\documentclass[twoside,12pt]{article}
\usepackage{epsfig}
\usepackage{lineno}
\usepackage{soul}
\newcommand{\bfq}{\mbox{\boldmath $q$}}
\newcommand{\bfP}{\mbox{\boldmath $P$}}
\newcommand{\bfS}{\mbox{\boldmath $S$}}
\newcommand{\bfs}{\mbox{\boldmath $s$}}
\newcommand{\bfx}{\mbox{\boldmath $x$}}
\newcommand{\bfz}{\mbox{\boldmath $z$}}
\newcommand{\bfZ}{\mbox{\boldmath $Z$}}
\newcommand{\bfy}{\mbox{\boldmath $y$}}

\newcommand{\bfL}{\mbox{\boldmath $L$}}
\newcommand{\bfA}{\mbox{\boldmath $A$}}
\newcommand{\bfB}{\mbox{\boldmath $B$}}

\newcommand{\bfk}{\mbox{\boldmath $k$}}
\newcommand{\bfb}{\mbox{\boldmath $b$}}
\newcommand{\bfr}{\mbox{\boldmath $r$}}
\newcommand{\bfR}{\mbox{\boldmath $R$}}
\newcommand{\bfh}{\mbox{\boldmath $h$}}
\newcommand{\bfD}{\mbox{\boldmath $\Delta$}}
\newcommand{\bfe}{\mbox{\boldmath $e$}}
\newcommand{\bfc}{\mbox{\boldmath $c$}}

\usepackage{amssymb}
\usepackage{xcolor}
\usepackage{braket}
\usepackage{amsmath}
\usepackage{hyperref}
\def\bkt{\bfk_\perp}
\def\bpt{\bfp_\perp}
\def\kt{k_\perp}
\def\pt{p_\perp}
\def\pup{p^\uparrow}
\newcommand{\ua}{\uparrow}
\newcommand{\da}{\downarrow}
\newcommand{\bfp}{\mbox{\boldmath $p$}}
\newcommand{\rd}{{\rm d}}
\def\xb{x_{_{\!B}}}

\newcommand{\be}{\begin{equation}}
\newcommand{\ee}{\end{equation}}
\newcommand{\bea}{\begin{eqnarray}}
\newcommand{\eea}{\end{eqnarray}}
\newcommand{\nn}{\nonumber}
\def\bbt{\bfb_\perp}
\def\bDt{\bfD_\perp}
\def\bzt{\bfz_\perp}
\def\avk{\langle k_\perp ^2\rangle}
\def\avp{\langle p_\perp ^2\rangle}
\def\avPT{\langle P_T^2\rangle}

\def\T{_{_T}}
\def\C{_{_C}}
\def\lsim{\mathrel{\rlap{\lower4pt\hbox{\hskip1pt$\sim$}}\raise1pt\hbox{$<$}}}
\def\gsim{\mathrel{\rlap{\lower4pt\hbox{\hskip1pt$\sim$}}\raise1pt\hbox{$>$}}}
\begin{document}

\title{Transverse spin effects in hard semi-inclusive collisions}
\author{M.\ Anselmino$^{1}$, A.\ Mukherjee$^2$ and A.\ Vossen$^{3,4}$ \\
$^1$Dipartimento di Fisica, Universit\`a di Torino and INFN Sezione di Torino, \\
Via P.~Giuria 1, I-10125 Torino, Italy \\
$^2$Department of Physics, Indian Institute of Technology Bombay,\\
Mumbai-400076, India\\
$^3$Duke University, Durham, North Carolina 27708, USA \\
$^4$Jefferson Lab, 12000 Jefferson Avenue, Newport News, VA 23606, USA}
\maketitle
\begin{abstract} 
The nucleons (protons and neutrons) are by far the most abundant form of matter 
in our visible Universe; they are composite particles made of quarks and gluons,
the fundamental quanta of Quantum Chromo Dynamics (QCD). The usual interpretation 
of the nucleon dynamics in high energy interactions is often limited to a simple 
one-dimensional picture of a fast moving nucleon as a collection of co-linearly 
moving quarks and gluons (partons), interacting accordingly to perturbative QCD 
rules. However, massive experimental evidence shows that, in particular when 
transverse spin dependent observables are involved, such a simple picture is not 
adequate. The intrinsic transverse motion of partons has to be taken into account; 
this opens the way to a new, truly 3-dimensional (3D) study of the nucleon 
structure. A review of the main experimental data, their interpretation and 
understanding in terms of new transverse momentum dependent partonic distributions, 
and the progress in building a 3D imaging of the nucleon is presented.      
\end{abstract}
\tableofcontents

\section{Introduction}

The nucleons -- protons and neutrons -- form the almost totality of the visible 
matter in the Universe. We know that they are composite objects, made of quarks 
and gluons (collectively denoted as partons), which interact according to the 
strong interactions rules of Quantum Chromo Dynamics (QCD), a fundamental 
relativistic quantum field theory. 
However, the full partonic description of the nucleons is still a 
very mysterious and fascinating open issue.
Because of this, 
the understanding of the internal structure of the nucleons, both in momentum 
and in coordinate space, is the ultimate goal of many ongoing or planned 
experiments and the focus of theoretical activities worldwide. 

The experiments are mainly high energy scatterings of point-like leptons off 
protons and neutrons, in which the lepton scatters off a single parton, or 
inelastic collisions between nucleons, like Drell-Yan processes in which a 
quark and an antiquark annihilate into a pair of leptons. Also the production 
of a single hadron or two hadrons in the high energy collision of two nucleons 
can be related to QCD elementary interactions among partons. The outcome of 
these experiments, when correctly interpreted, gives information on the internal 
nucleon composition. The theoretical scheme in which these processes are 
studied is QCD, both in its perturbative and non-perturbative aspects. 

The cross sections for the above processes are written, according to a 
factorisation theorem, as the convolution of elementary partonic interactions - 
known from perturbative calculations in the Standard Model of strong and 
electro-weak interactions - with Partonic Distribution and Fragmentation 
Functions (PDFs and FFs). These are These are not calculable using 
perturbative methods, but their evolution with the 
large-scale $Q^2$ of the process can be computed in QCD. By measuring the cross 
sections one learns about the PDFs and FFs at a certain scale, and can evolve 
them to other values of $Q^2$, thus achieving predicting power. Independent 
information on the FFs can be obtained from other processes, like the 
annihilation of $e^+$ and $e^-$ into pairs of hadrons.

For a long time, the PDFs and FFs were considered as collinear splitting 
processes, which corresponds to a 1-dimensional imaging of a fast nucleon as 
a simple set of co-linearly moving partons. Recently, it has become more and 
more clear that the understanding of many experimental results - in particular 
those involving spin degrees of freedom - must take into account the transverse 
degrees of freedom, that is the intrinsic motion of quarks and gluons inside 
the nucleons. This opens the way to the full study of the 3-dimensional (3D) 
structure of the nucleons.

The complete 3D information on the partonic momentum distributions has been 
encoded in Transverse Momentum Dependent Partonic Distribution Functions 
(TMD-PDFs). In experimental observables, they are often combined with 
Transverse Momentum Dependent Fragmentation Functions (TMD-FFs). Apart from 
perturbative QCD corrections, when integrated over transverse momentum the 
TMDs reduce to the collinear PDFs and FFs. A full knowledge of the partonic 
distributions must also include their dependence on hadronic and partonic spin, 
related to subtle spin-orbit correlations of the strong force. 

At leading order in $1/Q$ there are eight TMD-PDFs and, for spinless final 
hadrons, 2 TMD-FFs. Beside the TMD-PDFs and TMD-FFs, new objects - the 
Generalised Partonic Distributions, GPDs - offer information on the parton 
distribution in coordinate space. There are also eight leading order nucleon 
GPDs which give new information, like the correlation between the transverse 
position and the longitudinal momentum of partons, providing a 3D mapping of 
the nucleon. They are also related to the orbital momentum contribution of 
partons to the nucleon spin. The GPDs are off-diagonal matrix elements of 
quark and gluon operators between nucleon states and can be measured in hard 
exclusive processes such as the lepto-production of a photon or of a meson 
or the photo-production of a lepton pair. Like for the TMDs, the measured 
quantities are convolutions of GPDs with hard scattering amplitudes. In the 
diagonal limit the GPDs coincide with the PDFs.

Both the GPDs and the TMD-PDFs are particular limits of a vast class of 
functions, the so-called Wigner functions (or Generalised TMDs, GTMDs), which 
are the quantum mechanical version of the classical phase-space distributions. 
The really ultimate theoretical goal is that of reconstructing the nucleon 
Wigner functions; attempts to do that can be done, at the moment, by modelling 
the light-front nucleon wave functions. 

In the last 10-15 years the first measurements of azimuthal asymmetries in 
Semi Inclusive Deep Inelastic Scattering (SIDIS, lepton + nucleon $\to$ lepton 
+ hadron + X, $\ell \, N \to \ell \, h \, X$) processes by the HERMES (DESY, 
Germany), COMPASS (CERN) and Jefferson Laboratory (JLab, USA) Collaborations, 
together with the related theoretical analyses, have definitely revealed the 
role of the TMDs and allowed the first extraction of some of them. Similarly 
for the GPDs. Recent results by the Belle (KEK, Japan), BaBar (SLAC, USA)
and BES-III (BEPC, China) Collaborations in $e^+ e^- \to h_1 \, h_2 \, X$ 
processes have definitely shown the role of TMD-FFs. Important data are 
expected soon from the Drell-Yan (D-Y) processes at COMPASS and
possibly RHIC (BNL, USA), and from the 12 GeV upgrade of JLab. 
Great expectations are linked to the planned future Electron Ion Collider 
(EIC) in USA and the LHCb (polarised) fixed target experiment at CERN.

We have then reached a stage in which one should combine phenomenological 
studies of TMDs and GPDs with theoretical models of proton and neutron wave 
functions. It is the only way which may lead to a true 3D knowledge of the 
nucleon structure. The available data give the necessary (although not yet 
complete) information in modelling the 3D structure, while the soon expected 
new data will allow improvements of the models and tests of their predictions. 

In this review paper we focus on TMDs and inclusive processes, that is on the 
3D structure of nucleons in momentum space. The plan of the paper is the 
following. In Section 2 we summarise the experimental results which show and 
lead to the necessity of taking into account the transverse motion of partons 
inside the nucleons and the transverse momentum of hadrons in a parton 
hadronisation process. These are typically, but not exclusively, polarised 
interactions. We consider separately three kinds of processes: SIDIS, $\ell \, 
N \to \ell \, h \, X$; hard nucleon-nucleon interactions, $N\,N \to \ell^+ \, 
\ell^- \, X$, $N\,N \to h_1 \, h_2 \, X$ and $N\,N \to h_1 \, X$; hadron 
production in $e^+ \, e^-$ annihilations, $e^+ e^- \to h_1 \, h_2 \, X$ and 
$e^+ e^- \to \Lambda^\uparrow \, X$. Although the formal definition 
and discussion of TMDs will be presented in Section 3, some TMDs will already be 
mentioned in Section 2, when illustrating the experimental evidence for 
transverse motion. In particular, the Sivers TMD-PDF, that is the distribution
of unpolarised partons inside a transversely polarised proton, and the Collins 
TMD-FF, that is the transverse motion of a hadron within a jet generated by 
a transversely polarised quark.       

In Section 3 we present and discuss the TMD phenomenology; that is, after
introducing the TMD-PDFs and the TMD-FFs, we show how to relate them to physical 
observables, and how to extract TMD information from data, which is not a simple
procedure. This is mainly and explicitly done at leading order, again separately 
for the three kinds of processes described above. Some comments and full 
references to QCD corrections and TMD evolution are also given. 

In Section 4 we summarise our actual knowledge on some TMDs and their relevance 
towards a 3D imaging of the nucleon. Some specific issues, like the orbital
motion of quarks inside a nucleon and the universality of the TMDs, will 
only be
mentioned. The last part of this Section is amply devoted to the 
Wigner function, its importance and the ongoing attempts, mainly theoretical, 
to study it. 

In the Conclusions we summarise the content and the purpose of 
the paper, indicating open problems and possible further developments. The 
importance of new results from the running COMPASS and RHIC D-Y measurements, 
and from the operating 12 GeV JLab upgrade is discussed. Crucial improvements 
expected from the planned EIC facility are emphasised. 
        
Several excellent review papers related and complementary to the issues 
covered in this paper can be found in the literature~\cite{Barone:2001sp,
DAlesio:2007bjf,Barone:2010zz,Aidala:2012mv,Boglione:2015zyc,Avakian:2019drf}. 
A collections of topical contributions dedicated to the 3-dimensional nucleon 
structure can be found in Ref.~\cite{Anselmino:2015}, while the physics case 
of the Electron Ion Collider, a planned future machine devoted to the 
exploration of the nucleon structure, is discussed in Ref.~\cite{Accardi:2012qut}. 
This paper is focused on the phenomenological features of transverse spin 
physics and most technical aspects and subtleties of QCD, like TMD evolution, 
will not be discussed: a complete and fundamental introduction to a correct QCD 
description of high energy processes can be found in Ref.~\cite{Collins:2011zzd}.   

\section{Transverse spin effects and the parton transverse motion \label{sec2}}

In this Section we recall the experimental data which cannot be understood 
in the usual collinear QCD parton model scheme; they are mainly, but not uniquely,  
spin data. As usual, polarised experiments test a theory at a much deeper level 
than unpolarised quantities; in particular, Single Spin Asymmetries (SSAs)
originate from subtle Quantum Mechanical interference effects, which do 
not affect the unpolarised observables. In addition, if we consider parity 
conserving strong and electromagnetic interactions, only transverse 
SSAs are allowed by parity invariance; thus, they are the ideal probe to 
explore the transverse (with respect to the direction of motion) internal 
structure of hadrons.  

We consider high energy inclusive processes, which are usually described in 
terms of interactions among quarks and gluons. The relation between the measured 
hadronic quantities, the elementary QCD or QED partonic interactions, and the 
nucleon structure we wish to explore, is encoded in the factorisation scheme,
which we shall use and on which we shall comment in the next Section. We simply 
discuss, in this Section, the available data for the three kinds of processes 
mentioned in the Introduction: SIDIS, hard $N\,N$ collisions and 
$e^+e^-$ annihilations.    
 
\subsection{\it Spin effects in SIDIS \label{subsec2.1}}

Traditionally, since the end of the 60s, the exploration of the nucleon 
structure has been successfully performed via Deep Inelastic Scattering
(DIS, $\ell \, N \to \ell \, X$) in which a point-like lepton (typically 
electron, positron or muon) is scattered at high energy and large angle off 
a nucleon. The basic interpretation is that the lepton scatters off a quark,
via one virtual photon exchange, and the measurement of the final lepton 
energy and direction allows to learn about the longitudinal momentum fraction 
($x$) of the nucleon carried by the quark. The QCD corrections induce a 
dependence on the 4-momentum transfer squared of the lepton (the 4-momentum 
squared of the virtual photon, $q^2 = -Q^2$) which can be computed. Thus, one 
learns about the Parton Distribution Functions (PDFs), $f_q(x, Q^2)$, that is 
the number density of quarks $q$, carrying a fraction $x$ of the parent nucleon 
momentum, as seen at a distance $\sim 1/Q$. The correct prediction of the $Q^2$ 
dependence is one of the triumphs of perturbative QCD.

However, despite its great success, this study gives a one-dimensional (1D) 
picture of the nucleon, limited to the longitudinal degrees of freedom. This 
might be sufficient in many high energy experiments, where the transverse motion 
of partons inside the nucleon is negligible compared to the fast longitudinal
motion; indeed, many high energy cross sections are correctly predicted in 
several experiments. When introducing spin degrees of freedom this 1D 
picture allows to obtain information on the parton helicity distributions, 
that is the difference between the density number of partons with the same 
and opposite helicity as the parent proton: again, only longitudinal features
of the nucleon. The transversity distribution, that is the difference between 
the density number of partons with {\it transverse} spin parallel and antiparallel 
to the {\it transverse} spin of the parent proton, cannot be accessed in DIS. 
As it will be shown in the next Section, information on the transversity 
distributions can only be obtained by considering TMD effects. 

In general, transverse SSAs in hadronic processes cannot be understood in the 
simple collinear partonic picture of the nucleon. This is related to the 
fact that QCD or QED massless and parity conserving partonic interactions, 
do not allow transverse SSAs and a collinear fragmentation process cannot 
build up a transverse polarisation. Although PDFs have provided much 
information to shape our physical picture of the nucleon, they cannot answer 
key questions for understanding the structure of the nucleon, namely how its 
spin is apportioned between the spin of its constituents and their orbital 
angular momentum. We definitely need a 3D imaging of the nucleon, if we want
to understand its structure and to explain many experimental data.                         
 
So far, SIDIS processes ($\ell \, N \to \ell \, h \, X$) are the main probe 
exploited to explore the 3D structure of the nucleon. In such processes, 
differently from the DIS case in which one only detects the final 
lepton, the point-like lepton scatters off a quark, which, subsequently, 
fragments into an observed hadron. By looking at the hadron distribution one 
can get further information on the quark which generated it, its intrinsic 
motion and possible correlations between its spin, its motion and the spin 
of the nucleon.  
\begin{figure}[t]
\begin{center}
\includegraphics[width=15.truecm,angle=0]{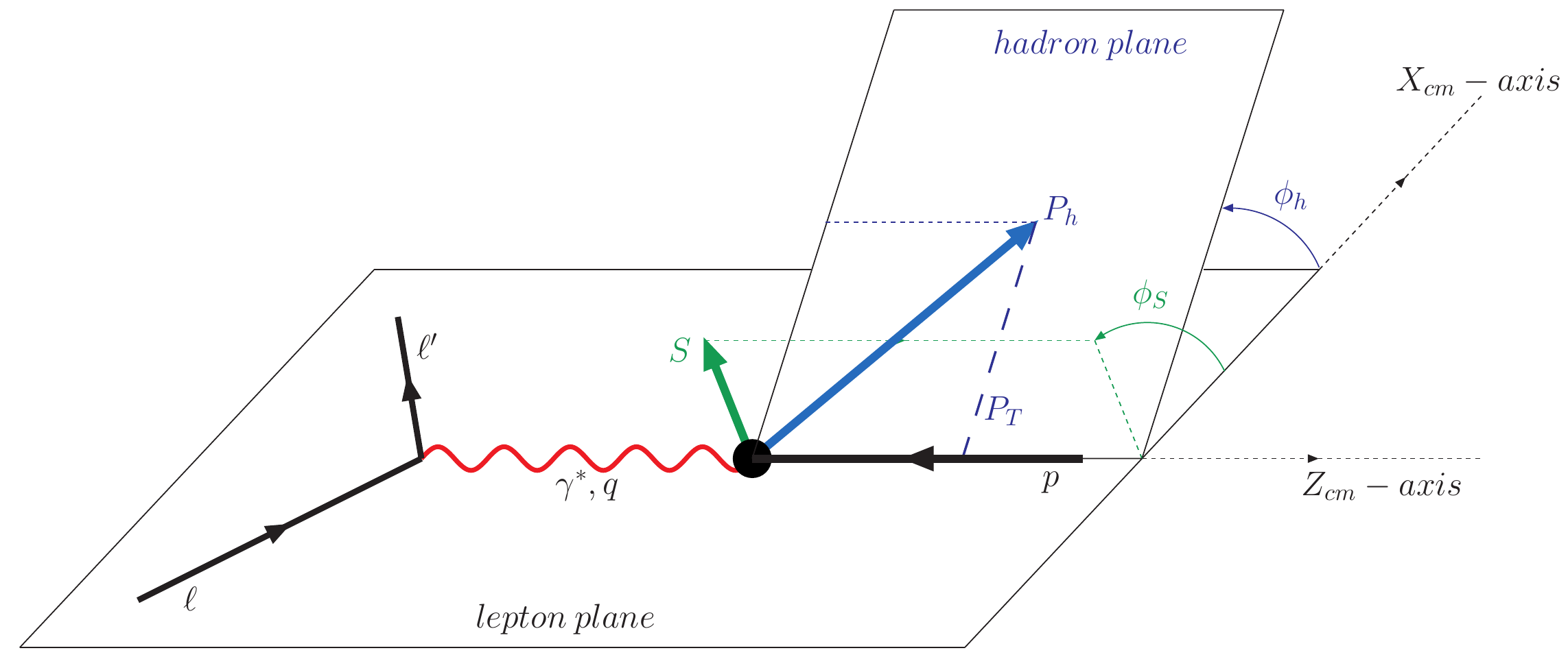}
\caption{Kinematics and definition of the variables of SIDIS processes in 
the $\gamma^* - N$ c.m. frame.}
\label{fig:SIDIS-kin}
\end{center}
\end{figure}

The kinematics of a typical SIDIS process, in the virtual photon-nucleon center 
of mass frame is shown in Fig.~\ref{fig:SIDIS-kin}, where the relevant kinematical
variables are defined. It is already clear from this figure that, in the simple 
leading order collinear parton model in which the $\gamma^*$ hits a quarks,
which bounces back and fragments co-linearly, one could not have a final hadron
with a transverse momentum $\bfP_T$. This could be generated by higher order 
QCD interactions, but, at leading order in the strong interaction coupling, a 
transverse momentum of the final hadron must be related to the intrinsic motion
of the quark in the nucleon and the transverse momentum of the hadron $h$
with respect to the momentum of the fragmenting quark.

The most general expression for the SIDIS cross section, with unpolarised
leptons and fully transversely polarised nucleons, assuming a single virtual 
photon exchange and neglecting masses, can be written 
as~\cite{Gourdin:1973qx,Kotzinian:1994dv,Mulders:1995dh,Bacchetta:2004zf, Diehl:2005pc,Bacchetta:2006tn,Anselmino:2011ch}:         
\bea
&& \hskip -20pt \frac{d\sigma^{\ell + p(S_T) \to \ell^\prime h X}}
{d\xb \, dQ^2 \, dz_h \, d^2 \bfP_T \, d\phi_S}
= \frac {2 \, \alpha^2}{Q^4} \times \label{dsigma}\\ [2pt] \nonumber \\
&\Bigg\{& \!\!\!\!\!
\frac{1+(1-y)^2}{2} F_{UU} + (2-y)\sqrt{1-y} \, \cos\phi_h \,
F_{UU}^{\cos\phi_h} + (1-y) \, \cos2\phi_h \, F_{UU}^{\cos2\phi_h}
\nonumber \\
&+& \!\! \bigg[ \, \frac{1+(1-y)^2}{2} \, \sin(\phi_h-\phi_S) \,
F_{UT}^{\sin(\phi_h-\phi_S)} + (1-y) \, \sin(\phi_h + \phi_S) \,
F_{UT}^{\sin(\phi_h + \phi_S)} \nonumber \\
&+& (1-y) \, \sin(3\phi_h-\phi_S) \,
F_{UT}^{\sin(3\phi_h-\phi_S)} + (2-y) \, \sqrt{1-y} \,
\Big( \sin\phi_S \, F_{UT}^{\sin\phi_S} +
\sin(2\phi_h-\phi_S) \, F_{UT}^{\sin(2\phi_h-\phi_S)} \Big) \bigg] 
\nonumber \Bigg\} \>,
\eea
where we have used the usual SIDIS variables:
\be
s = (\ell + p)^2 \quad\quad Q^2=-q^2 = -(\ell -\ell')^2 \quad\quad
\xb = \frac {Q^2}{2p \cdot q}\quad\quad
z_h = \frac{p \cdot P_h}{p \cdot q} \quad\quad
y =  \frac{p \cdot q}{p \cdot \ell} \> \cdot \label{sidisvar}
\ee 
The $F_{UU}$ and the $F_{UT}$ are structure functions which depend on the 
kinematical variables~(\ref{sidisvar}): the first index denotes the lepton 
polarisation state ($U$ = unpolarised) while the second one denotes the 
nucleon polarisation state (either $U$ = unpolarised or $T$ = transversely 
polarised). The full structure of the SIDIS cross section, with all lepton and 
nucleon polarisations, can be found in Refs.~\cite{Bacchetta:2006tn,
Anselmino:2011ch}; Eq.~(\ref{dsigma}) is the main source for all 
phenomenological SIDIS studies we discuss here.   

Obviously, the azimuthal modulations of the cross section require the detection 
of the transverse momentum $\bfP_T$ of the final hadron; by integration over
$\phi_h$ all terms, except that containing $F_{UU}$, would vanish. Notice also 
that the above SIDIS cross section can originate several transverse SSAs:
if one takes differences of cross sections with opposite nucleon transverse 
spins, $d\sigma(\phi_S) - d\sigma(\pi +\phi_S)$, many terms in 
Eq.~(\ref{dsigma}) survive. 

These asymmetries are often expressed through their azimuthal moments,
\be
A_{UT}^{W(\phi_h,\phi_S)} = 2 \,
\frac{\int\,d\phi_h \, d\phi_S\,\left[ d\sigma^\uparrow -
d\sigma^\downarrow\right]\,W(\phi_h,\phi_S)}
{\int\,d\phi_h \, d\phi_S\,\left[ d\sigma^\uparrow + d\sigma^\downarrow\right]}\,,
\label{eq:azi-mom}
\ee
where $W(\phi_h,\phi_S)$ is the appropriate azimuthal weight function
required in order to isolate the specific contribution of interest and
$d\sigma^{\uparrow,\downarrow}$ is the differential cross section of
Eq.~(\ref{dsigma}) with $S_T = \> \uparrow,\downarrow$ denoting,
respectively, a transverse polarisation with azimuthal angle $\phi_S$ and
$\phi_S + \pi$. 

For example, taking $W(\phi_h,\phi_S) = \sin(\phi_h - \phi_S)$, one obtains:
\be
A_{UT}^{\sin(\phi_h - \phi_S)} \equiv A_{\,Siv}^{\,p} =
\frac{F_{UT}^{\sin(\phi_h-\phi_S)}}{F_{UU}}\>, 
\label{eq:AUT-siv}
\ee 
while with $W(\phi_h,\phi_S) = \sin(\phi_h + \phi_S)$ one has
\be
A_{UT}^{\sin(\phi_h + \phi_S)} \equiv A_{\,Col}^{\,p} =
\frac{2(1-y)\,F_{UT}^{\sin(\phi_h+\phi_S)}}
{\bigl[ 1+(1-y)^2\bigr]\,F_{UU}}\> \cdot
\label{eq:AUT-coll}
\ee

\begin{figure}[t]
\begin{center}
\includegraphics[width=15.truecm,angle=0]{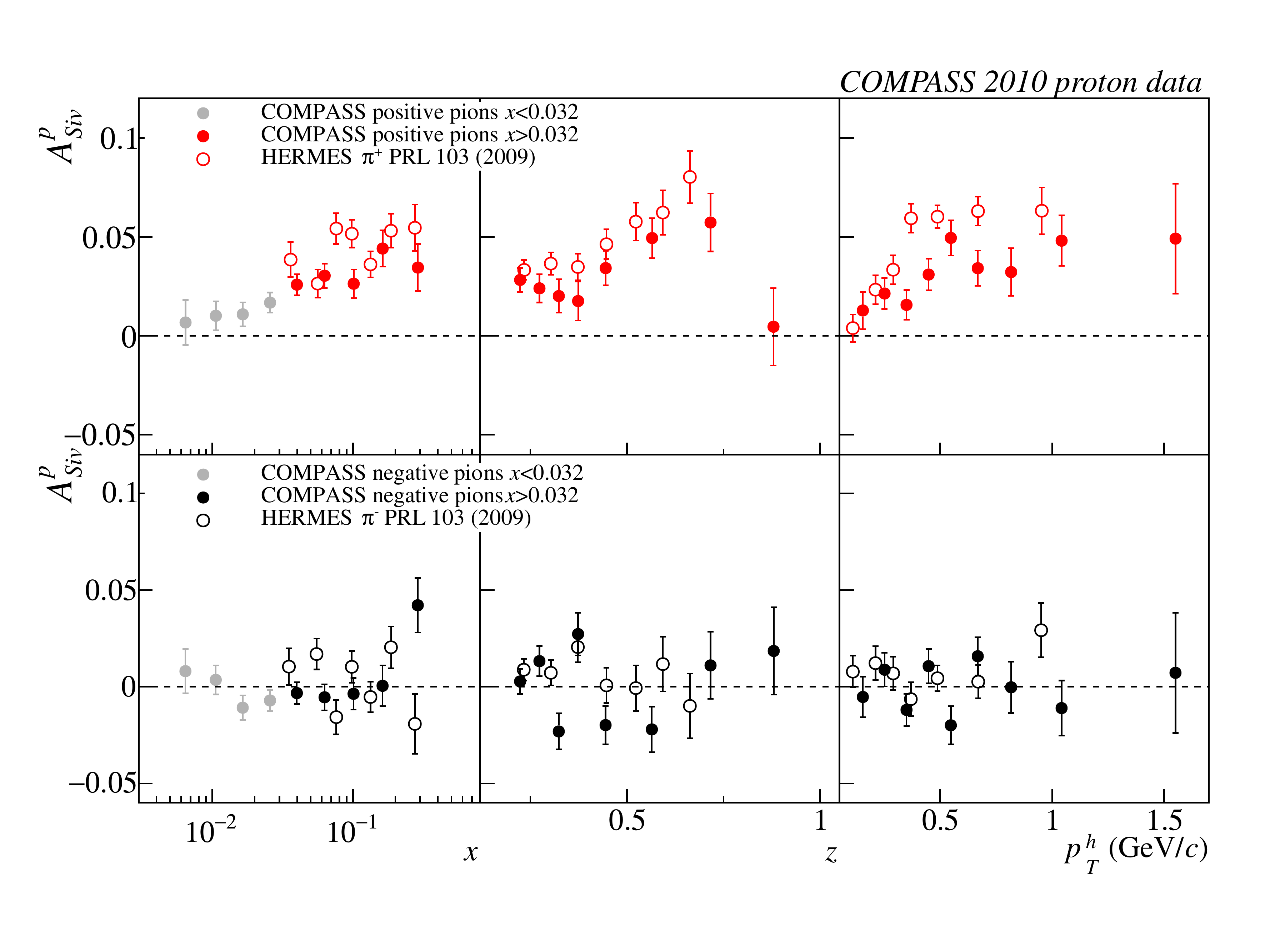}
\vskip -24pt
\caption{The weighted transverse SSA $A_{UT}^{\sin(\phi_h - \phi_S)}$, as 
measured by the COMPASS and Hermes Collaborations is shown as a function 
of its kinematical variables (notice that $x = \xb$, $z = z_h$ and $p^h_T = 
P_T$). This asymmetry is also denoted as $A_{\,Siv}^{\,p}$, because it will
be interpreted as related to a TMD-PDF introduced by Sivers. Figure reprinted from 
Ref.~\cite{Avakian:2019drf} with kind permission of Societ\`{a} Italiana di Fisica, 
\textcopyright Societ\`{a} Italiana di Fisica 2019.} 
\label{fig:C+H-Sivers}
\end{center}
\end{figure}

\begin{figure}[t]
\begin{center}
\includegraphics[width=15.truecm,angle=0]{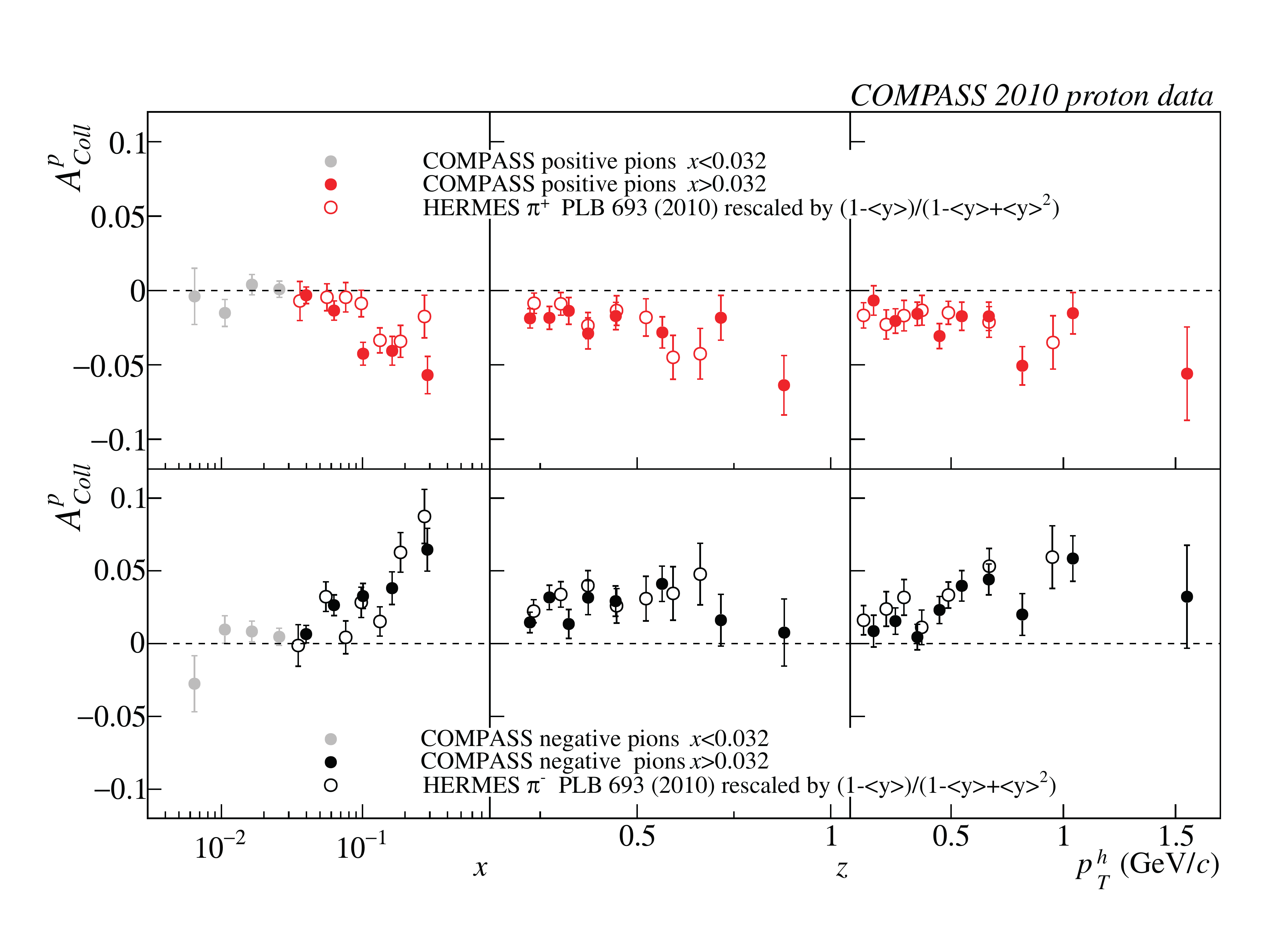}
\vskip -24pt
\caption{The weighted transverse SSA $A_{UT}^{\sin(\phi_h + \phi_S)}$, as 
measured by the COMPASS and Hermes Collaborations is shown as a function 
of its kinematical variables (notice that $x = \xb$, $z = z_h$ and $p^h_T = 
P_T$). This asymmetry is also denoted as $A_{\,Col}^{\,p}$, because it will
be interpreted as related to a TMD-FF introduced by Collins. Figure reprinted from 
Ref.~\cite{Avakian:2019drf} with kind permission of Societ\`{a} Italiana di Fisica, 
\textcopyright Societ\`{a} Italiana di Fisica 2019.} 
\label{fig:C+H-Collins}
\end{center}
\end{figure}

These SSAs have been observed by several experimental Collaborations: 
HERMES at HERA~\cite{Airapetian:2009ae,Airapetian:2010ds}, COMPASS at 
CERN~\cite{Adolph:2012sn,Adolph:2012sp,Adolph:2014zba,Adolph:2016dvl}, HALL A at 
JLab~\cite{Qian:2011py,Allada:2013nsw}. Some results from COMPASS and HERMES are 
shown in Figs.~\ref{fig:C+H-Sivers} and~\ref{fig:C+H-Collins}.

In Section~\ref{subsec3.2} we will interpret the SIDIS cross 
section, at least in limited kinematical regions, in terms of elementary 
lepton-quark interactions; at leading order in such interactions no SSAs 
is allowed and the spin effects must be originated by intrinsic non perturbative 
properties of the parton distributions and fragmentations, which will be 
encoded in the TMDs. Also the $P_T$ distribution of the unpolarised cross section 
will be related to TMDs.          

\subsection{\it Spin effects in hard $N\,N$ collisions \label{subsec2.2}}

The observation of transverse single spin asymmetries in hard $N\,N$ 
collisions played a pioneering role in the field of transverse spin physics.
Of special importance were the observation of a significant transverse 
polarisation of hyperons produced in the collision of a 300 GeV 
unpolarised proton beam with a Berillium target at 
Fermilab~\cite{Bunce:1976yb,Skubic:1978fi}, as well 
as the observation of left-right asymmetries, $A_N$, for pions produced in the 
forward direction of a polarised proton beam impinging on an unpolarised 
nuclear or proton target, first at the ZGS at 
ANL~\cite{Klem:1976ui,Dragoset:1978gg} and later at the AGS at 
BNL~\cite{Allgower:2002qi} and at the E704 experiment at 
Fermilab~\cite{Adams:1991rw, Adams:1991cs}, with beam energies ranging between 
6 and 200 GeV. As noted in the introduction, these results were in contradiction 
to the expectation that transverse spin effects are suppressed at high 
scales~\cite{Kane:1978nd} and therefore gave the first 
experimental hint of their importance in hard collisions. It can be argued that 
these early experiments did not reach high enough in the relevant momentum scale; 
however, recent data by the STAR experiment at the Relativistic Heavy Ion Collider 
(RHIC) shows that $A_N$ persists up to transverse momenta $p_T$ close to 
10 GeV~\cite{Pan:2016suv}.

These results provided first tantalising evidence for the 
importance of transverse spin and intrinsic transverse momenta of partons. 
In particular, the first attempt to explain the surprising left-right asymmetries 
observed in $p^\uparrow p \to \pi \, X$ processes prompted the first introduction 
of a TMD parton distribution (the Sivers distribution~\cite{Sivers:1989cc}), 
in the framework of a simple generalisation of the collinear factorisation 
scheme, subsequently denoted as the Generalised Parton Model (GPM). However, 
all these spin results in $p\,p$ single inclusive interactions are difficult 
to interpret in a partonic picture. As will be discussed later in 
Sec.~\ref{subsec3.3}, since they are one-scale processes, a collinear twist-3 
picture is applicable. The latter has non-trivial connections to the partonic 
TMD picture, {\it e.g.} via the Wandzura-Wilczec relations~\cite{Wandzura:1977qf}; 
however, it includes additional degrees of freedom parameterised by 
non-perturbative functions that would have to be measured as 
well~\cite{Gamberg:2017gle}. A significant recent development has been the 
experimental and theoretical effort put into observables where TMDs can be 
accessed directly such as di-hadron final states and two scale 
processes, like the production of hadrons in jets and the $W/Z$ 
production~\cite{Bacchetta:2004it,DAlesio:2010sag,Huang:2015vpy}.
  
To give an overview of the observables characterising the hadronic 
collisions discussed here, Table~\ref{tab:hadronicScales} shows, for each
process, the applicable framework and the large and small observable scales. 
The large scale is necessary for the factorisation of the hard scattering from 
the non-perturbative contributions to the cross section ($\gg M_p \simeq 1$ GeV), 
whereas the small scale provides sensitivity to the non-perturbative partonic 
structure and should therefore be of the order of the intrinsic transverse 
momentum in the nucleon ($\lsim 1$ GeV). Di-hadron correlations are a 
somewhat special case, since they can be used within a collinear framework.
The TMD approach for one scale $p\,p \rightarrow h \, X$ processes, assumes 
the validity of a Generalised Parton Model. 

\begin{table}
\begin{center}
\begin{tabular}{c|c|c|c|}
process & framework & hard scale & soft scale \\
\hline
$p\,p \rightarrow h \, X $ & twist-3&  $p_{T,h}$ & -\\
$p\,p \rightarrow h \, X $ & TMD, GPM&  $p_{T,h}$ & -\\
$p\,p\rightarrow h_1 \, h_2 \, X $ & collinear twist-2 & $p_{T,(h_1+h_2)}$ & -\\
$p\,p\rightarrow (jet+h) \, X $ &  TMD  &$p_{T,\textrm{jet}}$ & $k_{\perp h}$\\
$p\,p\rightarrow  \ell \, \ell' \, X$ & TMD & $M_{\ell-\ell'}$\ & $p_{T \gamma^*}$\\
$p\,p\rightarrow  W/Z^0 \, X$ & TMD &$M_{W/Z^0}$\ & $p_{T,W/Z^0}$\\
$p\, p\rightarrow  \gamma \, X$ &twist-3 & $p_{T,\gamma} $ &  - \\
\end{tabular}
\end{center}
\caption{\label{tab:hadronicScales}List of different processes sensitive to 
TMDs in hadronic collisions. The table shows the relevant hard scale (for the 
TMD factorisation and the twist-3 framework) and the soft scale (TMD framework 
only). The di-hadron production process is an outlier, since it can be 
described in a collinear, twist-2 framework due to the additional degrees of 
freedom in the final state. The TMD framework for $p\,p \to h \, X$
assumes the validity of the GPM. Symbols for the hard scales denote the hadron 
($p_{T,h}$), hadron-pair ($p_{T,(h_1+h_2)}$), jet ($p_{T,\textrm{jet}}$) and 
photon $p_{T,\gamma}$ transverse momenta as well as the masses of the di-lepton 
system in D-Y ($M_{\ell-\ell'}$) and the vector bosons in $W/Z$ production. 
Soft scales are given by the hadron transverse momentum within a jet, $k_{\perp h}$, 
for hadron in jet measurements and by the transverse momenta of the virtual 
photon and vector boson in D-Y and $W/Z^0$ production respectively. These quantities 
will be defined in more details when discussing the single processes 
in the following Sections.}
\end{table}

Richer final states, depending on several variables, that would  
allow a more direct access to partonic dynamics, as in SIDIS processes, were not 
possible at early $N\,N$ experiments due to limitations in energy, luminosity as 
well as detector capabilities. However, at RHIC, where longitudinal and 
transversely polarised protons can be collided with other protons or heavier 
nuclei, effects related to parton distribution functions can be accessed in 
two-scale processes at leading order and leading twist. 

Before exploring these cases in more detail, we give a short overview of the 
advantages of studying the nucleon structure in $N\,N $ collisions in addition 
to SIDIS. It should be noted that in the majority of cases discussed here 
both nuclei in the reactions are protons. This is natural, since this review 
is mainly concerned with the partonic structure of protons and for heavier 
nuclei additional nuclear effects would enter, which are not always well known. 
Furthermore, data on polarised nuclei beyond protons is quite rare as is data 
on heavier nuclei at center-of-mass energies beyond those within reach of fixed 
target experiments.

In hadronic collisions, there is no point-like probe; instead, at hard enough 
scales, the constituents of two composite particles scatter off each other. 
This inherently convolutes the structures of both particles into the 
physical observables as will be further detailed in Sec.~\ref{subsec3.3}. 
In particular, it is challenging to determine the kinematics of the partonic 
scattering. If the transverse energy in the event is sufficient, jets can be 
used as an approximation of the outgoing parton~\cite{Mukherjee:2012uz}, 
allowing the calculation of the kinematics of the underlying $2 \to 2$ 
partonic scattering in the Born approximation, leading to the expression in 
Eq.~(\ref{eq:diJetKinematics}) further below. Thus, $N\,N$ data is complementary
to SIDIS data and is crucial to complete our understanding of the proton 
structure. 

Differently from SIDIS processes, in hadronic collisions such as $p\,p$, 
gluons can be accessed at leading-order, since the probe is most often a 
color charged object as well. The presence of a color-charged probe and the 
associated difference in color flow allows to check the process dependence of 
interactions. An important examples for this is the predicted sign-change 
of transverse single spin asymmetries in SIDIS compared to D-Y 
measurements~\cite{Collins:2002kn} (see Section~\ref{subsec4.2}). This is an 
example of modified universality, where the modification is rather straightforward. 
Adding even more color-charges in the final state allows the existence of 
``entangled" gluon lines which is predicted to further complicate the process 
dependence~\cite{Rogers:2010dm}. However, it can be argued that a theoretical 
and experimental investigation of these effects is important for our full 
understanding of QCD.

While in SIDIS the coupling strength of the leptonic probe is given by $e_q^2$, 
where $e_q$ is the charge of the struck quark (in units of the proton charge), 
in contrast, in hadronic collisions the coupling strength is the same for all 
partons. Therefore complementary information on the flavour structure of the 
proton can be extracted. This is possible in SIDIS as well using effective 
neutron targets, but it requires additional running, {\it e.g.} with a 
deuterium or $^3He$ target. Nuclear effect in these targets can add complexity 
to the analysis.

\begin{figure}[t]
    \centering
    \vskip-36pt
    \includegraphics[width=0.45\textwidth]{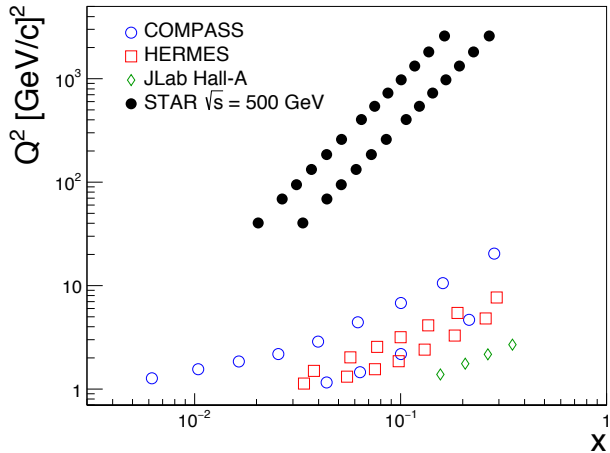}
    \vskip-64pt
    \caption{$x-Q^2$ coverage of the STAR Collaboration measuring 
    the Collins asymmetry for the production of hadrons in jets, compared with 
    the Collins asymmetry measurement in SIDIS experiments. Figure reprinted 
    from Ref.~\cite{Adamczyk:2017wld} and available under a 
    \href{https://creativecommons.org/licenses/by/4.0/legalcode}
    {Creative Commons Attribution 4.0 International}.}
\label{fig:hadronInJetKinCoverage}
\end{figure}

The last point we want to consider here is the extended kinematic coverage of 
hadronic collisions. Till the arrival of the EIC, polarised SIDIS 
experiments will be confined to a rather limited range in $Q^2$ due to their 
fixed target kinematics. In comparison, measurements at RHIC can reach values 
of $Q^2$ that are more than two orders of magnitude higher, as shown in 
Fig.~\ref{fig:hadronInJetKinCoverage}. 

RHIC is the first and only polarised $p\,p$ collider and naturally plays a 
prominent role in the study of transverse spin effects in hard $N\,N$ collisions. 
Even though we focus here on the RHIC results, it should be mentioned that 
measurements of interest for the TMD partonic structure of the nucleon are 
also conducted at other $N\,N$ machines, like the LHC. In these experiments 
there is obviously no access to observables that depend on polarisation in 
the initial state; however, measurements can be done that are sensitive 
{\it e.g.} to the intrinsic transverse momentum of partons in the nucleon 
by studying the $p_T$ spectrum of $W/Z$ bosons. For an 
overview, see Ref.~\cite{Angeles-Martinez:2015sea}.

Analogously to the SIDIS case, transverse spin and momentum dependent 
observables express themselves, given an appropriate reference system, 
in the dependence of the cross section on certain azimuthal angles which 
can be constructed from polarisation and momentum vectors. In the following 
we will summarise these observables for various processes in $N\,N$ collisions. 
The focus will be on observables that have been measured experimentally. 
Due to the added complexity in $N\,N$ collisions, the complete cross sections 
are rather lengthy, if they exist in the literature at all. Therefore they 
will not be reproduced here in their full length. The reader is referred to the 
appropriate given references. The phenomenology of hadronic collisions in 
terms of TMDs is discussed in detail in Sec.~\ref{sec3}. 

\subsubsection{Transverse single spin asymmetries in $p\,N\rightarrow h \, X$
processes}
\label{sec:anPions}
As described above, the transverse single spin asymmetry $A_N$ has a long 
history. For a forward moving transversely polarised beam, it is defined 
as:
\be
A_N \equiv \frac{\rd\sigma^\uparrow - \rd\sigma^\downarrow}
                {\rd\sigma^\uparrow + \rd\sigma^\downarrow} \>, \label{an}
\ee 
where $\rd\sigma$ is the differential cross section for the process 
$p \, N \to h(\bfp_h) +X$, and $\uparrow, \downarrow$ indicate opposite spin 
polarisation vectors perpendicular to the scattering plane. It is easy to see 
that, by rotational invariance, one has
\be
A_N \equiv \frac{\rd\sigma^\uparrow (\bfp_T) - \rd\sigma^\uparrow (-\bfp_T)}
                {\rd\sigma^\uparrow (\bfp_T) + \rd\sigma^\uparrow (-\bfp_T)} \>, \label{anlr}
\ee
where $\bfp_T$ is the component of the final hadron momentum $\bfp_h$ transverse 
to the polarised beam direction. That is, $A_N$ can also be simply seen as a 
left-right asymmetry in the inclusive production of a single hadron, while the 
beam polarisation remains fixed.      

Significant asymmetries have been observed in $p\,p$ collisions up to 
$\sqrt{s}= 500$ GeV~\cite{Pan:2016suv} for $\pi^0$. Data also exists for 
charged pions and kaons~\cite{Arsene:2008aa} as well as $\eta$ 
mesons~\cite{Adare:2014qzo,Li:2019iyt} and $J/\Psi$~\cite{Adare:2010bd}. 
Here we concentrate on the pseudo-scalar mesons. 
A common feature of the asymmetries is a rise with $x_F$, where the so-called 
Feynman-$x$ variable for a detected particle $A$ is defined as $x_F = p^{A}_L/ 
(p^A_{L \textrm{ max}})$. Here $p^A_L$ is its longitudinal momentum measured in a 
specific frame and $p^A_{L \textrm{ max}}$ the maximum longitudinal momentum 
that the particle can have in this frame. For $p\,p$ collisions with equal beam 
energies (in the c.m. system), $x_F$ reduces to $x_F = (2 p_L^A)/\sqrt{s}$ and 
$x_F=0$ corresponds to particles detected at an angle $\pi/2$ with respect to 
the beam axis in the lab frame.

\begin{figure}[t]
\begin{center}
\includegraphics[width=0.98\textwidth]{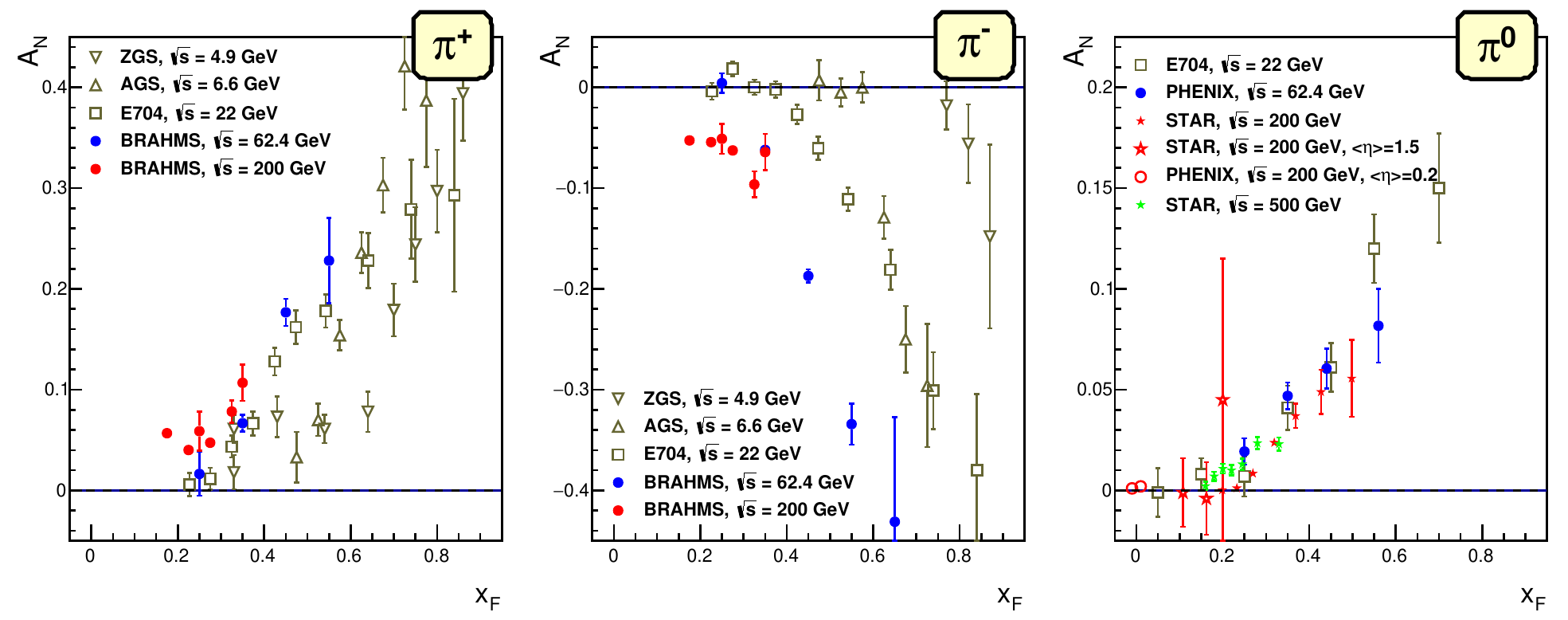}
\end{center}
\caption{Selection of world data on $A_N$ in $p\,p$ interactions for neutral 
and charged pions. In particular in the $\pi^0$ case, the so-called $x_F$ 
scaling is evident, which means that the asymmetry is almost independent of 
$\sqrt{s}$. In general, the dependence of $A_N$ on $x_F$ is almost linear.
Data compiled by Oleg Eyser~\cite{Aschenauer:2016our}.
\label{fig:A_N}}
\end{figure}

Fig.~\ref{fig:A_N} shows the world data on $A_N$ for c.m. energies $\sqrt{s}$ 
which go from 4.9~GeV at the ZGS up to 500 GeV at RHIC. As well as the rise 
in $x_F$, one can observe rising values of the asymmetries with the transverse 
momentum $p_T$ of the detected meson. Even at the highest c.m. energies available, 
no fall with $p_T$ was observed~\cite{Pan:2016suv}.
An interesting development has been the recent measurement of the nuclear 
dependence of $A_N$~\cite{Aidala:2019ctp}. The Phenix experiment observed a 
dependence of the asymmetries on the atomic number of the unpolarised beam, 
which has not been confirmed by the STAR experiment. These measurements might be 
sensitive to gluon saturation effects, which are not a focus of this review. 
We refer to Ref.~\cite{Benic:2018amn} for more details.

\subsubsection{Transverse single spin asymmetries in $\gamma^*,W/Z,\gamma$ 
production}
\label{sec:transvSpinEffectsInDY_W}
Closely related to the t-channel SIDIS process discussed earlier, are the 
corresponding $s$-channel processes in which the annihilation of a $q\,\bar{q}$ 
pair creates a virtual $\gamma^*$ or a real W/Z boson. For $q+\bar{q}\to 
\gamma^*\to \ell^+ \ell^-$, where $\ell^+ \ell^-$ is a final state lepton pair, 
this is the Drell-Yan process~\cite{Drell:1970wh}.  
Similarly to SIDIS, two non-perturbative objects enter the 
cross section of these processes, in this case two parton distribution functions, 
but no FF, due to the non-hadronic final state. This makes them relatively clean 
tools that allow a complementary access to TMD-PDFs. In particular, the Drell-Yan 
process with the possibility to measure transversity ``squared" in transverse 
double spin asymmetries without FF contribution~\cite{Anselmino:2004ki} as well 
as accessing the process dependence of the Sivers function~\cite{Collins:2002kn}, 
has attracted considerable attention in recent times. 

\begin{figure}[t]
  \centering
  \includegraphics[clip, trim=5cm 2cm 5cm 0, width=.48\textwidth]
  {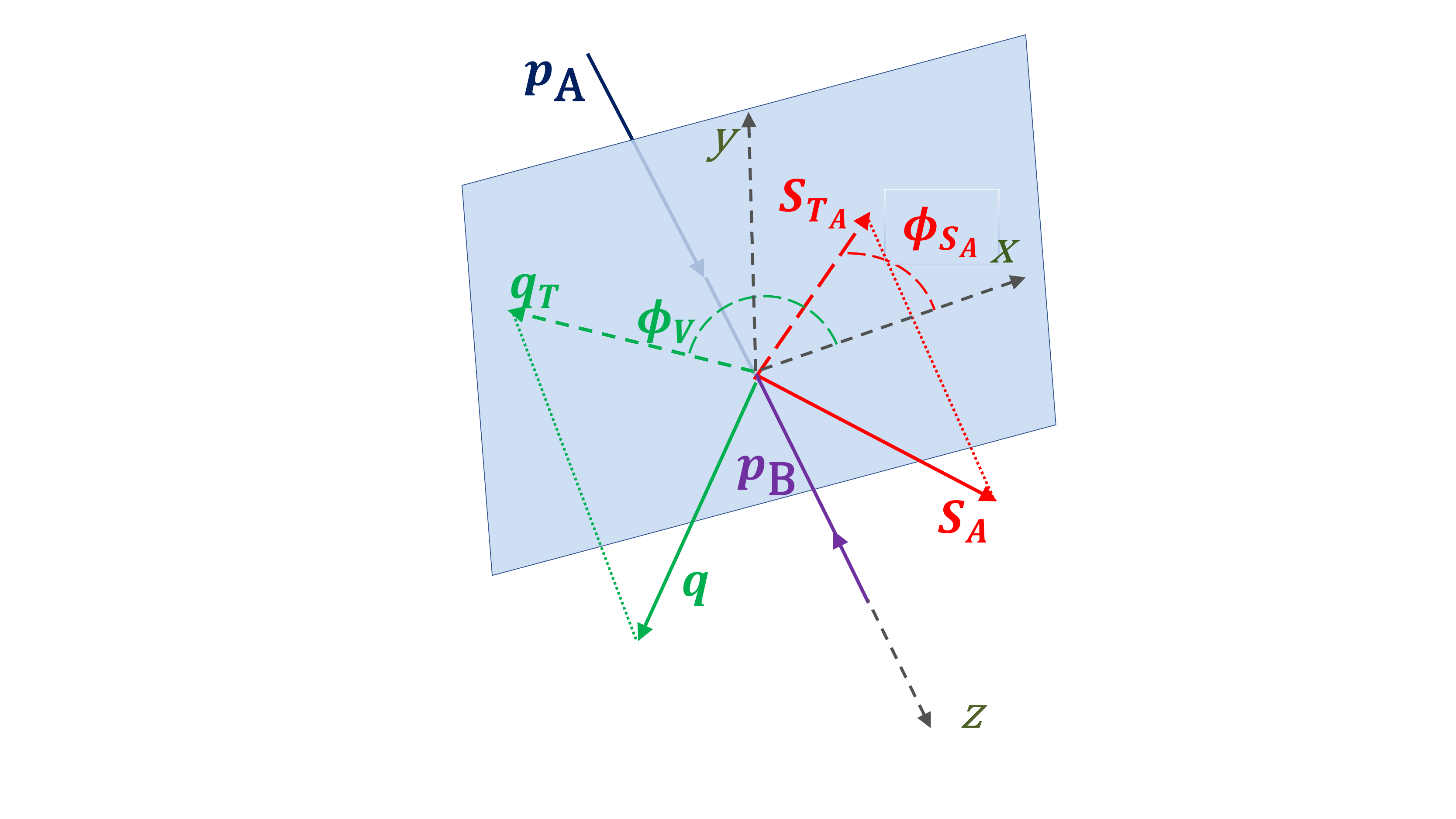}
  \includegraphics[clip, trim=4cm 0 5cm 0.2cm, width=.48\textwidth]
  {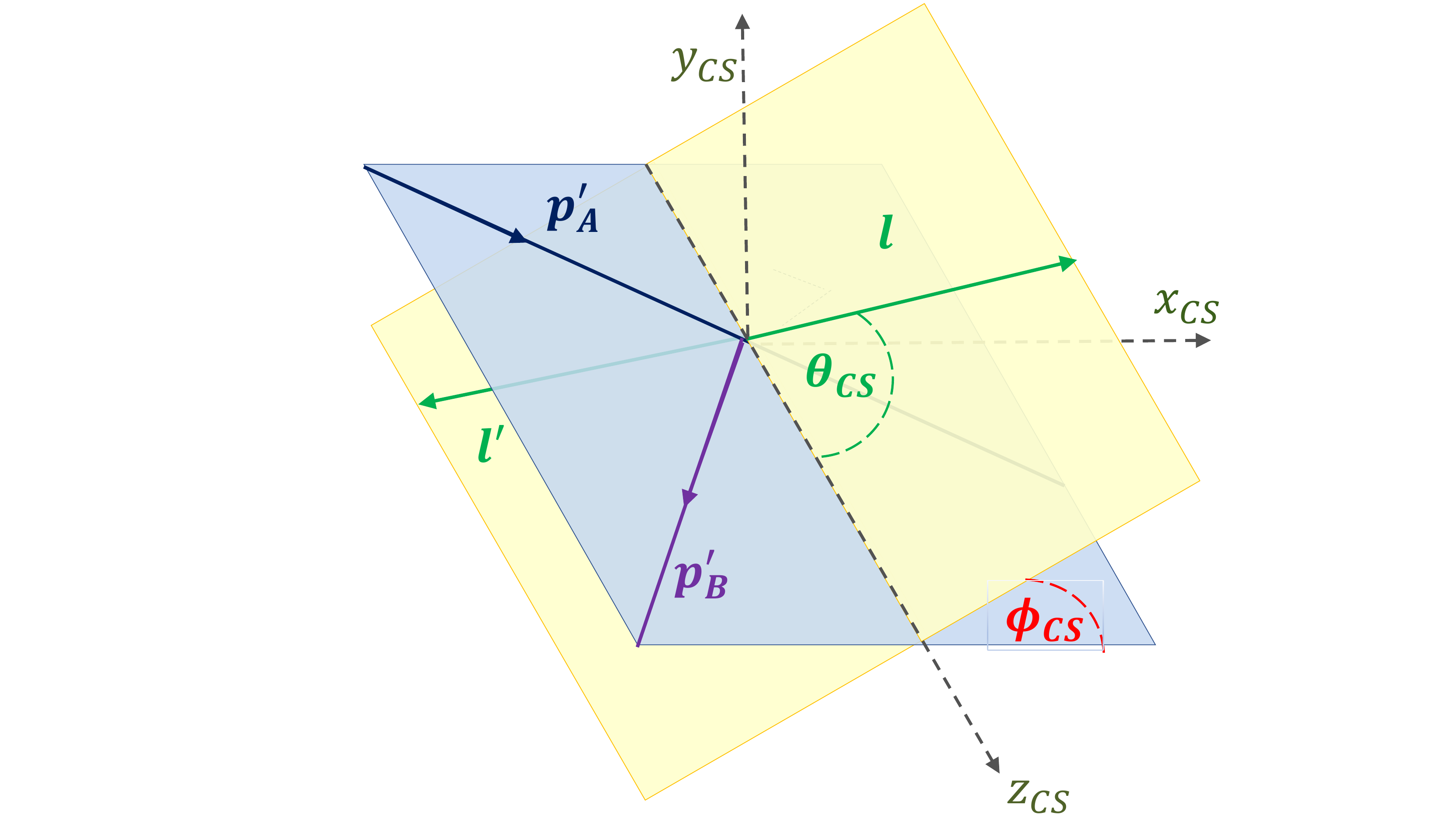}
  \caption{Two frames are commonly used in the analysis of the 
  Drell-Yan process. The target rest frame and the Collins-Soper frame.
  Since it is also used in the $W/Z$ case, here the center-of-mass frame 
  of the colliding hadrons is shown on the left side which is related to the 
  target rest frame by a boost along the $z$-axis. Therefore the azimuthal 
  angles entering the cross-section are the same. The Collins-Soper frame 
  is the lepton center-of-mass frame where the $z$-axis bisects the incoming 
  hadron momenta as shown in the figure on the right. The angle 
  $\phi_\textrm{CS}$ is then the azimuthal angle of the lepton plane with 
  respect to the hadron plane and the angle $\theta_\textrm{CS}$ is the angle 
  between the lepton direction and the $z$-axis.}    
  \label{fig:DY}
\end{figure}

When allowing for parton intrinsic motion, the TMDs express themselves in the 
dependence of the Drell-Yan cross-section on the azimuthal 
angles shown in Fig.~\ref{fig:DY}. The first ones, $\phi_V$ and $\phi_S$, are 
determined in the target rest frame and they are respectively the azimuthal 
angles of the momentum direction 
$\bfq$ of the vector boson -- the $\gamma^*$ in D-Y or the $W/Z$ in the case 
of weak boson production discussed further below -- and the transverse spin 
orientation of the beam. It is convenient to define them in the target 
rest frame as this is the natural frame for the experimental setup and it 
has a closer connection to the partonic picture~\cite{Arnold:2008kf}.
The remaining azimuthal angle $\phi_\textrm{CS}$ is customarily defined 
in a lepton pair center-of-mass frame. In this frame one also 
defines the polar angle $\theta_\textrm{CS}$. Here the subscript CS in $\theta_\textrm{CS}$ and $\phi_\textrm{CS}$ designates 
the Collins-Soper (CS) frame~\cite{PhysRevD.16.2219}. Another 
common frame, related to the CS system by a rotation, is the Gottfried-Jackson 
frame~\cite{Gottfried:1964nx}. 

The hard scale of the process is given by the virtuality $q^2$ of the 
$\gamma^*$ which can be determined from the invariant mass 
$M_{\ell-\ell'}$ of the $\ell^+\ell^-$ 
system. The TMD picture is valid for small transverse momenta $q_T$ of $\gamma^*$.
The cross section of the Drell-Yan process with one transversely polarised 
proton can be expressed in terms of azimuthal dependent structure functions 
analogous to the SIDIS process~\cite{Arnold:2008kf,Ji:2004xq}. 

The full expression for two polarized hadrons is quite lengthy (see {\it e.g.} 
Eq. (57) in Ref.~\cite{Arnold:2008kf}), as it contains various combinations of 
the TMDs of both hadrons. Therefore we will concentrate on two relevant cases 
here. First, considering the polarization of the hadrons (either longitudinal,
$S_L$, or transverse, $\bfS_T$), but integrating out the angles 
$\phi_\textrm{CS}$, $\theta_\textrm{CS}$ of the leptonic 
system~\cite{Arnold:2008kf,Huang:2015vpy}:
\bea
\label{dq:DY_StructFunctions}
\frac{d\sigma^{\rm DY}}{d^4q} &=& \frac{8\pi\alpha^2}{9s\,q^2}\bigg\{
F_{UU} + S_{AL} \, S_{BL} \, F_{LL}
\nonumber\\
&+& |\bfS_{AT}| \left[\sin(\phi_V - \phi_{S_A})\, 
F_{TU}^{\sin(\phi_V - \phi_{S_A})} \right]
+ |\bfS_{BT}| \left[\sin(\phi_V - \phi_{S_B})\,
F_{UT}^{\sin(\phi_V - \phi_{S_B})} \right]
\nonumber\\
&+& |\bfS_{AT}|\,S_{BL} \left[\cos(\phi_V - \phi_{S_A})\, 
F_{TL}^{\cos(\phi_V - \phi_{S_A})}\right]
+ S_{AL}\,|\bfS_{BT}| \left[ \cos(\phi_V - \phi_{S_B})\, 
F_{LT}^{\cos(\phi_V - \phi_{S_B})}\right]
\nonumber\\
&+& |\bfS_{AT}| \, |\bfS_{BT}| \left[
\cos(2\phi_V - \phi_{S_A} - \phi_{S_B})\,
F_{TT}^{\cos(2\phi_V - \phi_{S_A} - \phi_{S_B})}
+\cos(\phi_{S_A}-\phi_{S_B})\,F_{TT}^1 \right]\bigg\} \>.
\label{eq:DY}
\eea

%To unify the notation with the case of $W/Z$ production, we used here 
%$\phi_{S,V}$ instead of $\varphi_S$. 
As we will see in Sec~\ref{sec:phenoDY}, the $F_{TU}$ and $F_{UT}$ 
structure functions are sensitive to the Sivers functions of hadron 
$A$ and $B$, respectively, convoluted with the unpolarized PDF of the 
other hadron. Section~\ref{sec3} will explore the interpretation of the 
structure functions in terms of TMDs. 

Secondly, we consider the unpolarised cross-section in a di-lepton center 
of mass frame, {\it e.g.} the Collins-Soper frame. With $\Omega$ denoting the 
solid angle of the leptons, one can arrive for the angular distribution of the 
leptons at~\cite{Arnold:2008kf}:
\bea
\label{dq:DY_StructFunctions_upol}
\frac{d\sigma^{\rm DY}}{d^4q \, d\Omega} &=&
\frac{\alpha_{\textrm{em}}}{2s\,q^2}\bigg[ 
(1+\cos^2\theta_\textrm{CS})\,F_{UU}^1 \nn \\
&+& (1-\cos^2\theta_\textrm{CS})\,F_{UU}^2 + 
\sin2\theta_\textrm{CS}\,\cos\phi_\textrm{CS}\,F_{UU}^{\cos\phi_\textrm{CS}}+
\sin^2\theta_\textrm{CS}\,\cos2\phi_\textrm{CS}\,F_{UU}^{\cos2\phi_\textrm{CS}}\bigg],
\eea
where all angles are in the CS frame.
Defining
\be
\lambda=\frac{F_{UU}^1-F_{UU}^2}{F_{UU}^1+F_{UU}^2}\quad\quad
\mu=\frac{F_{UU}^{\cos\phi_\textrm{CS}}}{F_{UU}^1+F_{UU}^2}\quad\quad
\nu=\frac{2F_{UU}^{\cos2\phi_\textrm{CS}}}{F_{UU}^1+F_{UU}^2}
\ee
the cross section takes the form~\cite{Arnold:2008kf}
\begin{equation}
\frac{dN}{d\Omega}\equiv \frac{d\sigma^{\textrm{DY}}}{d^4q \, d\Omega}
\left/\frac{d\sigma^{\textrm{DY}}}{d^4 q}=\frac{3}{4\pi}\frac{1}{\lambda+3}
\left(1+\lambda \cos^2 \theta_\textrm{CS} +\mu \sin2\theta_\textrm{CS} 
\cos\phi_\textrm{CS} +\frac{\nu}{2}\sin^2 \theta_\textrm{CS}
\cos 2\phi_\textrm{CS} \right) \right. ,
\end{equation}
and the Lam-Tung relation~\cite{Lam:1978pu} can be written as $1-\lambda=2\nu$. 
It is the analogue to the Callan-Gross relation in SIDIS, since it is also a 
consequence of the interaction with point-like, spin-$\frac{1}{2}$ quarks. 
Unlike the Callan-Gross relation, the Lam-Tung relation holds at 
$\mathcal{O}(\alpha_s)$. Therefore, violations of the Lam-Tung relation can be 
seen as an indication of non-perturbative effects. Most notably, the 
Boer-Mulders function $h_1^\perp$, which will be introduced as 
one of the TMDs, leads to such a violation~\cite{Boer:1997nt}.

Albeit attractive, measuring spin-dependent asymmetries in the Drell-Yan 
process is challenging, since a large part of the cross-section is in a 
$M_{\ell-\ell'}$ region that receives significant contributions from 
resonances that decay into lepton pairs, like the $J/\Psi$. Recently, the 
COMPASS collaboration showed a first result on a Drell-Yan measurement using 
a polarised target to measure asymmetries related to the Sivers 
effect~\cite{Aghasyan:2017jop}. That measurement is unusual compared to other 
D-Y experiments, since it uses a pion beam, thus the pion PDFs enter in the 
relevant cross-sections. A Drell-Yan measurement in $p\,p$ interactions with 
a polarised target is also planned at the Fermilab experiment 
SpinQuest~\cite{Chen:2019hhx} 

The $\gamma^*$ in the D-Y process can be replaced by real W/Z bosons, which 
can then be detected via their hadronic or leptonic decay modes. Here the 
hard scale of the process is given by the mass of the weak boson.
Similar to the D-Y process, TMDs can be accessed in W/Z production through the 
dependence of the cross section on several azimuthal angles~\cite{Huang:2015vpy}. 
In this case it is convenient to consider the azimuthal angles of the 
polarisation vectors of the colliding beams in their center-of-mass system, 
$\varphi_{S_A}$ and $\varphi_{S_B}$ defined again relative to the azimuthal 
direction of the W/Z boson momentum in this system. Given that the azimuthal 
angles are invariant with respect to boosts along the $z$-axis, these are the 
same as the angles defined above for D-Y in the target rest frame.
The cross-section can then be written in terms of structure 
functions as ~\cite{Huang:2015vpy}:
\bea
\label{eq:W_StructFunctions}
\frac{d\sigma^W}{dy \, d^2\bfq_T} &=& \dfrac{\pi G_F M_W^2}{3 \sqrt{2}\,s} 
\bigg\{ F_{UU} + S_{AL}\,F_{LU} + S_{BL}\,F_{UL} + S_{AL}\,S_{BL}\,F_{LL}
\nonumber\\
&+& |\bfS_{AT}| \left[\sin(\phi_V - \phi_{S_A})\,F_{TU}^{\sin(\phi_V-\phi_{S_A})}
+ \cos(\phi_V - \phi_{S_A})\,F_{TU}^{\cos(\phi_V - \phi_{S_A})}
\right]
\nonumber\\
&+& |\bfS_{BT}| \left[\sin(\phi_V - \phi_{S_B})\,F_{UT}^{\sin(\phi_V - \phi_{S_B})}
+ \cos(\phi_V - \phi_{S_B})\,F_{UT}^{(\cos\phi_V - \phi_{S_B})}
\right]
\nonumber\\
&+& |\bfS_{AT}|\,S_{BL} \left[ \sin(\phi_V - \phi_{S_A})\,
F_{TL}^{\sin(\phi_V - \phi_{S_A})}
+ \cos(\phi_V - \phi_{S_A})\,F_{TL}^{\cos(\phi_V - \phi_{S_A})}
\right]
\\
&+& S_{AL}\,|\bfS_{BT}| \left[ \sin(\phi_V - \phi_{S_B})\, 
F_{LT}^{\sin(\phi_V - \phi_{S_B})}
+ \cos(\phi_V - \phi_{S_B})\,F_{LT}^{\cos(\phi_V - \phi_{S_B})}
\right]
\nonumber\\
&+& |\bfS_{AT}|\,|\bfS_{BT}| \left[
\cos(2\phi_V - \phi_{S_A} - \phi_{S_B})\,
F_{TT}^{\cos(2\phi_V - \phi_{S_A} - \phi_{S_B})}
+\cos(\phi_{S_A}-\phi_{S_B})\, F_{TT}^1\right.
\nonumber\\
&&\hspace{19mm}+\left.\sin(2\phi_V - \phi_{S_A} - \phi_{S_B}) 
F_{TT}^{\sin(2\phi_V - \phi_{S_A} - \phi_{S_B})}
+\sin(\phi_{S_A}-\phi_{S_B})\,F_{TT}^2\right]\bigg\} \>,
\nonumber
\label{eq:W}
\end{eqnarray}
where $y$ is the rapidity, which in terms of the four-momentum 
$q=(q_0,\bfq_T,q_L)$ is given by $y = \frac12 \ln \frac{q_0+q_L}{q_0-q_L}\cdot$

As explained in more detail in Ref.~\cite{Huang:2015vpy}, the transverse spin 
asymmetries in W/Z production differ from spin asymmetries in D-Y in two important 
ways.
    (1): Because the analogue of the decay-leptons is not accessible, the CS angles 
    $\phi$ and $\theta$ are effectively integrated out. This means that certain 
    TMDs that are accessible in D-Y are out of reach, {\it e.g.} the product of 
    Boer-Mulders and transversity.
    And (2): The parity violating nature of the weak interaction allows access 
    to ``wormgear" type TMDs in single spin asymmetries, which are not accessibly 
    in D-Y. This will be discussed further in Sec.~\ref{sec3}.

A pioneering measurement of transverse single spin asymmetries in $W$ and $Z^0$ 
production has been performed by the STAR experiment at RHIC~\cite{Adamczyk:2015gyk} 
with the main objective to extract Sivers type asymmetries.

A process that at first sight is similar to the D-Y and W/Z production processes 
described above, is the direct photon production, $p+p \rightarrow \gamma +X$.
However, since here a real photon is produced ($p_\gamma^2=0$), this is a single 
scale process that, analogous to the $A_N$ asymmetries discussed above, has better
to be treated in the twist-3 framework. Due to the relation of twist-3 functions 
to TMDs, this process can nevertheless be used to restrict TMDs like the Sivers 
function and has been suggested as another avenue to test the process dependence 
of the Sivers effect~\cite{Gamberg:2012iq} and to test the validity 
of the GPM approach which assumes TMD factorisation~\cite{Gamberg_2011}.

\subsubsection{Transverse single spin asymmetries in di-hadron production}
\label{sec:di-hadEffects}

Recent data from RHIC made the exploration of richer hadronic final states 
possible. We discuss two examples, di-hadrons in this Section and hadrons 
inside jets in the next Section. In both cases, to our knowledge, a complete 
set of structure functions does not exist in the literature; therefore, we 
will not reproduce the full cross-section, but will concentrate on the structure 
functions that have been explored experimentally, which in both cases are 
modulations sensitive to the transversity distribution and, in the case of jets, 
the Sivers distribution, as will be discussed in 
Sections~\ref{phenoDi-HadPP} and~\ref{sec:hadInJetsPheno}. Much more details on 
di-hadron fragmentation can be found in Ref.~\cite{Pisano:2015wnq}.

Considering only the practical relevant case of final states consisting of 
pseudo-scalar mesons, the hadronic tensor in di-hadron production can depend 
on an additional vector, the difference between the momenta of the outgoing 
hadrons $\bfR =\bfp_{h,1} - \bfp_{h,2}$. This additional vector allows 
sensitivity to the the transverse spin structure of the proton as noted by 
Collins, Heppelmann and Ladinsky~\cite{Collins:1993kq}. Using a co-ordinate 
system where the $z$-axis is given by the momentum vector of the hadron pair, the 
polarization sensitive part of the cross-section  
is usually parameterised with an azimuthal angle $\phi_R$, a polar angle 
$\theta$ and the invariant mass of the hadronic final state $M_h$. The 
angle $\phi_R$ is connected to the relative angular momentum of the final 
state with a quantisation axis transverse to $\bfp_h = \bfp_{h,1} + \bfp_{h,2}$, 
which can be seen as a proxy for the outgoing quark. Therefore, modulations of 
the di-hadron cross-section in $\phi_R$ are sensitive to the transverse spin 
structure of the proton. Well-known model calculations~\cite{Bacchetta:2002ux} 
for transverse polarisation dependent di-hadron fragmentation functions are 
based on the interference of hadron pairs in different partial waves~\footnote{It 
should be mentioned that there are alternative models based on string 
fragmentation~\cite{Matevosyan:2013aka,Matevosyan:2017uls}}. 

\begin{figure}[t]
    \centering
    \includegraphics[width=0.9\textwidth]{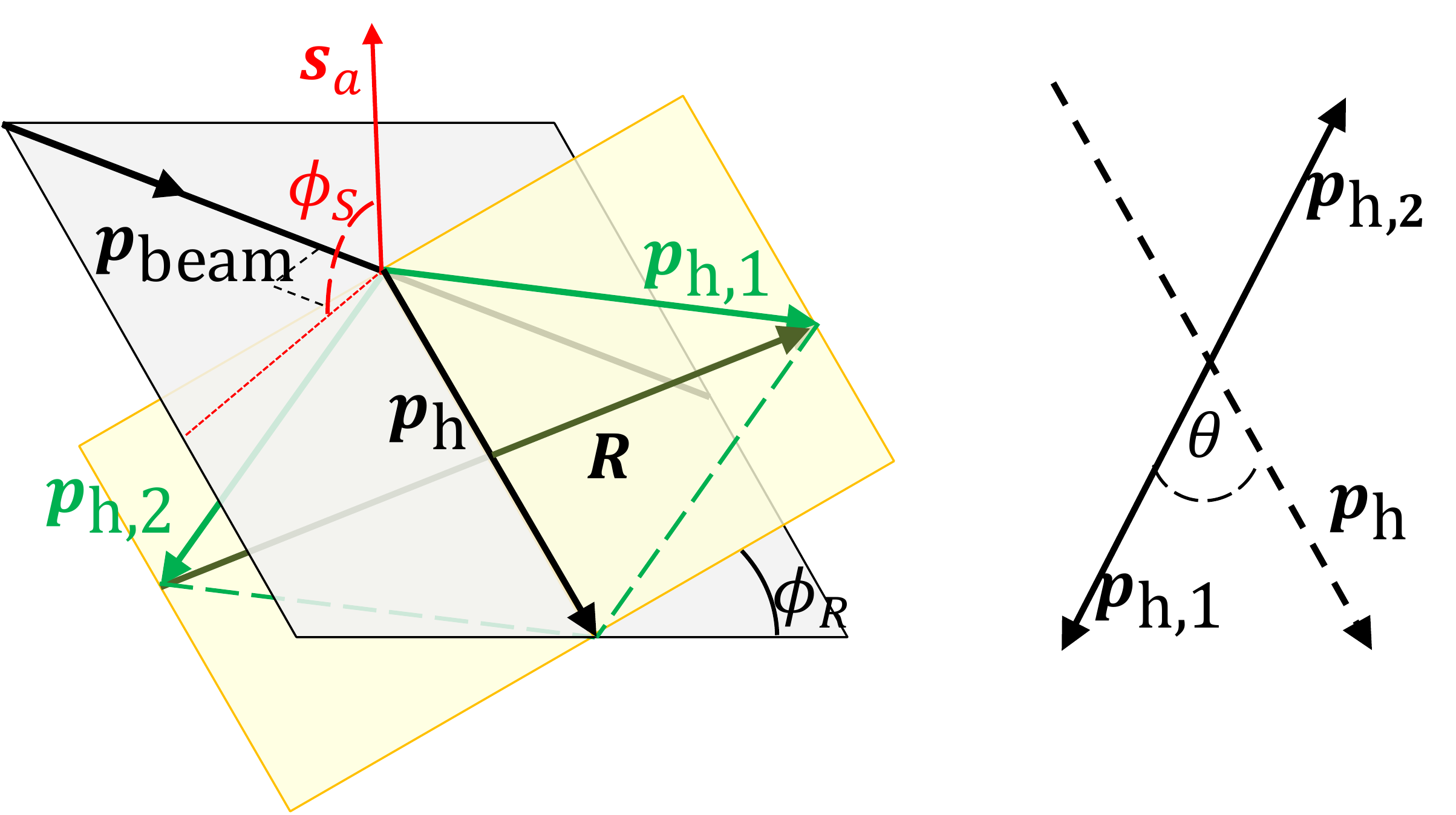}
    \caption{Left: Diagram of the azimuthal angles relevant for di-hadron 
    measurements in $p\,p$ interactions. Here $\bfp_{h,1(2)}$ is the momentum 
    of the positive (negative) pion, $\bfs_{a}$ is the beam polarization, and 
    $\phi_{R}$ is the angle between the scattering plane (gray) and the 
    di-hadron plane (yellow). The diagram on the right shows the polar angle 
    $\theta$ defined between the hadron direction in the center-of-mass system 
    of the hadron pair and the direction of the sum of the hadron momenta 
    $\bfp_h$ in the target rest frame.}
    \label{fig:IFF_pp_frame}
\end{figure}

For the parent quark polarisation dependent fragmentation into charged pions, 
the most relevant effect would come from the interference of an $s-$wave from 
non-resonant production and a $p-$wave from resonant production and subsequent 
decay of $\rho$ mesons. Expanded in partial waves, each interference term would 
then have a characteristic $\theta$ dependence with the most relevant $s-p$ 
interference term having a $\sin\theta$ dependence. Since the experimental 
acceptance usually peaks at $\sin\theta=1$ due to momentum cuts on the particles, 
current results in $p\,p$ integrate over the $\theta$ dependence, and only 
consider the dependence on the azimuthal angle $\phi_R$. The relevant quantities 
for this measurement are shown in Fig.~\ref{fig:IFF_pp_frame}. 
 
Experimental results from $p\,p$ interactions at RHIC have been published by the 
STAR Collaboration~\cite{Adamczyk:2015hri,Adamczyk:2017ynk} on the transverse 
single spin asymmetry $A_{UT}^{\sin\phi_{RS}}$
defined analogously to the SIDIS asymmetries in~\eqref{eq:azi-mom} as
\be
A_{UT}^{\sin \phi_{RS}} = 2 \,
\frac{\int\,d\phi_{RS} \,\left[ d\sigma^\uparrow -
d\sigma^\downarrow\right]\,\sin \phi_{RS}}
{\int\,d\phi_{RS}\,\left[ d\sigma^\uparrow + d\sigma^\downarrow\right]}\,,
\label{eq:A_UT}
\ee
where $\phi_{RS}=\phi_R-\phi_S$. 

As further discussed in Sec.~\ref{phenoDi-HadPP}, this asymmetry is sensitive 
to the contribution of the transversity PDF even after 
integrating over the transverse momentum degrees of freedom in the PDFs and 
FFs~\cite{Bacchetta:2004it}. The STAR measurement has been used for 
the first global extraction of transversity from SIDIS, $p\,p$ and $e^+e^-$ 
data~\cite{Radici:2018iag}.

\subsubsection{Transverse single spin asymmetries of jets and hadrons in jets}
\label{sec:hadInJets}
At high energies, hadronic final states in nuclear collisions are collimated 
into jets. Therefore, jets provide a connection to the initial state partonic 
kinematics. At leading order one can simply identify the parton direction 
with the jet~\cite{DAlesio:2013cfy}, but the connection can also be done at 
higher orders~\cite{Kang:2016mcy}.
\begin{figure}[t]
    \centering
    \includegraphics[angle=-90,width=0.9\textwidth]{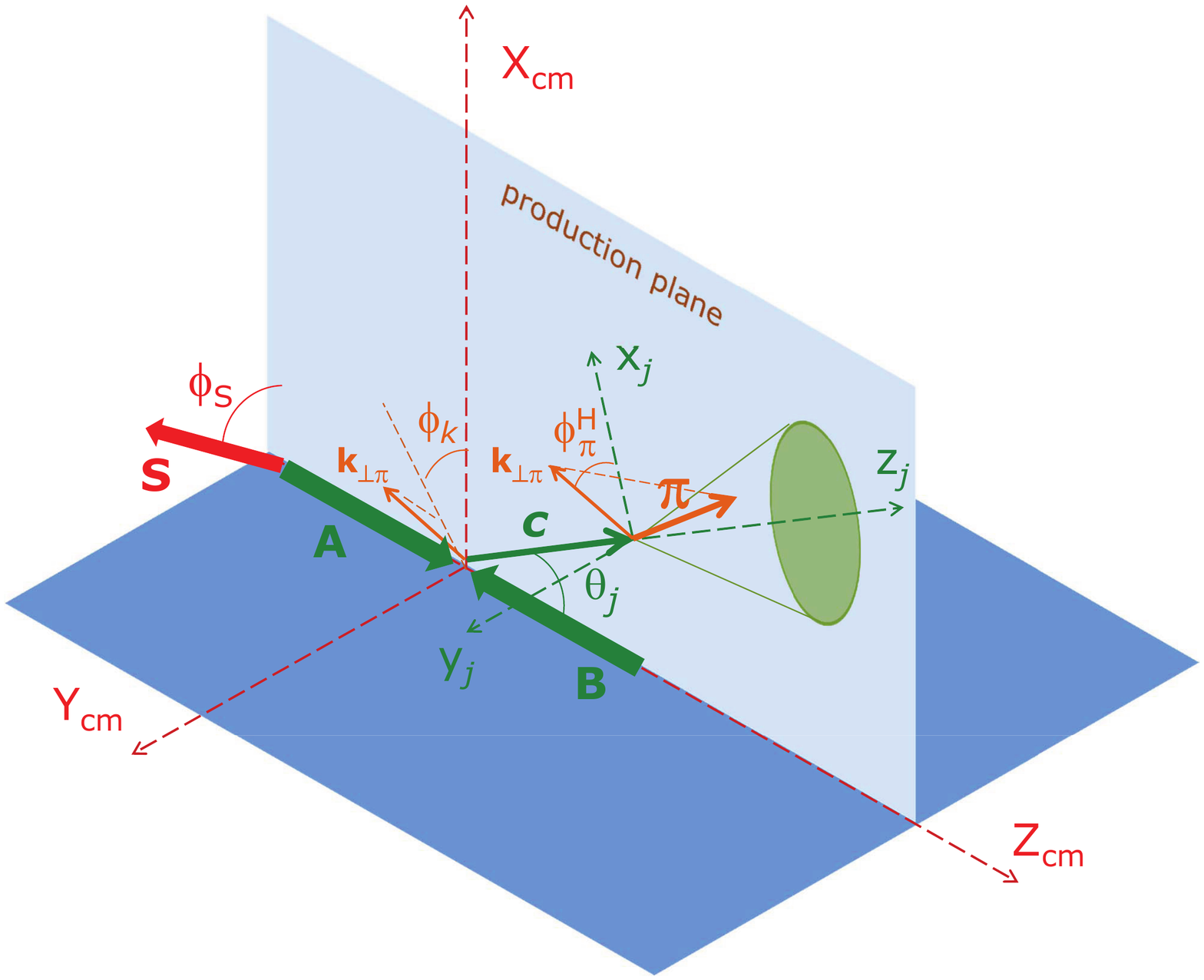}
    \caption{Diagram of the azimuthal angles relevant for hadron in jet 
    measurements in $p\,p$ interactions. The initial momenta of the colliding 
    hadrons are denoted $\bfA$ and $\bfB$, with $\bfA$ being assoicated with the 
    polarization vector $\bfS$. The outgoing parton momentum is denoted $\bfc$ 
    and has an azimuthal angle $\phi_k$. The azimuthal angle of $\bfS$ is 
    $\phi_S$ and the pion within the jet has a transverse momentum of 
    $k_{\perp,\pi}$ with respect to the jet axis. The azimuthal angle of 
    the pion around the jet axis is denoted $\phi_\pi^H$. Figure 
    from~\cite{DAlesio:2013cfy}. Reprinted by permission from  Springer 
    Nature Customer Service Centre GmbH, Springer Phys. Part. Nucl., 
    45(4):676-691, 2014, "Collins and Sivers effects in $\pup p \to$jet $\pi X$: 
    Universality and process dependence.", Umberto D'Alesio, Francesco Murgia, 
    and Cristian Pisano, Copyright 2014.}
    \label{fig:hadInJet_pp_frame}
\end{figure}

Jets are usually described by their transverse momentum $\bfp_T$, with respect
to the beam direction, as well as their position in $\eta-\phi$ space, where 
$\eta$ is the pseudo-rapidity and $\phi$ the azimuthal angle. Since differences 
in rapidity are boost invariant, a rule of thumb is that jets will cover about 
one unit in $\eta$ and $\phi$ regardless of their $x_F$. 
In practice, experimental requirements, such as detector uniformity, 
contributions from underlying events or the beam remnants, will often require 
the use of a smaller jet radius in reconstruction.
With the availability of high statistics datasets from the STAR experiment at 
RHIC, as well as the LHC experiments, interest in using jets to access proton 
structure has grown substantially. 

The challenge for jet physics is that the c.m. energy has to be high enough 
for jets to be created, and the jets must have high enough $p_T$ to provide 
a hard scale. At the same time, experiments have to have large enough 
acceptances to detect a jet. Given the rough size estimate of $1\times 1$ 
in $\eta-\phi$ space, this means usually full azimuthal coverage as well as 
a significant coverage in the polar angle. Forward detector, like early 
$N\,N$ experiments or current Fermilab experiments, are problematic
for jet physics, since the geometry means that jets in the acceptance have low 
$p_T$ and will often be contaminated by beam background. Therefore, most 
experimental input comes from the STAR experiment at RHIC as well as the 
LHC experiments.

Even though jets are among the most challenging observables in nuclear 
collisions, they are quite attractive, since they can be seen as proxies 
for the outgoing parton in the scattering. This makes an estimation of the 
underlying partonic kinematics in jet production possible; at LO we have  
$2 \to 2$ underlying processes. For example, the partonic $x$ can be calculated 
from the pseudo-rapidities $\eta$ and transverse momenta $p_T$ of the two jets 
as
\be
\label{eq:diJetKinematics}
x_1=(p_{T,1}\,e^{\eta_1} + p_{T,2}\,e^{\eta_2})/\sqrt{s}
\quad\quad   
x_2=(p_{T,1}\,e^{-\eta_1} + p_{T,2}\, e^{-\eta_2})/\sqrt{s}.
\ee

As it can be seen from the expression above, high $p_T$ jets preferentially 
select high $x$ partons. The fractional energy $z$ can be determined as the 
ratio of the energy of the detected hadron to the jet energy.
This measurement of $z$ is obviously also available in inclusive jet 
measurements. If only a single jet is detected, the formulae above to calculate 
$x_i$ at LO are not applicable; however, there is still a strong correlation 
between the $p_T$ and $\eta$ of the detected jet with the underlying partonic 
kinematics since the underlying $2 \to 2$ scattering in a high $p_T$ jet 
measurement is usually quite asymmetric in $x$. 

Compared to SIDIS processes, measurements of transverse spin phenomena in 
jets in $p\,p$ scattering has the advantage that, at least in the LO 
interpretation, the initial and final state $k_T$ dependences are separated, 
since presumably the jet is an approximation to the outgoing parton. However, 
if intrinsic transverse momenta are included, there are questions about the 
factorisation of this process~\cite{Rogers:2010dm}. For the Collins effect, 
{\it i.e.} the dependence of the transverse momentum of a hadron in a jet 
on the parent parton transverse polarisation, universality 
seems to hold~\cite{Yuan:2007nd,Kang:2017btw}.
 
Figure~\ref{fig:hadInJet_pp_frame} shows the relevant quantities for the 
process with one polarised proton. Polarisation dependent PDFs and FFs can 
be accessed by the dependence of the cross section on the azimuthal angles 
of the parent proton polarisation ($\phi_S)$, the azimuthal angle of the jet 
axis ($\phi_k$) and the hadron (in this case a pion) within the jet 
($\phi_\pi^H$). As will be discussed in~Section \ref{sec3}, in a LO treatment, 
the cross section has similarities with the SIDIS cross section when the 
lepton beam in the SIDIS case is replaced with a quark from the unpolarised 
proton. In particular, the transversity distribution can be accessed by 
measuring modulations in the azimuthal angle of hadrons around the 
jet axis ($\phi_\pi^H$) and the Sivers effect can be accessed by measuring the 
dependence of the cross section on the azimuthal angle of the jet axis ($\phi_k)$. 
Both of these measurements have been performed by the STAR 
Collaboration~\cite{Adamczyk:2012qj,Adamczyk:2017wld} and a significant 
evidence for the Collins signal, consistent with expectations from global 
fits~\cite{Kang:2017btw}, has been observed.

Transverse Single Spin Asymmetries (TSSAs) for jets measured at STAR do not 
show a significant signal, as expected for jets detected at mid-rapidity. The 
AnDY experiment performed a measurement for forward jets~\cite{Bland:2013pkt} 
which shows an indication of a non-vanishing asymmetry.
While for the observation of a hadron inside a jet the TMD picture is 
appropriate, due to the small transverse momentum of the hadron inside a jet, 
this is not so clear in single-jet measurements to access the Sivers effect. 
Here only a single hard scale, the $p_T$ of the jet, is observed, which 
makes the twist-3 picture more appropriate. However, the twist-3 measurement 
can be related to the Sivers function in the intermediate $p_T$ region, 
as discussed in Sec.~\ref{sec:hadInJetsPheno}. A measurement of the Sivers 
function that can be interpreted in the TMD picture are nearly back-to-back 
di-jets, where a spin dependence of the (small) relative transverse momentum 
of the two jets is observed. Such a measurement has been performed by the STAR 
Collaboration~\cite{Abelev:2007ii}, but the measurement was limited by statistics. 
In addition, factorization and universality are problematic in this case as 
will be further discussed in Sec.~\ref{sec:hadInJetsPheno}. It should be noted 
that the aforementioned factorization issues in back-to-back jet production 
can be avoided in SIDIS jet production, such as a future EIC. Here one of the 
jets is replaced conceptually by the outgoing lepton. Such a measurement would 
therefore combine some of the advantages of jet measurements, in particular the 
decoupling of the fragmentation functions for the TMD-PDF measurements and 
vice-versa with the advantage of the clean theoretical understanding of SIDIS.

\subsection{\it Spin effects in $e^+e^- \to h_1 \, h_2 \, X$ and 
$e^+e^- \to \Lambda^\uparrow \, X$ processes 
\label{subsec2.3}}

Let us consider the process in which unpolarised leptons and anti-leptons 
annihilate into two jets of particles, and we look at two hadrons belonging
to opposite jets. At first sight it might appear impossible to obtain spin 
effects in such a process: however, this is not the case~\cite{Boer:1997mf}. 
As usual, in $e^+e^-$ annihilation processes, the production
of hadrons goes via the subprocess $e^+e^- \to q \, \bar q$. The final $q$ 
and $\bar q$ are also not polarised, that is they can have spin ``up" or ``down",
with respect to the scattering plane, with the same probability (1/2); however 
there is a correlation between the spin of the quark and that of the anti-quark. 
Not necessarily if the first is up, the second is down, or viceversa. 
To be precise, with reference to the kinematical configuration of 
Fig.~\ref{fig:e+e-qqbar}, one has:
\begin{figure}[t]
\begin{center}
\vskip -48pt
\includegraphics[width=15.truecm,angle=0]{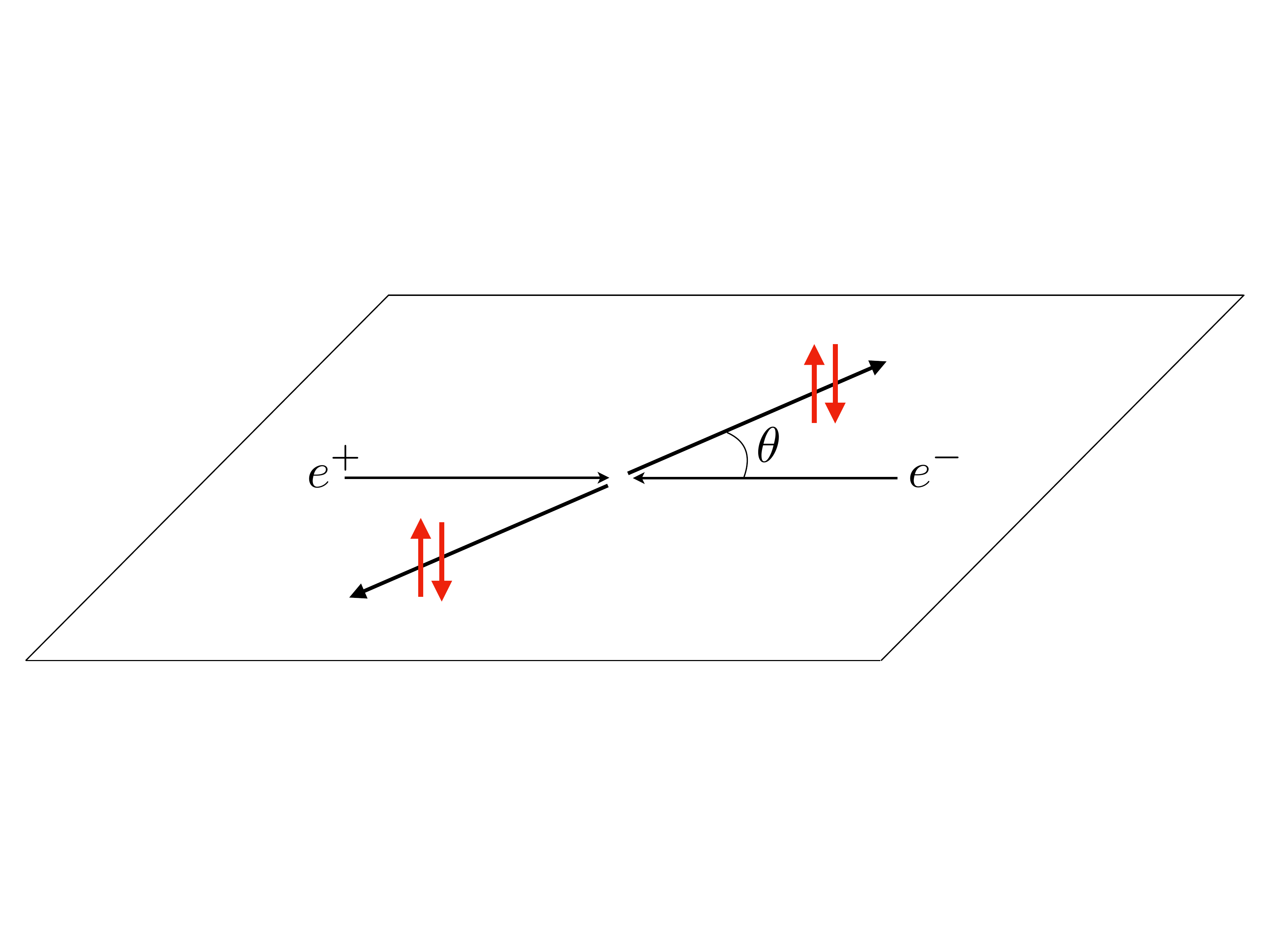}
\vskip -72pt 
\caption{Center of mass scattering plane for the $e^+e^- \to q \,\bar q$ process;
the spin polarisation vectors are perpendicular to the plane.}
\label{fig:e+e-qqbar}
\end{center}
\end{figure}
\be
\frac{d\sigma^{e^+e^- \to \, q^\uparrow \bar q^\uparrow}}{d\cos\theta} =
\frac{d\sigma^{e^+e^- \to \, q^\downarrow \bar q^\downarrow}}{d\cos\theta} =
\frac{3\pi\alpha^2}{4s} \, e_q^2 \cos^2\theta \quad\quad\quad 
\frac{d\sigma^{e^+e^- \to \, q^\uparrow \bar q^\downarrow}}{d\cos\theta} =
\frac{d\sigma^{e^+e^- \to \, q^\downarrow \bar q^\uparrow}}{d\cos\theta} =
\frac{3\pi\alpha^2}{4s} \, e_q^2 \label{e+e-qqbarcs} \>\cdot
\ee

The hadronisation process $q^{\uparrow, \downarrow} \to h \, X$, might have 
(and indeed has) a transverse spin dependence which affects the angular 
distribution of the produced hadron; then, if one looks in each same event 
$e^+e^- \to h_1 \, h_2 \, X$, at the correlation between the angular 
distributions of the two hadrons in the opposite jets, one can learn about 
such a transverse spin dependence. Again, by rotational invariance, no spin 
dependence would be possible in a collinear quark fragmentation process, which 
leads to the necessity of introducing transverse motions also in the 
phenomenological description of a quark fragmentation. 
\begin{figure}[t]
\begin{center}
\vskip -48pt
\includegraphics[width=15.truecm,angle=0]{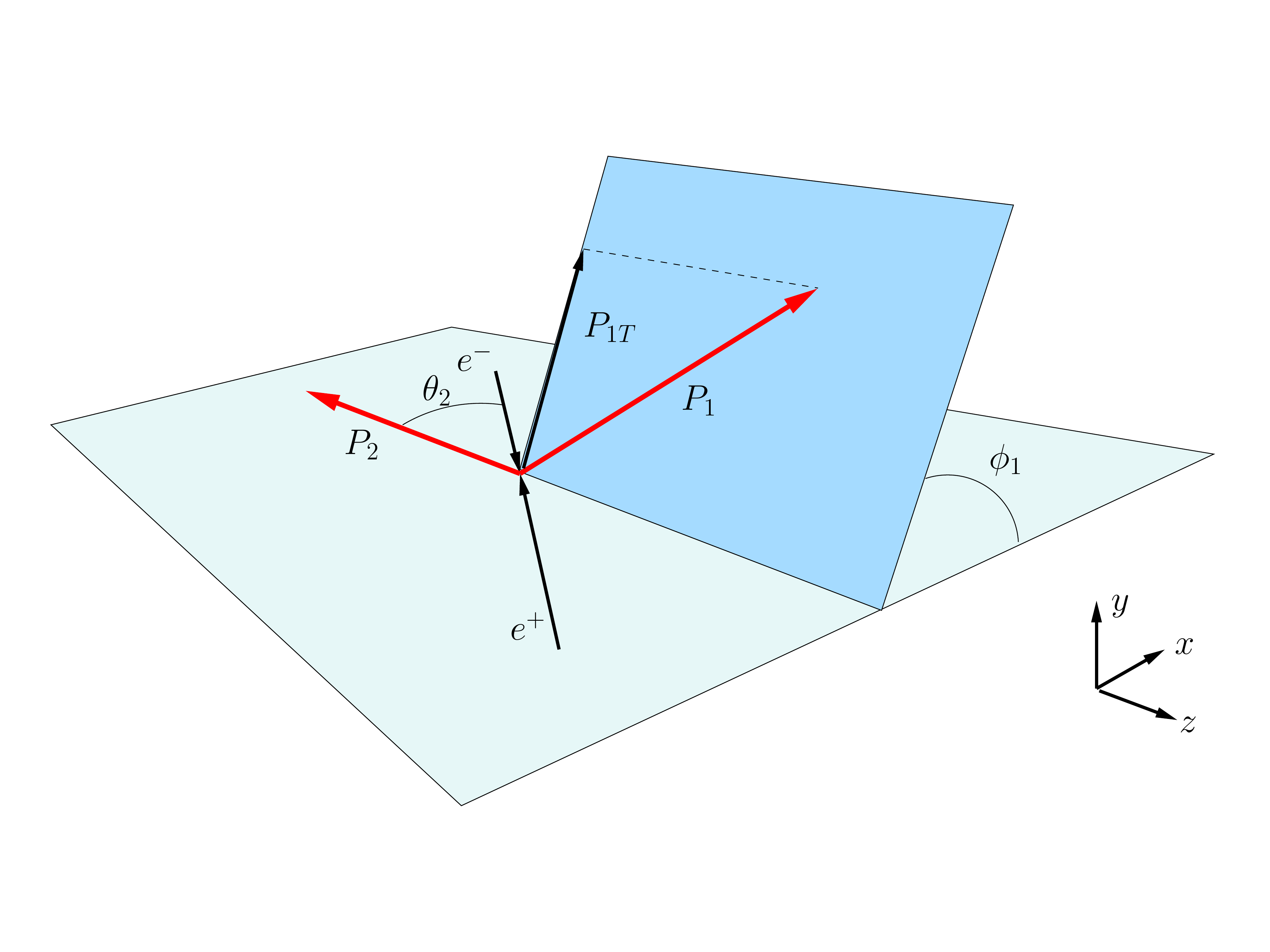}
\vskip -36pt 
\caption{Kinematical configuration and definition of variables for the process
$e^+e^- \to h_1 \, h_2 \, X$. $e^+$ and $e^-$ have opposite 
three-momenta, $\bfP_2$ is the momentum of a jet or of a fast hadron in a jet, 
and $\bfP_1$ is the momentum of a hadron belonging to the other jet.}
\label{e+e-P1}
\end{center}
\end{figure}

The general form of the cross section for the process $e^+e^- \to h_1 \, 
h_2 \, X$, via $e^+e^-$ annihilation into $q\,\bar q$, can be written in 
different ways, according to the reference frame used and the observables
one looks at. For example, using the reference frame and the kinematical
variables of Fig.~\ref{e+e-P1} one has~\cite{Boer:1997mf,Anselmino:2015sxa}:
\be
\frac{d\sigma^{e^+e^-\to h_1 h_2\,X}}{dz_1 \, dz_2 \, d^2\bfP_{1T} \, d\cos\theta_2}
= \frac{3\pi\alpha_{\rm em}^2}{2s}\,\left\{ D_{h_1h_2} + N_{h_1h_2}\,
\cos(2\phi_1)\right\}\,,
\label{eq:dsig-ee}
\ee
where $z_{1,2}$ are the light-cone momentum fractions of the hadrons $h_{1,2}$
resulting from the fragmentation of the quark and antiquark. They are essentially
the quark and antiquark energy fractions taken away by the hadrons.  

The cross-section~\eqref{eq:dsig-ee} has been measured by several experiments. The 
first measurement for $\pi^+\pi^-$ pairs was done by the Belle 
Collaboration~\cite{Abe:2005zx, Seidl:2008xc} and confirmed by 
BaBar~\cite{TheBABAR:2013yha}. BaBar also extended the measurement to the 
transverse momentum dependence and included charged kaons~\cite{Aubert:2015hha}. 
While Belle and BaBar measurements were done at a center-of-mass energy close to the 
mass of the $\Upsilon(4S)$, BES-III did a similar measurement at lower cms 
energies~\cite{Ablikim:2015pta}. Most recently, Belle published asymmetries 
of back-to-back hadrons including $\pi^0$ and $\eta$'s~\cite{Li:2019iyt}.

In the next Section the form factors $D_{h_1h_2}$ and $N_{h_1h_2}$ will be 
interpreted in terms of Fragmentation Functions, either collinear FFs or TMD-FFs. 
Eq.~(\ref{eq:dsig-ee}) will be used to relate experimental data to spin dependent 
TMD-FFs.  

%$e^+e^-$ annihilation processes, with the observation of one or two 
%final particles, are interesting in many respects. In addition to the case 
%discussed above, the process $e^+e^-\to h_1 \, h_2\,X$, with two hadrons inside 
%the same jet, can be used to get information on the di-hadron production and 
%the related di-hadron fragmentation function~\cite{Metz:2016swz},  
%discussed in Sections~\ref{sec:di-hadEffects} and ~\ref{phenoDi-HadPP}. 

The $e^+e^-$ annihilation processes, with the observation of one or two 
final particles, are interesting in many respects.
In addition to the case 
discussed above, the process $e^+e^-\to (h_{a1} \, h_{a2})\, (h_{b1}\,h_{b2})\,X$, with back-to-back pairs of hadrons inside 
the same jet, can be used to get information on
% the transverse spin dependent di-hadron production and 
the
%related 
polarized di-hadron fragmentation function~\cite{Metz:2016swz},  
as discussed in Section \ref{subsec3.4}.%}s~\ref{sec:di-hadEffects} and ~\ref{phenoDi-HadPP}. }

The measured transverse polarisation of $\Lambda$s and other hyperons, 
as mentioned in Section~\ref{subsec2.2}, was among the seminal data which prompted 
the study of transverse spin effects. These data was obtained from unpolarised 
$p\,p$ or $p\,N$ inclusive processes, and a first attempted explanation
was related to spin effects in the fragmentation of an unpolarised quark,
a TMD-FF, in the framework of the GPM~\cite{Anselmino:2000vs}, with subsequent
studies in SIDIS~\cite{Anselmino:2001js}. However, the usual 
reservations in using the GPM for a single scale $p\,p$ process, together 
with the scarcity or absence of data in other interactions, could not allow 
definite conclusions and alternative explanations were 
presented~\cite{Boer:2002ij}.    

Data on $\Lambda$ polarisation in $e^+e^- \to \Lambda^\uparrow \, X$, due to the 
simplicity of such a process, would offer an ideal occasion to understand 
whether the hadronisation process of an unpolarised quark might build up
a transverse polarisation; first data, although still limited, are becoming
available~\cite{Guan:2018ckx}.     

\section{Phenomenology of spin phenomena \label{sec3}}

We have discussed in the previous Section several examples of processes in which 
the simple -- yet in many cases successful -- 1D description of a fast moving
nucleon as an almost free set of co-linearly moving quarks and gluons cannot 
explain the experimental data. We have argued that the reason for this is
just the excessive simplicity of the 1D model; the neglected transverse degrees 
of freedom might play important roles, in particular, but not only, when such 
degrees of freedom are forced into existence by considering transversely 
polarised nucleons. Transverse always refers to the direction of motion of 
the nucleons. 

Much progress has been achieved in the last one or two decades by extending 
the collinear factorisation theorem, which allows to describe hadronic cross 
sections as convolutions of elementary quark and gluon interactions with
collinear PDFs or FFs. The PDFs give the number densities of partons inside a 
proton, while the FFs give the number densities of hadrons resulting in a 
parton hadronisation. The PDFs depend on the longitudinal momentum fraction  
$x$ of the proton carried by the parton and the FFs on the fraction $z$ of 
the parton momentum carried by the hadron; they both depend on the scale $Q^2$ 
of the process, and such a dependence can be computed in QCD (QCD evolution
of PDFs and FFs).  

The extension of the collinear factorisation theorem allows to describe 
cross sections again as convolutions: of elementary interactions with 
Transverse Momentum Dependent Parton Distribution Functions (TMD-PDFs) and
Fragmentation Functions (TMD-FFs). This TMD factorisation has been studied 
in a series of papers~\cite{Ji:2004xq,Collins:1984kg,Collins:1992kk,
Collins:2004nx,Ji:2004wu,Ji:2006ub,Ji:2006vf,Bacchetta:2008xw,Arnold:2008kf,
Anselmino:2009st} and proven to be valid for SIDIS, D-Y and 
$e^+e^- \to h_1 \, h_2 \, X$ processes; the situation is less clear for single 
and double inclusive hadronic processes, 
$p\,N \to h\, X$ and $p\,N \to h_1\,h_2 \, X$~\cite{Rogers:2010dm,Mulders:2011zt}.          
 
Throughout the paper we mainly adopt the TMD factorisation scheme and consider 
the TMDs, both TMD-PDFs and TMD-FFs, and their role in physical quantities at 
leading order. These TMDs have a partonic interpretation and we focus on their 
phenomenological extraction from data, which helps in trying to build a 3D imaging 
of the nucleon. In Section~\ref{subsec4.3} we look at the possibility of a real 
knowledge of the full 3D partonic structure of the nucleon, both in momentum and 
configuration space, with the Wigner function. We do not discuss the details of the 
QCD evolution of the TMDs (TMD evolution), which is studied at length in several 
papers~\cite{Collins:2011zzd,Aybat:2011zv,Aybat:2011ge,Bacchetta:2013pqa,
Scimemi:2019cmh}. The first phenomenological applications~\cite{Aybat:2011ta,Anselmino:2012aa,
Kang:2015msa,Bacchetta:2017gcc,Boglione:2018dqd} show that 
the TMD evolution will play an important role in future experiments, but does not 
significantly affect the TMD phenomenology of the actual available data.   
        
\subsection{\it Transverse Momentum Dependent Parton Distributions (TMD-PDFs) 
and Fragmentation Functions (TMD-FFs) \label{subsec3.1}}
\subsubsection {Quark TMD-PDFs}

\begin{figure}[t]
\begin{center}
\vskip -72pt
\includegraphics[width=15.truecm,angle=0]{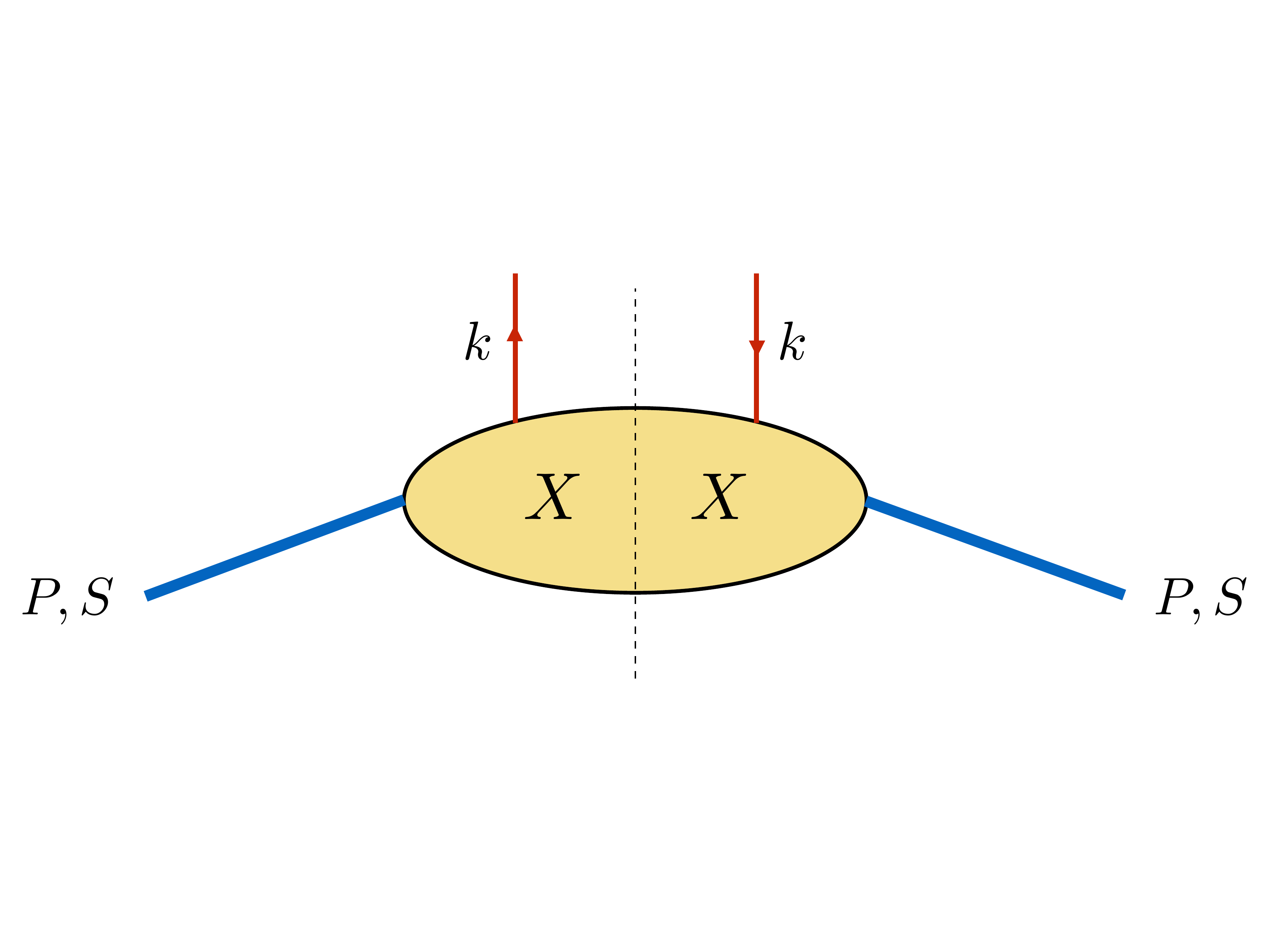}
\vskip -72pt 
\caption{Quark-quark correlator contributing to inclusive processes; the 
off-diagonal version, in which the initial and final nucleon momenta are different,
would contribute to amplitudes of exclusive processes.}
\label{correlator}
\end{center}
\end{figure}

The quark TMD-PDFs contribute to cross sections of inclusive processes in which 
one quark interacts with an external probe, like a point-like lepton or another 
parton; one can think of the quark as ``extracted" from the parent nucleon, which 
breaks up into unobserved particles. This is typically represented by the 
quark-quark correlator (handbag diagram) of Fig.~\ref{correlator}, which in Dirac
space is given by~\cite{Barone:2001sp}:
\bea
\Phi_{ij}(k; P,S) &=& \frac{1}{(2\pi)^4} \sum_X \int  
\frac{{\rm d}^3\bfP_X}{(2\pi)^3\,2E_X}\,(2\pi^4)\,
\delta^4(P - k - P_X)\, \langle PS|\overline \Psi_j(0)|X\rangle \langle X|
\Psi_i(0)|PS\rangle \nonumber \\
&=& \frac{1}{(2\pi)^4} \int {\rm d}^4\xi \, e^{ik \cdot \xi} \,
\langle PS| \overline \Psi_j(0)\Psi_i(\xi) |PS\rangle \>.\label{corr-def} 
\eea
In the collinear case, in which, apart from mass corrections, the quark 
momentum is just a fraction $x$ of the nucleon momentum, the most general 
dependence of the correlator on $x$ can be written as:

\be
\Phi(x,S) = \frac 12 \, \left[ f_1(x) \, \slash \!\!\! n_+ + 
S_L \, g_{1L}(x) \, \gamma^5 \, \slash \!\!\! n_+ +
h_{1T}(x)\, i \sigma_{\mu\nu} \, \gamma^5 \, n^\mu_+ \, S_T^\nu 
\right] \>. \label{corr-coll}
\ee

In the above equations $P$ and $S$ are respectively the four-momentum and the 
covariant spin vector of the nucleon, which can be longitudinally ($S_L$) or 
transversely ($S_T$) 
polarised. $n_+$ is a convenient light-like four-vector which, up to mass 
terms, is along the nucleon momentum. For completeness one should insert into
the definition of the correlator, Eq.~(\ref{corr-def}), a Wilson line, or gauge 
link, which guarantees the color gauge invariance of the correlator (see, for 
example, Refs.~\cite{Mulders:1995dh,Goeke:2005hb}). 

In Eq.~(\ref{corr-coll}), for each quark flavour, $f_1(x)$ is 
the unpolarised PDF, $g_{1L}(x)$ is the helicity distribution and $h_{1T}(x)$ 
the transversity distribution. These definitions conventionally refer to protons. 
These quantities sometimes appear in the literature with different names:
\be 
f_1(x) \equiv q(x) \quad\quad g_{1L}(x) \equiv \Delta q(x) \quad\quad
h_{1T}(x) \equiv h_1(x) \equiv \Delta_T q(x) \equiv \delta q(x) \>.
\ee  

When letting the quark have an intrinsic motion $\bkt$ inside the 
proton Eq.~(\ref{corr-coll}) gets more terms which vanish in the limit
$k_\perp \to 0$~\cite{Ralston:1979ys,Tangerman:1994eh,Mulders:1995dh,Boer:1997nt}:
\bea
\Phi(x,\bkt,S) &=& \frac 12 \, \left[ f_1 \, \slash \!\!\! n_+ + f_{1T}^\perp \, 
\frac {\epsilon_{\mu\nu\rho\sigma}\gamma^\mu n_+^\nu k_\perp^\rho S_T^\sigma}{M}
+ \left( S_L \, g_{1L} + \frac{\bkt \cdot \bfS_T}{M} \, g_{1T}^\perp \right)
\gamma^5 \, \slash \!\!\! n_+ \right.\nonumber \\
&+& h_{1T}\, i \sigma_{\mu\nu} \, \gamma^5 \, n^\mu_+ \, S_T^\nu +
\left.\left( S_L \, h_{1L}^\perp + \frac{\bkt \cdot \bfS_T}{M}\, 
h_{1T}^\perp \right)
\, \frac {i \sigma_{\mu\nu} \, \gamma^5 \, n^\mu_+ \, k_\perp^\nu }{M}
+ h_1^\perp \frac {\sigma_{\mu\nu} k_\perp^\mu n^\nu_+}{M} \right] \>.
\label{corr-8tmd}
\eea 

Eq.~(\ref{corr-8tmd}) gives the most general expression of the quark-quark 
correlator at leading twist (twist-2); it contains 8 independent functions, 
which are usually referred to as the 8 leading twist quark TMD-PDFs. The 
notations in which they are written require some comments: $f$, $g$ and $h$ 
indicate respectively unpolarised, longitudinally polarised and transversely 
polarised quarks; the subscript 1 refers to the fact that they are leading 
twist TMDs; the subscripts $L$ and $T$ show the polarisation, longitudinal 
or transverse, of the proton (no subscript stays for unpolarised nucleons);
the superscript $\perp$ appears for TMDs which do not contribute to the 
correlator in the collinear limit\footnote{Notice, however, that 
in the original literature~\cite{Tangerman:1994eh} the function $g_{1T}^\perp$ 
does not have the superscript $\perp$, as the function itself might not vanish 
in the limit $k_\perp \to 0$.}. 
One could also add at each TMD a superscript $q$ to identify the quark
flavour. $M$ is a mass parameter taken as the proton mass.   

\begin{figure}[t]
\begin{center}
\vskip -72pt
\includegraphics[width=15.truecm,angle=0]{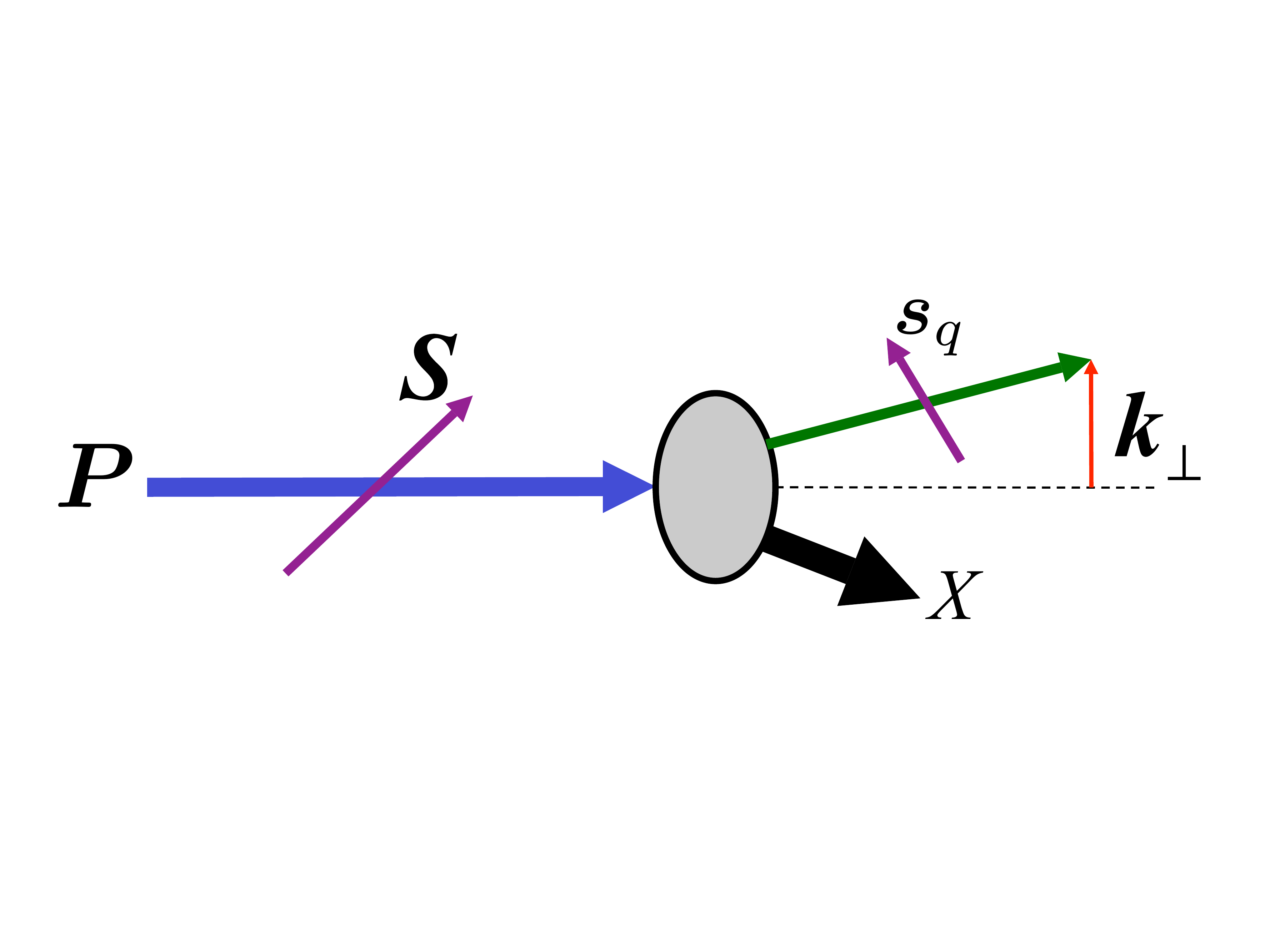}
\vskip -96pt 
\caption{Usual graphical representation of the TMD-PDFs of quarks with 
spin vector $\bfs_q$ and transverse intrinsic momentum $\bkt$ inside 
a proton with momentum $\bfP$ and spin vector $\bfS$.}
\label{tmd-pdf}
\end{center}
\end{figure}

Eq.~(\ref{corr-8tmd}) is the usual formal definition of the quark-quark correlator.
However, it is interesting to look at the TMDs in a simple and intuitive way, 
which clarifies their physical meaning and emphasises their partonic interpretation.
It is custumary to represent the PDFs and the TMDs as splitting processes
in which a proton breaks up into a quark + remnants ($X$), like in 
Fig.~\ref{tmd-pdf}.
 
We can ask how many independent combinations of the vectors ($\bfp, \bkt$) 
and pseudo-vectors ($\bfS, \bfs_q$) can make up, as required by QCD parity 
invariance, a scalar quantity, which can then appear as an independent term 
in the correlator. The answer can be written, in momentum space, 
as~\cite{Murgia:2010}:  
\bea
\widetilde\Phi(x, \hat{\bfP}, \hat{\bfk}_\perp, \bfS, \bfs_q) &=& 
\frac 12 \, \Big\{ f_{q/p}(x, k_\perp) \nonumber \\
&+& \Delta^N f_{q^\uparrow/p}(x, k_\perp)\, 
(\hat{\bfP} \times \hat{\bfk}_\perp) \cdot \bfs_q  
+ \frac 12 \, \Delta^N f_{q/p^\uparrow}(x, k_\perp)\, 
(\hat{\bfP} \times \hat{\bfk}_\perp) \cdot \bfS \nonumber \\
&+& \Delta^-f_{s_y/S_T}^q(x, k_\perp)\left[ \bfS \cdot \bfs_q - 
(\hat{\bfP} \cdot \bfS)(\hat{\bfP} \cdot \bfs_q) - 
(\hat{\bfk}_\perp \cdot \bfS)(\hat{\bfk}_\perp \cdot \bfs_q) \right]\nonumber \\
&+& \Delta f_{s_z/S_L}^q(x, k_\perp) (\hat{\bfP} \cdot \bfS)(\hat{\bfP} \cdot \bfs_q)
+   \Delta f_{s_x/S_L}^q(x, k_\perp) (\bfP \cdot \bfS)(\hat{\bfk}_\perp \cdot \bfs_q)
\nonumber \\
&+& \Delta f_{s_z/S_T}^q(x, k_\perp) (\hat{\bfk}_\perp \cdot \bfS)
(\hat{\bfP} \cdot \bfs_q)
+   \Delta f_{s_x/S_T}^q(x, k_\perp) (\hat{\bfk}_\perp \cdot \bfS)
(\hat{\bfk}_\perp \cdot \bfs_q) \Big\} \label{TMD-noi}
\eea 
and indeed contains 8 independent terms. These are written adopting the notation 
of Ref.~\cite{Anselmino:2011ch}; it has the advantage that the different functions 
are indeed polarised quark TMD-PDFs or differences between two of them with some 
opposite spin direction. The exact relation between the notations of 
Eqs.~(\ref{corr-8tmd}) and (\ref{TMD-noi}) is given in Eqs. (22)-(25) and (36), (37) 
of Ref.~\cite{Anselmino:2011ch}. 

As we said, Eq.~(\ref{TMD-noi}) can be read directly in terms of TMD-PDFs. If one 
averages over the proton spin $\bfS$ and sums over the emitted quark spin $\bfs_q$
one obtains: 
\be 
\frac 12 \, \sum_{\bfS, \bfs_q} \widetilde\Phi(x, \hat{\bfP}, 
\hat{\bfk}_\perp, \bfS, \bfs_q)= f_{q/p}(x, k_\perp) \label{PDF-unp}
\ee  
which is the unpolarised TMD-PDF of quark $q$. Similarly, summing over $\bfs_q$ 
only, one has the Sivers distribution~\cite{Sivers:1989cc,Sivers:1990fh} 
of unpolarised quarks inside a polarised proton:  
\be
\sum_{\bfs_q} \widetilde\Phi(x, \hat{\bfP}, \hat{\bfk}_\perp, \bfS, \bfs_q)
= f_{q/p}(x, k_\perp) + \frac 12 \, \Delta^N f_{q/p^\uparrow}(x, k_\perp) \,
(\hat{\bfP} \times \hat{\bfk}_\perp) \cdot \bfS \equiv f_{q/p^\uparrow}(x, \bkt)\>. \label{PDF-Sivers} 
\ee
Notice that, due to the scalar mixed product $(\hat{\bfP} \times \hat{\bfk}_\perp) 
\cdot \bfS$, only directions of $\bfS$ transverse to the proton momentum $\bfP$ 
contribute. 

Averaging over the proton spin $\bfS$ one has the Boer-Mulders 
distribution~\cite{Boer:1997nt} of polarised quarks inside an unpolarised 
proton,
\be
\frac 12 \, \sum_{\bfS} \widetilde\Phi(x, \hat{\bfP}, \hat{\bfk}_\perp, \bfS, \bfs_q)
= \frac 12 \, f_{q/p}(x, k_\perp) + \frac 12 \, \Delta^N f_{q^\uparrow/p}(x, k_\perp)
\, (\hat{\bfP} \times \hat{\bfk}_\perp) \cdot \bfs_q 
\equiv f_{q^\uparrow/p}(x, \bkt) \>. \label{PDF-BM} 
\ee

The other single terms in Eq.~(\ref{TMD-noi}) can be seen as differences of 
polarised quark distributions inside polarised protons. Notice that the proton 
is taken to move along the longitudinal $\hat{\bfZ}$-direction, $S_L = S_Z$, 
while the quark $\hat{\bfx}, \hat{\bfy}$ and $\hat{\bfz}$-axes are defined in 
the quark helicity rest frame~\cite{Anselmino:2011ch}.   

Let us summarise the meaning and different notations adopted for the 8 twist-2 
TMDs; in each of them one can think of the quark as carrying a momentum 
$\bfp_q = x \bfP + \bkt$. 
\begin{itemize} 
\item
The TMD-PDFs for unpolarised quarks of flavour $q$ inside an unpolarised proton 
are usually written as
\be f_1^q(x, \kt) \equiv f_{q/p}(x, k_\perp) \equiv q(x, k_\perp) 
\label{TMD-unp} 
\ee     
\item
The Sivers distribution of unpolarised quarks inside a transversely polarised 
proton is given in Eq.~(\ref{PDF-Sivers}). The function
\be
\Delta^N f_{q/p^\uparrow}(x, k_\perp) \equiv -\frac{2\kt}{M}\,
f_{1T}^{\perp q} (x, k_\perp)
\label{TMD-Siv}
\ee
is the Sivers function. Notice that, often, $f_{1T}^\perp$ alone is referred to as 
the Sivers function. 
\item
The Boer-Mulders distribution of polarised quarks inside an unpolarised 
proton is given in Eq.~(\ref{PDF-BM}). The function
\be
\Delta^N f_{q^\uparrow/p}(x, k_\perp) \equiv -\frac{\kt}{M}\,
h_{1}^{\perp q} (x, k_\perp)
\label{TMD-BM}
\ee
is the Boer-Mulders function. Notice that, often, $h_{1}^\perp$ alone is referred 
to as the Boer-Mulders function. 
\item
The TMD helicity distribution, that is the difference between the distributions
of quarks with positive and negative helicities, inside a positive helicity 
proton is given by:
\be
\Delta f_{s_z/S_L}^q(x, k_\perp) \equiv \Delta q(x, k_\perp)
\equiv g_{1L}^q(x, \kt)  
\label{TMD-hel}
\ee       
\item
The TMD transversity distribution, that is the difference between the 
distributions of quarks with opposite transverse spin, inside a proton with 
transverse spin is given by~\cite{Anselmino:2011ch}:
\be
\frac 12 \left( \Delta^-f_{s_y/S_T}^q(x, k_\perp)
+ \Delta f_{s_x/S_T}^q(x, k_\perp) \right) = h_{1T}^q(x, \kt) +
\frac{\kt^2}{2M^2} \, h_{1T}^{\perp q}(x, k_\perp) \equiv h_1^q(x, k_\perp)\label{TMD-tra}
\ee 
\item
The remaining three distributions in Eq.~(\ref{TMD-noi}) refer to differences of 
polarised TMDs with opposite quark polarisations inside a polarised proton, as 
indicated by the indexes: 
\bea 
&& \Delta f_{s_x/S_L}^q(x, \kt) \equiv \frac{\kt}{M} \, h_{1L}^{\perp q}(x, \kt) 
\quad\quad 
\Delta f_{s_z/S_T}^q(x, \kt) \equiv \frac{\kt}{M} \, g_{1T}^{\perp q}(x, \kt) \\
&& \Delta f_{s_x/S_T}^q(x, \kt) \equiv h_1^q(x, \kt) +
\frac{\kt^2}{2M^2}\,h_{1T}^{\perp q}(x, \kt) \>. \label{TMD-sS}
\eea
%$
$h_{1T}^{\perp}$ is also denoted ``pretzelosity", due to the typical shapes it 
produces in the proton rest frame~\cite{Miller:2003sa,Miller:2007ae}, while 
$h_{1L}^{\perp}$ and $g_{1T}^{\perp}$ are denoted ``worm-gear" 
TMDs~\cite{Bacchetta:2010uj}, as they relate 
quark and proton polarisations which are (almost) orthogonal.

In the collinear limit only the unpolarised, the helicity and the transversity 
TMDs do not vanish. Upon integration over ${\rm d}^2\bkt$ they give the usual 
collinear PDFs:
\bea
&&\int {\rm d}^2\bkt \, f_{q/p}(x, \kt) = f_{q/p}(x) 
\equiv q(x) \equiv f_1^q(x)\\
&&\int {\rm d}^2\bkt \, \Delta f_{s_z/S_L}^q(x, \kt) = \Delta q(x) 
\equiv g_{1L}(x) \\
&&\int {\rm d}^2\bkt \, h_1^q(x, \kt) = \Delta_Tq(x) \equiv h_1(x) \>. 
\eea
\end{itemize}

\subsubsection {Gluon TMD-PDFs}

We have so far only considered quarks TMDs, but similar quantities can 
be defined also for gluons~\cite{Mulders:2000sh,Anselmino:2005sh}. Again, there 
are 8 leading-twist TMDs, despite the fact that no transverse 
polarisation can exist for massless particles; its role is somewhat replaced 
by linear polarisation. We simply list here the gluon TMDs which have 
received more attention lately, pointing out several useful references, 
without any further discussion.
\begin{itemize} 
\item
The TMD-PDF for unpolarised gluons inside an unpolarised proton is usually 
written as
\be f_1^g(x, \kt) \equiv f_{g/p}(x, k_\perp) \equiv g(x, k_\perp) \>.
\label{TMD-unp-g} 
\ee     
\item
Analogously to the quark Sivers distribution one has the gluon Sivers distribution,
\be
\Delta^N f_{g/p^\uparrow}(x, k_\perp) \equiv -\frac{2\kt}{M}\,
f_{1T}^{\perp g} (x, k_\perp) \>.
\label{TMD-Siv-g}
\ee
A review paper on the status and future prospects of the gluon 
Sivers function can be found in Ref.~\cite{Boer:2015vso}. 
\item
The quantity denoted by $h_1^{\perp g}(x, \kt)$ is somewhat the analogous of the 
Boer-Mulders distribution $h_1^{\perp q}$: it is related to the distribution of 
linearly polarised gluons inside an unpolarised proton and it can 
lead to several azimuthal asymmetries in heavy quark pair production in 
unpolarised $e\,p$ and $p\,p$ collisions~\cite{Boer:2010zf,Pisano:2013cya,
Mukherjee:2016qxa,Kishore:2018ugo}, and 
to typical transverse momentum distributions of Higgs bosons at 
LHC~\cite{Boer:2011kf,Boer:2013fca,Boer:2014lka}.
\item
The gluon TMD helicity distribution is similar to the quark TMD helicity 
distribution:
\be
\Delta f_{s_z/S_L}^g(x, \kt) \equiv \Delta g(x, \kt) \>.
\ee
Its integrated collinear version $\Delta g(x)$, plays an important role in 
several processes with longitudinal polarisation, which are not considered here.    
\end{itemize}

\subsubsection {TMD-FFs} \label{sec:tmdFF}
Analogously to the distributions of quarks and gluons in a nucleon, also the 
fragmentation process of a parton into hadrons is not, in general, a collinear 
event; in many cases the transverse degrees of freedom can be safely integrated,
leading to a 1-dimensional fragmentation function, usually denoted as 
$D_1^q(z,Q^2)$ or $D_{h/q}(z,Q^2)$, which only depends, apart from the scale 
of the process, on the light-cone momentum fraction of the fragmenting quark 
taken away by the hadron. 

\begin{figure}[t]
\begin{center}
\vskip -72pt
\includegraphics[width=15.truecm,angle=0]{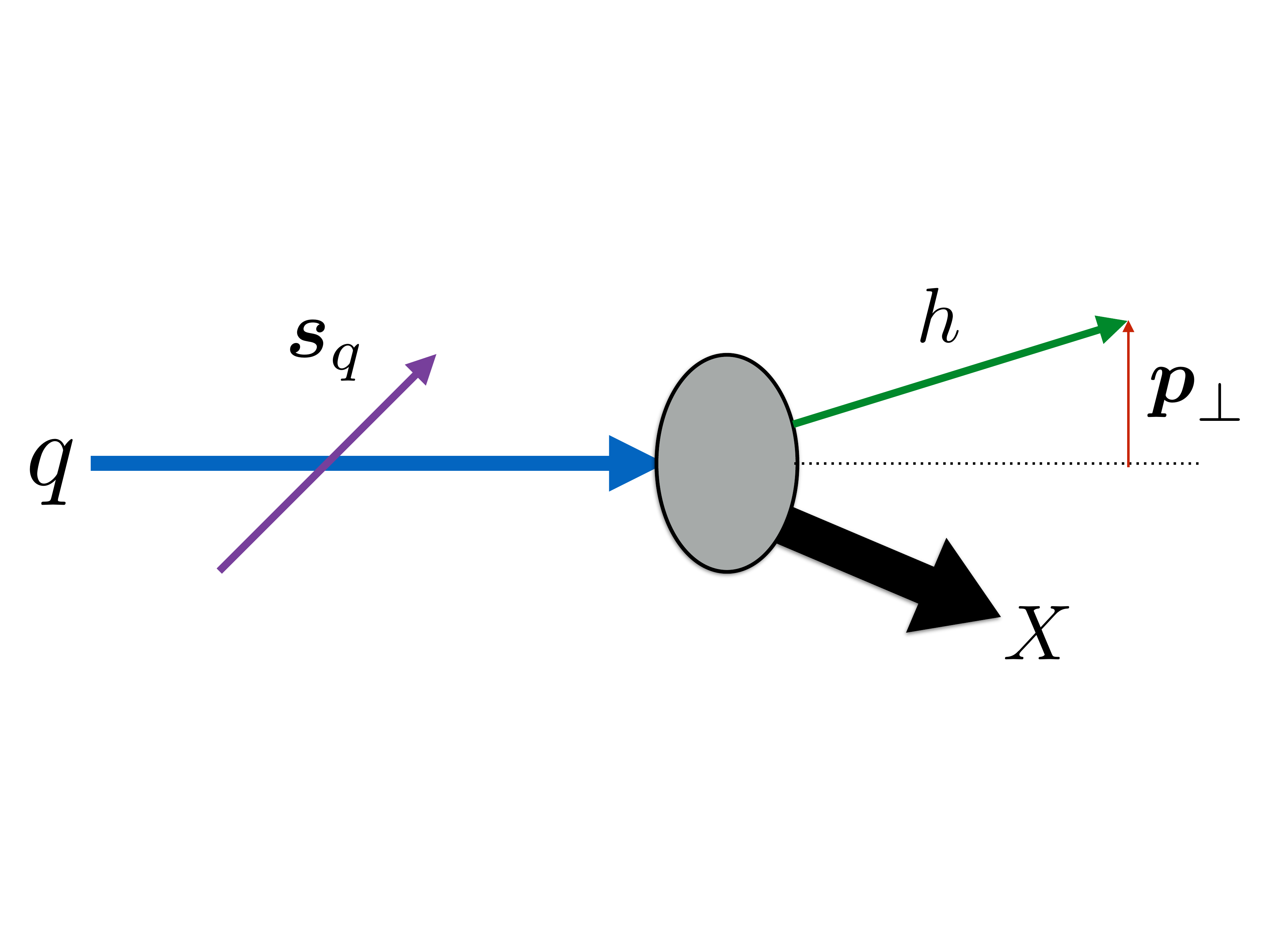}
\vskip -96pt 
\caption{Usual graphical representation of the TMD-FF for a quark with 
spin vector $\bfs_q$ which fragments into a hadron $h$ 
with transverse momentum $\bpt$ inside the jet, $\bfp_h = z\,\bfp_q + \bpt$.}
\label{tmd-ff}
\end{center}
\end{figure}

Introducing the transverse momentum $\bpt$ of the final hadron within the jet 
created in the quark hadronisation process, allows new transverse degrees of 
freedom. We consider the simple case of a final spinless or unpolarised hadron 
and refer to Fig.~\ref{tmd-ff} as a guide for the TMD-FFs, like we did 
in Eq.~(\ref{TMD-noi}). In this case we have less (pseudo)-vectors at our 
disposal and the most general, parity invariant expression for the TMD-FF can
be written as:
\bea
D_{h/q, \bfs_q}(z, \bpt; \bfs_q) &=& D_{h/q}(z,\pt) + \frac 12 \, 
\Delta^ND_{h/q^\uparrow} (z, \pt) \, \bfs_q \cdot 
(\hat{\bfp}_q \times \hat{\bfp}_\perp) \label{Coll-TO} \\
&=& D_{h/q}(z,\pt) + \frac {\pt}{z M_h}\,H_1^{\perp q} \, \bfs_q \cdot 
(\hat{\bfp}_q \times \hat{\bfp}_\perp) \label{Coll-Ams} \>.
\eea       
Eqs.~(\ref{Coll-TO}) and~(\ref{Coll-Ams}) show, in two different 
notations~\cite{Anselmino:2011ch,Mulders:1995dh}, the Collins 
distribution~\cite{Collins:1992kk}, which correlates the transverse spin of the 
fragmenting quark with the final hadron transverse momentum in the jet. 
The quantities $\Delta^ND_{h/q^\uparrow}$ or, often, $H_1^{\perp q}$ are 
referred to as the Collins function~(\ref{Coll-TO})~\cite{Anselmino:2011ch}.
A recent and most comprehensive review paper on parton FFs can be found in 
Ref.~\cite{Metz:2016swz}.    

In the case of the fragmentation of a quark into a spin 1/2 
hadron, one has, similarly to the 8 TMD-PDFs, 8 independent TMD-FFs (see 
Ref.~\cite{Metz:2016swz} and references therein). We only mention here the 
so-called ``polarising fragmentation function", describing the fragmentation 
of an unpolarised quark into a polarised spin 1/2 hadron, with spin $\bfS_h$.  
Analogously to the Collins distribution and two different 
notations~\cite{Anselmino:2000vs,Mulders:1995dh}, it is defined as  
\bea
D_{h,\bfS_h/q}(z, \bpt; \bfS_h) &=& \frac 12 \, D_{h/q}(z,\pt) + 
\frac 12 \, \Delta^ND_{h^\uparrow/q} (z, \pt) \, 
\bfS_h \cdot (\hat{\bfp}_q \times \hat{\bfp}_\perp) \label{pol-TO} \\
&=& \frac 12 \, D_{h/q}(z,\pt) + \frac {\pt}{z M_h}\,D_{1T}^{\perp h/q} \, 
\bfS_h \cdot (\hat{\bfp}_q \times \hat{\bfp}_\perp) \label{pol-Ams} \>.
\eea       

The polarising fragmentation function $\Delta^ND_{h^\uparrow/q}$, or
$D_{1T}^{\perp h/q}$, might be responsible for the polarisation of the $\Lambda$s 
observed in the interactions of unpolarised nucleons~\cite{Anselmino:2000vs}.
As briefly discussed in Section~\ref{subsec3.4}, it could be accessed in 
$e^+e^- \to \Lambda^\uparrow \, X$ processes.  
       
\subsection {\it How to interpret spin data in SIDIS \label{subsec3.2}}
In the previous Sections we have presented clear experimental evidence showing 
the necessity of considering, for the QCD partonic structure of nucleons 
in high energy inclusive processes, the full 3-dimensional motion of 
quarks and gluons. We do not address here the issue of the partonic distribution 
in coordinate space, which can be explored in exclusive 
processes~\cite{Diehl:2003ny}. We have 
introduced the TMD formalism appropriate to investigate and codify 
the 3D nucleon internal momentum structure. It is worth mentioning that the 
first parton model ideas date back exactly to 50 years ago~\cite{Feynman:1969ej}, 
and, since then, despite many QCD successes and tests, not much progress has been 
achieved in understanding the proton and neutron inner composition.

SIDIS processes are the main source of information on the nucleon structure
in momentum space. The SIDIS cross section has, at leading one photon exchange 
order, a well defined and rich expression, Eq.~(\ref{dsigma}). Thanks to the 
TMD factorisation scheme, in the energy region in which $P_T \simeq \kt \ll 
\sqrt {Q^2}$~\cite{Boglione:2016bph}, the form factors in Eq.~(\ref{dsigma}) 
can be written as convolutions of TMD-PDFs and TMD-FFs. Then, by comparing 
Eq.~(\ref{dsigma}), in which the TMDs have been inserted, with experimental 
data, one learns about the TMDs. Actually, it is very useful to consider the 
azimuthal moments of the spin asymmetries given in Eq.~(\ref{eq:azi-mom}), 
which isolate a single form factor at a time, like in Eqs.~(\ref{eq:AUT-siv})
and~(\ref{eq:AUT-coll}). The full expressions of the form factors $F_{UU}$ 
and $F_{UT}$ as TMD convolutions can be found in 
Refs.~\cite{Bacchetta:2006tn,Anselmino:2011ch}. Notice, that three of the form 
factors of Eq.~(\ref{dsigma}), that is $F_{UU}^{\cos\phi_h}$, 
$F_{UT}^{\sin\phi_S}$ and $F_{UU}^{\sin(2\phi_h - \phi_S)}$, are of 
${\cal O}(\kt/Q)$. Also, as we are only considering nucleons transversely  
polarised ($S_L = 0, \> S_T = 1$) and unpolarised leptons, the TMDs $g_{1L}$, 
$g_{1T}^{\perp}$ and $h_{1L}^{\perp}$ do not contribute to Eq.~(\ref{dsigma});
they contribute to SIDIS cross sections with longitudinally polarised leptons 
and/or nucleons.     

We concentrate here on the TMD interpretation of the form factors in 
Eq.~(\ref{dsigma}) which have been clearly found not to be negligible and have
large experimental support: they correspond to the unpolarised quark TMD, 
the Sivers function, the Collins function and the transversity distribution. 
Limited experimental information is also available for the Boer-Mulders function. 
Actually, some experimental data exist for all the other structure functions 
and TMDs~\cite{Kotzinian:2007uv,Parsamyan:2013ug}, but most of them are very 
small and compatible with zero. 

At ${\cal O}(\kt/Q$), the unpolarised cross section from 
Eq.~(\ref{dsigma}) is given by~\cite{Bacchetta:2006tn,Anselmino:2011ch}:
\bea
\frac{d\sigma^{\ell + p \to \ell^\prime h X}} {d\xb \, dQ^2 \, dz_h \, dP_T^2} 
&=& \frac{2 \pi^2 \alpha^2}{Q^4} \, \left[1 + (1-y)^2 \right] \, F_{UU} 
\nonumber \\
&=& \frac{2 \pi^2 \alpha^2}{Q^4} \, \left[ 1 + (1-y)^2 \right]
\nonumber \\
&\times& \sum_{q} e_q^2 \,
\int d^2\bkt \, d^2\bfp_\perp
\> \delta ^{(2)}\Big(\bfP_T - z_h\bfk_\perp -\bfp_\perp\Big)\,
f_{q/p} (x, k_{\perp}) \, D_{h/q}(z, p_{\perp}) \label{eq:SIDIS_FUU}
\eea
where $\xb = x, z_h =z$ and $\bpt = \bfP_T - z\bkt$.

Notice that this expression can be derived from the TMD parton model 
expression~\cite{Anselmino:2005nn}:
\be
\frac{d\sigma^{\ell + p \to \ell^\prime h X}} {d\xb \, dQ^2 \, dz_h \, d^2\bfP_T}
= \sum_q \int d^2 \bkt\;
f_{q/p}(x, \kt) \; \frac{d \hat\sigma ^{\ell q\to \ell q}}{dQ^2} \,
D_{h/q}(z, \pt) \>, \label{simple-unp}
\ee
which shows the explicit convolution of the TMD-PDFs, the TMD-FFs and the 
cross section for the elementary process:
\be
\frac{d \hat\sigma^{\ell q\to \ell q}}{d Q^2} = e_q^2 \,
\frac{2\pi \alpha^2}{Q^4}\,\left[ 1 + (1-y)^2 \right]  + {\cal O}(\kt/Q) \>.
\label{part-Xsec}
\ee
The terms of ${\cal O}(\kt/Q$) give a dependence on $\phi_h$ to 
Eq.~(\ref{simple-unp})~\cite{Anselmino:2005nn,Anselmino:2011ch}, which 
vanishes in Eq.~(\ref{eq:SIDIS_FUU}), which is integrated over $\phi_h$.  
Eq.~(\ref{eq:SIDIS_FUU}) will be used to relate data on hadron multiplicities in
unpolarised SIDIS with the unpolarised TMD-PDFs, $f_{q/p}(x, \kt)$, and 
TMD-FFs, $D_{h/q}(z, \pt)$. 

The Sivers effect, that is the correlation between the proton spin and the parton 
transverse momentum, is hidden in the $F_{UT}^{\sin(\phi_h-\phi_S)}$ term of 
Eq.~(\ref{dsigma}). One could use the $\sin(\phi_h-\phi_S)$ azimuthal moment
of Eq.~(\ref{eq:AUT-siv}), together with the expression of 
$F_{UT}^{\sin(\phi_h-\phi_S)}$ in ~\cite{Bacchetta:2006tn,Anselmino:2011ch} to 
relate the Sivers function with data. Or, one can simply use the analogous 
of Eq.~(\ref{simple-unp}) in case of a transversely polarised 
proton~\cite{Anselmino:2005nn}:  
\be
\frac{d\sigma^{\ell + \pup \to \ell \, h X}} {d\xb \, dQ^2 \, dz_h \, 
d^2\bfP_T \, d\phi_S}= \frac {1}{2\pi} \sum_q \int d^2 \bkt \,
f_{q/\pup}(x, \bkt) \; \frac{d \hat\sigma ^{\ell q\to \ell q}}{dQ^2} \,
D_{h/q}(z, \pt) \>, \label{simple-pol}
\ee
inserting the Sivers distribution $f_{q/\pup}(x, \bkt)$ given in 
Eq.~(\ref{PDF-Sivers}). Then, from Eq.~(\ref{eq:azi-mom}), one has:
\be
A^{\sin (\phi_h-\phi_S)}_{UT} = \label{aut-siv}
\frac{\displaystyle  \sum_q \int
{d\phi_S \, d\phi_h \, d^2 \bfk _\perp}\,
\Delta^N \! f_{q/\pup} (x,\kt) \, (\hat{\bfP}\times \hat{\bfk}_\perp) \cdot \bfS \;
\frac{d \hat\sigma ^{\ell q\to \ell q}}{dQ^2} \,
D_q^h(z,p_\perp) \sin (\phi_h -\phi_S) } 
{\displaystyle \sum_q \int {d\phi_S \,d\phi_h \, d^2 \bfk _\perp}\;
f_{q/p}(x,k _\perp) \; \frac{d \hat\sigma ^{\ell q\to \ell q}}{dQ^2} \;
 \; D_q^h(z,p_\perp) } \> \cdot 
\ee
The momentum dependence $(\hat{\bfP} \times \hat{\bfk}_\perp) \cdot \bfS$ of 
the Sivers distribution~(\ref{PDF-Sivers}) gives a factor $\sin(\varphi -\phi_S)$, 
where $\varphi$ is the azimuthal angle of $\bkt$. It is this factor which 
generates, when integrating ($d\sigma^\uparrow - d\sigma^\downarrow$) over 
$d^2\bkt$, the typical $\sin(\phi_h -\phi_S)$ dependence of the Sivers effect.   
Eq.~(\ref{aut-siv}), which could be further simplified, relates data on 
$A^{\sin (\phi_h-\phi_S)}_{UT}$ to a convolution of the Sivers function 
$\Delta^N \! f_{q/\pup} (x,\kt)$ with the unpolarised TMD-PDF. It will be 
used for extracting information on $\Delta^N \! f_{q/\pup}(x, \kt)$. 

In a similar way, one can extract from Eq.~(\ref{dsigma}) and the azimuthal 
moment~(\ref{eq:AUT-coll}) information on the structure function 
$F_{UT}^{\sin(\phi_h + \phi_S)}$ which depends on the Collins 
function~\cite{Bacchetta:2006tn,Anselmino:2011ch}. Equivalently, one can 
follow the approach of Refs.~\cite{Anselmino:2011ch,Anselmino:2007fs}
which start by writing the SIDIS cross section in terms of helicity amplitudes,  
keeping into account all phase factors which appear in the different
stages of the process. From Eqs.~(\ref{eq:azi-mom}) one
obtains~\cite{Anselmino:2011ch,Anselmino:2007fs}:        
\be
A^{\sin (\phi_h+\phi_S)}_{UT} = \label{sin-asym}
\frac{\displaystyle  \sum_q e_q^2 \int {d\phi_h \, d\phi_S \, d^2
\bfk _\perp}\,\Delta _T q (x,\kt) \,
\frac{d (\Delta {\hat \sigma})}{dQ^2}\,
\Delta^N D_{h/q^\ua}(z,\pt) \sin(\phi_S + \varphi +\phi_q^h)
\sin(\phi_h + \phi_S) } {\displaystyle \sum_q e_q^2 \, \int {d\phi_h
\,d\phi_S \, d^2 \bfk _\perp}\; f_{q/p}(x,k _\perp) \;
\frac{d\hat\sigma}{dQ^2}\; \; D_{h/q}(z,p_\perp) } \>, 
\ee
where 
\be
\frac{d (\Delta {\hat \sigma})}{dQ^2} \equiv
\frac{d \hat\sigma^{\ell q^\ua \to \ell q^\ua}}{d Q^2} -
\frac{d \hat\sigma^{\ell q^\ua \to \ell q^\da}}{d Q^2} = 
\frac{4\pi \alpha^2}{Q^4}\,(1-y)
\label{part-Xsec-tr}
\ee
is the transverse spin transfer in the elementary interaction. 

Eq.~(\ref{sin-asym}) describes the asymmetry in a process in which a transversely
polarised quark inside a transversely polarised protons scatters off an 
unpolarised lepton and then, with the remaining transverse polarisation, fragments 
into the final observed hadron $h$. In the TMD factorisation scheme it is a 
convolution of the TMD transversity with the spin transfer in the elementary 
interaction and the Collins TMD-FF. The azimuthal dependence 
$\sin(\phi_S + \varphi +\phi_q^h)$ is a subtle effect which arises from the 
combination of phase factors in the transversity distribution function, the 
elementary interaction and the Collins distribution~(\ref{Coll-TO}); $\phi_q^h$ 
can be written in terms of $\phi_h$ and $\varphi$~\cite{Anselmino:2011ch}.

Eq.~(\ref{sin-asym}) relates the measured asymmetry $A^{\sin (\phi_h+\phi_S)}_{UT}$
to a convolution of the transversity distribution and the Collins fragmentation; 
it has allowed the first ever extraction of the transversity distribution
$\Delta_Tq(x) = h_1^q(x)$~\cite{Anselmino:2007fs}, which is not accessible in DIS, 
due to its chiral odd nature. Independent information on the Collins function 
can be obtained in $e^+e^- \to h_1 \, h_2 \, X$ processes. 

Eqs.~(\ref{eq:SIDIS_FUU}), (\ref{aut-siv}) and~(\ref{sin-asym}) allow to get 
information on the unpolarised, the Sivers and the transversity TMDs. 
They are coupled to unpolarised TMD-FFs or to the Collins TMD-FF. Independent 
information on these quantities can be obtained from $e^+e^-$ annihilation
processes. 

Similar relations to SIDIS observables can be found for the other 
TMDs~\cite{Bacchetta:2006tn,Anselmino:2011ch}; in particular the Boer-Mulders 
function, which contributes to $F_{UU}^{\cos 2\phi_h}$, has been 
studied~\cite{Barone:2009hw,Christova:2019fbj} and some data are 
available~\cite{Airapetian:2012yg,Adolph:2014pwc,Kerbizi:2018bvk,
Moretti:2019lkw,Matousek:2019dlk}. We do not discuss here these 
TMDs; in Sections~\ref{subsec3.5}, \ref{subsec4.1}, \ref{subsec4.2}  
and~\ref{subsec4.3} we look in more details at our actual knowledge of the 
Collins, the unpolarised, the Sivers and the transversity TMDs, emphasising 
what we learn from them about the nucleon structure.    
 
\subsection{\it How to interpret spin data in hard $N\,N$ collisions
\label{subsec3.3}}

As anticipated in Sec~\ref{subsec2.2}, hadronic collisions are more 
challenging to interpret in the TMD framework than the SIDIS process.
Single scale processes, like the TSSA $A_N$ or the inclusive production of 
polarised $\Lambda$'s in unpolarised $p\,p$ scattering, can be 
described in a collinear framework. The mechanisms generating TSSA's in a 
collinear framework only enter at sub-leading twist, {\it i.e.} twist-3.
To get further insights into this statements, it is valuable to consider the 
necessary ingredients for the existence of transverse spin effects.

In particular TSSAs require one helicity-flip amplitude and a 
phase shift between two amplitudes in order not to vanish. This is because
TSSAs are related to matrix elements that are off-diagonal in the helicity 
basis due to decompositions of the kind:
\begin{equation}
\ket{\uparrow}/\ket{\downarrow} = \left(\ket{+}\pm\ket{-} \right)
/\sqrt{2} \Rightarrow \ket{\uparrow}\bra{\uparrow} - 
\ket{\downarrow}\bra{\downarrow}=\ket{+}\bra{-}+\ket{-}\bra{+}.
\end{equation} 
Therefore, non-zero TSSAs require a non-vanishing helicity-flip amplitude. 
It was realised early on that such helicity-flip amplitudes are 
suppressed, in QCD or QED high energy scatterings, by factors $m_q/\sqrt{Q^2}$. 
This leads to the well-known suggestion of the measurement of TSSAs as a test 
of QCD by Kane, Pumplin and Repko~\cite{Kane:1978nd}.
Furthermore, TSSAs are so-called naive T-odd observables, since they are T-odd 
if one does not consider the full final state, which is the case in semi-inclusive 
measurements. Only the imaginary part of the matrix element can contribute to 
T-odd effects. Since this needs the interference of amplitudes with a relative 
phase shift, another crucial ingredient to TSSAs are initial or final state 
interactions introducing such a shift. These interactions introduce a factor 
of $\alpha_s$, which is another reason to look for interactions in the 
non-perturbative regime for the sources of these asymmetries. 
It also shows that the TSSAs probe QCD at the amplitude level. At twist-3, 
the interference term can be generated by handbag diagrams, where an additional 
parton is exchanged on one side. For the initial state, an example 
of this is shown in Fig.~\ref{fig:correlatorQGG} which is discussed in the 
next Section along with the respective twist-3 correlator. In the case of the TMD 
picture, model calculations show that at a microscopic level, initial and final 
state interactions, in Drell-Yan and SIDIS processes respectively, 
indeed provide the necessary phase shift~\cite{Brodsky:2002cx,Brodsky:2002rv,
Boer:2003cm,Brodsky:2013oya}.

Intuitively, one can also expect the twist-3 framework and the TMD framework 
to be related since an integration over the intrinsic transverse momenta in 
the TMD framework should contribute to the asymmetries described in the 
twist-3 framework. And indeed, those exist in general via the Wandzura-Wilczek 
relations~\cite{Wandzura:1977qf} as well as more specific relations in the 
case of the Sivers function and the Collins asymmetries, as will be discussed 
later in this section. The di-hadron TSSAs described here, are an exception, 
in that they can be described in a leading twist collinear framework. 
They are not sensitive to TMDs but to the collinear transversity $h_1(x)$ 
coupling to a collinear FF. In this case, the additional degrees of freedom 
in the di-hadron final state plays a similar role as the polarisation degrees 
of freedom in semi-inclusive polarised hadron production, 
where the polarisation is measured with respect to the production plane and the 
necessary interference term comes from the amplitudes in which hadron pairs are 
produced with different relative angular momentum.

In $N\,N$ processes with a large and a small scale, the TMD factorisation 
picture can be appropriate. However, an important difference to the SIDIS 
process is that hard hadronic collisions have two color charges in the initial 
state. If the final state also contains a color charge, {\it i.e.} in the 
production of hadrons, TMD factorisation becomes more problematic, leading 
to modified universality~\cite{Rogers:2010dm,Mulders:2011zt} due to a phenomenon 
sometimes called ``color entanglement". However, for the ``hadron-in-jet" 
measurements discussed here, recent theoretical progress indicates that the 
TMD framework is applicable~\cite{Kang:2017glf}.
On the other side of the spectrum, Drell-Yan and $W/Z$ production can be 
seen as the crossed-channel analogue of the SIDIS process and can thus 
naturally be interpreted in the TMD framework. The difference in color flow 
is still present; however, the universality modification can be calculated 
and reduces to a sign flip of the observable.

Another complication in the interpretation of hadronic scattering observables 
should be mentioned. The initial partonic kinematics are not directly accessible, 
which means that observables are inherently convolutions over the initial 
kinematics. In the case of final state fragmentation, three non-perturbative 
functions are entering the cross-section. Since gluons enter the cross-section 
at leading order in hadronic collisions, the cross-sections can also contain 
convolutions over gluon PDFs and FFs which are less well known than their quark 
counterparts, injecting additional uncertainties. Keeping these points in mind, 
below the interpretation of the observables introduced in~Sec~\ref{subsec2.2} 
will be discussed. 

\subsubsection{Asymmetries in $p^\uparrow N\rightarrow h \, X$ processes}
\label{sec:pheno_AN}
The first attempts of explaining the large observed TSSAs in 
$p^\uparrow N \to h \, X$ were based on twist-3 matrix elements of the quark 
and gluon fields of the initial state by Efremov and 
Teryaev~\cite{Efremov:1984ip} and Qiu and Sterman~\cite{Qiu:1991pp,Qiu:1998ia}. 
\begin{figure}
    \centering
    \vskip -72pt
\includegraphics[width=15.truecm,angle=0]{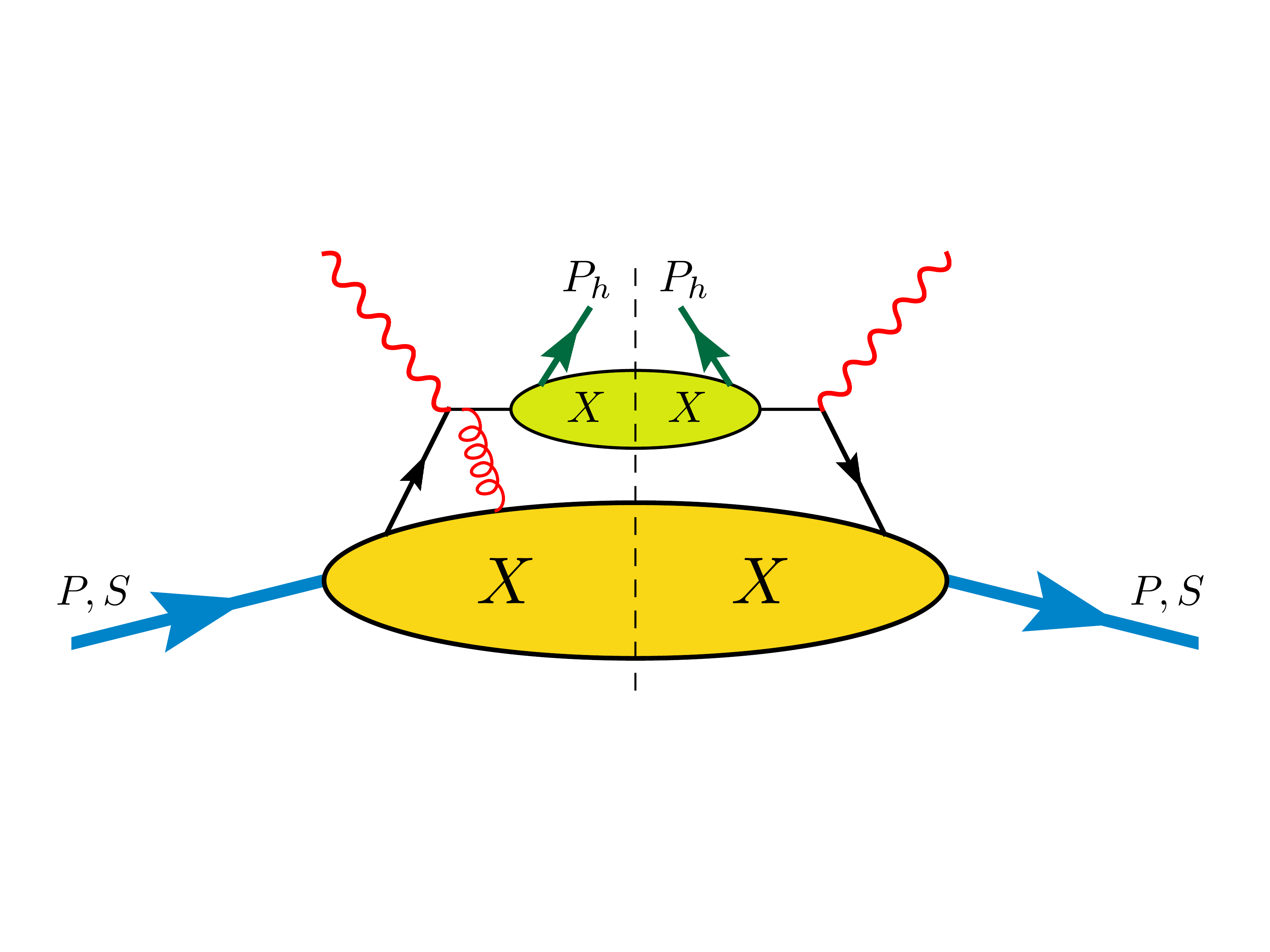}
\vskip -72pt 
\caption{
Quark-gluon-quark correlator contributing to inclusive processes at twist-3. 
The upper blob symbolizes the fragmentation process.}
    \label{fig:correlatorQGG}
\end{figure}
These twist-3 matrix elements encode three-parton correlations, {\it e.g.} 
quark-gluon-quark correlations represented by the handbag diagram shown in 
Fig.~\ref{fig:correlatorQGG}. In these handbag diagrams, an additional leg 
connects the outgoing quark to the blob representing the unobserved hadronic 
final state, $X$. Relevant here is the case when this leg is a gluon. 
The half of the handbag diagram with the extra leg therefore represents an 
amplitude with a final state interaction of the outgoing quark with $X$, 
which is related to model calculations of TMDs~\cite{Brodsky:2002cx}.  
Such an object would be expected to depend on the momentum fractions 
carried both by the gluon and by the quark; however, it turns out that the 
TSSAs receive their dominant contribution from the region in which the gluon 
is very soft. Therefore the gluon does not transmit momentum, but merely 
leads to the needed phase shift. These matrix elements are therefore known 
as the soft-gluon pole (SGP), or Efremov-Teryaev-Qiu-Sterman (ETQS) 
matrix elements $T_F(x,x)$. In general, twist-3 correlations are dependent 
on two partonic momentum fractions, but due to the softness of the exchanged 
gluon momentum, they can be taken as equal.
The phase shift imparted by the gluon bears resemblance to the model 
calculations for the Sivers effect of Ref.~\cite{Brodsky:2002cx} and therefore 
it might not be surprising that $T_F$ can be related to the Sivers 
function~\cite{Boer:2003cm,Ji:2006vf,Ji:2006ub,Ji:2006br} in the kinematic 
region where both formalism are valid, {\it i.e.} at intermediate $p_T$, 
$\Lambda_{\textrm{QCD}}< p_T < Q$, according to the relation:
 %~\cite{Kang:2011hk}
%
\begin{equation}
\label{eq:TfToSivers}
g\,T_F(x,x) = -\int d^2 \bkt \, \frac{k_\perp}{M} \, 
f_{1T}^\perp(x,k_\perp) \>.
\end{equation}

Once the Sivers function was extracted with sufficient precision from SIDIS 
data, it also became clear that the ETQS matrix elements were not the dominant 
mechanism behind the large TSSAs of hadrons~\cite{Kanazawa:2014dca}. 
There was even a ``sign mismatch" between the values of $T_F(x,x)$ 
requested by the $A_N$ data and the values obtained by inserting the Sivers 
function into Eq.~(\ref{eq:TfToSivers})~\cite{Kang:2011hk}. It should be 
mentioned, that SGPs can 
also come from $ggg$ correlations~\cite{Beppu:2013uda} and that another pole 
contribution in the initial state, the soft-fermionic pole (SFP), was considered 
as well~\cite{Koike:2009ge}, but both, the $ggg$ and the SFP 
contributions, also proved to be insufficient to explain the magnitude of 
$A_N$~\cite{Beppu:2013uda,Kanazawa:2010au,Kanazawa:2011bg}.
Other sizeable contributions might be expected from TMD effects 
in the fragmentation process.  
However, when fragmentation contributions do not play a role, such as 
in the production of prompt photons discussed in Sec.~\ref{sec:phenoDY}, 
the ETQS mechanism should be dominating and can shed light on the Sivers 
function in hadronic interactions.

An early attempt to include both initial state contributions as well as 
fragmentation contributions was performed in the framework of the so-called 
Generalized Parton Model (GPM)~\cite{Anselmino:2005sh,DAlesio:2004eso} which 
is essentially an extension of the SIDIS TMD framework to various observables in 
$p\,p$. A review on the interpretation of hadronic interactions in 
the GPM can be found in Refs.~\cite{DAlesio:2004eso,Aschenauer:2015ndk} and a 
study of the contribution of Collins and Sivers TMDs to $A_N$ in such a scheme 
can be found in Refs.~\cite{Anselmino:2012rq,Anselmino:2013rya}.

A modified version of the GPM, which takes into account colour 
factors due to initial and final state interaction, is the so-called colour 
gauge invariant (CGI) GPM~\cite{Bacchetta:2005rm,DAlesio:2011kkm}, which leads 
to non universality of the TMDs. This modified universality, {\it e.g.} of the 
Sivers function, was implemented via the respective color factors in 
Ref.~\cite{DAlesio:2011kkm}. However, by construction 
this model misses other twist-3 contributions to $A_N$ which do not have a 
leading twist TMD counterpart. Additionally, factorisation is only assumed.

The first calculation of the fragmentation contribution at twist-3 was performed 
in Ref~\cite{Metz:2012ct}. As other twist-3 functions, they can generally be 
decomposed into so-called intrinsic, kinematical and dynamical parts.
However, these parts are not independent from each other, so that the transverse 
spin dependent cross-section can be written in terms of $H_1^{\perp(1))}$ 
and $\tilde{H}$. Here $H_1^{\perp(1))}$  is the kinematical contribution, which 
can be written in terms of the $\bfk_T$ integrated Collins FF. The dynamical 
contribution $\tilde{H}$ describes quark-gluon-quark correlations. In terms of 
these functions, as well as the ETQS function $T_F$ describing the initial state 
spin-dependent dynamics at twist-3, the transverse polarisation dependent cross
section can be expressed as~\cite{Gamberg:2017gle}:
\bea
   && d\Delta\sigma^{pp\rightarrow \pi X}(\bfS_T)= 
  \frac{2|\bfP_{hT}|\,\alpha_s}{S} \\
  &&   \sum_{i,a,b,c}\int_{z_\textrm{min}}^1 \frac{dz}{z^3}
  \int_{x_\textrm{min}}\frac{dx}{x}\,\frac{1}{x'}\,\frac{1}{xS+U/z}\, f_1^b(x')
  \left[ M_h \, h_1^a(x) \, \mathcal{H}^{\pi/c,i}(x,x'z) + \frac{M}{\hat{u}}
  \, \mathcal{F}^{a,i}(x,x',z)\,D^{\pi,c}_1(z) \right] \nn.
\eea
Here $\Delta \sigma(\bfS_T)=\sigma(\bfS_T)-\sigma(-\bfS_T)$ and $\bfP_{hT}$ 
is chosen along the $x-$axis whereas $\bfS_T$ is chosen along the $y-$axis 
so that $\Delta\sigma$ is related to the observable 
$A_N=\Delta \sigma /(d\sigma(\bfS_T)+d\sigma(-\bfS_T))$.
The indices $a,b,c$ run over the partons participating in the hard scattering. 
Parton $a$ is taken from the polarised proton and parton $b$ from the 
unpolarised one, with parton $c$ fragmenting into the observed pion. The index 
$i$ runs over the different hard scattering processes. The 
elementary parton cross sections, which are dependent on the partonic 
Mandelstam variables $\hat{s} = x\,x'S$, $\hat{t}=x\,T/z$, $\hat{u}=x'\,U/z$ 
with $S=(P+P')^2$, $T=(P-P_h)^2$, $U=(P'-P_h)^2$, are encapsulated in the 
``hard factors" $F=\{H,H_1^\perp,T_F\}$, $S^i_F$ and $\tilde{S}_F^i$.
They are explicitly given in Ref.~\cite{Gamberg:2017gle}. Here $P,P'$ are the 
initial four-momenta of the protons involved in the scattering and $P_h$ the 
four-momentum of the outgoing hadron. The lower integration limits are 
$x_{\textrm{min}}=-(U/Z)(T/Z+S)$ and $z_{\textrm{min}}=-(T+U)/S$. The momentum 
fraction $x'$ of the unpolarized parton $b$ is related to a given momentum 
fraction $x$ of parton $a$ by $x'=-(xT/z)/(xS+U/z)$. The functions $\mathcal{F}$ 
and $\mathcal{H}$ describe the spin-dependent, non-perturbative dynamics of the 
initial and final state, respectively
\begin{equation}
    \mathcal{F}^{a}(x,x',z)=\pi \left[T_{F}^a(x,x) - 
    x\frac{dT_{F}^a(x,x)}{dx} \right] S_{T_F}^i
\end{equation}
and
\begin{equation}
    \mathcal{H}(x,x',z)=\left[H_1^{\perp(1)}(z) - 
    z\frac{dH_1^{\perp(1)}(z)}{dz} \right] \tilde{S}^i_{H_1^\perp}+
\left[-2H_1^{\perp (1)}(z)+\frac{1}{z} \tilde{H}(z) \right] \tilde{S}^i_H \>.
\end{equation}

\begin{figure}
    \centering
    \includegraphics[width=0.45\textwidth]{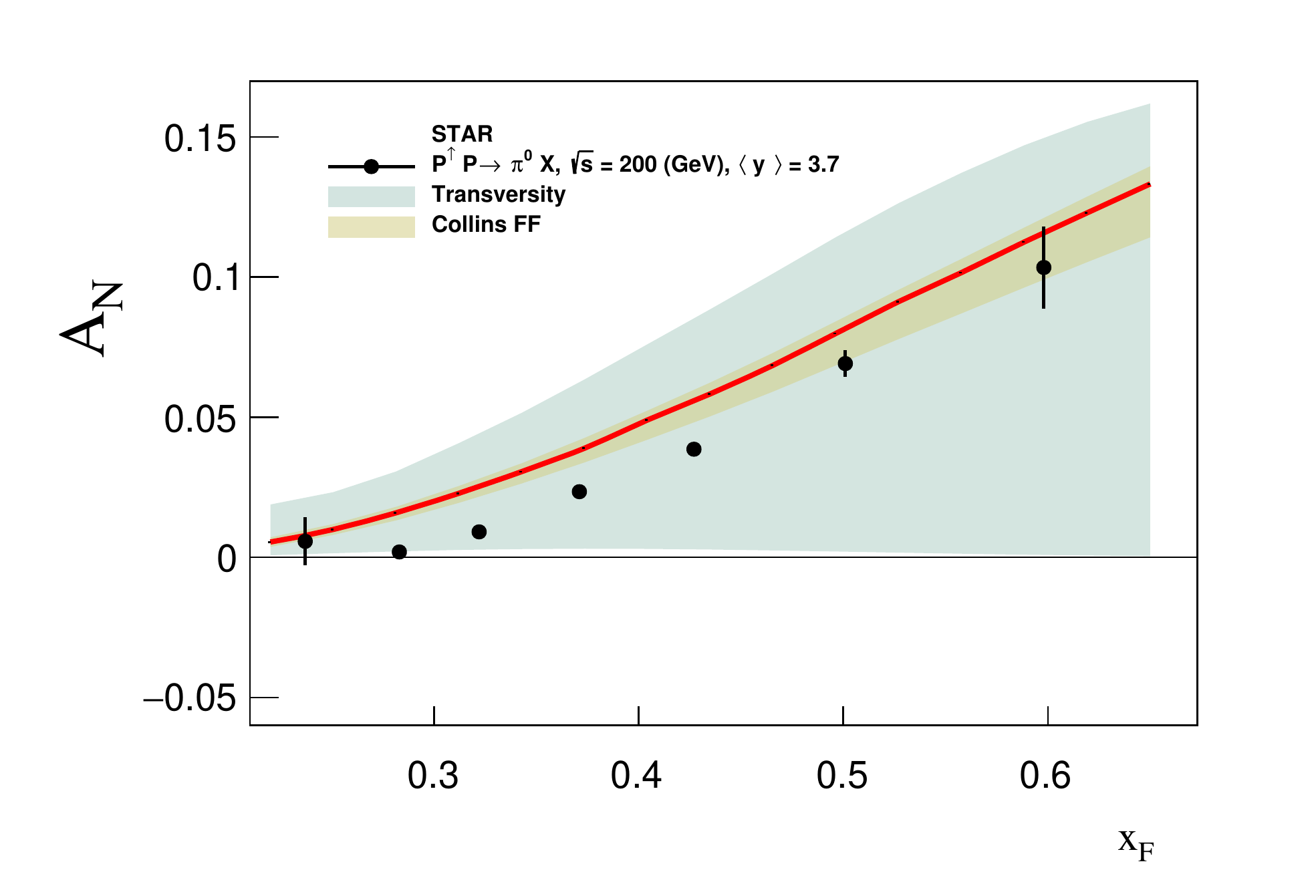}
    \includegraphics[width=0.45\textwidth]{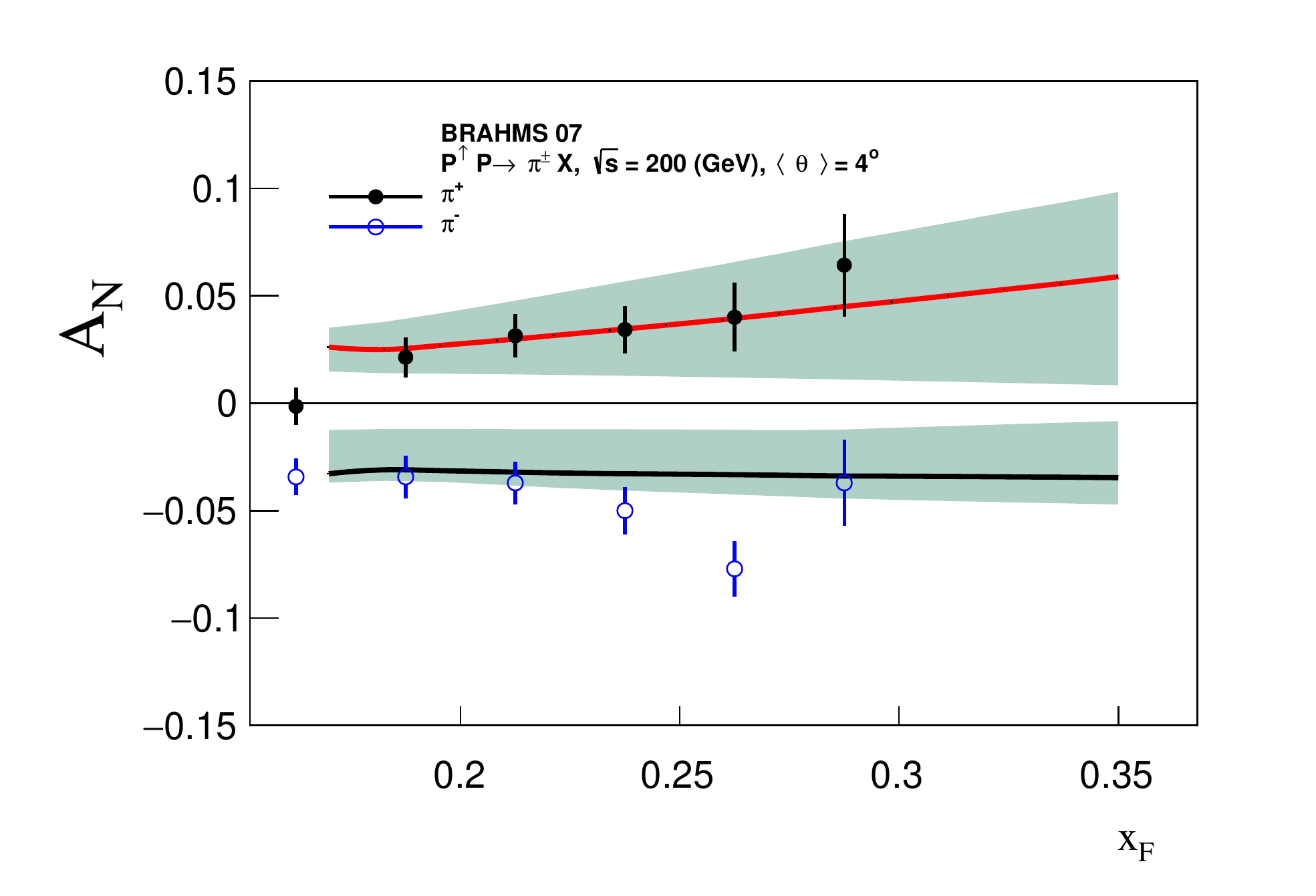}
    \caption{Calculations of $A_N$ for $\pi^0$ at $\sqrt{s}=200$~GeV 
    compared with STAR data (left) and for charged pions compared with Brahms 
    data (right)~\cite{Gamberg:2017gle}. The calculations show quite a good 
    agreement with the data. The main source of uncertainty comes from the poor 
    knowledge of $h_1$ and $H_1^\perp$ in the kinematic regime probed. 
    Figure from Ref.~\cite{Gamberg:2017gle}.
\label{fig:AN_fit}}
\end{figure}

The functions have been calculated using a Wandzura-Wilczek 
approximation~\cite{Gamberg:2017gle} and, with input for $H^{\perp(1)}$ 
and $h_1$  using a recent global fit~\cite{Kang:2015msa}, numerical estimates 
for $A_N$ have been obtained. Figure~\ref{fig:AN_fit} shows these estimates 
compared with some $A_N$ data from RHIC. Note that these curves are not fits, 
but estimates based on previous extractions of transversity and the Collins FF. 
As is shown in the figure, the leading uncertainty originates from the uncertainty 
on $h_1$ which is probed in a kinematic regime at high $x$ that is not well 
constrained yet from existing data. Conversely, this illustrates the opportunity 
to constrain transversity from existing $A_N$ data from RHIC. 

\subsubsection{Asymmetries in $\gamma^*$, $W/Z$, $\gamma$ production}
\label{sec:phenoDY}
Similar to the expansion of the structure functions in the SIDIS case, 
{\it e.g.} for $F_{UU}$ given in Eqs.~\eqref{eq:SIDIS_FUU}, the structure 
functions for Drell-Yan and $W$ production given in 
Eqs.~\eqref{dq:DY_StructFunctions},~\eqref{dq:DY_StructFunctions_upol} 
and~\eqref{eq:W_StructFunctions} can be expanded at $\mathcal{O}(\kt/Q^2)$ 
in terms of TMDs. The main difference to the SIDIS case is that the terms 
entering this expansion are now convolutions of the TMDs describing the 
parton distributions in the two interacting protons, while in the SIDIS case 
those are convolutions of PDFs and FFs.
This reflects the fact that D-Y and $W$-production are $s$-channel processes, 
while SIDIS is a $t-$channel process. The difference in color flow between 
them, coloured states in the initial vs. final state, leads to the predicted 
modified universality of TMD functions measured in D-Y or $W$-production 
compared to SIDIS, most notably the sign change of the Sivers 
function~\cite{Collins:2002kn,Brodsky:2002cx,Brodsky:2002rv}, Eq.~(\ref{ch-sign}). 

The notation used here will not distinguish between the TMDs measured 
in D-Y/W and SIDIS. Focusing first on the experimentally relevant transverse 
single spin asymmetry in the case where the angular distribution of the final 
state leptons is integrated out, the structure function 
$F_{UT}^{\sin(\phi_V-\phi_S)}$ can be written, in the aforementioned 
approximation, as:
\begin{equation}
\label{eq:F_UT_DY}
    F_{UT}^{\sin(\phi_V-\phi_S)}=\mathcal{C}\left[f_1 \, f_{1T}^\perp\right] (\textrm{DY})
\end{equation}
and
\begin{equation}
\label{eq:F_UT_W}
    F_{UT}^{\sin(\phi_V-\phi_S)}=\mathcal{C}^W \,2\left[f_1 \, f_{1T}^\perp\right] 
    (\textrm{W-production}).
\end{equation}
Here the Sivers function of the polarized hadron $A$ in the 
reaction is probed and the azimuthal angles are defined in 
Sec.~\ref{sec:transvSpinEffectsInDY_W}. The convolution over transverse momenta, 
again analogous to the SIDIS expression in~\eqref{eq:SIDIS_FUU}, is given by
\begin{multline}
\label{eq:dyConv}
    \mathcal{C}[w(\bfk_{a,\perp},\bfk_{b,\perp})\,f_1 \,\bar{f}_2]=\\
    \frac{1}{3}\sum_q e_q^2\int d^2 \bfk_{a,\perp}\,d^2\bfk_{b,\perp} 
    \, \delta^{(2)}(\bfq_\perp-\bfk_{a,\perp}-\bfk_{b,\perp})\,
    w(\bfk_{a,\perp},\bfk_{b,\perp})
    \left[f_1^q(x_a,\bfk_{a,\perp}) \, f_2^{\bar{q}}(x_b,\bfk_{b,\perp})+        
    q\leftrightarrow \bar{q}\right]\>,
\end{multline}
where $\bfq$ is the momentum of the virtual photon. The weight function 
$w$ is introduced for completeness and later use; in the case of the SSA 
that is sensitive to the Sivers function it is unity.
For $W$ production the convolution integral has to be modified to respect 
the coupling of the weak force. It reads:
\begin{multline}
    \mathcal{C}^W[w(\bfk_{a,\perp},\bfk_{b,\perp})\,f_1 \, \bar{f}_2]=\\
    \frac{1}{3}\,|V_{qq'}|\sum_{q,q'} \int d^2 \bfk_{a,\perp}d^2\,\bfk_{b,\perp}
    \, \delta^{(2)}(\bfq_\perp-\bfk_{a,\perp}-\bfk_{b,\perp})\,
    w(\bfk_{a,\perp},\bfk_{b,\perp}) \left[ f_1^q(x_a,\bfk_{a,\perp}) \, 
    f_2^{\bar{q'}}(x_b,\bfk_{b,\perp})+q\leftrightarrow \bar{q'}\right],
\end{multline}
where $V_{qq'}$ is the relevant entry in the CKM matrix. Also note that the 
sum now runs over all flavour combinations. $Z$ production is not considered 
here, but the corresponding expression can be found in Ref.~\cite{Huang:2015vpy}.
Eq.~\eqref{eq:F_UT_W} acquires a factor of two compared to~\eqref{eq:F_UT_DY} 
due to the different coupling constants of the weak force.

As discussed in Sec.~\ref{sec:transvSpinEffectsInDY_W}, further information 
about TMDs can be gained from the dependence of the unpolarised structure 
function on the decay angles of the leptons in the D-Y process. Here we use 
the angles in the Collins-Soper frame. In this frame, the $\cos2\phi_{CS}$ 
modulation of the cross-section is sensitive to the Boer-Mulders functions 
of the colliding hadrons as
\begin{equation}
    F_{UU}^{\cos2\phi_{CS}}=\mathcal{C}\left[\frac{(\bfh\cdot \bfk_{a,\perp})
    (\bfh\cdot\bfk_{b,\perp})-\bfk_{a,\perp}\cdot\bfk_{b,\perp}}{M_a\,M_b}
    \, h_1^\perp \, \bar{h}_1^\perp\right].
\end{equation}
The quantity $\bfh$ is a shorthand notation for the unit vector 
$\bfq_\perp/q_\perp$

In the D-Y limit $Q^2=0$, {\it i.e} for real photon production, the TMD picture 
cannot be used anymore; instead, the ETQS matrix elements discussed in 
Sec.~\ref{sec:pheno_AN} can be accessed in transverse single spin 
asymmetries~\cite{Kouvaris:2006zy,Kang:2011hk,Gamberg:2012iq,Kanazawa:2014nea}. 
The spin dependent part of the cross section for direct photon production, with 
$\bfP_\gamma$ and $E_\gamma$ denoting the photon momentum and energy, can be 
written as~\cite{Gamberg:2012iq}: 
\begin{eqnarray}
\label{eq:anPromptPhoton}
\frac{d\Delta\sigma^{pp\rightarrow \gamma X}}{d^3 \bfP_\gamma} &=& 
\epsilon_{\alpha \beta} \, \bfS_T^\alpha \, \bfP_{\gamma\perp}^\beta
\frac{\alpha_{\rm em}\,\alpha_s}{E_\gamma \, s}\sum_{a,b}
\int \frac{dx'}{x'}f_{b/B}(x')\int \frac{dx}{x}
\left[T^a_{F}(x, x) - x\frac{d}{dx}T^a_{F}(x, x)\right]
\nn \\
&\times&
\frac{1}{\hat{u}}\, S^{ab\to \gamma}(\hat s,\hat t,\hat u) \,
\delta\left(\hat s+\hat t+\hat u\right),
\end{eqnarray}
where $\Delta\sigma=[\sigma(\bfS_T)-\sigma(-\bfS_T)]/2$ 
and $S^{ab\to \gamma}$ is the hard factor of the process, which depends on 
the partonic Mandelstam variables and is explicitly given in 
Ref.~\cite{Gamberg:2012iq}. Fragmentation photons can also contribute to 
the asymmetries, {\it e.g.} via the Collins effect, but the contribution 
by direct photons sensitive to the Sivers effect dominates~\cite{Gamberg:2012iq}.
In this process, similar to D-Y production, only initial state interactions 
are responsible for the phase shift in the QCD amplitude, leading to the 
transverse single spin asymmetry. Due to the connection between the ETQS 
functions and the Sivers function shown in Eq.~\eqref{eq:TfToSivers}, the 
production of direct photons in hadronic collisions can therefore also be 
used to test the sign-change of the Sivers function from SIDIS.

\subsubsection{Asymmetries in di-hadron production}
\label{phenoDi-HadPP}
A detailed overview of Di-hadron FFs (DiFFs) can be found in 
Refs~\cite{Pisano:2015wnq} and~\cite{Metz:2016swz}. 
DiFFs can be introduced similarly to the TMD-FFs described in 
Sec.~\ref{sec:tmdFF}. If the production of two unpolarised hadrons $h_{1,2}$ 
is considered, where each of the hadrons carry the light-cone momentum fraction 
$z_i$ and the transverse momentum dependence is integrated over, the DiFFs 
depend on the quantities $z=z_1+z_2$, $\zeta=(z_1-z_2)/z$, $\bfR_T^2$ and 
$\phi_R$. For the $p\,p$ case discussed here, $\bfR$ and $\phi_R$ are defined 
in Sec.~\ref{sec:di-hadEffects}. The vector $\bfR_T$ is the transverse 
projection of $\bfR$. At large invariant masses $M_h$ of the two hadron system, 
the DiFFs can be calculated perturbatively and are connected to similar twist-3 
matrix elements as the Collins FF at large transverse momentum~\cite{Zhou:2011ba}. 
The relevant case to explore the partonic structure of the nucleon is the 
regime in which $M_h$ is small and the DiFFs are non-perturbative objects. 
Here, analogously to the TMD-FF regime where $\pt$ is small, polarisation 
quantum numbers of the parent quark are imprinted in the correlations of the 
two hadrons. In this regime $\zeta$ can be shown to be a linear polynomial 
in $\cos\theta$, with $\theta$ the polar angle defined in 
Sec.~\ref{sec:di-hadEffects}, and the DiFFs can be expanded in partial 
waves~\cite{Bacchetta:2002ux}. Different partial waves carry different 
angular momentum quantum numbers; therefore, di-hadron fragmentation can 
be treated analogously to single hadron fragmentation with the corresponding 
angular momentum~\cite{Gliske:2014wba}.

Similar to the expression for TMD-FFs in Eq.~(\ref{Coll-TO}), the di-hadron 
FF of a transversely polarised quark with polarisation vector $\bfs_q$ into 
unpolarised hadrons can then be written as
\begin{equation}
D_{h_1,h_2/q, \bfs_q}(z,\zeta,\bfR_T,\phi_R)= D_{h_1,h_2/q}+
H_1^{\sphericalangle q} \, \frac{\bfs_q \cdot ({\hat{\bfp}}_q \times 
\bfR_T)}{M_h} \>\cdot
\end{equation}
It is important to note, that, unlike the single-hadron case, the transverse 
polarisation dependence does not vanish upon integration over transverse 
momenta. The shift in the strong phase, which is needed to generate the 
TSSA (see Sec.~\ref{subsec3.3}) can be understood in terms of the partial 
wave expansion as being generated by the interference of different partial 
waves. There have been several suggestions as to the channels which interfere 
in $\pi^+\pi^-$ production~\cite{Jaffe:1997hf, Radici:2001na,Bacchetta:2006un}, 
leading to different predicted $M_h$ dependences. Data seems to prefer models 
similar to those of Refs.~\cite{Radici:2001na,Bacchetta:2006un}, where the 
dominant contribution at the energy of current experiments originates from 
the interference of pions produced in a relative $p-$wave from $\rho$ decay,
with the non-resonant production in a relative $s-$wave. It should be 
mentioned that a similar $M_h$ dependence can be generated from NJL 
models~\cite{Matevosyan:2017uls}.

Since for DiFFs the transverse polarisation dependence survives upon 
integration over intrinsic transverse momenta and thus a collinear framework 
can be recovered, they are a natural way to extract the transversity distribution 
in $p\,p$ collisions, where the partonic kinematics are not known a priori and 
can only be approximated from reconstructed jets. In $p\,p$ processes at 
leading order the transverse polarisation dependent di-hadron production for 
small $M_h$ can then be written as~\cite{Bacchetta:2004it}:
\begin{equation}
d\sigma_{UT}\propto \sum_{a,b,c,d} |\bfS| \, \sin(\phi_{RS})  
\int \frac{dx_a \, dx_b}{z_c} \, f_1^a(x_a) \, h_1^b \,
\frac{d\Delta \hat{\sigma}_{ab^\uparrow\rightarrow c^\uparrow d}}{d\hat{t}} 
\, \sin\theta \, H_1^\sphericalangle(z,\cos\theta,M_h)\>.
\end{equation}
Here, $a,b,c,d$ are the partons participating in the $2\rightarrow 2$ scattering 
and $\phi_{RS}$ is the angle between $\bfR_T$ and the polarisation of the 
proton, introduced in Sec.~\ref{sec:di-hadEffects}.
Since the di-hadron asymmetries in $p\,p$ can be interpreted in a collinear 
framework, where factorisation and evolution is presumably understood, they 
can be incorporated in a global analysis of transversity, which has been 
performed in Ref.~\cite{Radici:2018iag}, leading to the first global extraction 
of transversity from SIDIS, $p\,p$ and $e^+e^-$ annihilation data.

Because in $p\,p$ collisions the partons dominantly interact via 
flavour blind gluon exchange, $u-$quark dominance, important in the case of 
SIDIS data with a proton target, is much less of a concern. Therefore, 
$p\,p$ data can in principle give significant information on the $d-$quark 
transversity, which in current global extractions of transversity from SIDIS 
and $e^+e^-$ data is mainly constrained from data taken on deuterium targets, 
which is very limited. However, while the overall uncertainty on $h_1$ is 
reduced by the inclusion of $p\,p$ data, the essentially unknown unpolarised gluon 
DiFFs, which appears in the denominator of the spin asymmetries in $p\,p$, 
leads still to quite large systematic uncertainties on the extracted values. 

\subsubsection{Asymmetries of jets and hadrons in jets}
\label{sec:hadInJetsPheno}

Intuitively, azimuthal asymmetries of hadrons within jets, as described in 
Sec.~\ref{sec:hadInJets}, should be sensitive to the Collins effect as it has 
been suggested in Refs.~\cite{Yuan:2007nd,Yuan:2009dw}. If the jet axis can be 
identified with the outgoing quark direction, these measurements actually 
have an advantage compared to the corresponding SIDIS measurements, because 
the intrinsic quark momenta in the initial and the final state decouple. 
However, there are also challenges in the interpretation. On the one hand, 
as is typical in $p\,p$ processes, a considerable range in $x$ of the PDFs 
of the polarised and unpolarised proton is contributing. Jet reconstruction 
mitigates this issue somewhat, since the rapidity and the $p_T$ of the 
reconstructed jet impose some constraints on this distribution. Additionally, 
there were questions about the validity of TMD factorisation in the context 
of more than 2 color charges participating in a $2\rightarrow 2$ process, due 
to an effect dubbed ``color-entanglement"~\cite{Rogers:2010dm,Mulders:2011zt}, 
as already discussed in the introduction. However, a recent re-evaluation of 
the hadron in jet observables~\cite{Kang:2017glf} indeed showed that the TMD 
fragmentation functions entering the hadron in jet cross sections are universal. 
This supports the conclusion of Refs.~\cite{Yuan:2007nd,Yuan:2009dw}, which 
also argued that the Collins function would be the same as measured in SIDIS 
and $e^+e^-$. This finding is specific to TMD-FFs, where the separation of 
the color charges in the final state is large enough, so that there are no 
gluons reconnecting to the initial state TMDs.
 
A first study for the Collins and other asymmetries in jets 
has been presented in the context of the GPM~\cite{DAlesio:2010sag}, according 
to which the single polarised cross section for the $\pup p \to (h,{\rm jet})X$ 
process can be schematically written as:
\begin{eqnarray}
\label{eq:upolPPTMD}
\frac{d^6\sigma^{\pup p \to (h,\,{\rm jet})X}}{d^3\bfP_{{\rm jet}} \, 
dz \, d^2\bfP_{hT}} &\propto& 
\sum_{a,b,c,d} \int \frac{dx_a}{x_a} \, d^2 \bfk_{\perp a} 
\int \frac{dx_b}{x_b} \, d^2 \bfk_{\perp b} \, 
\delta(\hat{s} + \hat{t} + \hat{u}) \nonumber \\
&\times& \; \Big( \hat{\sigma}_{{\rm unp}}^{ab \to cd} \, 
f_{a/\pup}(x_a,\bfk_{\perp a}) \, f_1^{b}(x_b,k_{\perp b}) \, 
D_{h/c}(z,k_{\perp h})
\nonumber \\
& & \hspace{0.2cm} + \, \sin(\phi_S - \phi_h) \, \hat{\sigma}_{\textrm{pol}}^{ij 
\to kl} \, h_1^{a}(x_a,k_{\perp a}) \, f_1^{b}(x_b, k_{\perp b}) \, 
H_1^{\perp c}(z,k_{\perp h}) + \ldots \Big) \,,
\end{eqnarray}
where one recognises the TMD-PDFs and FFs defined in 
Eqs.~(\ref{PDF-Sivers}), (\ref{TMD-unp}, (\ref{TMD-tra}) and (\ref{Coll-Ams}).
$\phi_S$ and $\phi_h$ are the azimuthal angles respectively of the transverse 
spin vector of the polarised hadron and the transverse momentum of the hadron 
relative to the jet, as described in Sec.~\ref{sec:hadInJets}. Later work~\cite{Kang:2017btw} including TMD evolution agrees well with STAR data. 

If in Eq.~\ref{eq:upolPPTMD} the hadron $\bfP_{hT}$ is integrated over and 
only the dependence on $\phi_S$ is considered, sensitivity to the Sivers 
functions is obtained in the GPM~\cite{Anselmino:2013rya}. However, 
the applicability of this framework for $p\,p$ processes is limited. As we already 
commented, this is for instance due to questions about the validity of 
factorisation and universality. Additionally, in the case of the Sivers 
measurement in jets, only a single hard scale, the jet $P_T$, is measured. 
Then, it might be considered more natural to use the collinear twist-3 
approach; calculations within this framework~\cite{Gamberg:2013kla} seem to 
agree with the AnDY results~\cite{Bland:2013pkt}. Since the color factors in 
the twist-3 framework have a correspondence to the Wilson lines entering in 
the TMD framework, which lead to the modified universality (the sign change), 
the jet $A_N$ has been claimed to be the first evidence of this effect. 
The structure of the jet $A_N$ cross section is quite similar to the one in 
Eq.~\eqref{eq:anPromptPhoton} for prompt photons. They only differ in the color 
factors corresponding to the final state interactions, which, in the case of jet 
production, suppress the asymmetries.
It should be mentioned, that, within the cited uncertainties, the AnDY results also 
agree with the GPM calculations from Ref.~\cite{Anselmino:2013rya}. 

\subsection{\it How to interpret azimuthal correlations of back-to-back 
hadrons and the $\Lambda$ polarisation in semi-inclusive $e^+e^-$ 
annihilations \label{subsec3.4}}

This sub-section focuses on the sensitivity of the structure functions defined 
in Eq.~\eqref{eq:dsig-ee} in Sec.~\ref{subsec2.3} to the Collins FF, which 
is most relevant for this review. Additionally, a short interpretation of 
the back-to-back production of hadron pairs, sensitive to the di-hadron FF 
$H_1^\sphericalangle$ as well as the production of $\Lambda^\uparrow$ hyperons 
will be given, as these channels are relevant for the extraction of (TMD)-FFs 
entering in processes discussed in other sections in this review.

A detailed review of TMD FFs and their interpretation, including $H_1^\perp$, 
$H_1^\sphericalangle$ and the polarising FF $D_{1T}^{\perp\Lambda}$ can be 
found in Ref.~\cite{Metz:2016swz}. Coming back to 
Eq.~\eqref{eq:dsig-ee}, that is the production of two back-to-back hadrons in 
$e^+ e^- \to h_1 \, h_2 \, X$ processes, the cross section was worked out in 
Ref.~\cite{Boer:1997mf} for one photon annihilation. A complete discussion 
including electroweak and polarisation effects can be found in 
Ref~\cite{Pitonyak:2013dsu} and papers quoted therein.

At leading order in perturbation theory and in $1/Q$ the cross section 
reads~\cite{Boer:1997mf}
\begin{eqnarray}
\frac{d^5 \sigma^{e^+ e^- \to h_1 h_2 X}}
{d\cos\theta_2 \, dz_1 \, dz_2 \, d^2\bfP_{1T}} & = &
\frac{6 \pi \alpha_{\rm em}^2}{Q^2} \, z_1^2 \, z_2^2 \, \bigg( 
A(y) \, \mathcal{C}_{e^+ e^-} \big[ D_1 \bar{D}_1 \big] 
\nonumber \\
&+& \, B(y) \, \cos(2\phi_1) \,
\mathcal{C}_{e^+ e^-} \bigg[\frac{2 \, \hat{\bfh} \cdot \bfk_{1T} \, 
\hat{\bfh} \cdot \bfk_{2T} - \bfk_{1T} \cdot \bfk_{2T}}{M_1 \, M_2} \,
 H_1^\perp \,\bar{H}_1^\perp \bigg] \bigg) \,,
\label{eq:back2backHadrons}
\end{eqnarray}
where $\hat\bfh$ is the unit vector along $\bfP_{1\perp}$ and the 
angles $\theta_2$ and $\phi_1$ are defined in Fig.~\ref{e+e-P1}.
The convolution $\mathcal{C}$ is defined as
\begin{eqnarray} \label{e:convolution_epem}
{\cal C}_{e^+ e^-}[w D \bar{D}] & = & \sum_q e_q^2 \int d^2\bfk_{1T} \, 
d^2\bfk_{2T} \, \delta^{(2)}(\bfk_{1T} + \bfk_{2T} + \bfP_{1T}/z_1) 
\nonumber \\
& & \hspace{1.0cm} \times \, w(\bfk_{1T},\bfk_{2T}) \, 
D^q(z_1, \, z_1^2 \bfk_{1T}^{\,2}) \, 
\bar D^q(z_2, \, z_2^2 \bfk_{2T}^{\,2}) 
+ \{ q \leftrightarrow \bar{q} \}\,.
\end{eqnarray}
$A(y)$ and $B(y)$ are kinematic factors: 
\begin{equation} \label{e:y_dependence_epem}
A(y) = \frac{1}{2} - y + y^2 \,, \qquad \qquad B(y) = y(1-y) \,,
\end{equation}
with $y=(1 + \cos\theta_2)/2$. They are the analogue of the 
spin transfer coefficients in Eq.~\eqref{part-Xsec-tr} in the SIDIS case 
and can be interpreted as the projection of the transverse polarisation 
of the produced quarks. In the $e^+e^-$ CMS system, the maximal sensitivity 
to the transverse polarisation of the quarks is therefore reached at a 
direction normal to the beam axis, see Eqs.~\eqref{e+e-qqbarcs}.
Information on the Collins functions can be obtained by looking 
at the azimuthal modulation of the cross section, second line of 
Eq.~(\ref{eq:back2backHadrons}).

Comparing Eq.~\eqref{eq:back2backHadrons} to Eq.~\eqref{eq:dsig-ee}, 
the structure function $N_{h_1h_2}$ can then be identified with the term 
including the Collins functions,
\begin{equation}
4 \, z_1^2 \, z_2^2 \, B(y) \, \mathcal{C}_{e^+ e^-} \bigg[ 
\frac{2\hat{h} \cdot \bfk_{1T} \, \hat{h} \cdot \bfk_{2T} - \bfk_{1T} 
\cdot \bfk_{2T}}{M_1 \, M_2} \, H_1^\perp \,\bar{H}_1^\perp \bigg] \>,
\end{equation}
and $D_{h_1h_2}$ with the term containing the unpolarised FFs,
\begin{equation}
4 \, z_1^2 \, z_2^2 \, A(y) \, \mathcal{C}_{e^+ e^-} \big[ 
D_1 \bar{D}_1 \big] \>.
\end{equation}
Notice that the definitions of $\bfk_T$ and $D_1$ differ by factors $z$ from 
the definition of $\bpt$ and $D_{h/q}$.

For completeness, it should be mentioned that the Collins FF can also be 
accessed in a symmetric coordinate system, where the axis around which the 
azimuthal angles are measured is given by the thrust axis in the event. In 
this case, instead of the convolution of the Collins FFs of quark and anti-quark, 
the product of the $\bfk_T$ moments of the respective Collins FFs are 
measured~\cite{Boer:phd,Anselmino:2007fs}.

The term in the cross-section of back-to-back hadron production 
$e^+e^-\rightarrow h_1\,h_2\,X$ that is sensitive to the Collins FF $H_1^\perp$  
can be symbolically expressed as
\begin{equation}
\sum_q e^2_q \, H_1^{\perp \, h_a/q}(z_a,k_{aT}) 
\otimes H_1^{\perp \, h_b/\bar{q}}(z_b,k_{bT}) + \{q\leftrightarrow \bar{q}\},
\end{equation}
where the convolution over the transverse momenta is expressed as $\otimes$.
In a similar way, one can consider the process, in which two hadron pairs are 
created back-to-back:
\begin{equation}
    e^+e^-\rightarrow (h_{a1},h_{a2}) \, (h_{b1},h_{b2}) \, X.
\end{equation}
Here, the di-hadron FF $H_1^\sphericalangle$ is accessed, which can 
{\it e.g.} be used in the extraction of the transversity distribution 
from the di-hadron asymmetries measured in $p\,p$ and described in 
Sec~\ref{sec:di-hadEffects}. This is further discussed in 
Sec.~\ref{phenoDi-HadPP}. Instead of the angle $\phi_1$, which enters the 
asymmetries sensitive to the Collins effect, the di-hadron cross section 
depends on $\phi_R$, which is defined analogously to the $p\,p$ case (see 
Sec~\ref{sec:di-hadEffects}). Since $H_1^\sphericalangle$ does not vanish 
upon the integration over intrinsic transverse momenta, the amplitude of the 
$\sin(2\phi_R)$ modulation is then sensitive to the product of the di-hadron FFs:
\begin{equation}
\sum_q e^2_q \, H_1^{\sphericalangle \, h_{a1} h_{a2}/q}(z_a,M_{ha}) \, 
H_1^{\sphericalangle \, h_{b1} h_{b2}/\bar{q}}(z_b,M_{hb})\>.
\end{equation}

Finally, the transverse polarization of $\Lambda$ hyperons in $e^+e^-$ 
annihilation 
\begin{equation}
e^+e^-\rightarrow \Lambda^\uparrow X
\end{equation}
is sensitive to the term $\sum_q e^2_q \, D_{1T}^{\perp \Lambda/q}(z,\bpt)$, 
with the polarising fragmentation function $D_{1T}^{\perp \Lambda/q}$. 
In the case where another hadron is detected opposite to the $\Lambda$, 
some flavour sensitivity can be gained by the entrance of the unpolarised FF 
of the associated hadron in the expression of the cross section. The relevant 
term can then be expressed symbolically as 
\begin{equation}
\sum_q e^2_q \, D_{1T}^{\perp \Lambda/q}(z,\bpt) \, D_{\Lambda/\bar{q}}(z).
\end{equation}
Notice that the polarised $\Lambda$ production enjoyed 
increased relevance recently, due to the recent Belle 
measurements~\cite{Guan:2018ckx}, discussed earlier, and their possible 
interpretation in terms of Polarising Fragmentation 
Functions~\cite{Anselmino:2019cqd}.
The Belle experiment measured a two-scale process in which the hyperon 
polarisation was determined with respect to the plane spanned by the 
hyperon momentum as well as a proxy for the outgoing quark momentum. 
This was either the thrust axis or a hadron in the opposite hemisphere. 
Therefore the TMD picture is appropriate and has very recently 
been used for first extractions of the polarizing FF $D_{1T}^{\perp \Lambda/q}
(z,\bpt)$~\cite{DAlesio:2020wjq,Callos:2020qtu}. However, similar to the scalar 
hadron production processes discussed earlier, there is a related 
single-scale process, which can be treated in a twist-3 picture. 
Here the hyperon polarisation is measured with respect to the plane 
spanned by hyperon momentum and beam-axis. See 
Refs.~\cite{Koike:2017fxr,Gamberg:2018fwy} for recent work using this framework.

\subsection{\it The Collins fragmentation function \label{subsec3.5}}

Before discussing, in the next Section, our actual knowledge of the
TMD-PDFs and the nucleon 3-dimensional structure, let us comment on the 
information we have gathered so far on the Collins FF. This function is not 
directly related to the nucleon structure, but it is an essential piece of 
information which we need, as it often combines with TMF-PDFs into physical 
observables.

The TMD fragmentation function, Eqs.~(\ref{Coll-TO}) and~(\ref{Coll-Ams}), first 
introduced by Collins~\cite{Collins:1992kk}, embeds fundamental properties of the 
mysterious hadronisation process of quarks; it correlates the transverse spin of 
the fragmenting quark to the azimuthal distribution, around the quark direction, 
of the final hadrons. It is believed to be universal, the same in $e^+e^-$, SIDIS
and Drell-Yan processes~\cite{Collins:2004nx}, for which TMD factorisation holds.

As discussed in previous Sections, the Collins function, being chiral-odd, must 
couple, in physical observables, to another chiral-odd function; this can be the 
transversity distribution, in SIDIS, or another Collins function in $e^+e^-$ 
processes. The combined fit of azimuthal asymmetries in these two processes has 
indeed allowed an extraction of the Collins function~\cite{Anselmino:2007fs,
Anselmino:2008jk,Anselmino:2013vqa,Anselmino:2015sxa,Kang:2015msa}. 

\begin{figure}[t]
\begin{center}
\vskip -36 pt
\includegraphics[width=15.truecm,angle=0]{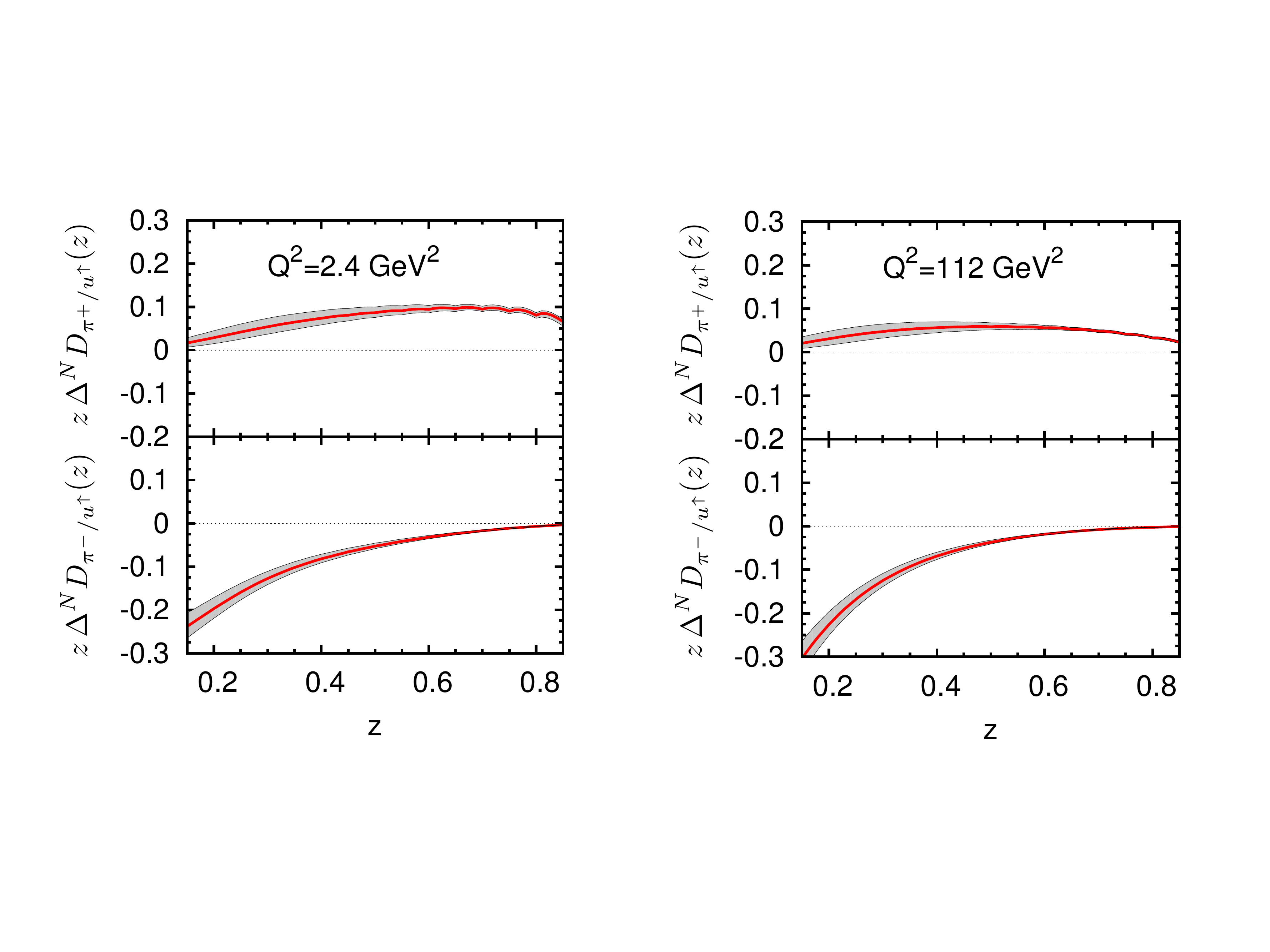}
\vskip -72pt 
\caption{The values of $z$ times the lowest $\pt$ moment of the favoured (upper
plots) and disfavoured (lower plots) Collins functions at $Q^2 = 2.4$ GeV$^2$ 
(left panel) and at $Q^2 = 112$ GeV$^2$ (right panel). The figure is adapted with 
permission from Ref.~\cite{Anselmino:2015sxa}. Copyrighted by the American 
Physical Society.}
\label{Collins}
\end{center}
\end{figure}

An example of the extracted Collins functions is shown in Fig.~\ref{Collins},
from Ref.~\cite{Anselmino:2015sxa}. The usual assumption, for pion production,  
is that of defining two kinds of functions: the favourite Collins functions, which 
is generated by a valence quark of the pion, like $\Delta^ND_{\pi^+/u^\uparrow}$,
and the disfavoured Collins function, which is generated by a sea quark of the 
pion, like $\Delta^ND_{\pi^-/u^\uparrow}$. The plots show the values of 
$z\,\Delta^ND_{h/q^\uparrow}(z,Q^2)$, where
\be
\Delta^ND_{h/q^\uparrow}(z,Q^2)= \int d^2\bpt \> 
\Delta^ND_{h/q^\uparrow}(x,\pt,Q^2) \>,      
\ee
at two different values of $Q^2$, which are the mean $Q^2$ 
of the SIDIS data and the $Q^2$ of the Belle $e^+e^-$ data; all details
can be found in Ref.~\cite{Anselmino:2015sxa}.

Let us stress once more that the data on the azimuthal distribution of two 
hadrons in $e^+e^- \to q \, \bar q \to h_1\,h_2\,X$ processes, discussed in 
Section~\ref{sec:hadInJetsPheno}, are a clear indication of the Collins effect 
at work. Such effects have been observed by the Belle~\cite{Abe:2005zx,Seidl:2008xc}, 
the BaBar~\cite{TheBABAR:2013yha} and the BESIII~\cite{Ablikim:2015pta} 
Collaborations.  

\section{From data to the 3-dimensional imaging of the nucleon \label{sec4}}

\subsection{\it The unpolarised TMD-PDFs \label{subsec4.1}}

The first information one can try to obtain on the transverse motion of quarks 
and gluons inside the proton is given by the unpolarised parton distribution 
$f_{a/p}(x,\kt)$. For quarks, this could be easily accessed in unpolarised SIDIS 
processes, from data on the cross section, Eqs.~(\ref{eq:SIDIS_FUU}) 
or~(\ref{simple-unp}). The unpolarised TMD-PDFs couple to the unpolarised 
TMD-FFs, $D_{h/q}(z,\pt)$. 

However, data on unpolarised SIDIS cross section are not so abundant. 
The first attempts to obtain information on the unpolarised TMDs were performed 
in Refs.~\cite{Anselmino:2005nn,Collins:2005ie,Schweitzer:2010tt}. 
A most simple factorised Gaussian parameterisation was assumed for the unknown 
TMDs:  
\bea
&&f_{q/p} (x,\kt)= f_{q/p} (x)\,\frac{e^{-\kt^2/\avk}}{\pi\avk}
\label{unp-dist}\\
&&D_{h/q}(z,\pt)=D_{h/q}(z)\,\frac{e^{-\pt^2/\avp}}{\pi\avp}\>,
\label{unp-frag}
\eea
where $f_{q/p}(x)$ and $D_{h/q}(z)$ are the usual collinear PDFs and FFs and 
where the Gaussian widths were assumed to be flavour independent and constant. 

Ref.~\cite{Anselmino:2005nn} exploited Fermilab~\cite{Adams:1993hs} 
and EMC data~\cite{Arneodo:1986cf} on $\langle \cos\phi_h \rangle$, a 
kinematical effect originated by the terms of ${\cal O}(\kt/Q)$ in the 
elementary interaction [see Eq.~(\ref{part-Xsec})], the so-called Cahn effect, 
contributing to $F_{UU}^{\cos\phi_h}$. Ref.~\cite{Collins:2005ie} best fitted
values of $\langle P_T \rangle$ from Ref.~\cite{Airapetian:2002mf}, while 
Ref.~\cite{Schweitzer:2010tt} compared with JLab data on cross 
section~\cite{Mkrtchyan:2007sr,Osipenko:2008aa} and HERMES data on 
$\langle P_T^2 \rangle$~\cite{Airapetian:2009jy}. The Gaussian model of the 
TMDs proved to be adequate to fit the data and the values of $\avk$ and $\avp$
resulting in these three cases are comparable, in the approximate range (in 
GeV$^2$): $0.25 \leq \avk \leq (0.38\pm 0.06)$ and $0.15 
\leq \avp \leq 0.20$.     

More recently, plenty of new data became available by the COMPASS 
and HERMES Collaborations, which measured, rather than the 
cross section, the hadron multiplicity. According to COMPASS 
notation~\cite{Adolph:2013stb,Aghasyan:2017ctw} 
the differential hadron multiplicity is defined as:
\be
\frac{d^2 n^h(\xb, Q^2, z_h, P_T^2)}{dz_h \, dP_T^2} \equiv
\frac{1}{\displaystyle{\frac{d^2 \sigma^{DIS} (\xb, Q^2)}{d\xb \, dQ^2}}} \>
\frac{d^4 \sigma (\xb, Q^2, z_h, P_T^2)}{d\xb \, dQ^2 \, dz_h \, dP_T^2} \>,
\label{mult-c}
\ee
while HERMES definition~\cite{Airapetian:2012ki} is: 
\be
M_n^h(\xb, Q^2,z_h, P_T)
\equiv
\frac{1}{\displaystyle{\frac{d^2 \sigma^{DIS} (\xb, Q^2)}{d\xb \, dQ^2}}} \>
\frac{d^4 \sigma (\xb, Q^2, z_h, P_T)}{d\xb \, dQ^2 \, dz_h \, dP_T}
\> ,\label{mult-h}
\ee
where the index $n$ denotes the kind of target and the Deep Inelastic Scattering 
(DIS) cross section has the usual leading order collinear expression,  
\be
\frac{d^2 \sigma^{DIS} (\xb, Q^2)}{d\xb \, dQ^2} = 
\frac {2 \, \pi \, \alpha^2}{Q^4} \, \left[ 1 + (1-y)^2 \right] \,
\sum_{q} e_q^2 \> f_{q/p} (\xb) \label{xs-DIS} \> \cdot
\ee

From Eqs.~(\ref{eq:SIDIS_FUU}) and~(\ref{mult-c})--(\ref{xs-DIS}) one simply has: 
\be
\frac{d^2 n^h(\xb, Q^2, z_h, P_T)}{dz_h \, dP_T^2} = 
\frac{1}{2P_T} M_n^h(\xb, Q^2,z_h, P_T) =
\frac{\pi \> F_{UU}} {\sum_{q} e_q^2 \> f_{q/p} (\xb)} \> \cdot
\label{mult-fuu}
\ee

The first analyses of data based on Eq.~(\ref{mult-fuu}) have been performed 
in Refs.~\cite{Signori:2013mda,Anselmino:2013lza} assuming for the TMDs the 
factorised and gaussian behaviour of Eqs.~(\ref{unp-dist}) and~(\ref{unp-frag}).   
In this case the expression of $F_{UU}$~(\ref{eq:SIDIS_FUU}) can be exactly 
calculated: 
\be
F_{UU}  =  \sum_{q} \, e_q^2 \,f_{q/p}(\xb)\,D_{h/q}(z_h)
\frac{e^{-P_T^2/\avPT}}{\pi\avPT} \>, \label{G-FUU}
\ee
where
\be
\avPT = \avp + z_h^2 \, \avk \>. \label{avPT}
\ee

As we said, in the simplest version of the gaussian model for the TMDs, they 
are assumed to be flavour independent and their widths $\avk$ and $\avp$ to be 
constant~\cite{Anselmino:2013lza}; this cannot be true for all values of $x$ 
and $z$, but might be adequate for analysing the available experimental data 
which cover limited kinematical regions. 
The possible flavour, $x$ and $z$ dependence of the gaussian widths was studied
in Refs.~\cite{Signori:2013mda,Bacchetta:2017gcc}.      

From this first study of the unpolarised TMDs, based on hadron multiplicity and 
the HERMES~\cite{Airapetian:2012ki} and 
COMPASS~\cite{Adolph:2013stb,Aghasyan:2017ctw} data, it turns out that the 
gaussian assumptions~(\ref{unp-dist}) and~(\ref{unp-frag}) -- possibly with 
flavour, $x$ and $z$ dependences -- are able to fit well most data. 

\begin{figure}[t]
\begin{center}
\includegraphics[width=15.truecm,angle=0]{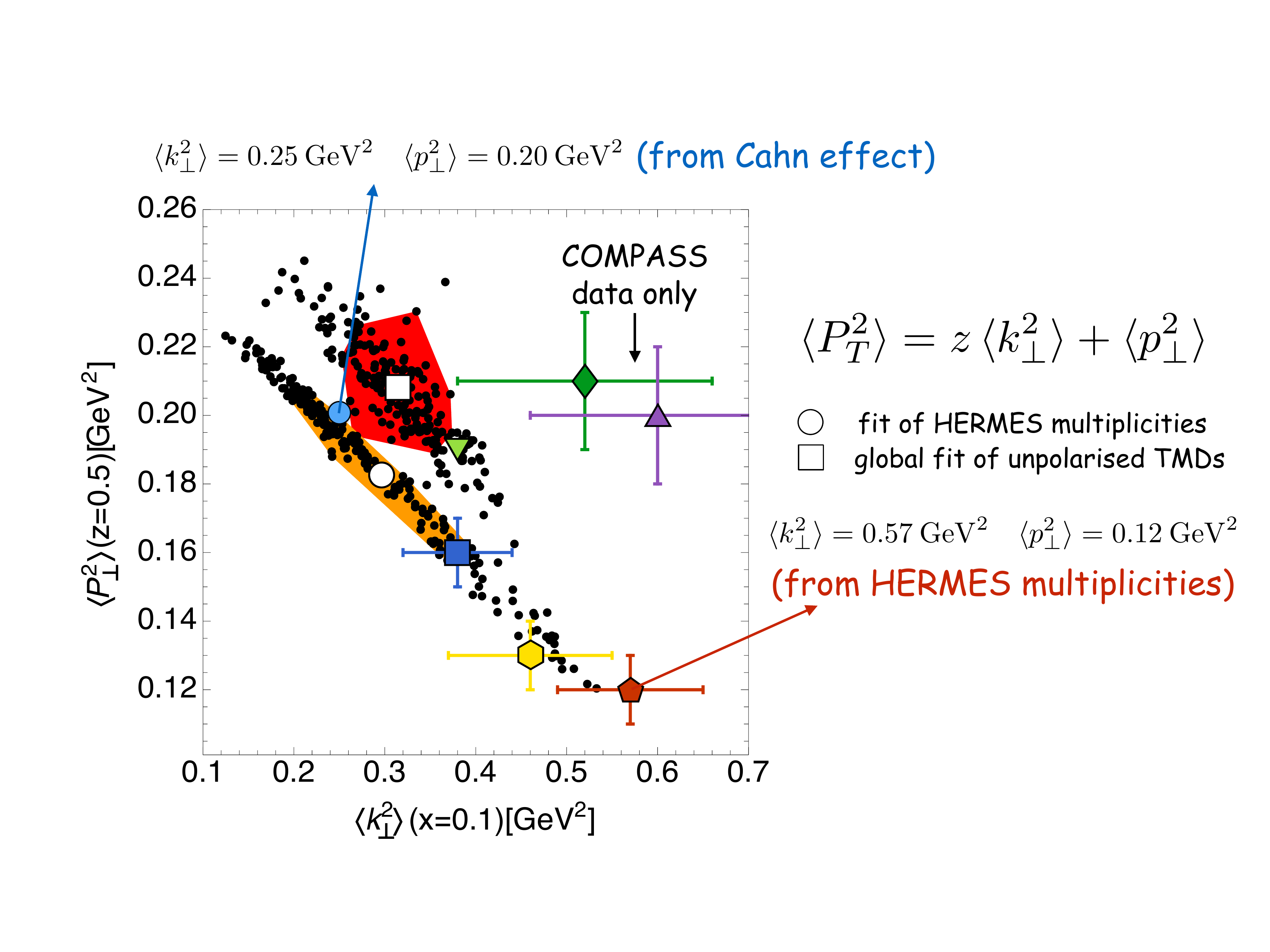}
\vskip -36pt 
\caption{Values of $\avk(x=0.1)$ and $\avp(z=0.5)$ obtained from different 
fits of the hadron multiplicities, as shown in the figure. For further details 
we refer to the text and to Ref.~\cite{Bacchetta:2017gcc} from which this figure 
has been taken and adjusted. Original figure available under a 
\href{https://creativecommons.org/licenses/by/4.0/legalcode}
{Creative Commons Attribution 4.0 International}.}
\label{kt-kp}
\end{center}
\end{figure}

However, the hadron multiplicy, as given by Eqs.~(\ref{mult-fuu})-(\ref{avPT}),
depends on the parameter $\avPT$, which, in turns, is a combinations of the 
two parameters $\avk$ and $\avp$. This correlations makes it difficult to 
extract separately and independently the gaussian widths of $f_{q/p}(x, \kt)$ 
and $D_{h/q}(z, \pt)$; similarly good fits of the multiplicities, corresponding
to different values of  $\avk$ and $\avp$, are possible. This is shown in 
Fig.~\ref{kt-kp}, which is adjusted from Ref.~\cite{Bacchetta:2017gcc}. 
The black dots around the central values are the outcome of different replicas 
of the fit and the coloured regions contains the 68\% of the replicas that are 
closest to the average value. The red region shows the results of the first 
global fit to unpolarised TMDs~\cite{Bacchetta:2017gcc}, which takes into account
SIDIS, Drell-Yan and Z boson production data, with TMD evolution. This study 
clearly shows the crucial role of TMDs in explaining the shape of the differential 
cross section for $Z$-boson production at Tevatron, Fig.~8 of 
Ref.~\cite{Bacchetta:2017gcc}.          

The study of the unpolarised TMDs, mainly from SIDIS, but also from Drell-Yan and
$Z$-boson production, clearly shows the need to introduce the transverse motions 
of partons inside the nucleon; this conclusion, based on experimental data, has 
solid theoretical grounds for processes in which one can identify two scales: a 
small one of the order of a few hundreds of MeV, and a large one, possibly of the
order of few or more GeV~\cite{Boglione:2016bph}. Thus, one can separate a soft
non perturbative region, which pertains to the nucleon structure, and a large 
energy region, described by perturbative Standard Model interactions. From data 
fitting, the resulting values of $\langle \kt \rangle$, are indeed of the order 
of few hundreds of MeV, giving us information on the quark intrinsic motion. 

The knowledge of $\avk$, even neglecting its uncertainty due to the strong 
correlation with $\avp$, is important, but far from resolving the momentum 
distribution of partons inside the nucleon; crucial questions like the spin and
the orbital momentum of quarks, and the correlation with the nucleon spin, 
remain open.    

\subsection{\it The Sivers function \label{subsec4.2}}

The Sivers distribution of unpolarised partons inside a transversely polarised 
proton has a long and interesting history. It was introduced in 
1990~\cite{Sivers:1989cc,Sivers:1990fh}, to explain the large and unexpected 
values of the SSA $A_N$ observed in $p \, N \to \pi \, X$ processes, as 
explained in Section~\ref{sec:anPions}. It was then criticised~\cite{Collins:1992kk} 
and dismissed as violating fundamental parity and time reversal properties 
of QCD. Then, a model calculation in Ref.~\cite{Brodsky:2002cx} showed explicitly
the possibility of having a non zero Sivers function in SIDIS processes, thanks 
to final state interactions of the scattered quark with the proton remnants.
The criticism of Ref.~\cite{Collins:1992kk} was reconsidered~\cite{Collins:2002kn}, 
taking into account the path-ordered exponential of the gluon field (gauge link) 
in the operator definition of parton densities, Eq.~(\ref{correlator}). This led,
rather than to a vanishing of the Sivers function, to the prediction that such 
a function should have opposite signs in SIDIS and Drell-Yan 
processes~\cite{Collins:2002kn}:             
\be
(f_{1T}^\perp)_{\rm SIDIS} = - (f_{1T}^\perp)_{\rm DY} \>. \label{ch-sign}
\ee
This prediction is considered as an important test of our understanding of 
the origin and nature of SSAs in SIDIS and Drell-Yan processes, within a QCD 
TMD factorisation scheme. First experimental results~\cite{Adamczyk:2015gyk} 
hint at a confirmation of the sign change~\cite{Huang:2015vpy}, but no definite 
conclusion can still be drawn~\cite{Anselmino:2016uie,Matousek:2018qqd}.

The quark Sivers function has been extracted from SIDIS data on the weighted 
asymmetry $A^{\sin(\phi_h-\phi_S)}_{UT}$ \cite{Airapetian:2009ae,
Airapetian:2010ds,Adolph:2012sn,Adolph:2012sp,Adolph:2014zba,Allada:2013nsw}, 
interpreted through Eq.~(\ref{aut-siv}), by several groups~\cite{Anselmino:2005an,
Anselmino:2005nn,Anselmino:2005ea,Vogelsang:2005cs,Collins:2005ie,
Collins:2005wb,Anselmino:2008sga,Bacchetta:2011gx,Anselmino:2015sxa,
Anselmino:2016uie}. This extraction requires some assumptions on the functional 
shape of the Sivers function and a choice of parameters. A most typical and simple 
assumption for $\Delta^N f_{q/\pup}(x, k_\perp)$, Eq.~(\ref{TMD-Siv}), is, again, a 
factorisation of the $x$ and $\kt$ dependences and a gaussian shape for the 
$\kt$ dependence.  

For example, a parameterisation of the Sivers function is 
given by~\cite{Anselmino:2005ea,Anselmino:2008sga}
\be
\Delta^N \! f_ {q/\pup}(x,\kt) = 2 \, {\cal N}_q(x) \, h(\kt) \, 
f_ {q/p} (x,\kt) \; , \label{sivfac}
\ee
with
\bea
&&{\cal N}_q(x) =  N_q \, x^{\alpha_q}(1-x)^{\beta_q} \,
\frac{(\alpha_q+\beta_q)^{(\alpha_q+\beta_q)}}
{\alpha_q^{\alpha_q} \beta_q^{\beta_q}}\; ,
\label{siversx} \\
&&h(\kt) = \sqrt{2e}\,\frac{k_\perp}{M_{S}}\,e^{-{k_\perp^2}/{M_{S}^2}}\; ,
\label{siverskt}
\eea
where $N_q$, $\alpha_q$, $\beta_q$ and $M_S$ (GeV/$c$) are free parameters
to be determined by fitting the experimental data. The functional shapes of 
${\cal N}_q(x)$ (with $-1 \leq N_q \leq 1$) and $h(\kt)$ are such that the
positivity bound for the Sivers function,
\be
\frac{|\Delta^N\!f_ {q/\pup}(x,\kt)|}{2 f_ {q/p} (x,\kt)}\le 1\>,
\label{pos}
\ee
is automatically fulfilled for any value of $x$ and $\kt$. The unpolarised TMD
$f_ {q/p} (x,\kt)$ is given in Eq.~(\ref{unp-dist}), so that: 
\bea
\Delta^N f_{q/\pup} (x,\kt) &=& 2 \, {\cal N}_q(x) \, f_ {q/p} (x) \,
\sqrt{2e}\,\frac{\kt}{M_S} \, e^{-\kt^2/M^2_S}\,
\frac{e^{-\kt^2/\avk}}{\pi\avk} \\
&\equiv& \Delta^N f_{q/\pup}(x) \; \sqrt{2e}\,\frac{\kt}{M_S} \;
\frac{e^{-\kt^2/\avk_S}}{\pi\avk}
\label{Siv-dist}
\eea
where
\be
\avk_S = \frac{\avk \, M^2_S}{\avk +  M^2_S} \> \cdot \label{S-2}
\ee

From Eqs.~(\ref{aut-siv}) and~(\ref{Siv-dist}) one 
obtains~\cite{Anselmino:2011ch,Anselmino:2018psi}:
\be
A^{\sin (\phi_h-\phi_S)}_{UT} =
\frac{\displaystyle 
\sum_q\,e_q^2\,\Delta^N f_{q/p^\uparrow}(x) \, D_{h/q}(z)
\times \sqrt{\frac{e}{2}}\,\frac{P_T}{M_S}\,\frac{z\,\langle k_\perp^2\rangle_S^2}
{\langle k_\perp^2\rangle}\,
\frac{e^{-P_T^2/\langle P_T^2\rangle_S}}{\pi\langle P_T^2\rangle_S^2}}
{\displaystyle \sum_q\,e_q^2\,f_{q/p}(x)\,D_{h/q}(z)\,
\frac{e^{-P_T^2/\langle P_T^2\rangle}}{\pi\langle P_T^2\rangle}} \>,
\label{eq:AUT-siv-fen}
\ee
where:
\be
\avPT = \avp + z^2 \avk
\quad\quad\quad
\avPT_S = \avp + z^2 \avk_S \,.
\label{eq:PT-S}
\ee

\begin{figure}[t]
\begin{center}
\vskip -36 pt
\includegraphics[width=17.truecm,angle=0]{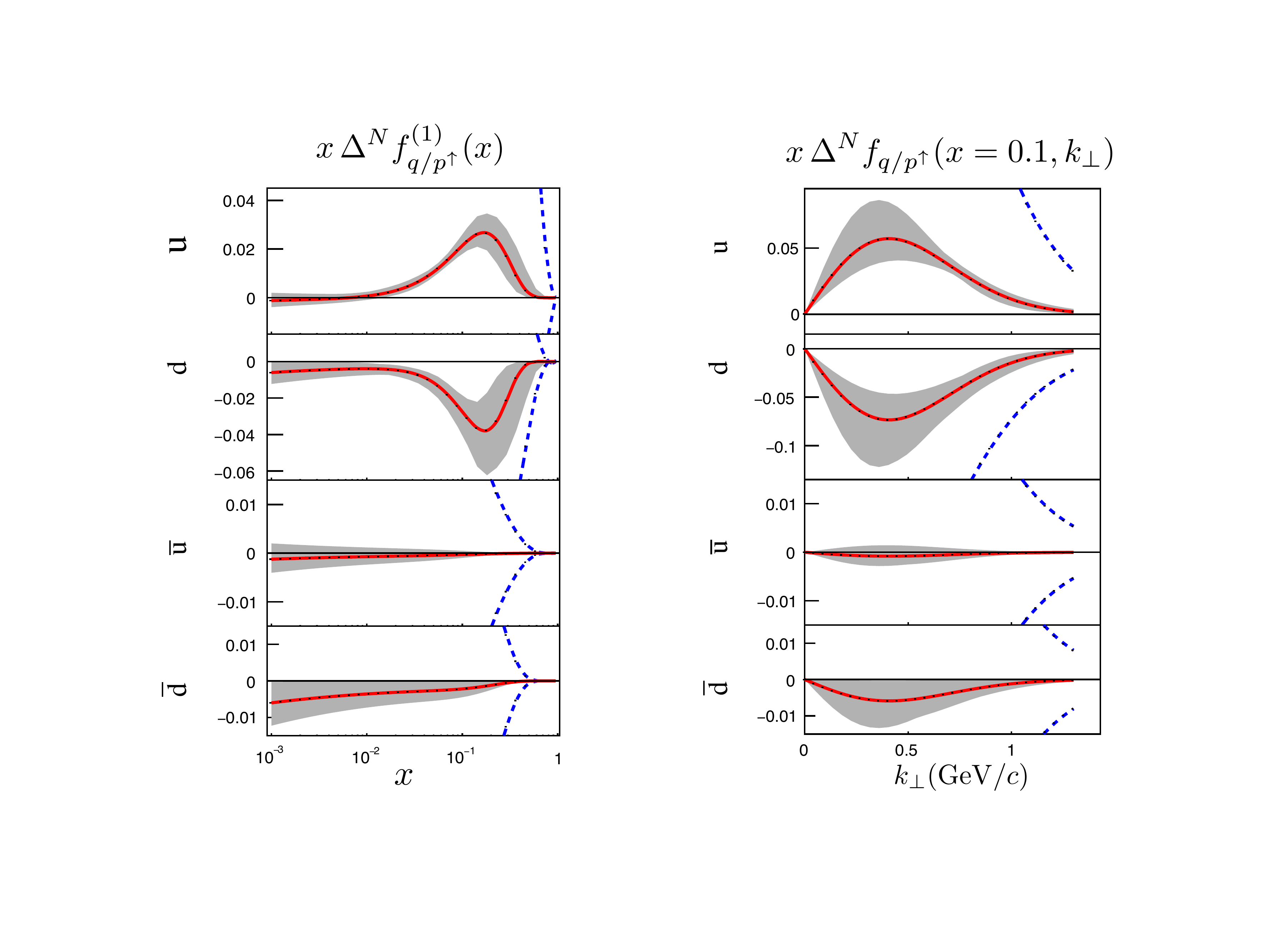}
\vskip -48pt 
\caption{The Sivers distributions for $u = u_v + \bar u$, $d = d_v + \bar d$, 
$\bar u$ and $\bar d$ as extracted in Ref.~\cite{Anselmino:2016uie}. 
Left panel: the $x$ dependence of the first moment of the Sivers functions, 
Eq.~(\ref{moment}) , multiplied by $x$. Right panel: the $\kt$
dependence of $x$ times the Sivers functions, at $x=0.1$. Details can 
be found in Ref.~\cite{Anselmino:2016uie}.
Figure from Ref.~\cite{Anselmino:2016uie} available under a 
\href{https://creativecommons.org/licenses/by/4.0/legalcode}
{Creative Commons Attribution 4.0 International}.}
\label{Sivers}
\end{center}
\end{figure}

\begin{figure}[t]
\begin{center}
\vskip -24 pt
\includegraphics[width=17.truecm,angle=0]{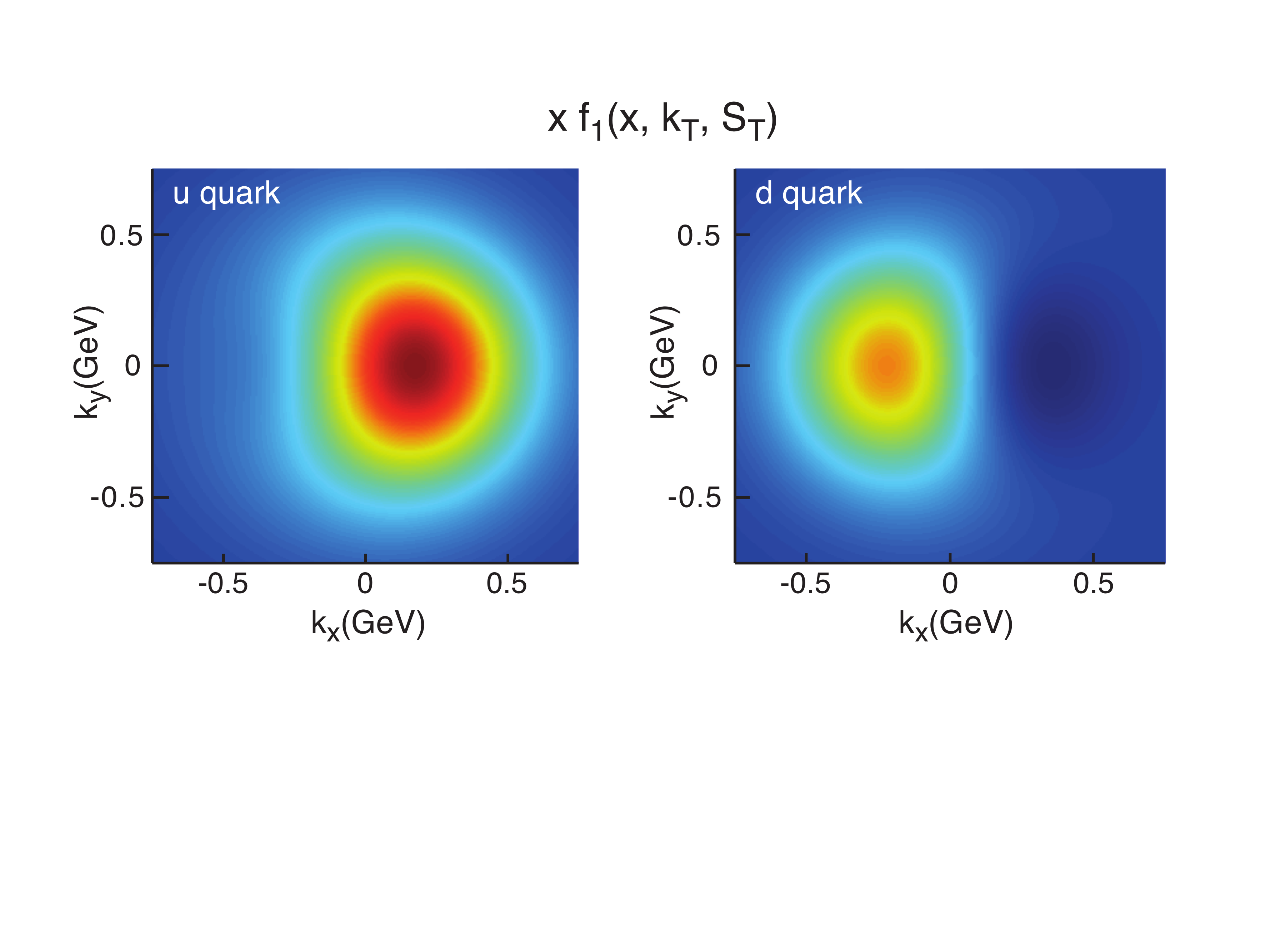}
\vskip -110pt 
\caption{The transverse-momentum distribution of $u$ (left) and $d$ (right) quarks  
with longitudinal momentum fraction $x= 0.1$ in a transversely polarised proton 
moving in the $\hat{\bfz}$-direction, while polarised in the $\hat{\bfy}$-direction.  
The corresponding Sivers distribution, $f_1(x,\kt,S_T) = f_{q/p^\uparrow}(x,\bkt)$, 
is evaluated using the Sivers function from Ref.~\cite{Anselmino:2010bs}. 
The color code indicates the probability of finding the quarks. Figure from 
Ref.~\cite{Accardi:2012qut} and available under a 
\href{https://creativecommons.org/licenses/by/4.0/legalcode}
{Creative Commons Attribution 4.0 International}.}
\label{Sivers-3D}
\end{center}
\end{figure}

The Sivers distribution, more than the unpolarised TMDs, offers a subtle and deeper 
way of probing the 3D structure of nucleons, as it couples the intrinsic motion of 
partons to a fundamental property of the the proton, its spin. The results so far 
reached deserve some comments.   
\begin{itemize} 
\item
Fig.~\ref{Sivers} shows a recent extraction of the quark Sivers functions from 
SIDIS data; they are taken from Ref.~\cite{Anselmino:2016uie} but do not differ,
qualitatively, from other extractions. The quantity shown on the left plot is  
the first moment of the Sivers function~\cite{Anselmino:2005an} (multiplied by $x$):
\be
\Delta^Nf_{q/p^\uparrow}^{(1)}(x)= \int d^2\bfk_\perp \frac{k_\perp}{4M} \, 
\Delta^Nf_{q/p^\uparrow}(x,k_\perp) = - f_{1T}^{\kt(1)q}(x) \>. \label{moment}
\ee 

At this stage, despite the simple and approximate interpretation of the data, 
Eq.~(\ref{eq:AUT-siv-fen}), one can definitely conclude that the Sivers effect, 
as origin of SSAs, at least in SIDIS processes, is well established. 
The corresponding Sivers functions need not be too large, 
and are well below the positivity limit~(\ref{pos}) (dashed blue lines in 
Fig.~\ref{Sivers}). The $u$ and $d$ quark Sivers functions are opposite in sign 
and peak at $x$-values in the valence region; in fact, the sea quark Sivers 
functions are compatible with zero. 
\item
The Burkardt sum rule for the Sivers distribution~\cite{Burkardt:2004ur}:
\be
\sum_a \int dx \, d^2 \bfk_\perp \, \bfk_\perp \, f_{a/\pup}(x, \bfk_\perp) 
\equiv  \sum_a \, \langle \bfk_\perp^a \rangle = 0 \>, \label{bsr}
\ee
is almost saturated by valence quarks alone ($a = u,d$)~\cite{Anselmino:2008sga}, 
leaving little room to a contribution from a gluon Sivers function, as confirmed 
by other studies~\cite{Anselmino:2006yq,Brodsky:2006ha,DAlesio:2015fwo}.

The gluon Sivers functions could be directly accessed in $p^\uparrow \, N$ 
interactions, in particular via the large $P_T$ inclusive production of charmed 
particles (like $D$ or $J/\psi$ mesons)~\cite{Anselmino:2004nk,
DAlesio:2017rzj,DAlesio:2019gnu} or prompt photons~\cite{DAlesio:2018rnv}. However,
for these processes, the simple TMD factorisation scheme might not be fully 
justified.    
\item
The Sivers distribution, Eq.~(\ref{PDF-Sivers}), induces a correlation between 
the parton intrinsic motion $\bkt$ and the parent nucleon polarisation $\bfS$, 
through the scalar expression $(\hat{\bfP} \times \hat{\bfk}_\perp) \cdot \bfS$. 
At fixed values of the proton momentum $\bfP$ the density number of partons inside
the transversely polarised proton is not isotropic in $\bkt$. The evidence of a 
non zero Sivers function has allowed the first 3-dimensional imaging of a proton.
An example is shown in Fig.~\ref{Sivers-3D} which shows the density of $u$ and $d$ 
quarks in the transverse momentum plane, for a proton moving along the 
$\hat{\bfz}$-axis and polarised along the $\hat{\bfy}$-axis. 
The $(\hat{\bfP} \times \hat{\bfk}_\perp) \cdot \bfS$ correlation induces a 
momentum deformation in $k_x$. Similar pictures can be drawn for different 
$x$-values, the so-called nucleon-tomography.
\item
Even if the Sivers function might contribute to SSAs with opposite signs
in SIDIS and Drell-Yan processes~\cite{Collins:2002kn}, one can think that it is 
related to fundamental features of the proton structure. Then, as it does not 
depend on the spin of the partons, which are unpolarised, but depends on the proton 
spin, it must be related to another pseudo-vector, that is the parton orbital 
angular momentum, say $\bfL_q$. It would be interesting to find a dependence on 
$\bfS \cdot \bfL_q$ embedded in the Sivers function; this was indeed proposed  
by Sivers~\cite{Sivers:2006rg,Sivers:2007pq}, as a possible normalisation of
$\Delta^Nf_{q/p^\uparrow}(x,k_\perp)$.         
\end{itemize}

\subsection{\it The transversity distribution \label{subsec4.3}}

The transversity distribution is one of the three basic PDFs, which survive 
in the collinear limit; however, differently from the unpolarised PDF and the
helicity distribution, it cannot be accessed in DIS processes, due to its chiral 
odd nature~\cite{Barone:2001sp}. It can be accessed in SIDIS processes, coupled to
the Collins TMD-FF, as detailed in Eq.~(\ref{sin-asym}). Independent information 
on the Collins function can be obtained from $e^+e^- \to h_1 \, h_2 \, X$ processes, 
as discussed in Section~\ref{subsec3.5}.

The combined extraction of the transversity and the Collins functions was performed
in Refs.~\cite{Anselmino:2007fs,Anselmino:2008jk,Anselmino:2013vqa,Anselmino:2015sxa,
Kang:2015msa}. As in the case of the Sivers function, a simple functional form 
was chosen for the unknown transversity and Collins TMDs: 
\be
\Delta_T q(x, \kt) =
\frac{1}{2} \, {\cal N}^{\T}_q(x)\,
\left[f_{q/p}(x)+\Delta q(x) \right] \;
\frac{e^{-{\kt^2}/{\avk}}}{\pi \avk} \label{tr-funct} \,, 
\ee
\be 
\Delta^N D_{h/q^\uparrow}(z,\pt) = 2\,{\cal N}^{\C}_q(z)\;
D_{h/q}(z)\;h(\pt)\,\frac{e^{-\pt^2/{\avp}}}{\pi \avp}\;,
\label{coll-funct}
\ee
with
\be
h(\pt)=\sqrt{2e}\,\frac{p_\perp}{M}\,e^{-{p_\perp^2}/{M^2}}\,.
\label{h-funct}
\ee
Simple parameterisations were adopted for ${\cal N}^{\T}_q(x)$ and
${\cal N}^{\C}_q(z)$~\cite{Anselmino:2013vqa,Anselmino:2015sxa} in such a way 
that the transversity distribution function automatically obeys the Soffer 
bound~\cite{Soffer:1994ww}
\be
|\Delta_T q(x)| \le \frac{1}{2}
\left[ f_{q/p}(x) + \Delta q(x)\right ]\,,
\label{soffer}
\ee
and the Collins function satisfies the positivity bound
\be
|\Delta^N D_{h/q^\uparrow}(z, p_\perp)| \le 2 D_{h/q}(z, p_\perp) \,.
\label{bound}
\ee

By insertion of the above expressions into Eq.~(\ref{sin-asym}) one 
obtains~\cite{Anselmino:2007fs,Kotzinian:1994dv,Mulders:1995dh}:
\be
A^{\sin (\phi_S+\phi_h)}_{_{UT}} =
\frac{\displaystyle \sum_q e_q^2 \; {\cal N}^{\T}_q(x)
\left[f_{q/p}(x)+\Delta q(x) \right]\;
{\cal N}^{\C}_q(z)\, D_{h/q}(z) \times
\frac{P_T}{M}\,\frac{1-y}{s\,x\,y^2}\,
\sqrt{2e} \, \frac{\avp ^2 \C}{\avp}
\, \frac{e^{-P_T^2/\avPT \C}}{\avPT ^2 \C}} 
{\displaystyle \sum_q e_q^2 \, f_{q/p}(x)\; D_{h/q}(z)\,
\frac{e^{-P_T^2/\avPT}}{\avPT} \,
\frac{[1+(1-y)^2]}{s\,x\,y^2}}\;,
\label{sin-asym-final}
\ee
where
\be
\avp \C= \frac{M^2 \avp}{M^2 +\avp} \quad\quad
 \avPT=\avp+z^2\avk \quad\quad
\avPT \C=\avp \C+z^2\avk\,.
\ee
          
\begin{figure}[t]
\begin{center}
\vskip -36 pt
\includegraphics[width=18.truecm,angle=0]{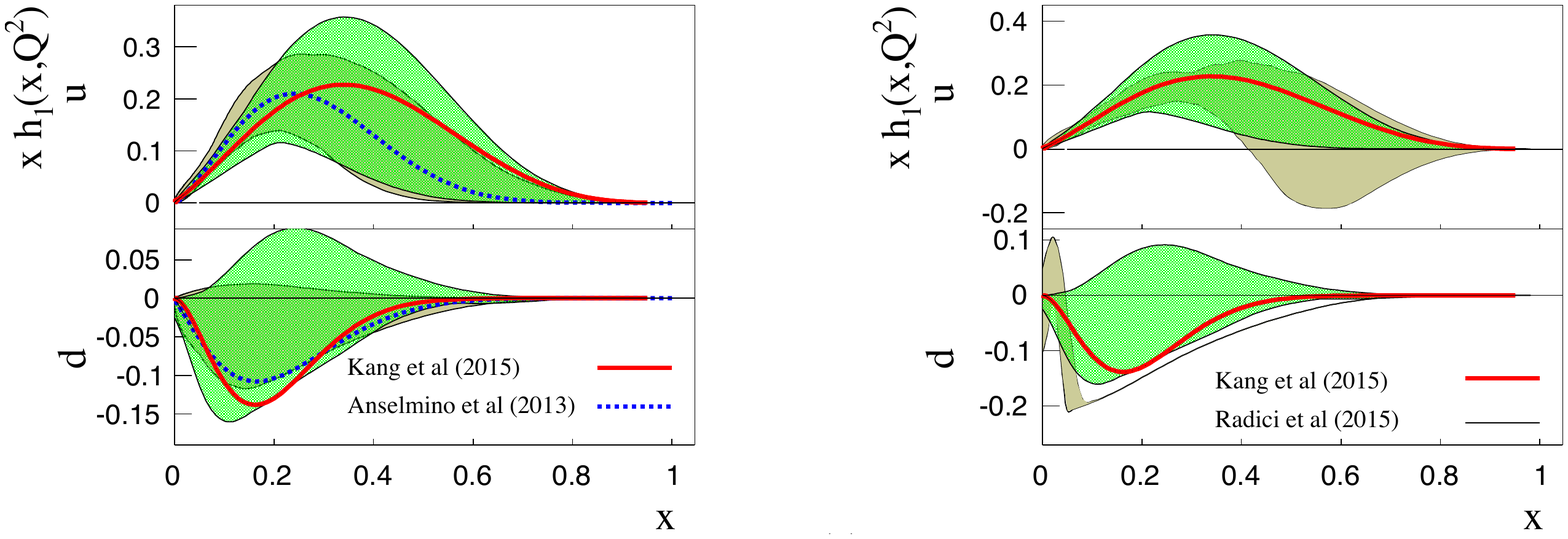}
\vskip -108pt 
\caption{Plots of $x$ times the transversity distribution, $x\,h_1^q = 
x\,\Delta q$, for $u$ and $d$ quarks, as obtained in Refs.~\cite{Kang:2015msa} 
(Kang et al. (2015)), Ref.~\cite{Anselmino:2013vqa} (Anselmino et al. (2013)) and 
Ref.~\cite{Radici:2015mwa} (Radici et al. (2015)). See text for further details.
Reprinted with permission from Zhong-Bo Kang, Alexei Prokudin, Peng Sun, 
and Feng Yuan, Phys. Rev., D93(1):014009, 2016. Copyright 2016 by the American Physical Society.}
\label{transv}
\end{center}
\end{figure}

Some results on the transversity distributions are shown in Fig.~\ref{transv}, 
obtained by different groups, with different best fitting procedures. The red 
solid lines, both in the left and right plots, have been obtained in 
Ref.~\cite{Kang:2015msa} by fitting SIDIS data from HERMES~\cite{Airapetian:2009ae},
COMPASS~\cite{Adolph:2014zba,Alekseev:2008aa} and Jlab~\cite{Qian:2011py}, 
together with $e^+e^-$ data from Belle~\cite{Seidl:2008xc} and 
BaBar~\cite{TheBABAR:2013yha}. The QCD evolution of the TMDs 
have been taken into account. The dotted blue line in the left plots is from 
Ref.~\cite{Anselmino:2013vqa}, obtained by a combined fitting of 
SIDIS~\cite{Airapetian:2010ds,Martin:2013eja} and $e^+e^-$ data~\cite{Seidl:2008xc}, 
without TMD evolution. The remaining line in the right plot is from 
Ref.~\cite{Radici:2015mwa}, which couples the chiral-odd transversity 
distribution with the chiral-odd di-hadron fragmentaion functions.         

Let us summarise our knowledge on the transversity distribution. 
\begin{itemize} 
\item
The combined fit of $A^{\sin (\phi_S+\phi_h)}_{_{UT}}$ from SIDIS data and 
azimuthal modulations in $e^+e^- \to h_1\,h_2\,X$ processes, has allowed the 
first extraction of the transversity distributions of $u$ and $d$ quarks. 
The results obtained by different groups are in good agreement, and show 
reasonable values of $\Delta_Tu(x) = h_1^u(x)$ and $\Delta_Td(x) = h_1^d(x)$, 
not far from the corresponding
helicity distributions, and peaked in the quark valence region. However, 
in SIDIS, the transversity distribution is coupled to the Collins function, 
and, in $e^+e^-$, one measures the product of two Collins functions: then, one 
cannot fix independently the signs of $\Delta_Tu(x)$ and $\Delta_Td(x)$, only 
their relative values, which turn out to be opposite. It is natural, following 
the helicity distributions or the $SU(6)$ spin-flavour symmetry, to assume 
$\Delta_Tu(x)$ to be positive and $\Delta_Td(x)$ to be negative, but this is not
determined by the extraction procedure.
\item
An alternative way of accessing the transversity distribution, by coupling it 
to the di-hadron fragmentation function~\cite{Bianconi:1999cd,Radici:2001na}, 
has been developed in Refs.~\cite{Bacchetta:2011ip,Bacchetta:2012ty,Radici:2015mwa}. 
Information on the di-hadron fragmentation function is obtained from $e^+e^-$ 
data~\cite{Vossen:2011fk}. This analysis yields results similar to those obtained 
from the combined fits of transversity and Collins functions, but it has the 
advantage that it can be used also in $p\,p$ 
interactions~\cite{Bacchetta:2004it,Radici:2018iag}, proving the universality 
of the transversity distribution~\cite{Radici:2016lam}. Also in this case one can 
only fix the relative sign of $u$ and $d$ quark transversities.
\item
The transversity distribution has the important feature that it is related to the 
tensor charge~\cite{Jaffe:1991kp},
\be 
\delta q \equiv \int_0^1 \!\!\! dx \, \left[ h_1^q(x) - h_1^{\bar q}(x) \right] \>,
\label{Tcharge}      
\ee
which can be computed on a lattice~\cite{Bhattacharya:2016zcn,Alexandrou:2017qyt,
Gupta:2018lvp}. The available data on $h_1^q(x)$ cover only a limited region in 
$x$, so that one needs some extrapolation in order to compute the full integral 
of Eq.~(\ref{Tcharge}); at the moment, there seems to be a discrepancies between 
the value of $\delta u$, as obtained from the extracted $h_1^u(x)$, and the 
lattice results~\cite{Radici:2018abr,Radici:2019myq,DAlesio:2020vtw,Benel:2019mcq}.
A very recent global analysis of data on Transverse Single Spin 
Asymmetries~\cite{Cammarota:2020qcw} includes, for the first time, results on 
the single spin asymmetry $A_N$ in $p^\uparrow p$ scattering and sees less tension 
with the lattice results. The inclusion of $A_N$ in this extraction is based on 
the connection of $A_N$, with transversity in the twist-three framework  
discussed in Sec.~\ref{sec:pheno_AN}.
\item
It is worth mentioning that the optimal access to the transversity distributions
could, in principle, be obtained by measuring double transverse spin asymmetries
in proton-antiproton Drell-Yan processes, $\pup \, \bar p^\uparrow \to \ell^+ \, 
\ell^- \, X$, which would mainly involve valence quark transversities. In order to 
enhance the amount of events, one could even think of measuring the di-lepton 
production at the $J/\psi$ peak, $\pup \, \bar p^\uparrow \to J/\psi \, X \to 
\ell^+ \, \ell^- \, X$~\cite{Anselmino:2004ki}. The availability of a polarised 
antiproton beams proves, however, to be a very difficult task~\cite{Barone:2005pu}.          
\end{itemize}
 
\section{The ultimate goal: the nucleon Wigner functions}

\begin{figure}[t]
\begin{center}
\vskip -36 pt
\includegraphics[width=18.truecm,angle=0]{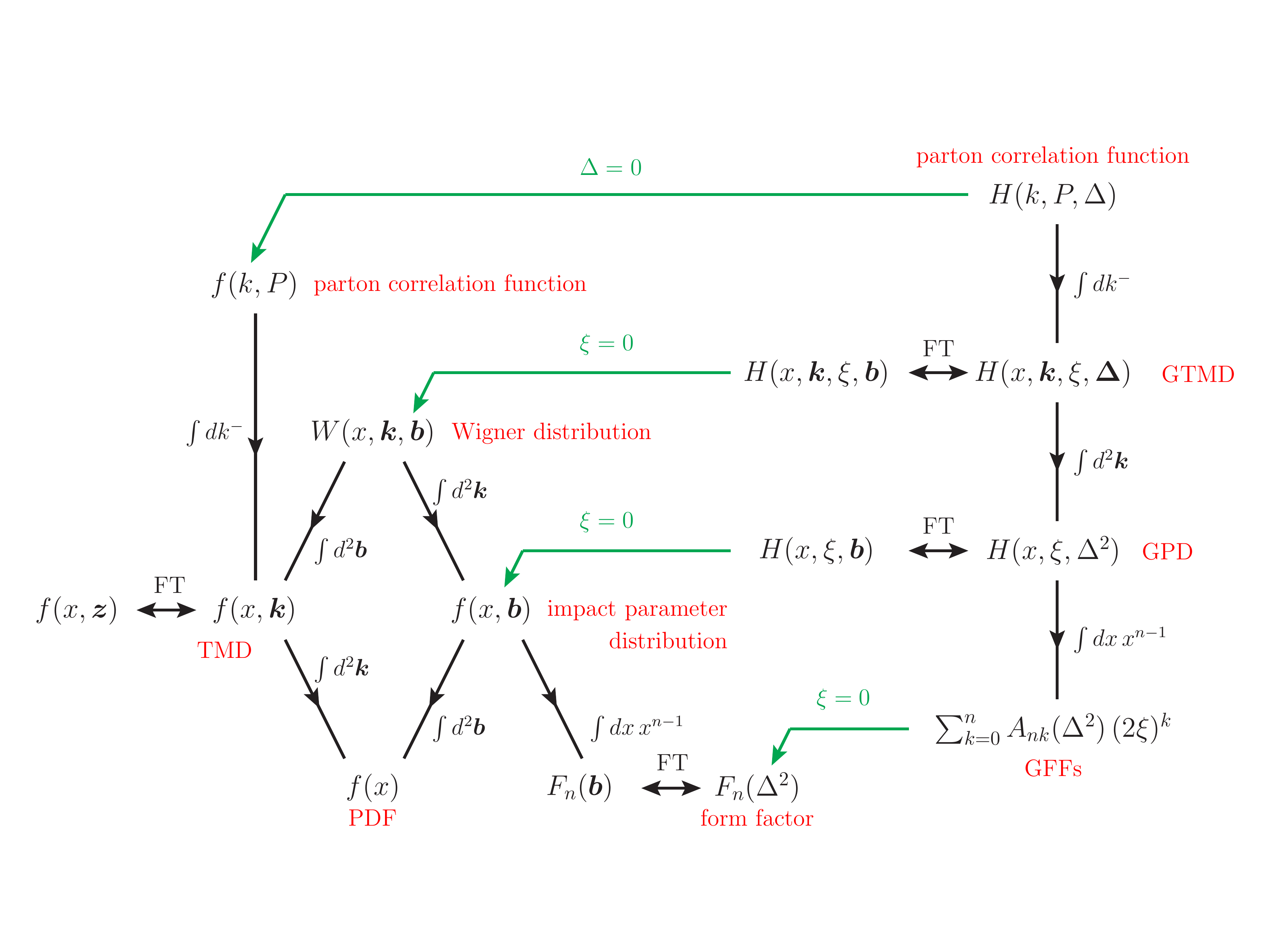}
\vskip -36pt 
\caption{Summary of the relations between different distributions and correlators. 
The Table is reprinted from Ref.~\cite{Diehl:2015uka}, where more details can be 
found. Notice that the vectors $\bfk$, $\bfb$ and $\bfD$ of the figure are 
defined respectively as $\bkt$, $\bbt$ and $\bDt$ in the text.
Reprinted with kind permission of The European Physical Journal (EPJ), Markus Diehl, 
"Introduction to GPDs and TMDs", Eur. Phys. J., A52(6):149, 2016, 
\textcopyright Societ\`{a} Italiana di Fisica/ Springer-Verlag 2016.}
\label{wigner}
\end{center}
\end{figure}

The TMDs give a three-dimensional momentum space information about the quarks and 
gluons inside the nucleon. They do not give any information about their spatial 
distribution. Spatial information about the quarks and gluons can be obtained in 
terms of nucleon form factors; the Fourier transform of the form factor gives the 
charge distribution of the nucleon. Form factors are obtained by taking moments 
of the Generalised Parton Distributions (GPDs), which are defined as off-forward 
matrix elements of quarks and gluon operators, in a generalisation of 
Eq.~(\ref{corr-def}) in which the initial and final proton momenta differ by an 
amount usually defined as $\xi$ (for the longitudinal direction) and 
$\bfD_\perp$ or $\bfq_\perp$ (for the transverse direction). GPDs can be accessed 
in exclusive processes and are not discussed here; a seminal review paper can be 
found in~\cite{Diehl:2003ny}. In Ref.~\cite{Burkardt:2000za} it was shown that a 
Fourier transform of the GPDs with respect to the transverse momentum transfer, 
$\bfD_\perp$, taken at $\xi = 0$, gives Impact Parameter Dependent Parton 
Distribution Functions (IPDPDFs or, shortly, IPDs)). The IPDs give the density 
of partons with light-cone momentum fraction $x$ and transverse impact parameter 
$\bfb_\perp$. However, the most general information about the quark and the
gluon distribution in the nucleon could be obtained in terms of the phase space 
or Wigner distributions, and the Generalised Transverse Momentum Dependent parton 
distributions (GTMDs), which we shall discuss in the next Sections. The Wigner 
function is related, by some integration and by taking particular limits or 
Fourier transforms, to TMDs, GPDs, PDFs, form factors, IPDs and GTMDs. A 
comprehensive table of all these connections can be found in 
Ref.~\cite{Diehl:2015uka} and is shown for convenience in Fig.~\ref{wigner}.

\subsection{\it Introduction to Wigner distributions for quarks and gluons}

In classical mechanics, the laws of dynamics are formulated in phase space, 
that describes the position $\bfr$ as well as the momentum $\bfp$ of each particle.  
In quantum mechanics, due to Heisenberg's uncertainly principle, the position and 
momentum of a particle cannot be determined simultaneously. In 1932, 
Wigner~\cite{Wigner:1932eb} introduced a quantum phase space distribution, which,
in one space and one momentum dimensions, reads: 
\be
W(x,p) = \int d \eta \, e^{i p \eta} \, \psi^*(x -\eta/2) \, \psi(x+\eta/2) \>.
\ee
$\psi(x)$ is the wave function of the system at the position $x$ and $p$ is the 
conjugate momentum. Wigner distributions are the analog of classical phase space 
distributions, to which they reduce in the classical limit. 
They have been used widely in different branches of physics, for example in 
quantum information, heavy ion physics, nonlinear dynamics, optics, image 
processing and so on~\cite{Balazs:1983hk,Hillery:1983ms,Lee}. The quantum 
mechanical Wigner distributions have also been measured in some particular 
cases~\cite{Vogel:1989zz, Smithey:1993zz, Banaszek:1999ya}. Wigner distributions 
by themselves are not positive definite, and do not have a probabilistic 
interpretation; however, upon integration over $p$ or $x$ they give respectively 
the probability density of the quantum system in $x$ or $p$ space. A few modified 
Wigner type distributions have been used in the literature, notably the Husimi 
distributions~\cite{Hagiwara:2014iya}, that are positive definite. 

Wigner type phase space distributions for nucleons were introduced in 
Refs.~\cite{Ji:2003ak,Belitsky:2003nz}. The impact parameter dependent parton 
distributions (IPDs) originally introduced by Burkardt had the skewness 
$\xi=0$, in which limit they have a probability interpretation. A generalisation 
to non-zero $\xi$ was discussed in \cite{Diehl:2002he}. 
Refs.~\cite{Brodsky:2006ku,Brodsky:2006in} discussed the Compton scattering 
amplitudes in longitudinal position space by taking their Fourier transform with 
respect to the skewness. The two dimensional position space picture of the 
IPDs was extended to three dimensions, in the rest frame of the nucleon, in 
Ref.~\cite{Ji:2003ak}. The Wigner operator for quarks can be defined as 
\be
\hat W^\Gamma (\bfr, k) = \int d^4 \eta ~e^{i k \cdot \eta}~ {\bar{\psi}} ( \bfr-\eta/2) 
\, \Gamma \, \psi ( \bfr +\eta/2) \>,
\ee
where $\Gamma$ is a Dirac matrix structure and $k$ is the $4$-momentum conjugate 
to the space-time separation $\eta$. A gauge link, not shown, has to be included 
for the color gauge invariance of QCD. 
The corresponding Wigner function for a non-relativistic system can 
be defined by taking an expectation value of the above operator for a state with 
the center-of-mass at ${\bf{R}} = 0$. As the proton is a relativistic object, 
recoil effects cannot be neglected, and the rest frame state cannot be uniquely 
defined. The Wigner distribution for the proton is defined in the Breit 
frame~\cite{Ji:2003ak,Belitsky:2003nz}. The most general Wigner distribution 
is a function of seven variables, three positions, and four momenta. Integrating 
out the light cone energy $k^-$ of the quarks, one gets the six dimensional 
reduced Wigner distribution. A five dimensional Wigner distribution was introduced 
in~\cite{Lorce:2011kd} in the infinite momentum frame or light-cone formalism by 
defining the Wigner operator for quarks at fixed light-cone time $z^+=0$:
\be
\hat W^\Gamma (\bbt, \bkt, x) = {1\over 2} \int \frac{dz^- \, d^2\bzt}
{(2 \pi)^3} \, e^{i (x p^+ z^- - \bkt \cdot \bzt)} \,
{{\bar{\psi}} ( y-z/2) \, \Gamma \, \mathcal{W} \, \psi (y+z/2)}\mid_{z^+=0} \>.
\label{W-op}
\ee
Here $y^\mu =\{ 0,0, \bbt \}$,  $x$ is the  average light-cone longitudinal 
momentum fraction of the nucleon carried by  the quark, $x = k^+/p^+$, 
$\Gamma= \gamma^+, \gamma^+ \gamma_5 , i \sigma^{j+} \gamma_5$ with $j=1,2$ at 
twist two; different choices of $\Gamma$ would give different phase space 
distribution. $\mathcal{W}$ is the gauge link or Wilson line. It should be noted 
that in the above expression $\bkt$ and $\bbt$ are not Fourier conjugate variables. 
$\bkt$ is the average transverse momentum of the active quark. The transverse 
impact parameter $\bbt$ is the conjugate of the momentum transfer $\bDt$ from 
the initial to the final state. Most studies of the Wigner functions in the 
literature consider the Drell-Yan-West frame, where $\Delta^+=0$. Including a 
non-zero $\Delta^+$ would spoil the semi-classical probabilistic interpretation 
of the Wigner functions. 
 
The Wigner distribution is defined, for a nucleon state with polarisation $\bfS$,
as:
\be 
\rho^\Gamma (\bbt, \bkt, x , \bfS) = \int {d^2 \bDt \over {( 2 \pi)}^2 } 
\, e^{-i \bDt \cdot \bbt} \, W^\Gamma ( \bDt, \bkt, x, \bfS) \>,
\label{wigner_def}
\ee
where,
\be
W^\Gamma (\bDt, \bkt, x, \bfS)= \langle p^+, {\bDt/2}, \bfS \mid \hat W^\Gamma 
({\bf{0}}_\perp, \bkt, x)   \mid p^+,-{\bDt/2}, \bfS \rangle \>.
\label{W-dist}
\ee
As  transverse boosts are Galilean in light-front framework, one can construct 
a nucleon state localised in transverse coordinates, with its transverse 
center-of-momentum fixed. Thus, Eq.~(\ref{wigner_def}) is consistent with 
special relativity. Different choices of $\Gamma$ matrix give different Wigner 
distributions, that probe various combinations of the quark and nucleon 
polarisation. 

\subsection{\it Wigner distributions and Generalized Transverse Momentum Dependent 
parton distribution functions (GTMDs)} 

\begin{figure}[t]
\begin{center}
\vskip -120 pt
\includegraphics[width=18.truecm,angle=0]{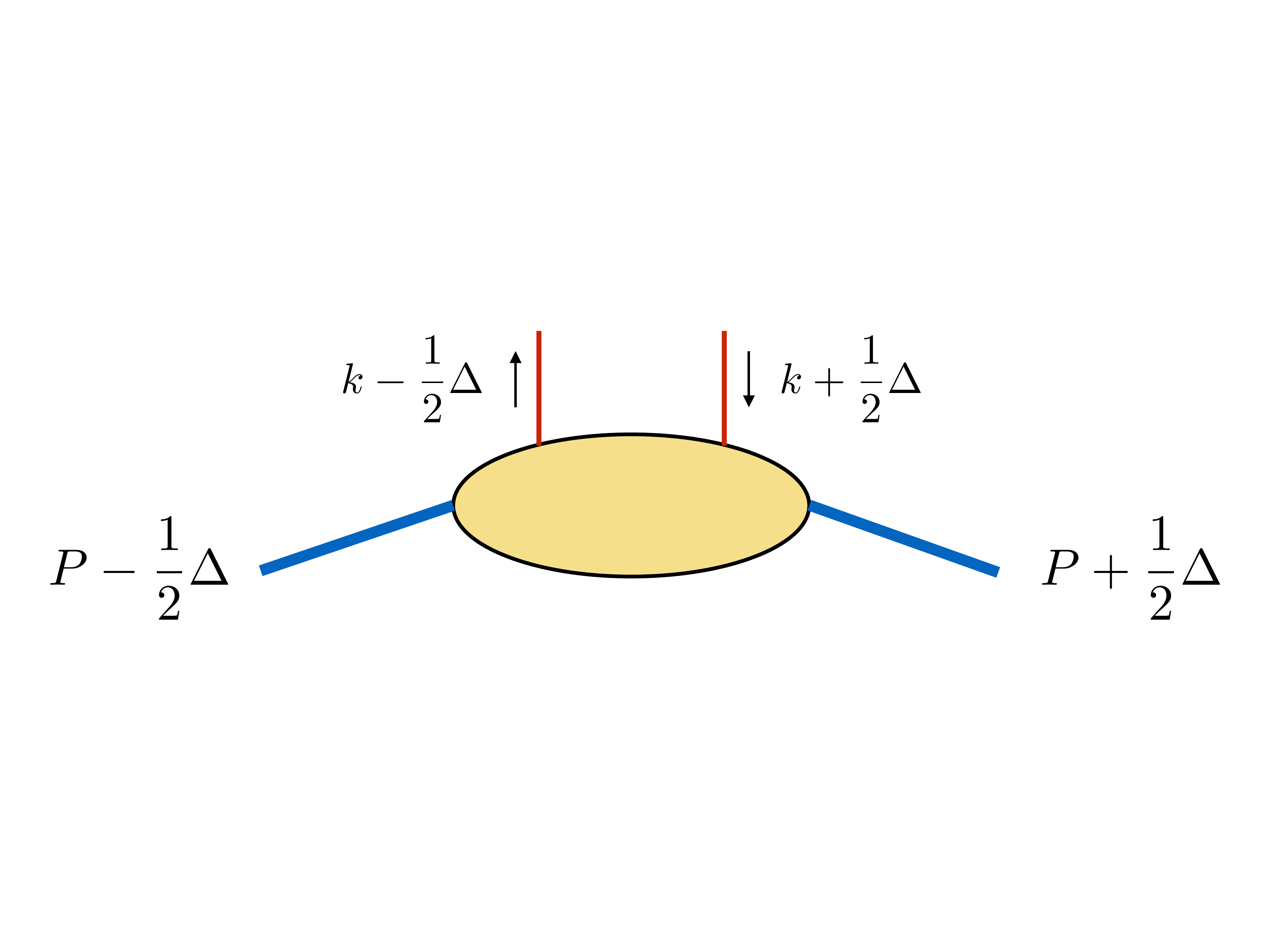}
\vskip -130pt 
\caption{The most general off-forward quark-quark correlator; $P$, $\Delta$ and $k$
are unintegrated 4-momenta. When $\Delta = 0$ one recovers the correlator, for 
inclusive processes, of Fig.~\ref{correlator}.}
\label{corr-gen}
\end{center}
\end{figure}

There is a direct relation between the Wigner distributions and the most general 
parton correlation functions~\cite{Meissner:2009ww}. These are the completely 
unintegrated parton correlators of the nucleon, described in 
Fig.~\ref{corr-gen} and defined as~\cite{Diehl:2015uka}: 
\be
H^\Gamma(k, P, \Delta) = (2\pi)^{-4} \int d^4z \, e^{ikz} \>
\langle p(P + \frac 12 \Delta)| \, \bar q(-\frac 12 z) \, \Gamma \, q(\frac 12 z) 
\, |p(P + \frac 12 \Delta) \rangle \>,
\ee
where the Dirac matrix $\Gamma$ selects the parton spin degrees of freedom; the 
proton spin labels and the necessary gauge link, or Wilson operator, between the 
quark fields have been omitted.  

Integrating $H^\Gamma$ over the light-cone energy of the quark, $k^-$, see 
Fig.~\ref{wigner}, one gets the Generalized Transverse Momentum Dependent 
parton distributions (GTMDs).  
The Wigner distributions are Fourier transforms of the GTMDs. The most general 
parametrization of the GTMD correlator for a nucleon is given in 
Refs.~\cite{Meissner:2009ww,Lorce:2013pza}. At leading twist,  there are 16 GTMDs 
depending on the choice of $\Gamma$. The GTMDs are in general complex quantities.
They are constrained by Hermiticity and time reversal \cite{Lorce:2013pza} 
properties.  The GTMDs can be written as
\be
X(P, k, \Delta, \eta) =  X^e(P, k, \Delta) +i X^o (P, k, \Delta, \eta) \>,
\ee
where $\eta$ denotes the direction of the gauge link (plus or minus). 
$X^e$ is even under time reversal and $X^o$ is odd.
Due to the Hermiticity property of the GTMDs, the Wigner distributions, which are 
2D Fourier transforms of the GTMDs, are real, although they can also be separated 
into a T-even and T-odd part. In the limit $\bDt=0$, the GTMDs reduce to the TMDs.
It can be shown that due to the Hermiticity contraint all GTMDs which are odd 
in $\bDt$ vanish in the TMD limit. Upon integration over the parton momentum 
$\bkt$, they are related to the GPDs. In this case all effects of the T-odd part 
of the GTMDs vanish and there is no dependence on $\eta$. Thus, the GTMDs and 
the Wigner functions can be thought of as ``mother distributions", providing 
all information about the quark and gluon distributions of the nucleon, and 
beyond. A complete list of the relations between the various GTMDs and the 
corresponding TMDS and GPDs upto twist-four is given in \cite{Meissner:2009ww}. 
Some nontrivial relations between  GPDs and TMDs can be understood in terms of 
GTMDs. Three model independent relations exist at twist-two level between the 
GPDs and TMDs, these connect the GPD $H(x,0,0)$ to the moment of $f_1(x, k_T^2)$, 
$\tilde H(x,0,0)$ to the moment of $g_{1L}(x, k_T^2)$ and $H_T(x,0,0)$ to a 
linear combination of moments of $h_{1T} (x, k_T^2)$ and $h_{1T}^\perp(x, k_T^2)$. 
These are model independent relations as the corresponding GPDs and TMDs are 
related to the same GTMDs. On the other hand, in spectator type models relations 
have been derived connecting the Sivers function to the GPD $E$ and Boer-Mulders 
function to a combination of  GPDs $E_T$ and $\tilde H_T$. These are called 
relations of the second kind and cannot be derived in a model independent way 
as these GPDs and TMDs are not related to the same GTMDs \cite{Meissner:2009ww}. 
There are also relations of the third kind for example connecting the T-even 
pretzelosity distribution to the GPD $\tilde H_T$. This is also a model 
dependent relation as these are related to different GTMDs. At twist three 
and twist four there are  no non-trivial model independent relations between 
GPDs and TMDs.

Unlike quark GTMDs, gluon GTMDs and unintegrated gluon distributions are known 
in the literature since a long time: for example, they were discussed in the 
small $x$ regime in~\cite{Martin:1999wb} in the context of diffractive vector 
meson production process and in~\cite{Khoze:2000cy} for Higgs production. 
Parametrizations of gluon GTMDs are given in Ref.~\cite{Lorce:2013pza}. 

\subsection{\it Definitions}

The Wigner distributions are not positive definite. Integrating out one or more 
variables they can be related to better known objects: integrating over $\bbt$ 
effectively sets $\bDt$ to zero and one obtains the standard TMD correlator; on 
the other hand, by integration over $\bkt$, the Wigner distributions reduce to the 
impact parameter dependent PDFs (IPDs), which are the Fourier transforms of 
the GPD correlations. TMDs and IPDs can be interpreted as densities in 
momentum and transverse position space respectively. New distributions are 
obtained from $\rho^\Gamma (\bbt, \bkt, x , \bfS)$ by integrating over 
$k_x$ and $b_y$ or $k_y$ and $b_x$; these are not related to any known TMDs 
or GPDs, and therefore carry further information beyond what can be probed by 
the TMDs and GPDs. At leading twist one can define 16 independent Wigner 
distributions corresponding to different quark and proton polarisations ($U$, $L$
and $T$)~\cite{Lorce:2011kd,Liu:2015eqa}. In the list below the first subscript 
refers to the proton and the second to the quark; $\hat\bfe_j$ ($j=1,2,3$ or 
$x,y,z$) denote the unit vectors along the coordinate axes and the nucleon moves 
along $\hat\bfe_z$.    

\subsubsection{Unpolarized target and different quark polarizations}

\noindent
The unpolarized Wigner distribution: 
\bea
\rho_{UU}(\bbt, \bkt, x)&=&\frac1{2}\Big[\rho^{[\gamma^+]}(\bbt, \bkt, x,\hat \bfe_z)
+\rho^{[\gamma^+]}(\bbt, \bkt, x,-\hat \bfe_z)\Big] \>.
\label{rhouu} 
\eea
The unpolarized-longitudinally polarized Wigner distribution:
\bea
\rho_{UL}(\bbt, \bkt, x)&=&\frac1{2}\Big[\rho^{[\gamma^+\gamma^5]}(\bbt, \bkt, x,
\hat \bfe_z)+\rho^{[\gamma^+\gamma^5]}(\bbt, \bkt, x,-\hat \bfe_z)\Big] \>.
\eea
The unpolarized-transversely polarized Wigner distribution ($j = 1,2$):
\bea
\rho^j_{UT}(\bbt, \bkt, x)&=&\frac1{2}\Big[\rho^{[i \sigma^{+j}\gamma^5]}(\bbt, \bkt, x,\hat \bfe_z)+\rho^{[i \sigma^{+j}\gamma^5]}(\bbt, \bkt, x,-\hat \bfe_z)\Big] \>.
\eea

\subsubsection{Longitudinally polarized target and different quark polarizations}

\noindent
The longitudinal-unpolarized Wigner distribution:
\bea
\rho_{LU}(\bbt, \bkt, x)&=&\frac1{2}\Big[\rho^{[\gamma^+]}(\bbt, \bkt, x,\hat \bfe_z)
-\rho^{[\gamma^+]}(\bbt, \bkt, x,-\hat \bfe_z)\Big] \>.
\eea
The longitudinal Wigner distribution:
\bea
\rho_{LL}(\bbt, \bkt, x)&=&\frac1{2}\Big[\rho^{[\gamma^+\gamma^5]}(\bbt, \bkt, x,
\hat \bfe_z)-\rho^{[\gamma^+\gamma^5]}(\bbt, \bkt, x,-\hat \bfe_z)\Big] \>.
\label{rholl}
\eea
The longitudinal-transversely polarized Wigner distribution ($j = 1,2$):
\bea
\rho^j_{LT}(\bbt, \bkt, x)&=&\frac1{2}\Big[\rho^{[i \sigma^{+j}\gamma^5]}(\bbt, \bkt, 
x,\hat \bfe_z)-\rho^{[i \sigma^{+j}\gamma^5]}(\bbt, \bkt, x,-\hat \bfe_z)\Big] \>.
\eea

\subsubsection{Transversely polarized target and different quark polarizations}

\noindent
The transverse-unpolarized Wigner distribution ($j = 1,2$):
\bea
\rho^j_{TU}(\bbt, \bkt, x)&=&\frac1{2}\Big[\rho^{[\gamma^+]}(\bbt, \bkt, x,
\hat \bfe_j)-\rho^{[\gamma^+]}(\bbt, \bkt, x,-\hat \bfe_j)\Big] \>.
\eea
The transverse-longitudinally polarized  Wigner distribution ($j = 1,2$):
\bea
\rho^j_{TL}(\bbt, \bkt, x)&=&\frac1{2}\Big[\rho^{[\gamma^+\gamma^5]}(\bbt, \bkt, x,
\hat \bfe_j)-\rho^{[\gamma^+\gamma^5]}(\bbt, \bkt, x,-\hat \bfe_j)\Big] \>.
\eea
The transversely polarised  Wigner distribution ($j,k = 1,2$):
\bea
\rho_{TT}^{jk}(\bbt, \bkt, x)&=&\frac1{2}\delta_{jk}\Big[\rho^{[i \sigma^{+k}
\gamma^5]}(\bbt, \bkt, x,\hat \bfe_j)-\rho^{[i \sigma^{+k}\gamma^5]}(\bbt, \bkt, x,
-\hat \bfe_j)\Big] \>.
\label{rhott}
\eea
The pretzelous Wigner distribution ($j,k = 1,2$): 
\bea
\label{pretz}
\rho^{\perp jk}_{TT}(\bbt,\bkt, x)&=& \frac1{2}\epsilon_{jk}\Big[\rho^{[i \sigma^{+k}
\gamma^5]}(\bbt, \bkt,x,\hat \bfe_j)-\rho^{[i \sigma^{+k}\gamma^5]}(\bbt, \bkt, x,
-\hat \bfe_j)\Big] \>.
\eea

\subsubsection{Properties: connection to GPDs and TMDs and orbital angular momentum}

The Wigner distributions could, in principle, offer a complete information on 
the nucleon structure and a few comments might help in understanding their 
importance and their relations with TMDs. 
\begin{itemize}
\item
In the above definitions $\bfS = +\hat \bfe_z$ ($- \hat \bfe_z$) corresponds 
to helicity $+$ ($-$) of the target state. 
$\pm\hat \bfe_j$ ($j = 1,2$) correspond to transversity states and can be 
expressed in terms of helicity states. Notice that in the pretzelous Wigner 
function the transverse polarisations of the quark and of the proton are in 
orthogonal directions. 
\item
By integration over $x$ one gets the Wigner distributions in $\bbt$ and $\bkt$ 
space. $\rho_{UU}(\bbt, \bkt)$ probes unpolarized quarks in unpolarized nucleon. 
Any distortion in the $\bbt$ or $\bkt$ space is a measure of the effect of the 
orbital motion of the quarks. 

$\rho_{LU}$ gives the distortion in the $\bbt$ 
and $\bkt$ space in the distribution of the unpolarised quarks due to the 
longitudinal polarisation of the proton. In fact, this is related to the quark 
Orbital Angular Momentum (OAM) \cite{Lorce:2011kd,Lorce:2011ni,Hatta:2011ku}:
\be
l^q_z= \int \! dx \int \! d^2 \bbt \int \! d^2 \bkt \, (\bbt \times \bkt)_z 
\, \rho^q _{LU} (\bbt, \bkt, x) \>. \label{oam}
\ee
At the density level, {\it i.e.} without integrating over $x$, the above expression 
gives the canonical OAM of the quark, when the gauge link in the  Wigner 
distribution [see Eqs.~(\ref{wigner_def}), (\ref{W-dist}) and (\ref{W-op})]
is staple-like, irrespective of its direction, future pointing or past 
pointing~\cite{Leader:2013jra}.   
This reduces to the Jaffe-Manohar OAM \cite{Jaffe:1989jz} in the light-cone gauge.   
On the other hand, for a straight line gauge link, Eq.~(\ref{oam}) gives the 
kinetic OAM, which is related to the GPD $E$ through Ji's relation \cite{Ji:1996nm}. 
The kinetic and canonical OAM are related through a potential term 
(see Refs.~\cite{Leader:2013jra} and \cite{Wakamatsu:2014zza} for a review). 
The kinetic OAM is related to the pretzelosity distribution in some models.
The relation of quark and gluon OAM to unintegrated correlators was first 
discussed in Ref.~\cite{Hagler:2003jw}. 

$\rho_{UL}$ probes longitudinally polarised quarks in unpolarised proton, and 
this is related to the correlation between quark spin and OAM,
\be
C^q_z= \int \! dx \int \! d^2 \bbt \int \! d^2 \bkt \, (\bbt \times \bkt)_z 
\, \rho^q _{UL} (\bbt, \bkt, x) \>. \label{soam}
\ee
\item 
By integration over $d^2\bbt$, as shown in Fig.~\ref{wigner}, 8 of the 
16 Wigner distributions, (\ref{rhouu})--(\ref{pretz}), reduce to the 
leading twist distributions of Eq.~(\ref{TMD-noi}), while the other 8 vanish. 
Similarly, by integration over $d^2\bkt$ one recovers the eight leading twist 
IPDs, which have not been discussed here.

In particular, $\rho_{UU}$ is related to the unpolarised quark TMD, 
$f_1(x,\kt)$. $\rho_{LL}$ gives the correlation of the longitudinal spin 
of the quark in a longitudinally polarised proton and in the TMD limit it  
is related to the helicity distribution $g_{1L}(x,\kt)$. The TMD limit of 
$\rho_{TU}$ and $\rho_{UT}$ are respectively connected to the T-odd Sivers 
function, $f_{1T}^\perp$, and Boer-Mulders function, $h_1^\perp$.

$\rho_{LT}$ gives the correlations between a transversely polarized quark 
in a longitudinally polarised proton. In the TMD limit, $\rho_{LT}$ it is 
related to the longitudinal-transversity worm-gear TMD $h_{1L}^\perp$. 
Similarly, $\rho_{TL}$ describes the correlation between the longitudinally 
polarized quark in a transversely polarised proton and in the TMD limit  
is linked to the transverse-helicity worm-gear TMD $g_{1T}^\perp$.
$\rho_{TT}$ and $\rho_{TT}^\perp$ are related, in the TMD limit, to the 
transversity and pretzelosity TMDs, $h_{1T}$ and $h_{1T}^\perp$.

Notice that by integration over $d^2\bkt$ or $d^2\bbt$ both $\rho_{LU}$ and 
$\rho_{UL}$ vanish: they do not have a TMD or GPD limit. Thus, they carry 
completely new information about the nucleon structure, as shown in 
Eqs.~(\ref{oam}) and (\ref{soam}). 

Multipole decomposition of the Wigner distributions were investigated 
in Ref.~\cite{Lorce:2015sqe}.
\end{itemize}       

\subsection{\it Model Calculations} 
\begin{figure}[t]
\begin{center}
\vskip -18 pt
\includegraphics[width=18.truecm,angle=0]{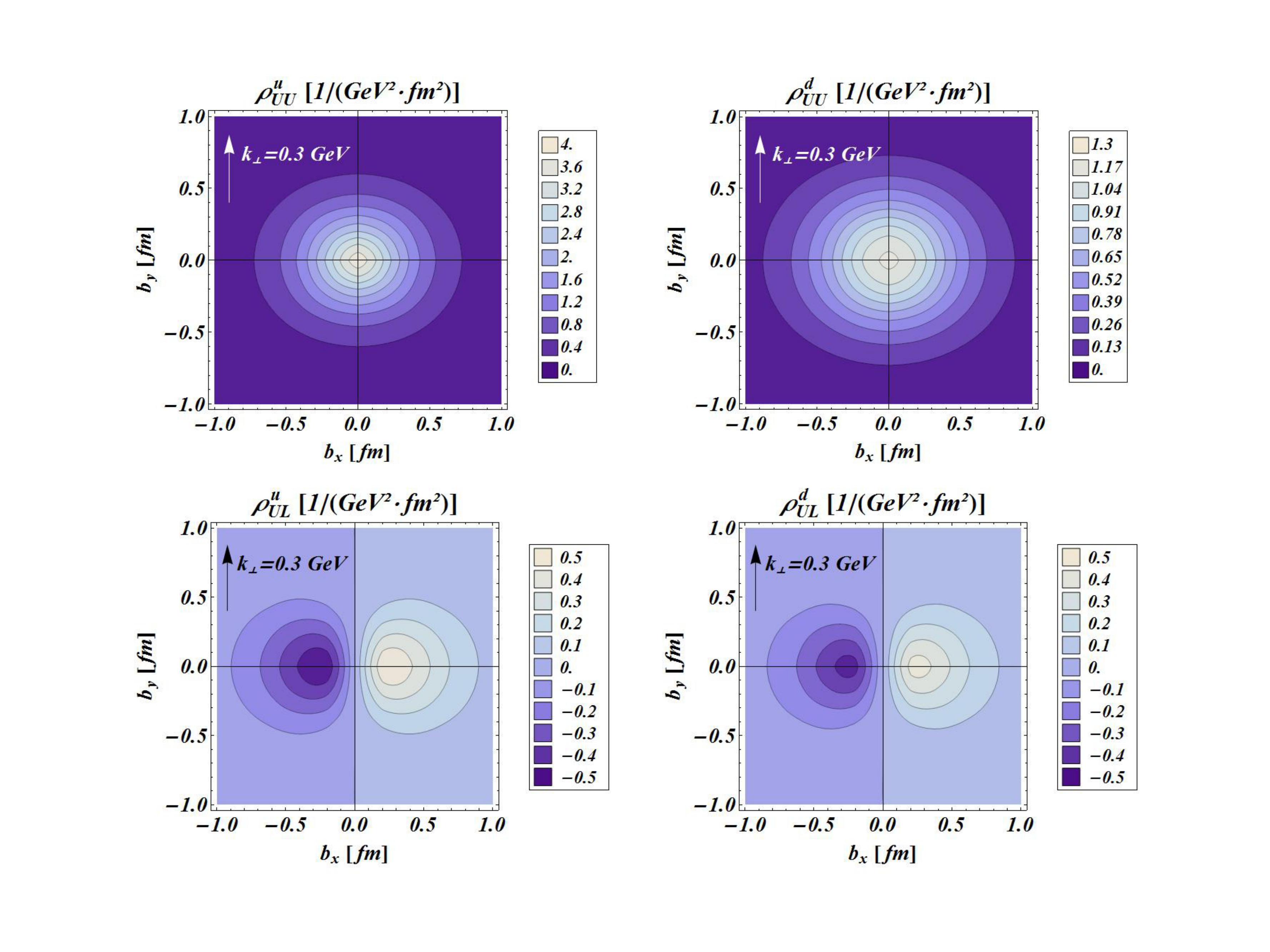}
\vskip -36pt 
\caption{%(colour online) 
Wigner distributions for $u$ and $d $ quarks in $\bbt$ space, 
at fixed $\bkt = \kt \hat\bfe_y$ with $\kt = 0.3$ GeV, in a constituent 
quark model, from Ref.~\cite{Lorce:2011kd}. The upper panels refer to unpolarised 
protons and the lower ones to longitudinally polarised ones. Reprinted with 
permission from C. Lorc\'{e} and B. Pasquini, Phys. Rev., D84:014015, 2011. 
Copyright 2011 by the American Physical Society.}
\label{Wigner-model}
\end{center}
\end{figure}
Model calculations of the Wigner distributions and the GTMDs are important as 
so far they have not been extracted from experimental data. In 
Ref.~\cite{Lorce:2011kd} the quark Wigner distributions for unpolarised as well 
as longitudinally polarised protons were investigated in a light-cone constituent 
quark model as well as in a light-cone chiral quark soliton model, using the 
valence light-front wave function (LFWF). The results for these two models are
rather similar and some of them are shown as examples of 3D imaging of the proton. 

The two upper plots in Fig.~\ref{Wigner-model} show the transverse phase space 
distribution $\rho_{UU} (\bbt,\bkt)$ obtained by integrating the Wigner 
distribution over $x$, for unpolarised $u$ and $d$ quarks inside an unpolarised 
proton. The results are given in the $b_x$ and $b_y$ plane, at fixed value of 
$\bkt = \kt \hat\bfe_y$ with $\kt = 0.3$ GeV.   
It is seen that the distribution is spread in the direction 
$\bbt \perp \bkt$ more than $\bbt \parallel \bkt$. This is expected in a model 
with confinement, where the radial momentum of a quark decreases at the periphery  
of the $\bbt$ space and the polar momentum dominates due to the OAM.
 
The two lower plots in Fig.~\ref{Wigner-model} show the same $\bbt$ distribution
for a longitudinally ({\it i.e.} along $\hat\bfe_z$) polarised proton. 
In this case $\rho_{UL}(\bbt,\bkt)$ shows the effect of the net OAM,
which for $u$ quarks tend to align with the nucleon spin whereas for
$d$ quarks is anti-aligned. The Jaffe-Manohar OAM and Ji OAM have also
been computed in this Ref.~\cite{Lorce:2011kd}. Both definitions of the OAM 
give the same result for the total quark ($u+d$) OAM in the two models, although
contributions from each flavour varies. This is expected as there is no gluon in 
these models. 

All the 16 leading twist quark Wigner distributions for a nucleon have been 
calculated in Ref.~\cite{Liu:2015eqa} in a light cone spectator model, with 
the inclusion of both scalar and axial vector diquarks.  Dipole and 
quadrupole structures are seen  for the Wigner distributions
that are related to the correlations between the spin and OAM of the
quarks and the nucleon, as well as the transverse momentum. Results in
this model show some difference compared to the models in
\cite{Lorce:2011kd} in impact  parameter space. Both for $u$ and $d$
quarks, the spin-orbit correlation $C^q_z$ is negative in this model
which is opposite to what is observed in light cone constituent quark
model \cite{Lorce:2011kd}.  This is  a model dependent result. The 16
leading twist Wigner distributions were investigated in a diquark model
in \cite{Chakrabarti:2017teq} including both scalar and axial vector
diquarks, the analytic form of the light-front wave function (LFWF) was
obtained using soft wall ADS/QCD prediction. Unlike the other models,
there is no favored configuration between $\bbt$ and $\bkt$ here. In this
model $C^q_z$ is negative, meaning  the quark OAM is anti-aligned to
quark spin which agrees with the observation in scalar diquark model
\cite{Liu:2015eqa,Chakrabarti:2016yuw}. $\rho_{UT}$ and $\rho_{TU}$ show a 
dipolar behaviour which is similar to that of Ref.~\cite{Liu:2015eqa}. 
The above phenomenological models do not have gluonic degrees of freedom  
and the calculations describe the T-even part of the Wigner distributions.  

The quark OAM through GTMDs was calculated in Ref.~\cite{Courtoy:2016des} in the
MIT bag model. In Ref.~\cite{Hagiwara:2016kam} gluon Wigner, Husimi distributions 
and GTMDs were investigated in color glass condensate in the small $x$ regime 
taking into account the gluon saturation effects. The Husimi distribution is 
obtained from the Wigner distribution by using a Gaussian smearing in $\bbt$ 
and $\bkt$. As a result the Husimi distributions do not reduce to GPDs or TMDs 
after integration over $\bkt$ or $\bbt$. However, the Husimi distributions 
are positive and unlike Wigner distributions can be interpreted as probability
distributions in phase space.   

The above phenomenological models for the nucleon do not include gluonic 
degrees of freedom. In 
Refs.~\cite{Mukherjee:2014nya, Mukherjee:2015aja, More:2017zqq, More:2017zqp} 
the quark and gluon Wigner distributions and GTMDs were calculated for a spin-1/2 
system of a quark dressed with a gluon, which may be thought of as a 
relativistic composite system of a quark and a gluon. The quark and gluon OAM 
and spin-orbit correlations were also calculated in this model. A similar model 
was also used in Ref.~\cite{Kanazawa:2014nha} for the GTMDs $F_{14}$ and $G_{11}$ 
and the results agree. The results for the quark and the gluon OAM and spin-orbit 
correlations were found to depend on the quark mass parameter, and the helicity 
sum rule is found to be satisfied. The Jaffe-Manohar OAM is different from 
Ji's OAM in this model. In Refs.~\cite{Mukherjee:2014nya, Mukherjee:2015aja, 
More:2017zqq, More:2017zqp} the gauge link was taken to be unity and as a result 
only the T-even part of the GTMDs were calculated. The T-odd part of the Wigner 
functions and the GTMDs were calculated in Ref.~\cite{Chakrabarti:2019wjx} by 
incorporating the final state interactions at the level of one gluon exchange 
in the LFWF. In a recent work \cite{Mueller:2019gjj} an ab initio world line 
approach was used to construct the phase space distributions of systems with 
internal symmetries. 
  
\subsection{\it Experiments to access the Wigner Distributions and GTMDs} 
  
Although several model calculations exist for both quark and gluon GTMDs 
and Wigner distributions, accessing them in experiments is still a challenge. 
Quite a few theoretical studies are available in the literature on how to probe  
the GTMDs in specific processes. In Ref.~\cite{Bhattacharya:2017bvs} it was 
shown that the quark GTMDs in the ERBL region $(-\xi < x < \xi) $ can be accessed  
in exclusive double Drell-Yan process in pion-nucleon scattering where one 
detects the two di-lepton pairs and the nucleon. As this process involve a 
staple-like Wilson line, the GTMDs here can probe the canonical or Jaffe-Manohar 
OAM. In Ref.~\cite{Bhattacharya:2018lgm} it was shown that the gluon GTMDs can 
be accessed in exclusive production of double pseudo-scalar quarkonium in 
nucleon-nucleon collision. Gluon Wigner distributions at small $x$ can be 
accessed  through the corresponding GTMDs in hard diffractive di-jet production  
at the future EIC~\cite{Hatta:2016dxp}. This process probes the dipole gluon 
GTMDs. The longitudinal SSA in this process is a direct probe of the gluon 
OAM and helicity at small $x$~\cite{Hatta:2016aoc} as well as moderate 
$x$~\cite{Ji:2016jgn}. Ref.~\cite{Hagiwara:2017fye} suggested to probe the 
gluon Wigner distributions in ultra-peripheral $pA$ collisions.  

\section{Conclusions} 

We have discussed our present knowledge of the 3-dimensional nucleon structure,
mainly in momentum space; that is, within the QCD parton model, we have summarised
the attempts, both experimental and theoretical, to understand the intrinsic 
motion of quarks and gluons inside the proton. 

First, we have presented plenty of experimental evidence, which requires the 
extension of the simple 1-dimensional picture of a fast moving proton as a set 
of almost free partons co-linearly moving, to a 3-dimensional picture including 
the transverse motion of partons. Such evidence is particularly striking when 
looking at experiments with transversely polarised protons; many transverse spin 
effects, for long time wrongly considered irrelevant or unnecessary, can only be 
understood in terms of the intrinsic 3D motion of quarks.      
 
Then, we have described the phenomenological approach which allows to gather 
3D information on the nucleon structure. The TMD-PDFs give the number densities 
of partons inside a proton, taking into account their longitudinal and transverse
momenta, together with the spin degrees of freedom. They often couple with 
TMD-FFs, giving the number densities of hadrons resulting in the fragmentation
of a quark or gluon: also in this case, the transverse momentum of the hadron 
with respect to the parton direction and the transverse spin degrees of freedom 
have to be taken into account. The TMD-PDFs and TMD-FFs appear in the 
theoretical expressions of several physical observables, which can be measured; 
thus, their extraction from data is possible. 

Some of the TMD-PDFs and FFs have been found to play crucial roles in SIDIS and 
Drell-Yan processes, as well as in $e^+e^-$ annihilations. In particular the 
effects resulting from the intrinsic motion of unpolarised quarks inside both 
unpolarised and transversely polarised protons have been well established and
explored. Similarly for the fragmentation of a transversely polarised quark 
into spinless hadrons. A first attempt at a 3D imaging of the proton in 
momentum space is possible, as shown in Fig.~\ref{Sivers-3D}. Notice that a 3D 
imaging in coordinate space is possible through the study of particular 
exclusive processes, which are not discussed in this paper.  
       
Finally, a full 3D imaging of the nucleon would be possible only through a 
knowledge of the Wigner distribution, the quantum analog of the classical 
phase-space distribution. This is a much more demanding task, although it 
should be the ultimate goal of a complete study of the nucleon structure.  
So far, models for the light-cone wave function of the proton have been used
to construct model Wigner functions.       

The introduction of the new concepts of TMD-PDFs and TMD-FFs has allowed 
a much better description of many, otherwise mysterious, transverse spin data;
 our way of exploring the nucleon structure has deeply changed, 
with a lot of encouraging and crucial results.
Much more work remains to be done. The validity 
of the TMD factorisation scheme, crucial for the phenomenological extraction 
of the TMDs from data, seems to be limited to few processes and particular 
kinematical ranges. The huge amount of data on the hyperon polarisation in 
unpolarised inclusive $p\,N$ interactions still lack a fundamental precise 
explanation, capable of reliable predictions. The origin of some TMDs, like 
the Sivers distribution, from basic QCD interactions, which leads to the 
prediction of the sign change of the Sivers function in SIDIS and D-Y 
processes, still has to be definitely tested. 

New precise data from the ``safe" processes -- SIDIS, D-Y and 
$e^+e^- \to h_1 \, h_2\, X$ -- will allow a much better determination of the 
TMDs and their parameterisation; in parallel, the theoretical study of the TMD 
evolution will improve and allow more precise predictions. New data are expected 
from the ongoing JLab experiments at 12 GeV and the next COMPASS run with a
transversely polarised deuteron target, from RHIC and the $e^+e^-$ facilities.
All these efforts should be combined with models of the proton wave function
and the Wigner distribution, from which TMDs and other observables can be 
computed and compared with data. In particular, the orbital angular momentum 
of quarks and gluons, which is a fundamental piece of information, is still 
unknown; its knowledge should finally allow the complete understanding of the 
proton spin. 

Great expectations, among the international hadron and nuclear physics community,
are linked to the planned Electron Ion Collider, a facility with good hopes of 
being built in the USA. Its scientific program~\cite{Accardi:2012qut} has a 
large part and strong motivations devoted to the study of the 3D nucleon 
structure, both in momentum and position space. Much larger kinematical ranges 
of $x$ and $Q^2$ could be explored, with high luminosity, providing a great 
amount of new data, and leading to a much improved understanding and imaging of 
the nucleon structure.      

\vskip 24pt 
\noindent
{\bf Acknowledgements}

M.A. would like to thank the Physics Department of the Indian Institute of 
Technology in Mumbai (IIT Bombay), where part of the paper was written, for 
hospitality and support. This material is based upon work supported by the 
U.S. Department of Energy, Office of Science, Office of Nuclear Physics under 
Award Numbers DE-SC0019230 and DE-AC05-06OR23177. 

\bibliographystyle{unsrturl}
\bibliography{main}

\begin{thebibliography}{100}

\bibitem{Barone:2001sp}
Vincenzo Barone, Alessandro Drago, and Philip~G. Ratcliffe.
\newblock {Transverse polarisation of quarks in hadrons}.
\newblock {\em Phys. Rept.}, 359:1--168, 2002.
\newblock \href {http://arxiv.org/abs/hep-ph/0104283}
  {\path{arXiv:hep-ph/0104283}}, \href
  {http://dx.doi.org/10.1016/S0370-1573(01)00051-5}
  {\path{doi:10.1016/S0370-1573(01)00051-5}}.

\bibitem{DAlesio:2007bjf}
U.~D'Alesio and F.~Murgia.
\newblock {Azimuthal and Single Spin Asymmetries in Hard Scattering Processes}.
\newblock {\em Prog. Part. Nucl. Phys.}, 61:394--454, 2008.
\newblock \href {http://arxiv.org/abs/0712.4328} {\path{arXiv:0712.4328}},
  \href {http://dx.doi.org/10.1016/j.ppnp.2008.01.001}
  {\path{doi:10.1016/j.ppnp.2008.01.001}}.

\bibitem{Barone:2010zz}
Vincenzo Barone, Franco Bradamante, and Anna Martin.
\newblock {Transverse-spin and transverse-momentum effects in high-energy
  processes}.
\newblock {\em Prog. Part. Nucl. Phys.}, 65:267--333, 2010.
\newblock \href {http://arxiv.org/abs/1011.0909} {\path{arXiv:1011.0909}},
  \href {http://dx.doi.org/10.1016/j.ppnp.2010.07.003}
  {\path{doi:10.1016/j.ppnp.2010.07.003}}.

\bibitem{Aidala:2012mv}
Christine~A. Aidala, Steven~D. Bass, Delia Hasch, and Gerhard~K. Mallot.
\newblock {The Spin Structure of the Nucleon}.
\newblock {\em Rev. Mod. Phys.}, 85:655--691, 2013.
\newblock \href {http://arxiv.org/abs/1209.2803} {\path{arXiv:1209.2803}},
  \href {http://dx.doi.org/10.1103/RevModPhys.85.655}
  {\path{doi:10.1103/RevModPhys.85.655}}.

\bibitem{Boglione:2015zyc}
Mariaelena Boglione and Alexei Prokudin.
\newblock {Phenomenology of transverse spin: past, present and future}.
\newblock {\em Eur. Phys. J.}, A52(6):154, 2016.
\newblock \href {http://arxiv.org/abs/1511.06924} {\path{arXiv:1511.06924}},
  \href {http://dx.doi.org/10.1140/epja/i2016-16154-6}
  {\path{doi:10.1140/epja/i2016-16154-6}}.

\bibitem{Avakian:2019drf}
H.~Avakian, B.~Parsamyan, and A.~Prokudin.
\newblock {Spin orbit correlations and the structure of the nucleon}.
\newblock {\em Riv. Nuovo Cim.}, 42(1):1--48, 2019.
\newblock \href {http://dx.doi.org/10.1393/ncr/i2019-10155-3}
  {\path{doi:10.1393/ncr/i2019-10155-3}}.

\bibitem{Anselmino:2015}
Editors, M.~Anselmino, M.~Guidal, and P.~Rossi.
\newblock {The 3-dimensional nucleon structure}.
\newblock {\em Eur. Phys. J.}, A52(6):149--164, 2016.

\bibitem{Accardi:2012qut}
A.~Accardi et~al.
\newblock {Electron Ion Collider: The Next QCD Frontier}.
\newblock {\em Eur. Phys. J.}, A52(9):268, 2016.
\newblock \href {http://arxiv.org/abs/1212.1701} {\path{arXiv:1212.1701}},
  \href {http://dx.doi.org/10.1140/epja/i2016-16268-9}
  {\path{doi:10.1140/epja/i2016-16268-9}}.

\bibitem{Collins:2011zzd}
John Collins.
\newblock {Foundations of perturbative QCD}.
\newblock {\em Camb. Monogr. Part. Phys. Nucl. Phys. Cosmol.}, 32:1--624, 2011.

\bibitem{Gourdin:1973qx}
M.~Gourdin.
\newblock {Semiinclusive reactions induced by leptons}.
\newblock {\em Nucl. Phys.}, B49:501--512, 1972.
\newblock \href {http://dx.doi.org/10.1016/0550-3213(72)90615-3}
  {\path{doi:10.1016/0550-3213(72)90615-3}}.

\bibitem{Kotzinian:1994dv}
Aram Kotzinian.
\newblock New quark distributions and semiinclusive electroproduction on the
  polarized nucleons.
\newblock {\em Nucl. Phys.}, B441:234--248, 1995.
\newblock \href {http://arxiv.org/abs/hep-ph/9412283}
  {\path{arXiv:hep-ph/9412283}}.

\bibitem{Mulders:1995dh}
P.~J. Mulders and R.~D. Tangerman.
\newblock {The Complete tree level result up to order 1/Q for polarized deep
  inelastic leptoproduction}.
\newblock {\em Nucl. Phys.}, B461:197--237, 1996.
\newblock [Erratum: Nucl. Phys.B484,538(1997)].
\newblock \href {http://arxiv.org/abs/hep-ph/9510301}
  {\path{arXiv:hep-ph/9510301}}, \href
  {http://dx.doi.org/10.1016/S0550-3213(96)00648-7,
  10.1016/0550-3213(95)00632-X} {\path{doi:10.1016/S0550-3213(96)00648-7,
  10.1016/0550-3213(95)00632-X}}.

\bibitem{Bacchetta:2004zf}
Alessandro Bacchetta, Piet~J. Mulders, and Fetze Pijlman.
\newblock {New observables in longitudinal single-spin asymmetries in
  semi-inclusive DIS}.
\newblock {\em Phys. Lett.}, B595:309--317, 2004.
\newblock \href {http://arxiv.org/abs/hep-ph/0405154}
  {\path{arXiv:hep-ph/0405154}}, \href
  {http://dx.doi.org/10.1016/j.physletb.2004.06.052}
  {\path{doi:10.1016/j.physletb.2004.06.052}}.

\bibitem{Diehl:2005pc}
M.~Diehl and S.~Sapeta.
\newblock {On the analysis of lepton scattering on longitudinally or
  transversely polarized protons}.
\newblock {\em Eur. Phys. J.}, C41:515--533, 2005.
\newblock \href {http://arxiv.org/abs/hep-ph/0503023}
  {\path{arXiv:hep-ph/0503023}}, \href
  {http://dx.doi.org/10.1140/epjc/s2005-02242-9}
  {\path{doi:10.1140/epjc/s2005-02242-9}}.

\bibitem{Bacchetta:2006tn}
Alessandro Bacchetta, Markus Diehl, Klaus Goeke, Andreas Metz, Piet~J. Mulders,
  and Marc Schlegel.
\newblock {Semi-inclusive deep inelastic scattering at small transverse
  momentum}.
\newblock {\em JHEP}, 02:093, 2007.
\newblock \href {http://arxiv.org/abs/hep-ph/0611265}
  {\path{arXiv:hep-ph/0611265}}, \href
  {http://dx.doi.org/10.1088/1126-6708/2007/02/093}
  {\path{doi:10.1088/1126-6708/2007/02/093}}.

\bibitem{Anselmino:2011ch}
M.~Anselmino, M.~Boglione, U.~D'Alesio, S.~Melis, F.~Murgia, E.~R. Nocera, and
  A.~Prokudin.
\newblock {General Helicity Formalism for Polarized Semi-Inclusive Deep
  Inelastic Scattering}.
\newblock {\em Phys. Rev.}, D83:114019, 2011.
\newblock \href {http://arxiv.org/abs/1101.1011} {\path{arXiv:1101.1011}},
  \href {http://dx.doi.org/10.1103/PhysRevD.83.114019}
  {\path{doi:10.1103/PhysRevD.83.114019}}.

\bibitem{Airapetian:2009ae}
A.~Airapetian et~al.
\newblock {Observation of the Naive-T-odd Sivers Effect in Deep-Inelastic
  Scattering}.
\newblock {\em Phys. Rev. Lett.}, 103:152002, 2009.
\newblock \href {http://arxiv.org/abs/0906.3918} {\path{arXiv:0906.3918}},
  \href {http://dx.doi.org/10.1103/PhysRevLett.103.152002}
  {\path{doi:10.1103/PhysRevLett.103.152002}}.

\bibitem{Airapetian:2010ds}
A.~Airapetian et~al.
\newblock {Effects of transversity in deep-inelastic scattering by polarized
  protons}.
\newblock {\em Phys. Lett.}, B693:11--16, 2010.
\newblock \href {http://arxiv.org/abs/1006.4221} {\path{arXiv:1006.4221}},
  \href {http://dx.doi.org/10.1016/j.physletb.2010.08.012}
  {\path{doi:10.1016/j.physletb.2010.08.012}}.

\bibitem{Adolph:2012sn}
C.~Adolph et~al.
\newblock {Experimental investigation of transverse spin asymmetries in muon-p
  SIDIS processes: Collins asymmetries}.
\newblock {\em Phys. Lett.}, B717:376--382, 2012.
\newblock \href {http://arxiv.org/abs/1205.5121} {\path{arXiv:1205.5121}},
  \href {http://dx.doi.org/10.1016/j.physletb.2012.09.055}
  {\path{doi:10.1016/j.physletb.2012.09.055}}.

\bibitem{Adolph:2012sp}
C.~Adolph et~al.
\newblock {II - Experimental investigation of transverse spin asymmetries in
  muon-p SIDIS processes: Sivers asymmetries}.
\newblock {\em Phys. Lett.}, B717:383--389, 2012.
\newblock \href {http://arxiv.org/abs/1205.5122} {\path{arXiv:1205.5122}},
  \href {http://dx.doi.org/10.1016/j.physletb.2012.09.056}
  {\path{doi:10.1016/j.physletb.2012.09.056}}.

\bibitem{Adolph:2014zba}
C.~Adolph et~al.
\newblock {Collins and Sivers asymmetries in muonproduction of pions and kaons
  off transversely polarised protons}.
\newblock {\em Phys. Lett.}, B744:250--259, 2015.
\newblock \href {http://arxiv.org/abs/1408.4405} {\path{arXiv:1408.4405}},
  \href {http://dx.doi.org/10.1016/j.physletb.2015.03.056}
  {\path{doi:10.1016/j.physletb.2015.03.056}}.

\bibitem{Adolph:2016dvl}
C~Adolph et~al.
\newblock {Sivers asymmetry extracted in SIDIS at the hard scales of the
  Drell-Yan process at COMPASS}.
\newblock {\em Phys. Lett.}, B770:138--145, 2017.
\newblock \href {http://arxiv.org/abs/1609.07374} {\path{arXiv:1609.07374}},
  \href {http://dx.doi.org/10.1016/j.physletb.2017.04.042}
  {\path{doi:10.1016/j.physletb.2017.04.042}}.

\bibitem{Qian:2011py}
X.~Qian et~al.
\newblock {Single Spin Asymmetries in Charged Pion Production from
  Semi-Inclusive Deep Inelastic Scattering on a Transversely Polarized $^3$He
  Target}.
\newblock {\em Phys. Rev. Lett.}, 107:072003, 2011.
\newblock \href {http://arxiv.org/abs/1106.0363} {\path{arXiv:1106.0363}},
  \href {http://dx.doi.org/10.1103/PhysRevLett.107.072003}
  {\path{doi:10.1103/PhysRevLett.107.072003}}.

\bibitem{Allada:2013nsw}
K.~Allada et~al.
\newblock {Single spin asymmetries of inclusive hadrons produced in electron
  scattering from a transversely polarized $^3$He target}.
\newblock {\em Phys. Rev.}, C89(4):042201, 2014.
\newblock \href {http://arxiv.org/abs/1311.1866} {\path{arXiv:1311.1866}},
  \href {http://dx.doi.org/10.1103/PhysRevC.89.042201}
  {\path{doi:10.1103/PhysRevC.89.042201}}.

\bibitem{Bunce:1976yb}
G.~Bunce et~al.
\newblock {$\Lambda_0$ Hyperon Polarization in Inclusive Production by 300-GeV
  Protons on Beryllium.}
\newblock {\em Phys. Rev. Lett.}, 36:1113--1116, 1976.
\newblock \href {http://dx.doi.org/10.1103/PhysRevLett.36.1113}
  {\path{doi:10.1103/PhysRevLett.36.1113}}.

\bibitem{Skubic:1978fi}
P.~Skubic et~al.
\newblock {Neutral Strange Particle Production by 300-GeV Protons}.
\newblock {\em Phys. Rev.}, D18:3115--3144, 1978.
\newblock \href {http://dx.doi.org/10.1103/PhysRevD.18.3115}
  {\path{doi:10.1103/PhysRevD.18.3115}}.

\bibitem{Klem:1976ui}
R.~D. Klem, J.~E. Bowers, H.~W. Courant, H.~Kagan, M.~L. Marshak, E.~A.
  Peterson, K.~Ruddick, W.~H. Dragoset, and J.~B. Roberts.
\newblock {Measurement of Asymmetries of Inclusive Pion Production in Proton
  Proton Interactions at 6-GeV/c and 11.8-GeV/c}.
\newblock {\em Phys. Rev. Lett.}, 36:929--931, 1976.
\newblock \href {http://dx.doi.org/10.1103/PhysRevLett.36.929}
  {\path{doi:10.1103/PhysRevLett.36.929}}.

\bibitem{Dragoset:1978gg}
W.~H. Dragoset, J.~B. Roberts, J.~E. Bowers, H.~W. Courant, H.~Kagan, M.~L.
  Marshak, E.~A. Peterson, K.~Ruddick, and R.~D. Klem.
\newblock {Asymmetries in Inclusive Proton-Nucleon Scattering at 11.75-GeV/c}.
\newblock {\em Phys. Rev.}, D18:3939--3954, 1978.
\newblock \href {http://dx.doi.org/10.1103/PhysRevD.18.3939}
  {\path{doi:10.1103/PhysRevD.18.3939}}.

\bibitem{Allgower:2002qi}
C.~E. Allgower et~al.
\newblock {Measurement of analyzing powers of $\pi^+$ and $\pi^-$ produced on a
  hydrogen and a carbon target with a 22-GeV/c incident polarized proton beam}.
\newblock {\em Phys. Rev.}, D65:092008, 2002.
\newblock \href {http://dx.doi.org/10.1103/PhysRevD.65.092008}
  {\path{doi:10.1103/PhysRevD.65.092008}}.

\bibitem{Adams:1991rw}
D.~L. Adams et~al.
\newblock {Comparison of spin asymmetries and cross-sections in $\pi^0$
  production by 200-GeV polarized anti-protons and protons}.
\newblock {\em Phys. Lett.}, B261:201--206, 1991.
\newblock \href {http://dx.doi.org/10.1016/0370-2693(91)91351-U}
  {\path{doi:10.1016/0370-2693(91)91351-U}}.

\bibitem{Adams:1991cs}
D.~L. Adams et~al.
\newblock {Analyzing power in inclusive $\pi^+$ and $\pi^-$ production at high
  $x_F$ with a 200-GeV polarized proton beam}.
\newblock {\em Phys. Lett.}, B264:462--466, 1991.
\newblock \href {http://dx.doi.org/10.1016/0370-2693(91)90378-4}
  {\path{doi:10.1016/0370-2693(91)90378-4}}.

\bibitem{Kane:1978nd}
Gordon~L. Kane, J.~Pumplin, and W.~Repko.
\newblock {Transverse Quark Polarization in Large $p_T$ Reactions, e+ e- Jets,
  and Leptoproduction: A Test of QCD}.
\newblock {\em Phys. Rev. Lett.}, 41:1689, 1978.
\newblock \href {http://dx.doi.org/10.1103/PhysRevLett.41.1689}
  {\path{doi:10.1103/PhysRevLett.41.1689}}.

\bibitem{Pan:2016suv}
Yuxi Pan.
\newblock {Transverse Single Spin Asymmetries of Forward $\pi^0$ and Jet-like
  Events in $\sqrt s$ = 500 GeV Polarized Proton Collisions at STAR}.
\newblock {\em Int. J. Mod. Phys. Conf. Ser.}, 40(01):1660037, 2016.
\newblock \href {http://dx.doi.org/10.1142/S2010194516600375}
  {\path{doi:10.1142/S2010194516600375}}.

\bibitem{Sivers:1989cc}
Dennis~W. Sivers.
\newblock {Single Spin Production Asymmetries from the Hard Scattering of
  Point-Like Constituents}.
\newblock {\em Phys. Rev.}, D41:83, 1990.
\newblock \href {http://dx.doi.org/10.1103/PhysRevD.41.83}
  {\path{doi:10.1103/PhysRevD.41.83}}.

\bibitem{Wandzura:1977qf}
S.~Wandzura and Frank Wilczek.
\newblock {Sum Rules for Spin Dependent Electroproduction: Test of Relativistic
  Constituent Quarks}.
\newblock {\em Phys. Lett.}, 72B:195--198, 1977.
\newblock \href {http://dx.doi.org/10.1016/0370-2693(77)90700-6}
  {\path{doi:10.1016/0370-2693(77)90700-6}}.

\bibitem{Gamberg:2017gle}
Leonard Gamberg, Zhong-Bo Kang, Daniel Pitonyak, and Alexei Prokudin.
\newblock {Phenomenological constraints on $A_N$ in $p^\uparrow p\to \pi\, X$
  from Lorentz invariance relations}.
\newblock {\em Phys. Lett.}, B770:242--251, 2017.
\newblock \href {http://arxiv.org/abs/1701.09170} {\path{arXiv:1701.09170}},
  \href {http://dx.doi.org/10.1016/j.physletb.2017.04.061}
  {\path{doi:10.1016/j.physletb.2017.04.061}}.

\bibitem{Bacchetta:2004it}
Alessandro Bacchetta and Marco Radici.
\newblock {Dihadron interference fragmentation functions in proton-proton
  collisions}.
\newblock {\em Phys. Rev.}, D70:094032, 2004.
\newblock \href {http://arxiv.org/abs/hep-ph/0409174}
  {\path{arXiv:hep-ph/0409174}}, \href
  {http://dx.doi.org/10.1103/PhysRevD.70.094032}
  {\path{doi:10.1103/PhysRevD.70.094032}}.

\bibitem{DAlesio:2010sag}
Umberto D'Alesio, Francesco Murgia, and Cristian Pisano.
\newblock {Azimuthal asymmetries for hadron distributions inside a jet in
  hadronic collisions}.
\newblock {\em Phys. Rev.}, D83:034021, 2011.
\newblock \href {http://arxiv.org/abs/1011.2692} {\path{arXiv:1011.2692}},
  \href {http://dx.doi.org/10.1103/PhysRevD.83.034021}
  {\path{doi:10.1103/PhysRevD.83.034021}}.

\bibitem{Huang:2015vpy}
Jin Huang, Zhong-Bo Kang, Ivan Vitev, and Hongxi Xing.
\newblock {Spin asymmetries for vector boson production in polarized p+p
  collisions}.
\newblock {\em Phys. Rev.}, D93(1):014036, 2016.
\newblock \href {http://arxiv.org/abs/1511.06764} {\path{arXiv:1511.06764}},
  \href {http://dx.doi.org/10.1103/PhysRevD.93.014036}
  {\path{doi:10.1103/PhysRevD.93.014036}}.

\bibitem{Mukherjee:2012uz}
Asmita Mukherjee and Werner Vogelsang.
\newblock {Jet production in (un)polarized pp collisions: dependence on jet
  algorithm}.
\newblock {\em Phys. Rev.}, D86:094009, 2012.
\newblock \href {http://arxiv.org/abs/1209.1785} {\path{arXiv:1209.1785}},
  \href {http://dx.doi.org/10.1103/PhysRevD.86.094009}
  {\path{doi:10.1103/PhysRevD.86.094009}}.

\bibitem{Collins:2002kn}
John~C. Collins.
\newblock {Leading twist single transverse-spin asymmetries: Drell-Yan and deep
  inelastic scattering}.
\newblock {\em Phys. Lett.}, B536:43--48, 2002.
\newblock \href {http://arxiv.org/abs/hep-ph/0204004}
  {\path{arXiv:hep-ph/0204004}}, \href
  {http://dx.doi.org/10.1016/S0370-2693(02)01819-1}
  {\path{doi:10.1016/S0370-2693(02)01819-1}}.

\bibitem{Rogers:2010dm}
Ted~C. Rogers and Piet~J. Mulders.
\newblock {No Generalized TMD-Factorization in Hadro-Production of High
  Transverse Momentum Hadrons}.
\newblock {\em Phys. Rev.}, D81:094006, 2010.
\newblock \href {http://arxiv.org/abs/1001.2977} {\path{arXiv:1001.2977}},
  \href {http://dx.doi.org/10.1103/PhysRevD.81.094006}
  {\path{doi:10.1103/PhysRevD.81.094006}}.

\bibitem{Adamczyk:2017wld}
Leszek Adamczyk et~al.
\newblock {Azimuthal transverse single-spin asymmetries of inclusive jets and
  charged pions within jets from polarized-proton collisions at $\sqrt{s} =
  500$ GeV}.
\newblock {\em Phys. Rev.}, D97(3):032004, 2018.
\newblock \href {http://arxiv.org/abs/1708.07080} {\path{arXiv:1708.07080}},
  \href {http://dx.doi.org/10.1103/PhysRevD.97.032004}
  {\path{doi:10.1103/PhysRevD.97.032004}}.

\bibitem{Angeles-Martinez:2015sea}
R.~Angeles-Martinez et~al.
\newblock {Transverse Momentum Dependent (TMD) parton distribution functions:
  status and prospects}.
\newblock {\em Acta Phys. Polon.}, B46(12):2501--2534, 2015.
\newblock \href {http://arxiv.org/abs/1507.05267} {\path{arXiv:1507.05267}},
  \href {http://dx.doi.org/10.5506/APhysPolB.46.2501}
  {\path{doi:10.5506/APhysPolB.46.2501}}.

\bibitem{Arsene:2008aa}
I.~Arsene et~al.
\newblock {Single Transverse Spin Asymmetries of Identified Charged Hadrons in
  Polarized p+p Collisions at s**(1/2) = 62.4-GeV}.
\newblock {\em Phys. Rev. Lett.}, 101:042001, 2008.
\newblock \href {http://arxiv.org/abs/0801.1078} {\path{arXiv:0801.1078}},
  \href {http://dx.doi.org/10.1103/PhysRevLett.101.042001}
  {\path{doi:10.1103/PhysRevLett.101.042001}}.

\bibitem{Adare:2014qzo}
A.~Adare et~al.
\newblock {Cross section and transverse single-spin asymmetry of $\eta$ mesons
  in $p^{\uparrow}+p$ collisions at $\sqrt{s}=200$ GeV at forward rapidity}.
\newblock {\em Phys. Rev.}, D90(7):072008, 2014.
\newblock \href {http://arxiv.org/abs/1406.3541} {\path{arXiv:1406.3541}},
  \href {http://dx.doi.org/10.1103/PhysRevD.90.072008}
  {\path{doi:10.1103/PhysRevD.90.072008}}.

\bibitem{Li:2019iyt}
H.~Li et~al.
\newblock {Azimuthal asymmetries of back-to-back $\pi^\pm-(\pi^0,\eta,\pi^\pm)$
  pairs in $e^+e^-$ annihilation}.
\newblock 2019.
\newblock \href {http://arxiv.org/abs/1909.01857} {\path{arXiv:1909.01857}}.

\bibitem{Adare:2010bd}
A.~Adare et~al.
\newblock {Measurement of Transverse Single-Spin Asymmetries for $J/\psi$
  Production in Polarized $p+p$ Collisions at $\sqrt{s} = 200$ GeV}.
\newblock {\em Phys. Rev.}, D82:112008, 2010.
\newblock [Erratum: Phys. Rev.D86,099904(2012)].
\newblock \href {http://arxiv.org/abs/1009.4864} {\path{arXiv:1009.4864}},
  \href {http://dx.doi.org/10.1103/PhysRevD.82.112008,
  10.1103/PhysRevD.86.099904} {\path{doi:10.1103/PhysRevD.82.112008,
  10.1103/PhysRevD.86.099904}}.

\bibitem{Aidala:2019ctp}
C.~Aidala et~al.
\newblock {Nuclear dependence of the transverse single-spin asymmetry in the
  production of charged hadrons at forward rapidity in polarized $p+p$, $p+$Al,
  and $p+$Au collisions at $\sqrt{s_{_{NN}}}=200$ GeV}.
\newblock {\em Phys. Rev. Lett.}, 123(12):122001, 2019.
\newblock \href {http://arxiv.org/abs/1903.07422} {\path{arXiv:1903.07422}},
  \href {http://dx.doi.org/10.1103/PhysRevLett.123.122001}
  {\path{doi:10.1103/PhysRevLett.123.122001}}.

\bibitem{Benic:2018amn}
Sanjin Benic and Yoshitaka Hatta.
\newblock {Single spin asymmetry in forward $pA$ collisions: Phenomenology at
  RHIC}.
\newblock {\em Phys. Rev.}, D99(9):094012, 2019.
\newblock \href {http://arxiv.org/abs/1811.10589} {\path{arXiv:1811.10589}},
  \href {http://dx.doi.org/10.1103/PhysRevD.99.094012}
  {\path{doi:10.1103/PhysRevD.99.094012}}.

\bibitem{Aschenauer:2016our}
Elke-Caroline Aschenauer et~al.
\newblock {The RHIC Cold QCD Plan for 2017 to 2023: A Portal to the EIC}.
\newblock 2016.
\newblock \href {http://arxiv.org/abs/1602.03922} {\path{arXiv:1602.03922}}.

\bibitem{Drell:1970wh}
S.~D. Drell and Tung-Mow Yan.
\newblock {Massive Lepton Pair Production in Hadron-Hadron Collisions at
  High-Energies}.
\newblock {\em Phys. Rev. Lett.}, 25:316--320, 1970.
\newblock [Erratum: Phys. Rev. Lett.25,902(1970)].
\newblock \href {http://dx.doi.org/10.1103/PhysRevLett.25.316,
  10.1103/PhysRevLett.25.902.2} {\path{doi:10.1103/PhysRevLett.25.316,
  10.1103/PhysRevLett.25.902.2}}.

\bibitem{Anselmino:2004ki}
M.~Anselmino, V.~Barone, A.~Drago, and N.~N. Nikolaev.
\newblock {Accessing transversity via $J/\psi$ production in $p^\uparrow \,
  \bar p^\uparrow$ interactions}.
\newblock {\em Phys. Lett.}, B594:97--104, 2004.
\newblock \href {http://arxiv.org/abs/hep-ph/0403114}
  {\path{arXiv:hep-ph/0403114}}, \href
  {http://dx.doi.org/10.1016/j.physletb.2004.05.029}
  {\path{doi:10.1016/j.physletb.2004.05.029}}.

\bibitem{Arnold:2008kf}
S.~Arnold, A.~Metz, and M.~Schlegel.
\newblock {Dilepton production from polarized hadron hadron collisions}.
\newblock {\em Phys. Rev.}, D79:034005, 2009.
\newblock \href {http://arxiv.org/abs/0809.2262} {\path{arXiv:0809.2262}},
  \href {http://dx.doi.org/10.1103/PhysRevD.79.034005}
  {\path{doi:10.1103/PhysRevD.79.034005}}.

\bibitem{PhysRevD.16.2219}
John~C. Collins and Davison~E. Soper.
\newblock Angular distribution of dileptons in high-energy hadron collisions.
\newblock {\em Phys. Rev. D}, 16:2219--2225, Oct 1977.
\newblock URL: \url{https://link.aps.org/doi/10.1103/PhysRevD.16.2219}, \href
  {http://dx.doi.org/10.1103/PhysRevD.16.2219}
  {\path{doi:10.1103/PhysRevD.16.2219}}.

\bibitem{Gottfried:1964nx}
K.~Gottfried and John~David Jackson.
\newblock {On the Connection between production mechanism and decay of
  resonances at high-energies}.
\newblock {\em Nuovo Cim.}, 33:309--330, 1964.
\newblock \href {http://dx.doi.org/10.1007/BF02750195}
  {\path{doi:10.1007/BF02750195}}.

\bibitem{Ji:2004xq}
Xiang-dong Ji, Jian-Ping Ma, and Feng Yuan.
\newblock {QCD factorization for spin-dependent cross sections in DIS and
  Drell-Yan processes at low transverse momentum}.
\newblock {\em Phys. Lett.}, B597:299--308, 2004.
\newblock \href {http://arxiv.org/abs/hep-ph/0405085}
  {\path{arXiv:hep-ph/0405085}}, \href
  {http://dx.doi.org/10.1016/j.physletb.2004.07.026}
  {\path{doi:10.1016/j.physletb.2004.07.026}}.

\bibitem{Lam:1978pu}
C.~S. Lam and Wu-Ki Tung.
\newblock {A Systematic Approach to Inclusive Lepton Pair Production in
  Hadronic Collisions}.
\newblock {\em Phys. Rev.}, D18:2447, 1978.
\newblock \href {http://dx.doi.org/10.1103/PhysRevD.18.2447}
  {\path{doi:10.1103/PhysRevD.18.2447}}.

\bibitem{Boer:1997nt}
Daniel Boer and P.~J. Mulders.
\newblock {Time reversal odd distribution functions in leptoproduction}.
\newblock {\em Phys. Rev.}, D57:5780--5786, 1998.
\newblock \href {http://arxiv.org/abs/hep-ph/9711485}
  {\path{arXiv:hep-ph/9711485}}, \href
  {http://dx.doi.org/10.1103/PhysRevD.57.5780}
  {\path{doi:10.1103/PhysRevD.57.5780}}.

\bibitem{Aghasyan:2017jop}
M.~Aghasyan et~al.
\newblock {First measurement of transverse-spin-dependent azimuthal asymmetries
  in the Drell-Yan process}.
\newblock {\em Phys. Rev. Lett.}, 119(11):112002, 2017.
\newblock \href {http://arxiv.org/abs/1704.00488} {\path{arXiv:1704.00488}},
  \href {http://dx.doi.org/10.1103/PhysRevLett.119.112002}
  {\path{doi:10.1103/PhysRevLett.119.112002}}.

\bibitem{Chen:2019hhx}
Andrew Chen et~al.
\newblock {Probing nucleon's spin structures with polarized Drell-Yan in the
  Fermilab SpinQuest experiment}.
\newblock In {\em {23rd International Symposium on Spin Physics (SPIN 2018)
  Ferrara, Italy, September 10-14, 2018}}, 2019.
\newblock URL:
  \url{http://lss.fnal.gov/archive/2018/conf/fermilab-conf-18-412-e.pdf}, \href
  {http://arxiv.org/abs/1901.09994} {\path{arXiv:1901.09994}}.

\bibitem{Adamczyk:2015gyk}
L.~Adamczyk et~al.
\newblock {Measurement of the transverse single-spin asymmetry in $p^\uparrow+p
  \to W^{\pm}/Z^0$ at RHIC}.
\newblock {\em Phys. Rev. Lett.}, 116(13):132301, 2016.
\newblock \href {http://arxiv.org/abs/1511.06003} {\path{arXiv:1511.06003}},
  \href {http://dx.doi.org/10.1103/PhysRevLett.116.132301}
  {\path{doi:10.1103/PhysRevLett.116.132301}}.

\bibitem{Gamberg:2012iq}
Leonard Gamberg and Zhong-Bo Kang.
\newblock {Single transverse spin asymmetry of prompt photon production}.
\newblock {\em Phys. Lett.}, B718:181--188, 2012.
\newblock \href {http://arxiv.org/abs/1208.1962} {\path{arXiv:1208.1962}},
  \href {http://dx.doi.org/10.1016/j.physletb.2012.10.002}
  {\path{doi:10.1016/j.physletb.2012.10.002}}.

\bibitem{Gamberg_2011}
Leonard Gamberg and Zhong-Bo Kang.
\newblock Process dependent sivers function and implication for single spin
  asymmetry in inclusive hadron production.
\newblock {\em Physics Letters B}, 696(1-2):109?118, Jan 2011.
\newblock URL: \url{http://dx.doi.org/10.1016/j.physletb.2010.11.066}, \href
  {http://dx.doi.org/10.1016/j.physletb.2010.11.066}
  {\path{doi:10.1016/j.physletb.2010.11.066}}.

\bibitem{Pisano:2015wnq}
Silvia Pisano and Marco Radici.
\newblock {Di-hadron fragmentation and mapping of the nucleon structure}.
\newblock {\em Eur. Phys. J.}, A52(6):155, 2016.
\newblock \href {http://arxiv.org/abs/1511.03220} {\path{arXiv:1511.03220}},
  \href {http://dx.doi.org/10.1140/epja/i2016-16155-5}
  {\path{doi:10.1140/epja/i2016-16155-5}}.

\bibitem{Collins:1993kq}
John~C. Collins, Steve~F. Heppelmann, and Glenn~A. Ladinsky.
\newblock {Measuring transversity densities in singly polarized hadron hadron
  and lepton - hadron collisions}.
\newblock {\em Nucl. Phys.}, B420:565--582, 1994.
\newblock \href {http://arxiv.org/abs/hep-ph/9305309}
  {\path{arXiv:hep-ph/9305309}}, \href
  {http://dx.doi.org/10.1016/0550-3213(94)90078-7}
  {\path{doi:10.1016/0550-3213(94)90078-7}}.

\bibitem{Bacchetta:2002ux}
Alessandro Bacchetta and Marco Radici.
\newblock {Partial wave analysis of two hadron fragmentation functions}.
\newblock {\em Phys. Rev.}, D67:094002, 2003.
\newblock \href {http://arxiv.org/abs/hep-ph/0212300}
  {\path{arXiv:hep-ph/0212300}}, \href
  {http://dx.doi.org/10.1103/PhysRevD.67.094002}
  {\path{doi:10.1103/PhysRevD.67.094002}}.

\bibitem{Matevosyan:2013aka}
Hrayr~H. Matevosyan, Anthony~W. Thomas, and Wolfgang Bentz.
\newblock {Dihadron fragmentation functions within the Nambu-Jona-Lasinio jet
  model}.
\newblock {\em Phys. Rev.}, D88(9):094022, 2013.
\newblock \href {http://arxiv.org/abs/1310.1917} {\path{arXiv:1310.1917}},
  \href {http://dx.doi.org/10.1103/PhysRevD.88.094022}
  {\path{doi:10.1103/PhysRevD.88.094022}}.

\bibitem{Matevosyan:2017uls}
Hrayr~H. Matevosyan, Aram Kotzinian, and Anthony~W. Thomas.
\newblock {Dihadron fragmentation functions in the quark-jet model:
  Transversely polarized quarks}.
\newblock {\em Phys. Rev.}, D97(1):014019, 2018.
\newblock \href {http://arxiv.org/abs/1709.08643} {\path{arXiv:1709.08643}},
  \href {http://dx.doi.org/10.1103/PhysRevD.97.014019}
  {\path{doi:10.1103/PhysRevD.97.014019}}.

\bibitem{Adamczyk:2015hri}
L.~Adamczyk et~al.
\newblock {Observation of Transverse Spin-Dependent Azimuthal Correlations of
  Charged Pion Pairs in $p^\uparrow+p$ at $\sqrt{s}=200$ GeV}.
\newblock {\em Phys. Rev. Lett.}, 115:242501, 2015.
\newblock \href {http://arxiv.org/abs/1504.00415} {\path{arXiv:1504.00415}},
  \href {http://dx.doi.org/10.1103/PhysRevLett.115.242501}
  {\path{doi:10.1103/PhysRevLett.115.242501}}.

\bibitem{Adamczyk:2017ynk}
L.~Adamczyk et~al.
\newblock {Transverse spin-dependent azimuthal correlations of charged pion
  pairs measured in p$^\uparrow$+p collisions at $\sqrt{s}$ = 500 GeV}.
\newblock {\em Phys. Lett.}, B780:332--339, 2018.
\newblock \href {http://arxiv.org/abs/1710.10215} {\path{arXiv:1710.10215}},
  \href {http://dx.doi.org/10.1016/j.physletb.2018.02.069}
  {\path{doi:10.1016/j.physletb.2018.02.069}}.

\bibitem{Radici:2018iag}
Marco Radici and Alessandro Bacchetta.
\newblock {First Extraction of Transversity from a Global Analysis of
  Electron-Proton and Proton-Proton Data}.
\newblock {\em Phys. Rev. Lett.}, 120(19):192001, 2018.
\newblock \href {http://arxiv.org/abs/1802.05212} {\path{arXiv:1802.05212}},
  \href {http://dx.doi.org/10.1103/PhysRevLett.120.192001}
  {\path{doi:10.1103/PhysRevLett.120.192001}}.

\bibitem{DAlesio:2013cfy}
Umberto D'Alesio, Francesco Murgia, and Cristian Pisano.
\newblock {Collins and sivers effects in $p^\uparrow p \to$ jet $\pi X$:
  Universality and process dependence}.
\newblock {\em Phys. Part. Nucl.}, 45(4):676--691, 2014.
\newblock \href {http://arxiv.org/abs/1307.4880} {\path{arXiv:1307.4880}},
  \href {http://dx.doi.org/10.1134/S1063779614040054}
  {\path{doi:10.1134/S1063779614040054}}.

\bibitem{Kang:2016mcy}
Zhong-Bo Kang, Felix Ringer, and Ivan Vitev.
\newblock {The semi-inclusive jet function in SCET and small radius resummation
  for inclusive jet production}.
\newblock {\em JHEP}, 10:125, 2016.
\newblock \href {http://arxiv.org/abs/1606.06732} {\path{arXiv:1606.06732}},
  \href {http://dx.doi.org/10.1007/JHEP10(2016)125}
  {\path{doi:10.1007/JHEP10(2016)125}}.

\bibitem{Yuan:2007nd}
Feng Yuan.
\newblock {Azimuthal asymmetric distribution of hadrons inside a jet at hadron
  collider}.
\newblock {\em Phys. Rev. Lett.}, 100:032003, 2008.
\newblock \href {http://arxiv.org/abs/0709.3272} {\path{arXiv:0709.3272}},
  \href {http://dx.doi.org/10.1103/PhysRevLett.100.032003}
  {\path{doi:10.1103/PhysRevLett.100.032003}}.

\bibitem{Kang:2017btw}
Zhong-Bo Kang, Alexei Prokudin, Felix Ringer, and Feng Yuan.
\newblock {Collins azimuthal asymmetries of hadron production inside jets}.
\newblock {\em Phys. Lett.}, B774:635--642, 2017.
\newblock \href {http://arxiv.org/abs/1707.00913} {\path{arXiv:1707.00913}},
  \href {http://dx.doi.org/10.1016/j.physletb.2017.10.031}
  {\path{doi:10.1016/j.physletb.2017.10.031}}.

\bibitem{Adamczyk:2012qj}
L.~Adamczyk et~al.
\newblock {Longitudinal and transverse spin asymmetries for inclusive jet
  production at mid-rapidity in polarized $p+p$ collisions at $\sqrt{s}=200$
  GeV}.
\newblock {\em Phys. Rev.}, D86:032006, 2012.
\newblock \href {http://arxiv.org/abs/1205.2735} {\path{arXiv:1205.2735}},
  \href {http://dx.doi.org/10.1103/PhysRevD.86.032006}
  {\path{doi:10.1103/PhysRevD.86.032006}}.

\bibitem{Bland:2013pkt}
L.~C. Bland et~al.
\newblock {Cross Sections and Transverse Single-Spin Asymmetries in Forward Jet
  Production from Proton Collisions at $\sqrt{s}=500$ GeV}.
\newblock {\em Phys. Lett.}, B750:660--665, 2015.
\newblock \href {http://arxiv.org/abs/1304.1454} {\path{arXiv:1304.1454}},
  \href {http://dx.doi.org/10.1016/j.physletb.2015.10.001}
  {\path{doi:10.1016/j.physletb.2015.10.001}}.

\bibitem{Abelev:2007ii}
B.~I. Abelev et~al.
\newblock {Measurement of transverse single-spin asymmetries for di-jet
  production in proton-proton collisions at s**(1/2) = 200-GeV}.
\newblock {\em Phys. Rev. Lett.}, 99:142003, 2007.
\newblock \href {http://arxiv.org/abs/0705.4629} {\path{arXiv:0705.4629}},
  \href {http://dx.doi.org/10.1103/PhysRevLett.99.142003}
  {\path{doi:10.1103/PhysRevLett.99.142003}}.

\bibitem{Boer:1997mf}
Daniel Boer, R.~Jakob, and P.~J. Mulders.
\newblock {Asymmetries in polarized hadron production in e+ e- annihilation up
  to order 1/Q}.
\newblock {\em Nucl. Phys.}, B504:345--380, 1997.
\newblock \href {http://arxiv.org/abs/hep-ph/9702281}
  {\path{arXiv:hep-ph/9702281}}, \href
  {http://dx.doi.org/10.1016/S0550-3213(97)00456-2}
  {\path{doi:10.1016/S0550-3213(97)00456-2}}.

\bibitem{Anselmino:2015sxa}
M.~Anselmino, M.~Boglione, U.~D'Alesio, J.~O. Gonzalez~Hernandez, S.~Melis,
  F.~Murgia, and A.~Prokudin.
\newblock {Collins functions for pions from SIDIS and new $e^+e^-$ data: a
  first glance at their transverse momentum dependence}.
\newblock {\em Phys. Rev.}, D92(11):114023, 2015.
\newblock \href {http://arxiv.org/abs/1510.05389} {\path{arXiv:1510.05389}},
  \href {http://dx.doi.org/10.1103/PhysRevD.92.114023}
  {\path{doi:10.1103/PhysRevD.92.114023}}.

\bibitem{Abe:2005zx}
Kazuo Abe et~al.
\newblock {Measurement of azimuthal asymmetries in inclusive production of
  hadron pairs in e+ e- annihilation at Belle}.
\newblock {\em Phys. Rev. Lett.}, 96:232002, 2006.
\newblock \href {http://arxiv.org/abs/hep-ex/0507063}
  {\path{arXiv:hep-ex/0507063}}, \href
  {http://dx.doi.org/10.1103/PhysRevLett.96.232002}
  {\path{doi:10.1103/PhysRevLett.96.232002}}.

\bibitem{Seidl:2008xc}
R.~Seidl et~al.
\newblock {Measurement of Azimuthal Asymmetries in Inclusive Production of
  Hadron Pairs in e+e- Annihilation at $\sqrt s$ = 10.58 GeV}.
\newblock {\em Phys. Rev.}, D78:032011, 2008.
\newblock [Erratum: Phys. Rev.D86,039905(2012)].
\newblock \href {http://arxiv.org/abs/0805.2975} {\path{arXiv:0805.2975}},
  \href {http://dx.doi.org/10.1103/PhysRevD.78.032011,
  10.1103/PhysRevD.86.039905} {\path{doi:10.1103/PhysRevD.78.032011,
  10.1103/PhysRevD.86.039905}}.

\bibitem{TheBABAR:2013yha}
J.~P. Lees et~al.
\newblock {Measurement of Collins asymmetries in inclusive production of
  charged pion pairs in $e^+e^-$ annihilation at BABAR}.
\newblock {\em Phys. Rev.}, D90(5):052003, 2014.
\newblock \href {http://arxiv.org/abs/1309.5278} {\path{arXiv:1309.5278}},
  \href {http://dx.doi.org/10.1103/PhysRevD.90.052003}
  {\path{doi:10.1103/PhysRevD.90.052003}}.

\bibitem{Aubert:2015hha}
J.~P. Lees et~al.
\newblock {Collins asymmetries in inclusive charged $KK$ and $K\pi$ pairs
  produced in $e^+e^-$ annihilation}.
\newblock {\em Phys. Rev.}, D92(11):111101, 2015.
\newblock \href {http://arxiv.org/abs/1506.05864} {\path{arXiv:1506.05864}},
  \href {http://dx.doi.org/10.1103/PhysRevD.92.111101}
  {\path{doi:10.1103/PhysRevD.92.111101}}.

\bibitem{Ablikim:2015pta}
M.~Ablikim et~al.
\newblock {Measurement of azimuthal asymmetries in inclusive charged dipion
  production in $e^+e^-$ annihilations at $\sqrt{s}$ = 3.65 GeV}.
\newblock {\em Phys. Rev. Lett.}, 116(4):042001, 2016.
\newblock \href {http://arxiv.org/abs/1507.06824} {\path{arXiv:1507.06824}},
  \href {http://dx.doi.org/10.1103/PhysRevLett.116.042001}
  {\path{doi:10.1103/PhysRevLett.116.042001}}.

\bibitem{Metz:2016swz}
Andreas Metz and Anselm Vossen.
\newblock {Parton Fragmentation Functions}.
\newblock {\em Prog. Part. Nucl. Phys.}, 91:136--202, 2016.
\newblock \href {http://arxiv.org/abs/1607.02521} {\path{arXiv:1607.02521}},
  \href {http://dx.doi.org/10.1016/j.ppnp.2016.08.003}
  {\path{doi:10.1016/j.ppnp.2016.08.003}}.

\bibitem{Anselmino:2000vs}
M.~Anselmino, Daniel Boer, U.~D'Alesio, and F.~Murgia.
\newblock {Lambda polarization from unpolarized quark fragmentation}.
\newblock {\em Phys. Rev.}, D63:054029, 2001.
\newblock \href {http://arxiv.org/abs/hep-ph/0008186}
  {\path{arXiv:hep-ph/0008186}}, \href
  {http://dx.doi.org/10.1103/PhysRevD.63.054029}
  {\path{doi:10.1103/PhysRevD.63.054029}}.

\bibitem{Anselmino:2001js}
M.~Anselmino, Daniel Boer, U.~D'Alesio, and F.~Murgia.
\newblock {Transverse lambda polarization in semiinclusive DIS}.
\newblock {\em Phys. Rev.}, D65:114014, 2002.
\newblock \href {http://arxiv.org/abs/hep-ph/0109186}
  {\path{arXiv:hep-ph/0109186}}, \href
  {http://dx.doi.org/10.1103/PhysRevD.65.114014}
  {\path{doi:10.1103/PhysRevD.65.114014}}.

\bibitem{Boer:2002ij}
Daniel Boer and Adrian Dumitru.
\newblock {Polarized hyperons from pA scattering in the gluon saturation
  regime}.
\newblock {\em Phys. Lett.}, B556:33--40, 2003.
\newblock \href {http://arxiv.org/abs/hep-ph/0212260}
  {\path{arXiv:hep-ph/0212260}}, \href
  {http://dx.doi.org/10.1016/S0370-2693(03)00081-9}
  {\path{doi:10.1016/S0370-2693(03)00081-9}}.

\bibitem{Guan:2018ckx}
Y.~Guan et~al.
\newblock {Observation of Transverse $\Lambda/\bar{\Lambda}$ Hyperon
  Polarization in $e^+e^-$ Annihilation at Belle}.
\newblock {\em Phys. Rev. Lett.}, 122(4):042001, 2019.
\newblock \href {http://arxiv.org/abs/1808.05000} {\path{arXiv:1808.05000}},
  \href {http://dx.doi.org/10.1103/PhysRevLett.122.042001}
  {\path{doi:10.1103/PhysRevLett.122.042001}}.

\bibitem{Collins:1984kg}
John~C. Collins, Davison~E. Soper, and George Sterman.
\newblock {Transverse Momentum Distribution in Drell-Yan Pair and W and Z Boson
  Production}.
\newblock {\em Nucl. Phys.}, B250:199, 1985.
\newblock \href {http://dx.doi.org/10.1016/0550-3213(85)90479-1}
  {\path{doi:10.1016/0550-3213(85)90479-1}}.

\bibitem{Collins:1992kk}
John~C. Collins.
\newblock Fragmentation of transversely polarized quarks probed in transverse
  momentum distributions.
\newblock {\em Nucl. Phys.}, B396:161--182, 1993.

\bibitem{Collins:2004nx}
John~C. Collins and Andreas Metz.
\newblock {Universality of soft and collinear factors in hard- scattering
  factorization}.
\newblock {\em Phys. Rev. Lett.}, 93:252001, 2004.
\newblock \href {http://arxiv.org/abs/hep-ph/0408249}
  {\path{arXiv:hep-ph/0408249}}, \href
  {http://dx.doi.org/10.1103/PhysRevLett.93.252001}
  {\path{doi:10.1103/PhysRevLett.93.252001}}.

\bibitem{Ji:2004wu}
Xiangdong Ji, Jian-Ping Ma, and Feng Yuan.
\newblock Qcd factorization for semi-inclusive deep-inelastic scattering at low
  transverse momentum.
\newblock {\em Phys. Rev.}, D71:034005, 2005.
\newblock \href {http://arxiv.org/abs/hep-ph/0404183}
  {\path{arXiv:hep-ph/0404183}}.

\bibitem{Ji:2006ub}
Xiangdong Ji, Jian-Wei Qiu, Werner Vogelsang, and Feng Yuan.
\newblock {A unified picture for single transverse-spin asymmetries in hard
  processes}.
\newblock {\em Phys. Rev. Lett.}, 97:082002, 2006.
\newblock \href {http://arxiv.org/abs/hep-ph/0602239}
  {\path{arXiv:hep-ph/0602239}}, \href
  {http://dx.doi.org/10.1103/PhysRevLett.97.082002}
  {\path{doi:10.1103/PhysRevLett.97.082002}}.

\bibitem{Ji:2006vf}
Xiangdong Ji, Jian-wei Qiu, Werner Vogelsang, and Feng Yuan.
\newblock {Single Transverse-Spin Asymmetry in Drell-Yan Production at Large
  and Moderate Transverse Momentum}.
\newblock {\em Phys. Rev.}, D73:094017, 2006.
\newblock \href {http://arxiv.org/abs/hep-ph/0604023}
  {\path{arXiv:hep-ph/0604023}}, \href
  {http://dx.doi.org/10.1103/PhysRevD.73.094017}
  {\path{doi:10.1103/PhysRevD.73.094017}}.

\bibitem{Bacchetta:2008xw}
Alessandro Bacchetta, Daniel Boer, Markus Diehl, and Piet~J. Mulders.
\newblock {Matches and mismatches in the descriptions of semi-inclusive
  processes at low and high transverse momentum}.
\newblock {\em JHEP}, 08:023, 2008.
\newblock \href {http://arxiv.org/abs/0803.0227} {\path{arXiv:0803.0227}},
  \href {http://dx.doi.org/10.1088/1126-6708/2008/08/023}
  {\path{doi:10.1088/1126-6708/2008/08/023}}.

\bibitem{Anselmino:2009st}
M.~Anselmino et~al.
\newblock {Sivers effect in Drell-Yan processes}.
\newblock {\em Phys. Rev.}, D79:054010, 2009.
\newblock \href {http://arxiv.org/abs/0901.3078} {\path{arXiv:0901.3078}},
  \href {http://dx.doi.org/10.1103/PhysRevD.79.054010}
  {\path{doi:10.1103/PhysRevD.79.054010}}.

\bibitem{Mulders:2011zt}
P.~J. Mulders and T.~C. Rogers.
\newblock {Gauge Links, TMD-Factorization, and TMD-Factorization Breaking}.
\newblock 2011.
\newblock \href {http://arxiv.org/abs/1102.4569} {\path{arXiv:1102.4569}}.

\bibitem{Aybat:2011zv}
S.~Mert Aybat and Ted~C. Rogers.
\newblock {TMD Parton Distribution and Fragmentation Functions with QCD
  Evolution}.
\newblock {\em Phys. Rev.}, D83:114042, 2011.
\newblock \href {http://arxiv.org/abs/1101.5057} {\path{arXiv:1101.5057}},
  \href {http://dx.doi.org/10.1103/PhysRevD.83.114042}
  {\path{doi:10.1103/PhysRevD.83.114042}}.

\bibitem{Aybat:2011ge}
S.~Mert Aybat, John~C. Collins, Jian-Wei Qiu, and Ted~C. Rogers.
\newblock {The QCD Evolution of the Sivers Function}.
\newblock {\em Phys. Rev.}, D85:034043, 2012.
\newblock \href {http://arxiv.org/abs/1110.6428} {\path{arXiv:1110.6428}},
  \href {http://dx.doi.org/10.1103/PhysRevD.85.034043}
  {\path{doi:10.1103/PhysRevD.85.034043}}.

\bibitem{Bacchetta:2013pqa}
Alessandro Bacchetta and Alexei Prokudin.
\newblock {Evolution of the helicity and transversity
  Transverse-Momentum-Dependent parton distributions}.
\newblock {\em Nucl. Phys.}, B875:536--551, 2013.
\newblock \href {http://arxiv.org/abs/1303.2129} {\path{arXiv:1303.2129}},
  \href {http://dx.doi.org/10.1016/j.nuclphysb.2013.07.013}
  {\path{doi:10.1016/j.nuclphysb.2013.07.013}}.

\bibitem{Scimemi:2019cmh}
Ignazio Scimemi and Alexey Vladimirov.
\newblock {Non-perturbative structure of semi-inclusive deep-inelastic and
  Drell-Yan scattering at small transverse momentum}.
\newblock 2019.
\newblock \href {http://arxiv.org/abs/1912.06532} {\path{arXiv:1912.06532}}.

\bibitem{Aybat:2011ta}
S.~Mert Aybat, Alexei Prokudin, and Ted~C. Rogers.
\newblock {Calculation of TMD Evolution for Transverse Single Spin Asymmetry
  Measurements}.
\newblock {\em Phys. Rev. Lett.}, 108:242003, 2012.
\newblock \href {http://arxiv.org/abs/1112.4423} {\path{arXiv:1112.4423}},
  \href {http://dx.doi.org/10.1103/PhysRevLett.108.242003}
  {\path{doi:10.1103/PhysRevLett.108.242003}}.

\bibitem{Anselmino:2012aa}
M.~Anselmino, M.~Boglione, and S.~Melis.
\newblock {A Strategy towards the extraction of the Sivers function with TMD
  evolution}.
\newblock {\em Phys. Rev.}, D86:014028, 2012.
\newblock \href {http://arxiv.org/abs/1204.1239} {\path{arXiv:1204.1239}},
  \href {http://dx.doi.org/10.1103/PhysRevD.86.014028}
  {\path{doi:10.1103/PhysRevD.86.014028}}.

\bibitem{Kang:2015msa}
Zhong-Bo Kang, Alexei Prokudin, Peng Sun, and Feng Yuan.
\newblock {Extraction of Quark Transversity Distribution and Collins
  Fragmentation Functions with QCD Evolution}.
\newblock {\em Phys. Rev.}, D93(1):014009, 2016.
\newblock \href {http://arxiv.org/abs/1505.05589} {\path{arXiv:1505.05589}},
  \href {http://dx.doi.org/10.1103/PhysRevD.93.014009}
  {\path{doi:10.1103/PhysRevD.93.014009}}.

\bibitem{Bacchetta:2017gcc}
Alessandro Bacchetta, Filippo Delcarro, Cristian Pisano, Marco Radici, and
  Andrea Signori.
\newblock {Extraction of partonic transverse momentum distributions from
  semi-inclusive deep-inelastic scattering, Drell-Yan and Z-boson production}.
\newblock {\em JHEP}, 06:081, 2017.
\newblock [Erratum: JHEP06,051(2019)].
\newblock \href {http://arxiv.org/abs/1703.10157} {\path{arXiv:1703.10157}},
  \href {http://dx.doi.org/10.1007/JHEP06(2017)081, 10.1007/JHEP06(2019)051}
  {\path{doi:10.1007/JHEP06(2017)081, 10.1007/JHEP06(2019)051}}.

\bibitem{Boglione:2018dqd}
M.~Boglione, U.~D'Alesio, C.~Flore, and J.~O. Gonzalez-Hernandez.
\newblock {Assessing signals of TMD physics in SIDIS azimuthal asymmetries and
  in the extraction of the Sivers function}.
\newblock {\em JHEP}, 07:148, 2018.
\newblock \href {http://arxiv.org/abs/1806.10645} {\path{arXiv:1806.10645}},
  \href {http://dx.doi.org/10.1007/JHEP07(2018)148}
  {\path{doi:10.1007/JHEP07(2018)148}}.

\bibitem{Goeke:2005hb}
K.~Goeke, A.~Metz, and M.~Schlegel.
\newblock {Parameterization of the quark-quark correlator of a spin-1/2
  hadron}.
\newblock {\em Phys. Lett.}, B618:90--96, 2005.
\newblock \href {http://arxiv.org/abs/hep-ph/0504130}
  {\path{arXiv:hep-ph/0504130}}, \href
  {http://dx.doi.org/10.1016/j.physletb.2005.05.037}
  {\path{doi:10.1016/j.physletb.2005.05.037}}.

\bibitem{Ralston:1979ys}
John~P. Ralston and Davison~E. Soper.
\newblock {Production of Dimuons from High-Energy Polarized Proton Proton
  Collisions}.
\newblock {\em Nucl. Phys.}, B152:109, 1979.
\newblock \href {http://dx.doi.org/10.1016/0550-3213(79)90082-8}
  {\path{doi:10.1016/0550-3213(79)90082-8}}.

\bibitem{Tangerman:1994eh}
R.~D. Tangerman and P.~J. Mulders.
\newblock {Intrinsic transverse momentum and the polarized Drell-Yan process}.
\newblock {\em Phys. Rev.}, D51:3357--3372, 1995.
\newblock \href {http://arxiv.org/abs/hep-ph/9403227}
  {\path{arXiv:hep-ph/9403227}}, \href
  {http://dx.doi.org/10.1103/PhysRevD.51.3357}
  {\path{doi:10.1103/PhysRevD.51.3357}}.

\bibitem{Murgia:2010}
U.~D'Alesio and F.~Murgia.
\newblock {Private communication}.
\newblock {\em Unpublished}.

\bibitem{Sivers:1990fh}
Dennis~W. Sivers.
\newblock {Hard scattering scaling laws for single spin production
  asymmetries}.
\newblock {\em Phys. Rev.}, D43:261--263, 1991.
\newblock \href {http://dx.doi.org/10.1103/PhysRevD.43.261}
  {\path{doi:10.1103/PhysRevD.43.261}}.

\bibitem{Miller:2003sa}
Gerald~A. Miller.
\newblock {Shapes of the proton}.
\newblock {\em Phys. Rev.}, C68:022201, 2003.
\newblock \href {http://arxiv.org/abs/nucl-th/0304076}
  {\path{arXiv:nucl-th/0304076}}, \href
  {http://dx.doi.org/10.1103/PhysRevC.68.022201}
  {\path{doi:10.1103/PhysRevC.68.022201}}.

\bibitem{Miller:2007ae}
Gerald~A. Miller.
\newblock {Densities, Parton Distributions, and Measuring the Non-Spherical
  Shape of the Nucleon}.
\newblock {\em Phys. Rev.}, C76:065209, 2007.
\newblock \href {http://arxiv.org/abs/0708.2297} {\path{arXiv:0708.2297}},
  \href {http://dx.doi.org/10.1103/PhysRevC.76.065209}
  {\path{doi:10.1103/PhysRevC.76.065209}}.

\bibitem{Bacchetta:2010uj}
Alessandro Bacchetta.
\newblock {Transverse-momentum-dependent parton distributions (TMDs)}.
\newblock {\em AIP Conf. Proc.}, 1374(1):29--34, 2011.
\newblock \href {http://arxiv.org/abs/1012.2315} {\path{arXiv:1012.2315}},
  \href {http://dx.doi.org/10.1063/1.3647094} {\path{doi:10.1063/1.3647094}}.

\bibitem{Mulders:2000sh}
P.~J. Mulders and J.~Rodrigues.
\newblock {Transverse momentum dependence in gluon distribution and
  fragmentation functions}.
\newblock {\em Phys. Rev.}, D63:094021, 2001.
\newblock \href {http://arxiv.org/abs/hep-ph/0009343}
  {\path{arXiv:hep-ph/0009343}}, \href
  {http://dx.doi.org/10.1103/PhysRevD.63.094021}
  {\path{doi:10.1103/PhysRevD.63.094021}}.

\bibitem{Anselmino:2005sh}
M.~Anselmino, M.~Boglione, U.~D'Alesio, E.~Leader, S.~Melis, and F.~Murgia.
\newblock {The general partonic structure for hadronic spin asymmetries}.
\newblock {\em Phys. Rev.}, D73:014020, 2006.
\newblock \href {http://arxiv.org/abs/hep-ph/0509035}
  {\path{arXiv:hep-ph/0509035}}, \href
  {http://dx.doi.org/10.1103/PhysRevD.73.014020}
  {\path{doi:10.1103/PhysRevD.73.014020}}.

\bibitem{Boer:2015vso}
Daniel Boer, C\'edric Lorc\'e, Cristian Pisano, and Jian Zhou.
\newblock {The gluon Sivers distribution: status and future prospects}.
\newblock {\em Adv. High Energy Phys.}, 2015:371396, 2015.
\newblock \href {http://arxiv.org/abs/1504.04332} {\path{arXiv:1504.04332}},
  \href {http://dx.doi.org/10.1155/2015/371396}
  {\path{doi:10.1155/2015/371396}}.

\bibitem{Boer:2010zf}
Daniel Boer, Stanley~J. Brodsky, Piet~J. Mulders, and Cristian Pisano.
\newblock {Direct Probes of Linearly Polarized Gluons inside Unpolarized
  Hadrons}.
\newblock {\em Phys. Rev. Lett.}, 106:132001, 2011.
\newblock \href {http://arxiv.org/abs/1011.4225} {\path{arXiv:1011.4225}},
  \href {http://dx.doi.org/10.1103/PhysRevLett.106.132001}
  {\path{doi:10.1103/PhysRevLett.106.132001}}.

\bibitem{Pisano:2013cya}
Cristian Pisano, Daniel Boer, Stanley~J. Brodsky, Maarten G.~A. Buffing, and
  Piet~J. Mulders.
\newblock {Linear polarization of gluons and photons in unpolarized collider
  experiments}.
\newblock {\em JHEP}, 10:024, 2013.
\newblock \href {http://arxiv.org/abs/1307.3417} {\path{arXiv:1307.3417}},
  \href {http://dx.doi.org/10.1007/JHEP10(2013)024}
  {\path{doi:10.1007/JHEP10(2013)024}}.

\bibitem{Mukherjee:2016qxa}
Asmita Mukherjee and Sangem Rajesh.
\newblock {$J/\psi $ production in polarized and unpolarized ep collision and
  Sivers and $\cos 2\phi $ asymmetries}.
\newblock {\em Eur. Phys. J.}, C77(12):854, 2017.
\newblock \href {http://arxiv.org/abs/1609.05596} {\path{arXiv:1609.05596}},
  \href {http://dx.doi.org/10.1140/epjc/s10052-017-5406-4}
  {\path{doi:10.1140/epjc/s10052-017-5406-4}}.

\bibitem{Kishore:2018ugo}
Raj Kishore and Asmita Mukherjee.
\newblock {Accessing linearly polarized gluon distribution in $J/\psi$
  production at the electron-ion collider}.
\newblock {\em Phys. Rev.}, D99(5):054012, 2019.
\newblock \href {http://arxiv.org/abs/1811.07495} {\path{arXiv:1811.07495}},
  \href {http://dx.doi.org/10.1103/PhysRevD.99.054012}
  {\path{doi:10.1103/PhysRevD.99.054012}}.

\bibitem{Boer:2011kf}
Daniel Boer, Wilco~J. den Dunnen, Cristian Pisano, Marc Schlegel, and Werner
  Vogelsang.
\newblock {Linearly Polarized Gluons and the Higgs Transverse Momentum
  Distribution}.
\newblock {\em Phys. Rev. Lett.}, 108:032002, 2012.
\newblock \href {http://arxiv.org/abs/1109.1444} {\path{arXiv:1109.1444}},
  \href {http://dx.doi.org/10.1103/PhysRevLett.108.032002}
  {\path{doi:10.1103/PhysRevLett.108.032002}}.

\bibitem{Boer:2013fca}
Daniel Boer, Wilco~J. den Dunnen, Cristian Pisano, and Marc Schlegel.
\newblock {Determining the Higgs spin and parity in the diphoton decay
  channel}.
\newblock {\em Phys. Rev. Lett.}, 111(3):032002, 2013.
\newblock \href {http://arxiv.org/abs/1304.2654} {\path{arXiv:1304.2654}},
  \href {http://dx.doi.org/10.1103/PhysRevLett.111.032002}
  {\path{doi:10.1103/PhysRevLett.111.032002}}.

\bibitem{Boer:2014lka}
Daniel Boer and Cristian Pisano.
\newblock {Impact of gluon polarization on Higgs boson plus jet production at
  the LHC}.
\newblock {\em Phys. Rev.}, D91(7):074024, 2015.
\newblock \href {http://arxiv.org/abs/1412.5556} {\path{arXiv:1412.5556}},
  \href {http://dx.doi.org/10.1103/PhysRevD.91.074024}
  {\path{doi:10.1103/PhysRevD.91.074024}}.

\bibitem{Diehl:2003ny}
M.~Diehl.
\newblock {Generalized parton distributions}.
\newblock {\em Phys. Rept.}, 388:41--277, 2003.
\newblock \href {http://arxiv.org/abs/hep-ph/0307382}
  {\path{arXiv:hep-ph/0307382}}, \href
  {http://dx.doi.org/10.1016/j.physrep.2003.08.002,
  10.3204/DESY-THESIS-2003-018} {\path{doi:10.1016/j.physrep.2003.08.002,
  10.3204/DESY-THESIS-2003-018}}.

\bibitem{Feynman:1969ej}
Richard~P. Feynman.
\newblock {Very high-energy collisions of hadrons}.
\newblock {\em Phys. Rev. Lett.}, 23:1415--1417, 1969.
\newblock \href {http://dx.doi.org/10.1103/PhysRevLett.23.1415}
  {\path{doi:10.1103/PhysRevLett.23.1415}}.

\bibitem{Boglione:2016bph}
M.~Boglione, J.~Collins, L.~Gamberg, J.~O. Gonzalez-Hernandez, T.~C. Rogers,
  and N.~Sato.
\newblock {Kinematics of Current Region Fragmentation in Semi-Inclusive Deeply
  Inelastic Scattering}.
\newblock {\em Phys. Lett.}, B766:245--253, 2017.
\newblock \href {http://arxiv.org/abs/1611.10329} {\path{arXiv:1611.10329}},
  \href {http://dx.doi.org/10.1016/j.physletb.2017.01.021}
  {\path{doi:10.1016/j.physletb.2017.01.021}}.

\bibitem{Kotzinian:2007uv}
Aram Kotzinian.
\newblock {Beyond Collins and Sivers: Further measurements of the target
  transverse spin-dependent azimuthal asymmetries in semi-inclusive DIS from
  COMPASS}.
\newblock In {\em {Proceedings, 15th International Workshop on Deep-inelastic
  scattering and related subjects (DIS 2007). Vol. 1 and 2: Munich, Germany,
  April 16-20, 2007}}, pages 647--650, 2007.
\newblock \href {http://arxiv.org/abs/0705.2402} {\path{arXiv:0705.2402}},
  \href {http://dx.doi.org/10.3204/proc07-01/107}
  {\path{doi:10.3204/proc07-01/107}}.

\bibitem{Parsamyan:2013ug}
Bakur Parsamyan.
\newblock {Six 'beyond Collins and Sivers' transverse spin asymmetries at
  COMPASS}.
\newblock {\em Phys. Part. Nucl.}, 45:158--162, 2014.
\newblock \href {http://arxiv.org/abs/1301.6615} {\path{arXiv:1301.6615}},
  \href {http://dx.doi.org/10.1134/S106377961401078X}
  {\path{doi:10.1134/S106377961401078X}}.

\bibitem{Anselmino:2005nn}
M.~Anselmino, M.~Boglione, U.~D'Alesio, A.~Kotzinian, F.~Murgia, and
  A.~Prokudin.
\newblock {The Role of Cahn and sivers effects in deep inelastic scattering}.
\newblock {\em Phys. Rev.}, D71:074006, 2005.
\newblock \href {http://arxiv.org/abs/hep-ph/0501196}
  {\path{arXiv:hep-ph/0501196}}, \href
  {http://dx.doi.org/10.1103/PhysRevD.71.074006}
  {\path{doi:10.1103/PhysRevD.71.074006}}.

\bibitem{Anselmino:2007fs}
M.~Anselmino, M.~Boglione, U.~D'Alesio, A.~Kotzinian, F.~Murgia, A.~Prokudin,
  and C.~Turk.
\newblock {Transversity and Collins functions from SIDIS and e+ e- data}.
\newblock {\em Phys. Rev.}, D75:054032, 2007.
\newblock \href {http://arxiv.org/abs/hep-ph/0701006}
  {\path{arXiv:hep-ph/0701006}}, \href
  {http://dx.doi.org/10.1103/PhysRevD.75.054032}
  {\path{doi:10.1103/PhysRevD.75.054032}}.

\bibitem{Barone:2009hw}
Vincenzo Barone, Stefano Melis, and Alexei Prokudin.
\newblock {The Boer-Mulders effect in unpolarized SIDIS: An Analysis of the
  COMPASS and HERMES data on the cos 2 phi asymmetry}.
\newblock {\em Phys. Rev.}, D81:114026, 2010.
\newblock \href {http://arxiv.org/abs/0912.5194} {\path{arXiv:0912.5194}},
  \href {http://dx.doi.org/10.1103/PhysRevD.81.114026}
  {\path{doi:10.1103/PhysRevD.81.114026}}.

\bibitem{Christova:2019fbj}
E.~Christova, D.~Kotlorz, and E.~Leader.
\newblock {Towards a model independent extraction of the Boer-Mulders
  function}.
\newblock 2019.
\newblock \href {http://arxiv.org/abs/1909.08218} {\path{arXiv:1909.08218}}.

\bibitem{Airapetian:2012yg}
A.~Airapetian et~al.
\newblock {Azimuthal distributions of charged hadrons, pions, and kaons
  produced in deep-inelastic scattering off unpolarized protons and deuterons}.
\newblock {\em Phys. Rev.}, D87(1):012010, 2013.
\newblock \href {http://arxiv.org/abs/1204.4161} {\path{arXiv:1204.4161}},
  \href {http://dx.doi.org/10.1103/PhysRevD.87.012010}
  {\path{doi:10.1103/PhysRevD.87.012010}}.

\bibitem{Adolph:2014pwc}
C.~Adolph et~al.
\newblock {Measurement of azimuthal hadron asymmetries in semi-inclusive deep
  inelastic scattering off unpolarised nucleons}.
\newblock {\em Nucl. Phys.}, B886:1046--1077, 2014.
\newblock \href {http://arxiv.org/abs/1401.6284} {\path{arXiv:1401.6284}},
  \href {http://dx.doi.org/10.1016/j.nuclphysb.2014.07.019}
  {\path{doi:10.1016/j.nuclphysb.2014.07.019}}.

\bibitem{Kerbizi:2018bvk}
Albi Kerbizi.
\newblock {Interpretation of the unpolarized azimuthal asymmetries in SIDIS}.
\newblock {\em PoS}, SPIN2018:053, 2018.
\newblock \href {http://arxiv.org/abs/1812.07477} {\path{arXiv:1812.07477}},
  \href {http://dx.doi.org/10.22323/1.346.0053}
  {\path{doi:10.22323/1.346.0053}}.

\bibitem{Moretti:2019lkw}
Andrea Moretti.
\newblock {Measurement of azimuthal asymmetries in SIDIS on unpolarized
  protons}.
\newblock {\em PoS}, SPIN2018:052, 2019.
\newblock \href {http://arxiv.org/abs/1901.01773} {\path{arXiv:1901.01773}},
  \href {http://dx.doi.org/10.22323/1.346.0052}
  {\path{doi:10.22323/1.346.0052}}.

\bibitem{Matousek:2019dlk}
J.~Matoušek.
\newblock {Measurement of the azimuthal modulations of hadrons in unpolarised
  SIDIS}.
\newblock {\em PoS}, DIS2019:189, 2019.
\newblock \href {http://arxiv.org/abs/1907.08851} {\path{arXiv:1907.08851}},
  \href {http://dx.doi.org/10.22323/1.352.0189}
  {\path{doi:10.22323/1.352.0189}}.

\bibitem{Brodsky:2002cx}
Stanley~J. Brodsky, Dae~Sung Hwang, and Ivan Schmidt.
\newblock {Final state interactions and single spin asymmetries in
  semiinclusive deep inelastic scattering}.
\newblock {\em Phys. Lett.}, B530:99--107, 2002.
\newblock \href {http://arxiv.org/abs/hep-ph/0201296}
  {\path{arXiv:hep-ph/0201296}}, \href
  {http://dx.doi.org/10.1016/S0370-2693(02)01320-5}
  {\path{doi:10.1016/S0370-2693(02)01320-5}}.

\bibitem{Brodsky:2002rv}
Stanley~J. Brodsky, Dae~Sung Hwang, and Ivan Schmidt.
\newblock {Initial state interactions and single spin asymmetries in Drell-Yan
  processes}.
\newblock {\em Nucl. Phys.}, B642:344--356, 2002.
\newblock \href {http://arxiv.org/abs/hep-ph/0206259}
  {\path{arXiv:hep-ph/0206259}}, \href
  {http://dx.doi.org/10.1016/S0550-3213(02)00617-X}
  {\path{doi:10.1016/S0550-3213(02)00617-X}}.

\bibitem{Boer:2003cm}
Daniel Boer, P.~J. Mulders, and F.~Pijlman.
\newblock {Universality of T odd effects in single spin and azimuthal
  asymmetries}.
\newblock {\em Nucl. Phys.}, B667:201--241, 2003.
\newblock \href {http://arxiv.org/abs/hep-ph/0303034}
  {\path{arXiv:hep-ph/0303034}}, \href
  {http://dx.doi.org/10.1016/S0550-3213(03)00527-3}
  {\path{doi:10.1016/S0550-3213(03)00527-3}}.

\bibitem{Brodsky:2013oya}
Stanley~J. Brodsky, Dae~Sung Hwang, Yuri~V. Kovchegov, Ivan Schmidt, and
  Matthew~D. Sievert.
\newblock {Single-Spin Asymmetries in Semi-inclusive Deep Inelastic Scattering
  and Drell-Yan Processes}.
\newblock {\em Phys. Rev.}, D88(1):014032, 2013.
\newblock \href {http://arxiv.org/abs/1304.5237} {\path{arXiv:1304.5237}},
  \href {http://dx.doi.org/10.1103/PhysRevD.88.014032}
  {\path{doi:10.1103/PhysRevD.88.014032}}.

\bibitem{Kang:2017glf}
Zhong-Bo Kang, Xiaohui Liu, Felix Ringer, and Hongxi Xing.
\newblock {The transverse momentum distribution of hadrons within jets}.
\newblock {\em JHEP}, 11:068, 2017.
\newblock \href {http://arxiv.org/abs/1705.08443} {\path{arXiv:1705.08443}},
  \href {http://dx.doi.org/10.1007/JHEP11(2017)068}
  {\path{doi:10.1007/JHEP11(2017)068}}.

\bibitem{Efremov:1984ip}
A.~V. Efremov and O.~V. Teryaev.
\newblock {QCD Asymmetry and Polarized Hadron Structure Functions}.
\newblock {\em Phys. Lett.}, 150B:383, 1985.
\newblock \href {http://dx.doi.org/10.1016/0370-2693(85)90999-2}
  {\path{doi:10.1016/0370-2693(85)90999-2}}.

\bibitem{Qiu:1991pp}
Jian-wei Qiu and George~F. Sterman.
\newblock {Single transverse spin asymmetries}.
\newblock {\em Phys. Rev. Lett.}, 67:2264--2267, 1991.
\newblock \href {http://dx.doi.org/10.1103/PhysRevLett.67.2264}
  {\path{doi:10.1103/PhysRevLett.67.2264}}.

\bibitem{Qiu:1998ia}
Jian-wei Qiu and George~F. Sterman.
\newblock {Single transverse spin asymmetries in hadronic pion production}.
\newblock {\em Phys. Rev.}, D59:014004, 1999.
\newblock \href {http://arxiv.org/abs/hep-ph/9806356}
  {\path{arXiv:hep-ph/9806356}}, \href
  {http://dx.doi.org/10.1103/PhysRevD.59.014004}
  {\path{doi:10.1103/PhysRevD.59.014004}}.

\bibitem{Ji:2006br}
Xiangdong Ji, Jian-Wei Qiu, Werner Vogelsang, and Feng Yuan.
\newblock {Single-transverse spin asymmetry in semi-inclusive deep inelastic
  scattering}.
\newblock {\em Phys. Lett.}, B638:178--186, 2006.
\newblock \href {http://arxiv.org/abs/hep-ph/0604128}
  {\path{arXiv:hep-ph/0604128}}, \href
  {http://dx.doi.org/10.1016/j.physletb.2006.05.044}
  {\path{doi:10.1016/j.physletb.2006.05.044}}.

\bibitem{Kanazawa:2014dca}
Koichi Kanazawa, Yuji Koike, Andreas Metz, and Daniel Pitonyak.
\newblock {Towards an explanation of transverse single-spin asymmetries in
  proton-proton collisions: the role of fragmentation in collinear
  factorization}.
\newblock {\em Phys. Rev.}, D89(11):111501, 2014.
\newblock \href {http://arxiv.org/abs/1404.1033} {\path{arXiv:1404.1033}},
  \href {http://dx.doi.org/10.1103/PhysRevD.89.111501}
  {\path{doi:10.1103/PhysRevD.89.111501}}.

\bibitem{Kang:2011hk}
Zhong-Bo Kang, Jian-Wei Qiu, Werner Vogelsang, and Feng Yuan.
\newblock {An Observation Concerning the Process Dependence of the Sivers
  Functions}.
\newblock {\em Phys. Rev.}, D83:094001, 2011.
\newblock \href {http://arxiv.org/abs/1103.1591} {\path{arXiv:1103.1591}},
  \href {http://dx.doi.org/10.1103/PhysRevD.83.094001}
  {\path{doi:10.1103/PhysRevD.83.094001}}.

\bibitem{Beppu:2013uda}
Hiroo Beppu, Koichi Kanazawa, Yuji Koike, and Shinsuke Yoshida.
\newblock {Three-gluon contribution to the single spin asymmetry for light
  hadron production in pp collision}.
\newblock {\em Phys. Rev.}, D89(3):034029, 2014.
\newblock \href {http://arxiv.org/abs/1312.6862} {\path{arXiv:1312.6862}},
  \href {http://dx.doi.org/10.1103/PhysRevD.89.034029}
  {\path{doi:10.1103/PhysRevD.89.034029}}.

\bibitem{Koike:2009ge}
Yuji Koike and Tetsuya Tomita.
\newblock {Soft-fermion-pole contribution to single-spin asymmetry for pion
  production in pp collisions}.
\newblock {\em Phys. Lett.}, B675:181--189, 2009.
\newblock \href {http://arxiv.org/abs/0903.1923} {\path{arXiv:0903.1923}},
  \href {http://dx.doi.org/10.1016/j.physletb.2009.04.017}
  {\path{doi:10.1016/j.physletb.2009.04.017}}.

\bibitem{Kanazawa:2010au}
Koichi Kanazawa and Yuji Koike.
\newblock {New Analysis of the Single Transverse-Spin Asymmetry for Hadron
  Production at RHIC}.
\newblock {\em Phys. Rev.}, D82:034009, 2010.
\newblock \href {http://arxiv.org/abs/1005.1468} {\path{arXiv:1005.1468}},
  \href {http://dx.doi.org/10.1103/PhysRevD.82.034009}
  {\path{doi:10.1103/PhysRevD.82.034009}}.

\bibitem{Kanazawa:2011bg}
Koichi Kanazawa and Yuji Koike.
\newblock {A phenomenological study on single transverse-spin asymmetry for
  inclusive light-hadron productions at RHIC}.
\newblock {\em Phys. Rev.}, D83:114024, 2011.
\newblock \href {http://arxiv.org/abs/1104.0117} {\path{arXiv:1104.0117}},
  \href {http://dx.doi.org/10.1103/PhysRevD.83.114024}
  {\path{doi:10.1103/PhysRevD.83.114024}}.

\bibitem{DAlesio:2004eso}
Umberto D'Alesio and Francesco Murgia.
\newblock {Parton intrinsic motion in inclusive particle production:
  Unpolarized cross sections, single spin asymmetries and the Sivers effect}.
\newblock {\em Phys. Rev.}, D70:074009, 2004.
\newblock \href {http://arxiv.org/abs/hep-ph/0408092}
  {\path{arXiv:hep-ph/0408092}}, \href
  {http://dx.doi.org/10.1103/PhysRevD.70.074009}
  {\path{doi:10.1103/PhysRevD.70.074009}}.

\bibitem{Aschenauer:2015ndk}
E.~C. Aschenauer, U.~D'Alesio, and F.~Murgia.
\newblock {TMDs and SSAs in hadronic interactions}.
\newblock {\em Eur. Phys. J.}, A52(6):156, 2016.
\newblock \href {http://arxiv.org/abs/1512.05379} {\path{arXiv:1512.05379}},
  \href {http://dx.doi.org/10.1140/epja/i2016-16156-4}
  {\path{doi:10.1140/epja/i2016-16156-4}}.

\bibitem{Anselmino:2012rq}
M.~Anselmino, M.~Boglione, U.~D'Alesio, E.~Leader, S.~Melis, F.~Murgia, and
  A.~Prokudin.
\newblock {On the role of Collins effect in the single spin asymmetry $A_N$ in
  $p^\uparrow p \to h X$ processes}.
\newblock {\em Phys. Rev.}, D86:074032, 2012.
\newblock \href {http://arxiv.org/abs/1207.6529} {\path{arXiv:1207.6529}},
  \href {http://dx.doi.org/10.1103/PhysRevD.86.074032}
  {\path{doi:10.1103/PhysRevD.86.074032}}.

\bibitem{Anselmino:2013rya}
M.~Anselmino, M.~Boglione, U.~D'Alesio, S.~Melis, F.~Murgia, and A.~Prokudin.
\newblock {Sivers effect and the single spin asymmetry $A_{N}$ in $p^{\uparrow}
  p \to hX$ processes}.
\newblock {\em Phys. Rev.}, D88(5):054023, 2013.
\newblock \href {http://arxiv.org/abs/1304.7691} {\path{arXiv:1304.7691}},
  \href {http://dx.doi.org/10.1103/PhysRevD.88.054023}
  {\path{doi:10.1103/PhysRevD.88.054023}}.

\bibitem{Bacchetta:2005rm}
A.~Bacchetta, C.~J. Bomhof, P.~J. Mulders, and F.~Pijlman.
\newblock {Single spin asymmetries in hadron-hadron collisions}.
\newblock {\em Phys. Rev.}, D72:034030, 2005.
\newblock \href {http://arxiv.org/abs/hep-ph/0505268}
  {\path{arXiv:hep-ph/0505268}}, \href
  {http://dx.doi.org/10.1103/PhysRevD.72.034030}
  {\path{doi:10.1103/PhysRevD.72.034030}}.

\bibitem{DAlesio:2011kkm}
Umberto D'Alesio, Leonard Gamberg, Zhong-Bo Kang, Francesco Murgia, and
  Cristian Pisano.
\newblock {Testing the process dependence of the Sivers function via hadron
  distributions inside a jet}.
\newblock {\em Phys. Lett.}, B704:637--640, 2011.
\newblock \href {http://arxiv.org/abs/1108.0827} {\path{arXiv:1108.0827}},
  \href {http://dx.doi.org/10.1016/j.physletb.2011.09.067}
  {\path{doi:10.1016/j.physletb.2011.09.067}}.

\bibitem{Metz:2012ct}
A.~Metz and D.~Pitonyak.
\newblock {Fragmentation contribution to the transverse single-spin asymmetry
  in proton-proton collisions}.
\newblock {\em Phys. Lett.}, B723:365--370, 2013.
\newblock [Erratum: Phys. Lett.B762,549(2016)].
\newblock \href {http://arxiv.org/abs/1212.5037} {\path{arXiv:1212.5037}},
  \href {http://dx.doi.org/10.1016/j.physletb.2013.05.043,
  10.1016/j.physletb.2016.10.011} {\path{doi:10.1016/j.physletb.2013.05.043,
  10.1016/j.physletb.2016.10.011}}.

\bibitem{Kouvaris:2006zy}
Chris Kouvaris, Jian-Wei Qiu, Werner Vogelsang, and Feng Yuan.
\newblock {Single transverse-spin asymmetry in high transverse momentum pion
  production in pp collisions}.
\newblock {\em Phys. Rev.}, D74:114013, 2006.
\newblock \href {http://arxiv.org/abs/hep-ph/0609238}
  {\path{arXiv:hep-ph/0609238}}, \href
  {http://dx.doi.org/10.1103/PhysRevD.74.114013}
  {\path{doi:10.1103/PhysRevD.74.114013}}.

\bibitem{Kanazawa:2014nea}
K.~Kanazawa, Y.~Koike, A.~Metz, and D.~Pitonyak.
\newblock {Transverse single-spin asymmetries in $p^\uparrow p \to \gamma X$
  from quark-gluon-quark correlations in the proton}.
\newblock {\em Phys. Rev.}, D91(1):014013, 2015.
\newblock \href {http://arxiv.org/abs/1410.3448} {\path{arXiv:1410.3448}},
  \href {http://dx.doi.org/10.1103/PhysRevD.91.014013}
  {\path{doi:10.1103/PhysRevD.91.014013}}.

\bibitem{Zhou:2011ba}
Jian Zhou and Andreas Metz.
\newblock {Dihadron fragmentation functions for large invariant mass}.
\newblock {\em Phys. Rev. Lett.}, 106:172001, 2011.
\newblock \href {http://arxiv.org/abs/1101.3273} {\path{arXiv:1101.3273}},
  \href {http://dx.doi.org/10.1103/PhysRevLett.106.172001}
  {\path{doi:10.1103/PhysRevLett.106.172001}}.

\bibitem{Gliske:2014wba}
Stephen Gliske, Alessandro Bacchetta, and Marco Radici.
\newblock {Production of two hadrons in semi-inclusive deep inelastic
  scattering}.
\newblock {\em Phys. Rev.}, D90(11):114027, 2014.
\newblock [Erratum: Phys. Rev.D91,no.1,019902(2015)].
\newblock \href {http://arxiv.org/abs/1408.5721} {\path{arXiv:1408.5721}},
  \href {http://dx.doi.org/10.1103/PhysRevD.90.114027,
  10.1103/PhysRevD.91.019902} {\path{doi:10.1103/PhysRevD.90.114027,
  10.1103/PhysRevD.91.019902}}.

\bibitem{Jaffe:1997hf}
R.~L. Jaffe, Xue-min Jin, and Jian Tang.
\newblock {Interference fragmentation functions and the nucleon's
  transversity}.
\newblock {\em Phys. Rev. Lett.}, 80:1166--1169, 1998.
\newblock \href {http://arxiv.org/abs/hep-ph/9709322}
  {\path{arXiv:hep-ph/9709322}}, \href
  {http://dx.doi.org/10.1103/PhysRevLett.80.1166}
  {\path{doi:10.1103/PhysRevLett.80.1166}}.

\bibitem{Radici:2001na}
Marco Radici, Rainer Jakob, and Andrea Bianconi.
\newblock {Accessing transversity with interference fragmentation functions}.
\newblock {\em Phys. Rev.}, D65:074031, 2002.
\newblock \href {http://arxiv.org/abs/hep-ph/0110252}
  {\path{arXiv:hep-ph/0110252}}, \href
  {http://dx.doi.org/10.1103/PhysRevD.65.074031}
  {\path{doi:10.1103/PhysRevD.65.074031}}.

\bibitem{Bacchetta:2006un}
Alessandro Bacchetta and Marco Radici.
\newblock {Modeling dihadron fragmentation functions}.
\newblock {\em Phys. Rev.}, D74:114007, 2006.
\newblock \href {http://arxiv.org/abs/hep-ph/0608037}
  {\path{arXiv:hep-ph/0608037}}, \href
  {http://dx.doi.org/10.1103/PhysRevD.74.114007}
  {\path{doi:10.1103/PhysRevD.74.114007}}.

\bibitem{Yuan:2009dw}
Feng Yuan and Jian Zhou.
\newblock {Collins Fragmentation and the Single Transverse Spin Asymmetry}.
\newblock {\em Phys. Rev. Lett.}, 103:052001, 2009.
\newblock \href {http://arxiv.org/abs/0903.4680} {\path{arXiv:0903.4680}},
  \href {http://dx.doi.org/10.1103/PhysRevLett.103.052001}
  {\path{doi:10.1103/PhysRevLett.103.052001}}.

\bibitem{Gamberg:2013kla}
Leonard Gamberg, Zhong-Bo Kang, and Alexei Prokudin.
\newblock {Indication on the process-dependence of the Sivers effect}.
\newblock {\em Phys. Rev. Lett.}, 110(23):232301, 2013.
\newblock \href {http://arxiv.org/abs/1302.3218} {\path{arXiv:1302.3218}},
  \href {http://dx.doi.org/10.1103/PhysRevLett.110.232301}
  {\path{doi:10.1103/PhysRevLett.110.232301}}.

\bibitem{Pitonyak:2013dsu}
D.~Pitonyak, M.~Schlegel, and A.~Metz.
\newblock {Polarized hadron pair production from electron-positron
  annihilation}.
\newblock {\em Phys. Rev.}, D89(5):054032, 2014.
\newblock \href {http://arxiv.org/abs/1310.6240} {\path{arXiv:1310.6240}},
  \href {http://dx.doi.org/10.1103/PhysRevD.89.054032}
  {\path{doi:10.1103/PhysRevD.89.054032}}.

\bibitem{Boer:phd}
D.~Boer.
\newblock {\em {Ph.D.~thesis, Vrije U., Amsterdam, 1998}}.

\bibitem{Anselmino:2019cqd}
M.~Anselmino, R.~Kishore, and A.~Mukherjee.
\newblock {Polarizing fragmentation function and the $\Lambda$ polarization in
  $e^+e^-$ processes}.
\newblock {\em Phys. Rev.}, D100(1):014029, 2019.
\newblock \href {http://arxiv.org/abs/1905.02777} {\path{arXiv:1905.02777}},
  \href {http://dx.doi.org/10.1103/PhysRevD.100.014029}
  {\path{doi:10.1103/PhysRevD.100.014029}}.

\bibitem{DAlesio:2020wjq}
Umberto D'Alesio, Francesco Murgia, and Marco Zaccheddu.
\newblock {First extraction of the $\Lambda$ polarising fragmentation function
  from Belle $e^+e^-$ data}.
\newblock 2020.
\newblock \href {http://arxiv.org/abs/2003.01128} {\path{arXiv:2003.01128}}.

\bibitem{Callos:2020qtu}
Daniel Callos, Zhong-Bo Kang, and John Terry.
\newblock {Extracting the Transverse Momentum Dependent Polarizing
  Fragmentation Functions}.
\newblock 2020.
\newblock \href {http://arxiv.org/abs/2003.04828} {\path{arXiv:2003.04828}}.

\bibitem{Koike:2017fxr}
Yuji Koike, Andreas Metz, Daniel Pitonyak, Kenta Yabe, and Shinsuke Yoshida.
\newblock {Twist-3 fragmentation contribution to polarized hyperon production
  in unpolarized hadronic collisions}.
\newblock {\em Phys. Rev.}, D95(11):114013, 2017.
\newblock \href {http://arxiv.org/abs/1703.09399} {\path{arXiv:1703.09399}},
  \href {http://dx.doi.org/10.1103/PhysRevD.95.114013}
  {\path{doi:10.1103/PhysRevD.95.114013}}.

\bibitem{Gamberg:2018fwy}
Leonard Gamberg, Zhong-Bo Kang, Daniel Pitonyak, Marc Schlegel, and Shinsuke
  Yoshida.
\newblock {Polarized hyperon production in single-inclusive electron-positron
  annihilation at next-to-leading order}.
\newblock {\em JHEP}, 01:111, 2019.
\newblock \href {http://arxiv.org/abs/1810.08645} {\path{arXiv:1810.08645}},
  \href {http://dx.doi.org/10.1007/JHEP01(2019)111}
  {\path{doi:10.1007/JHEP01(2019)111}}.

\bibitem{Anselmino:2008jk}
M.~Anselmino, M.~Boglione, U.~D'Alesio, A.~Kotzinian, F.~Murgia, A.~Prokudin,
  and S.~Melis.
\newblock {Update on transversity and Collins functions from SIDIS and e+ e-
  data}.
\newblock {\em Nucl. Phys. Proc. Suppl.}, 191:98--107, 2009.
\newblock \href {http://arxiv.org/abs/0812.4366} {\path{arXiv:0812.4366}},
  \href {http://dx.doi.org/10.1016/j.nuclphysbps.2009.03.117}
  {\path{doi:10.1016/j.nuclphysbps.2009.03.117}}.

\bibitem{Anselmino:2013vqa}
M.~Anselmino, M.~Boglione, U.~D'Alesio, S.~Melis, F.~Murgia, and A.~Prokudin.
\newblock {Simultaneous extraction of transversity and Collins functions from
  new SIDIS and e+e- data}.
\newblock {\em Phys. Rev.}, D87:094019, 2013.
\newblock \href {http://arxiv.org/abs/1303.3822} {\path{arXiv:1303.3822}},
  \href {http://dx.doi.org/10.1103/PhysRevD.87.094019}
  {\path{doi:10.1103/PhysRevD.87.094019}}.

\bibitem{Collins:2005ie}
J.~C. Collins, A.~V. Efremov, K.~Goeke, S.~Menzel, A.~Metz, and P.~Schweitzer.
\newblock {Sivers effect in semi-inclusive deeply inelastic scattering}.
\newblock {\em Phys. Rev.}, D73:014021, 2006.
\newblock \href {http://arxiv.org/abs/hep-ph/0509076}
  {\path{arXiv:hep-ph/0509076}}, \href
  {http://dx.doi.org/10.1103/PhysRevD.73.014021}
  {\path{doi:10.1103/PhysRevD.73.014021}}.

\bibitem{Schweitzer:2010tt}
P.~Schweitzer, T.~Teckentrup, and A.~Metz.
\newblock {Intrinsic transverse parton momenta in deeply inelastic reactions}.
\newblock {\em Phys. Rev.}, D81:094019, 2010.
\newblock \href {http://arxiv.org/abs/1003.2190} {\path{arXiv:1003.2190}},
  \href {http://dx.doi.org/10.1103/PhysRevD.81.094019}
  {\path{doi:10.1103/PhysRevD.81.094019}}.

\bibitem{Adams:1993hs}
M.~R. Adams et~al.
\newblock {Perturbative QCD effects observed in 490-GeV deep inelastic muon
  scattering}.
\newblock {\em Phys. Rev.}, D48:5057--5066, 1993.
\newblock \href {http://dx.doi.org/10.1103/PhysRevD.48.5057}
  {\path{doi:10.1103/PhysRevD.48.5057}}.

\bibitem{Arneodo:1986cf}
M.~Arneodo et~al.
\newblock {Measurement of Hadron Azimuthal Distributions in Deep Inelastic Muon
  Proton Scattering}.
\newblock {\em Z. Phys.}, C34:277, 1987.
\newblock \href {http://dx.doi.org/10.1007/BF01548808}
  {\path{doi:10.1007/BF01548808}}.

\bibitem{Airapetian:2002mf}
A.~Airapetian et~al.
\newblock {Measurement of single spin azimuthal asymmetries in semiinclusive
  electroproduction of pions and kaons on a longitudinally polarized deuterium
  target}.
\newblock {\em Phys. Lett.}, B562:182--192, 2003.
\newblock \href {http://arxiv.org/abs/hep-ex/0212039}
  {\path{arXiv:hep-ex/0212039}}, \href
  {http://dx.doi.org/10.1016/S0370-2693(03)00566-5}
  {\path{doi:10.1016/S0370-2693(03)00566-5}}.

\bibitem{Mkrtchyan:2007sr}
H.~Mkrtchyan et~al.
\newblock {Transverse momentum dependence of semi-inclusive pion production}.
\newblock {\em Phys. Lett.}, B665:20--25, 2008.
\newblock \href {http://arxiv.org/abs/0709.3020} {\path{arXiv:0709.3020}},
  \href {http://dx.doi.org/10.1016/j.physletb.2008.05.047}
  {\path{doi:10.1016/j.physletb.2008.05.047}}.

\bibitem{Osipenko:2008aa}
M.~Osipenko et~al.
\newblock {Measurement of unpolarized semi-inclusive pi+ electroproduction off
  the proton}.
\newblock {\em Phys. Rev.}, D80:032004, 2009.
\newblock \href {http://arxiv.org/abs/0809.1153} {\path{arXiv:0809.1153}},
  \href {http://dx.doi.org/10.1103/PhysRevD.80.032004}
  {\path{doi:10.1103/PhysRevD.80.032004}}.

\bibitem{Airapetian:2009jy}
A.~Airapetian et~al.
\newblock {Transverse momentum broadening of hadrons produced in semi-inclusive
  deep-inelastic scattering on nuclei}.
\newblock {\em Phys. Lett.}, B684:114--118, 2010.
\newblock \href {http://arxiv.org/abs/0906.2478} {\path{arXiv:0906.2478}},
  \href {http://dx.doi.org/10.1016/j.physletb.2010.01.020}
  {\path{doi:10.1016/j.physletb.2010.01.020}}.

\bibitem{Adolph:2013stb}
C.~Adolph et~al.
\newblock {Hadron Transverse Momentum Distributions in Muon Deep Inelastic
  Scattering at 160 GeV/$c$}.
\newblock {\em Eur. Phys. J.}, C73:2531, 2013.
\newblock \href {http://arxiv.org/abs/1305.7317} {\path{arXiv:1305.7317}},
  \href {http://dx.doi.org/10.1140/epjc/s10052-013-2531-6}
  {\path{doi:10.1140/epjc/s10052-013-2531-6}}.

\bibitem{Aghasyan:2017ctw}
M.~Aghasyan et~al.
\newblock {Transverse-momentum-dependent Multiplicities of Charged Hadrons in
  Muon-Deuteron Deep Inelastic Scattering}.
\newblock {\em Phys. Rev.}, D97(3):032006, 2018.
\newblock \href {http://arxiv.org/abs/1709.07374} {\path{arXiv:1709.07374}},
  \href {http://dx.doi.org/10.1103/PhysRevD.97.032006}
  {\path{doi:10.1103/PhysRevD.97.032006}}.

\bibitem{Airapetian:2012ki}
A.~Airapetian et~al.
\newblock {Multiplicities of charged pions and kaons from semi-inclusive
  deep-inelastic scattering by the proton and the deuteron}.
\newblock {\em Phys. Rev.}, D87:074029, 2013.
\newblock \href {http://arxiv.org/abs/1212.5407} {\path{arXiv:1212.5407}},
  \href {http://dx.doi.org/10.1103/PhysRevD.87.074029}
  {\path{doi:10.1103/PhysRevD.87.074029}}.

\bibitem{Signori:2013mda}
Andrea Signori, Alessandro Bacchetta, Marco Radici, and Gunar Schnell.
\newblock {Investigations into the flavor dependence of partonic transverse
  momentum}.
\newblock {\em JHEP}, 11:194, 2013.
\newblock \href {http://arxiv.org/abs/1309.3507} {\path{arXiv:1309.3507}},
  \href {http://dx.doi.org/10.1007/JHEP11(2013)194}
  {\path{doi:10.1007/JHEP11(2013)194}}.

\bibitem{Anselmino:2013lza}
M.~Anselmino, M.~Boglione, J.~O. Gonzalez~Hernandez, S.~Melis, and A.~Prokudin.
\newblock {Unpolarised Transverse Momentum Dependent Distribution and
  Fragmentation Functions from SIDIS Multiplicities}.
\newblock {\em JHEP}, 04:005, 2014.
\newblock \href {http://arxiv.org/abs/1312.6261} {\path{arXiv:1312.6261}},
  \href {http://dx.doi.org/10.1007/JHEP04(2014)005}
  {\path{doi:10.1007/JHEP04(2014)005}}.

\bibitem{Anselmino:2016uie}
M.~Anselmino, M.~Boglione, U.~D'Alesio, F.~Murgia, and A.~Prokudin.
\newblock {Study of the sign change of the Sivers function from STAR
  Collaboration W/Z production data}.
\newblock {\em JHEP}, 04:046, 2017.
\newblock \href {http://arxiv.org/abs/1612.06413} {\path{arXiv:1612.06413}},
  \href {http://dx.doi.org/10.1007/JHEP04(2017)046}
  {\path{doi:10.1007/JHEP04(2017)046}}.

\bibitem{Matousek:2018qqd}
J.~Matoušek.
\newblock {Weighted transverse spin asymmetries in 2015 COMPASS Drell-Yan
  data}.
\newblock {\em PoS}, SPIN2018:038, 2018.
\newblock \href {http://arxiv.org/abs/1812.08505} {\path{arXiv:1812.08505}},
  \href {http://dx.doi.org/10.22323/1.346.0038}
  {\path{doi:10.22323/1.346.0038}}.

\bibitem{Anselmino:2005an}
M.~Anselmino et~al.
\newblock {Comparing extractions of Sivers functions}.
\newblock In {\em {Transversity. Proceedings, Workshop, Como, Italy, September
  7-10, 2005}}, pages 236--243, 2005.
\newblock \href {http://arxiv.org/abs/hep-ph/0511017}
  {\path{arXiv:hep-ph/0511017}}, \href
  {http://dx.doi.org/10.1142/9789812773272_0028}
  {\path{doi:10.1142/9789812773272_0028}}.

\bibitem{Anselmino:2005ea}
M.~Anselmino, M.~Boglione, U.~D'Alesio, A.~Kotzinian, F.~Murgia, and
  A.~Prokudin.
\newblock {Extracting the Sivers function from polarized SIDIS data and making
  predictions}.
\newblock {\em Phys. Rev.}, D72:094007, 2005.
\newblock [Erratum: Phys. Rev.D72,099903(2005)].
\newblock \href {http://arxiv.org/abs/hep-ph/0507181}
  {\path{arXiv:hep-ph/0507181}}, \href
  {http://dx.doi.org/10.1103/PhysRevD.72.094007, 10.1103/PhysRevD.72.099903}
  {\path{doi:10.1103/PhysRevD.72.094007, 10.1103/PhysRevD.72.099903}}.

\bibitem{Vogelsang:2005cs}
Werner Vogelsang and Feng Yuan.
\newblock {Single-transverse spin asymmetries: From DIS to hadronic
  collisions}.
\newblock {\em Phys. Rev.}, D72:054028, 2005.
\newblock \href {http://arxiv.org/abs/hep-ph/0507266}
  {\path{arXiv:hep-ph/0507266}}, \href
  {http://dx.doi.org/10.1103/PhysRevD.72.054028}
  {\path{doi:10.1103/PhysRevD.72.054028}}.

\bibitem{Collins:2005wb}
J.~C. Collins, A.~V. Efremov, K.~Goeke, M.~Grosse~Perdekamp, S.~Menzel,
  B.~Meredith, A.~Metz, and P.~Schweitzer.
\newblock {Sivers effect in semi-inclusive deeply inelastic scattering and
  Drell-Yan}.
\newblock In {\em {Transversity. Proceedings, Workshop, Como, Italy, September
  7-10, 2005}}, pages 212--219, 2005.
\newblock \href {http://arxiv.org/abs/hep-ph/0510342}
  {\path{arXiv:hep-ph/0510342}}, \href
  {http://dx.doi.org/10.1142/9789812773272_0025}
  {\path{doi:10.1142/9789812773272_0025}}.

\bibitem{Anselmino:2008sga}
M.~Anselmino, M.~Boglione, U.~D'Alesio, A.~Kotzinian, S.~Melis, F.~Murgia,
  A.~Prokudin, and C.~Turk.
\newblock {Sivers Effect for Pion and Kaon Production in Semi-Inclusive Deep
  Inelastic Scattering}.
\newblock {\em Eur. Phys. J.}, A39:89--100, 2009.
\newblock \href {http://arxiv.org/abs/0805.2677} {\path{arXiv:0805.2677}},
  \href {http://dx.doi.org/10.1140/epja/i2008-10697-y}
  {\path{doi:10.1140/epja/i2008-10697-y}}.

\bibitem{Bacchetta:2011gx}
Alessandro Bacchetta and Marco Radici.
\newblock {Constraining quark angular momentum through semi-inclusive
  measurements}.
\newblock {\em Phys. Rev. Lett.}, 107:212001, 2011.
\newblock \href {http://arxiv.org/abs/1107.5755} {\path{arXiv:1107.5755}},
  \href {http://dx.doi.org/10.1103/PhysRevLett.107.212001}
  {\path{doi:10.1103/PhysRevLett.107.212001}}.

\bibitem{Anselmino:2018psi}
M.~Anselmino, M.~Boglione, U.~D'Alesio, F.~Murgia, and A.~Prokudin.
\newblock {Role of transverse momentum dependence of unpolarized parton
  distribution and fragmentation functions in the analysis of azimuthal spin
  asymmetries}.
\newblock {\em Phys. Rev.}, D98(9):094023, 2018.
\newblock \href {http://arxiv.org/abs/1809.09500} {\path{arXiv:1809.09500}},
  \href {http://dx.doi.org/10.1103/PhysRevD.98.094023}
  {\path{doi:10.1103/PhysRevD.98.094023}}.

\bibitem{Burkardt:2004ur}
Matthias Burkardt.
\newblock {Sivers mechanism for gluons}.
\newblock {\em Phys. Rev.}, D69:091501, 2004.
\newblock \href {http://arxiv.org/abs/hep-ph/0402014}
  {\path{arXiv:hep-ph/0402014}}, \href
  {http://dx.doi.org/10.1103/PhysRevD.69.091501}
  {\path{doi:10.1103/PhysRevD.69.091501}}.

\bibitem{Anselmino:2006yq}
M.~Anselmino, U.~D'Alesio, S.~Melis, and F.~Murgia.
\newblock {Constraints on the gluon Sivers distribution via transverse single
  spin asymmetries at mid-rapidity in $p^\uparrow p \to \pi^0 X$ processes at
  RHIC}.
\newblock {\em Phys. Rev.}, D74:094011, 2006.
\newblock \href {http://arxiv.org/abs/hep-ph/0608211}
  {\path{arXiv:hep-ph/0608211}}, \href
  {http://dx.doi.org/10.1103/PhysRevD.74.094011}
  {\path{doi:10.1103/PhysRevD.74.094011}}.

\bibitem{Brodsky:2006ha}
Stanley~J. Brodsky and Susan Gardner.
\newblock {Evidence for the Absence of Gluon Orbital Angular Momentum in the
  Nucleon}.
\newblock {\em Phys. Lett.}, B643:22--28, 2006.
\newblock \href {http://arxiv.org/abs/hep-ph/0608219}
  {\path{arXiv:hep-ph/0608219}}, \href
  {http://dx.doi.org/10.1016/j.physletb.2006.10.024}
  {\path{doi:10.1016/j.physletb.2006.10.024}}.

\bibitem{DAlesio:2015fwo}
U.~D'Alesio, F.~Murgia, and C.~Pisano.
\newblock {Towards a first estimate of the gluon Sivers function from A$_{N}$
  data in pp collisions at RHIC}.
\newblock {\em JHEP}, 09:119, 2015.
\newblock \href {http://arxiv.org/abs/1506.03078} {\path{arXiv:1506.03078}},
  \href {http://dx.doi.org/10.1007/JHEP09(2015)119}
  {\path{doi:10.1007/JHEP09(2015)119}}.

\bibitem{Anselmino:2004nk}
M.~Anselmino, M.~Boglione, U.~D'Alesio, E.~Leader, and F.~Murgia.
\newblock {Accessing Sivers gluon distribution via transverse single spin
  asymmetries in $p^\uparrow p \to D + X$ processes at RHIC}.
\newblock {\em Phys. Rev.}, D70:074025, 2004.
\newblock \href {http://arxiv.org/abs/hep-ph/0407100}
  {\path{arXiv:hep-ph/0407100}}, \href
  {http://dx.doi.org/10.1103/PhysRevD.70.074025}
  {\path{doi:10.1103/PhysRevD.70.074025}}.

\bibitem{DAlesio:2017rzj}
Umberto D'Alesio, Francesco Murgia, Cristian Pisano, and Pieter Taels.
\newblock {Probing the gluon Sivers function in $p^\uparrow p\to J/\psi\,X$ and
  $p^\uparrow p \to D\,X$}.
\newblock {\em Phys. Rev.}, D96(3):036011, 2017.
\newblock \href {http://arxiv.org/abs/1705.04169} {\path{arXiv:1705.04169}},
  \href {http://dx.doi.org/10.1103/PhysRevD.96.036011}
  {\path{doi:10.1103/PhysRevD.96.036011}}.

\bibitem{DAlesio:2019gnu}
Umberto D'Alesio, Francesco Murgia, Cristian Pisano, and Sangem Rajesh.
\newblock {Single-spin asymmetries in $p^\uparrow p \to J/\psi + X$ within a
  TMD approach: role of the color octet mechanism}.
\newblock 2019.
\newblock \href {http://arxiv.org/abs/1910.09640} {\path{arXiv:1910.09640}}.

\bibitem{DAlesio:2018rnv}
Umberto D'Alesio, Carlo Flore, Francesco Murgia, Cristian Pisano, and Pieter
  Taels.
\newblock {Unraveling the Gluon Sivers Function in Hadronic Collisions at
  RHIC}.
\newblock {\em Phys. Rev.}, D99(3):036013, 2019.
\newblock \href {http://arxiv.org/abs/1811.02970} {\path{arXiv:1811.02970}},
  \href {http://dx.doi.org/10.1103/PhysRevD.99.036013}
  {\path{doi:10.1103/PhysRevD.99.036013}}.

\bibitem{Sivers:2006rg}
Dennis Sivers.
\newblock {Single-Spin Observables and Orbital Structures in Hadronic
  Distributions}.
\newblock {\em Phys. Rev.}, D74:094008, 2006.
\newblock \href {http://arxiv.org/abs/hep-ph/0609080}
  {\path{arXiv:hep-ph/0609080}}, \href
  {http://dx.doi.org/10.1103/PhysRevD.74.094008}
  {\path{doi:10.1103/PhysRevD.74.094008}}.

\bibitem{Sivers:2007pq}
Dennis Sivers.
\newblock {Chiral dynamics and single-spin asymmetries}.
\newblock In {\em {Proceedings, XII Advanced Research Workshop on High Energy
  Spin Physics, DSPIN-07, Dubna, Russia, 3-7 Sep, 2007}}, pages 161--165, 2007.
\newblock \href {http://arxiv.org/abs/0711.3185} {\path{arXiv:0711.3185}}.

\bibitem{Anselmino:2010bs}
M.~Anselmino, M.~Boglione, U.~D'Alesio, S.~Melis, F.~Murgia, and A.~Prokudin.
\newblock {New insight on the Sivers transverse momentum dependent distribution
  function}.
\newblock {\em J. Phys. Conf. Ser.}, 295:012062, 2011.
\newblock \href {http://arxiv.org/abs/1012.3565} {\path{arXiv:1012.3565}},
  \href {http://dx.doi.org/10.1088/1742-6596/295/1/012062}
  {\path{doi:10.1088/1742-6596/295/1/012062}}.

\bibitem{Soffer:1994ww}
Jacques Soffer.
\newblock {Positivity constraints for spin dependent parton distributions}.
\newblock {\em Phys. Rev. Lett.}, 74:1292--1294, 1995.
\newblock \href {http://arxiv.org/abs/hep-ph/9409254}
  {\path{arXiv:hep-ph/9409254}}, \href
  {http://dx.doi.org/10.1103/PhysRevLett.74.1292}
  {\path{doi:10.1103/PhysRevLett.74.1292}}.

\bibitem{Radici:2015mwa}
Marco Radici, A.~Courtoy, Alessandro Bacchetta, and Marco Guagnelli.
\newblock {Improved extraction of valence transversity distributions from
  inclusive dihadron production}.
\newblock {\em JHEP}, 05:123, 2015.
\newblock \href {http://arxiv.org/abs/1503.03495} {\path{arXiv:1503.03495}},
  \href {http://dx.doi.org/10.1007/JHEP05(2015)123}
  {\path{doi:10.1007/JHEP05(2015)123}}.

\bibitem{Alekseev:2008aa}
M.~Alekseev et~al.
\newblock {Collins and Sivers asymmetries for pions and kaons in muon-deuteron
  DIS}.
\newblock {\em Phys. Lett.}, B673:127--135, 2009.
\newblock \href {http://arxiv.org/abs/0802.2160} {\path{arXiv:0802.2160}},
  \href {http://dx.doi.org/10.1016/j.physletb.2009.01.060}
  {\path{doi:10.1016/j.physletb.2009.01.060}}.

\bibitem{Martin:2013eja}
Anna Martin.
\newblock {COMPASS results on Collins and Sivers asymmetries for charged
  hadrons}.
\newblock {\em Phys. Part. Nucl.}, 45:141--145, 2014.
\newblock \href {http://arxiv.org/abs/1303.2076} {\path{arXiv:1303.2076}},
  \href {http://dx.doi.org/10.1134/S1063779614010079}
  {\path{doi:10.1134/S1063779614010079}}.

\bibitem{Bianconi:1999cd}
A.~Bianconi, S.~Boffi, R.~Jakob, and M.~Radici.
\newblock {Two hadron interference fragmentation functions. Part 1. General
  framework}.
\newblock {\em Phys. Rev.}, D62:034008, 2000.
\newblock \href {http://arxiv.org/abs/hep-ph/9907475}
  {\path{arXiv:hep-ph/9907475}}, \href
  {http://dx.doi.org/10.1103/PhysRevD.62.034008}
  {\path{doi:10.1103/PhysRevD.62.034008}}.

\bibitem{Bacchetta:2011ip}
Alessandro Bacchetta, Aurore Courtoy, and Marco Radici.
\newblock {First glances at the transversity parton distribution through
  dihadron fragmentation functions}.
\newblock {\em Phys. Rev. Lett.}, 107:012001, 2011.
\newblock \href {http://arxiv.org/abs/1104.3855} {\path{arXiv:1104.3855}},
  \href {http://dx.doi.org/10.1103/PhysRevLett.107.012001}
  {\path{doi:10.1103/PhysRevLett.107.012001}}.

\bibitem{Bacchetta:2012ty}
Alessandro Bacchetta, A.~Courtoy, and Marco Radici.
\newblock {First extraction of valence transversities in a collinear
  framework}.
\newblock {\em JHEP}, 03:119, 2013.
\newblock \href {http://arxiv.org/abs/1212.3568} {\path{arXiv:1212.3568}},
  \href {http://dx.doi.org/10.1007/JHEP03(2013)119}
  {\path{doi:10.1007/JHEP03(2013)119}}.

\bibitem{Vossen:2011fk}
A.~Vossen et~al.
\newblock {Observation of transverse polarization asymmetries of charged pion
  pairs in $e^+e^-$ annihilation near $\sqrt{s}=10.58$ GeV}.
\newblock {\em Phys. Rev. Lett.}, 107:072004, 2011.
\newblock \href {http://arxiv.org/abs/1104.2425} {\path{arXiv:1104.2425}},
  \href {http://dx.doi.org/10.1103/PhysRevLett.107.072004}
  {\path{doi:10.1103/PhysRevLett.107.072004}}.

\bibitem{Radici:2016lam}
Marco Radici, Alessandro~M. Ricci, Alessandro Bacchetta, and Asmita Mukherjee.
\newblock {Exploring universality of transversity in proton-proton collisions}.
\newblock {\em Phys. Rev.}, D94(3):034012, 2016.
\newblock \href {http://arxiv.org/abs/1604.06585} {\path{arXiv:1604.06585}},
  \href {http://dx.doi.org/10.1103/PhysRevD.94.034012}
  {\path{doi:10.1103/PhysRevD.94.034012}}.

\bibitem{Jaffe:1991kp}
R.~L. Jaffe and Xiang-Dong Ji.
\newblock {Chiral odd parton distributions and polarized Drell-Yan}.
\newblock {\em Phys. Rev. Lett.}, 67:552--555, 1991.
\newblock \href {http://dx.doi.org/10.1103/PhysRevLett.67.552}
  {\path{doi:10.1103/PhysRevLett.67.552}}.

\bibitem{Bhattacharya:2016zcn}
Tanmoy Bhattacharya, Vincenzo Cirigliano, Saul Cohen, Rajan Gupta, Huey-Wen
  Lin, and Boram Yoon.
\newblock {Axial, Scalar and Tensor Charges of the Nucleon from 2+1+1-flavor
  Lattice QCD}.
\newblock {\em Phys. Rev.}, D94(5):054508, 2016.
\newblock \href {http://arxiv.org/abs/1606.07049} {\path{arXiv:1606.07049}},
  \href {http://dx.doi.org/10.1103/PhysRevD.94.054508}
  {\path{doi:10.1103/PhysRevD.94.054508}}.

\bibitem{Alexandrou:2017qyt}
C.~Alexandrou et~al.
\newblock {Nucleon scalar and tensor charges using lattice QCD simulations at
  the physical value of the pion mass}.
\newblock {\em Phys. Rev.}, D95(11):114514, 2017.
\newblock [erratum: Phys. Rev.D96,no.9,099906(2017)].
\newblock \href {http://arxiv.org/abs/1703.08788} {\path{arXiv:1703.08788}},
  \href {http://dx.doi.org/10.1103/PhysRevD.96.099906,
  10.1103/PhysRevD.95.114514} {\path{doi:10.1103/PhysRevD.96.099906,
  10.1103/PhysRevD.95.114514}}.

\bibitem{Gupta:2018lvp}
Rajan Gupta, Boram Yoon, Tanmoy Bhattacharya, Vincenzo Cirigliano, Yong-Chull
  Jang, and Huey-Wen Lin.
\newblock {Flavor diagonal tensor charges of the nucleon from (2+1+1)-flavor
  lattice QCD}.
\newblock {\em Phys. Rev.}, D98(9):091501, 2018.
\newblock \href {http://arxiv.org/abs/1808.07597} {\path{arXiv:1808.07597}},
  \href {http://dx.doi.org/10.1103/PhysRevD.98.091501}
  {\path{doi:10.1103/PhysRevD.98.091501}}.

\bibitem{Radici:2018abr}
Marco Radici.
\newblock {First extraction of transversity from data on lepton-hadron
  scattering and hadronic collisions}.
\newblock {\em PoS}, SPIN2018:044, 2019.
\newblock \href {http://arxiv.org/abs/1810.00496} {\path{arXiv:1810.00496}},
  \href {http://dx.doi.org/10.22323/1.346.0044}
  {\path{doi:10.22323/1.346.0044}}.

\bibitem{Radici:2019myq}
Marco Radici.
\newblock {Update on phenomenological extraction of the proton tensor charge}.
\newblock {\em PoS}, DIS2019:199, 2019.
\newblock \href {http://dx.doi.org/10.22323/1.352.0199}
  {\path{doi:10.22323/1.352.0199}}.

\bibitem{DAlesio:2020vtw}
Umberto D'Alesio, Carlo Flore, and Alexei Prokudin.
\newblock {Role of the Soffer bound in determination of transversity and the
  tensor charge}.
\newblock 2020.
\newblock \href {http://arxiv.org/abs/2001.01573} {\path{arXiv:2001.01573}}.

\bibitem{Benel:2019mcq}
J.~Benel, A.~Courtoy, and R.~Ferro-Hernandez.
\newblock {Constrained fit of the valence transversity distributions from
  dihadron production}.
\newblock 2019.
\newblock \href {http://arxiv.org/abs/1912.03289} {\path{arXiv:1912.03289}}.

\bibitem{Cammarota:2020qcw}
Justin Cammarota, Leonard Gamberg, Zhong-Bo Kang, Joshua~A. Miller, Daniel
  Pitonyak, Alexei Prokudin, Ted~C. Rogers, and Nobuo Sato.
\newblock {The origin of single transverse-spin asymmetries in high-energy
  collisions}.
\newblock 2020.
\newblock \href {http://arxiv.org/abs/2002.08384} {\path{arXiv:2002.08384}}.

\bibitem{Barone:2005pu}
Vincenzo Barone et~al.
\newblock {Antiproton-proton scattering experiments with polarization}.
\newblock 2005.
\newblock \href {http://arxiv.org/abs/hep-ex/0505054}
  {\path{arXiv:hep-ex/0505054}}.

\bibitem{Burkardt:2000za}
Matthias Burkardt.
\newblock {Impact parameter dependent parton distributions and off forward
  parton distributions for $\zeta \to 0$}.
\newblock {\em Phys. Rev.}, D62:071503, 2000.
\newblock [Erratum: Phys. Rev.D66,119903(2002)].
\newblock \href {http://arxiv.org/abs/hep-ph/0005108}
  {\path{arXiv:hep-ph/0005108}}, \href
  {http://dx.doi.org/10.1103/PhysRevD.62.071503, 10.1103/PhysRevD.66.119903}
  {\path{doi:10.1103/PhysRevD.62.071503, 10.1103/PhysRevD.66.119903}}.

\bibitem{Diehl:2015uka}
Markus Diehl.
\newblock {Introduction to GPDs and TMDs}.
\newblock {\em Eur. Phys. J.}, A52(6):149, 2016.
\newblock \href {http://arxiv.org/abs/1512.01328} {\path{arXiv:1512.01328}},
  \href {http://dx.doi.org/10.1140/epja/i2016-16149-3}
  {\path{doi:10.1140/epja/i2016-16149-3}}.

\bibitem{Wigner:1932eb}
Eugene~P. Wigner.
\newblock {On the quantum correction for thermodynamic equilibrium}.
\newblock {\em Phys. Rev.}, 40:749--760, 1932.
\newblock \href {http://dx.doi.org/10.1103/PhysRev.40.749}
  {\path{doi:10.1103/PhysRev.40.749}}.

\bibitem{Balazs:1983hk}
N.~L. Balazs and B.~K. Jennings.
\newblock {Wigner's Function and Other Distribution Functions in Mock Phase
  Spaces}.
\newblock {\em Phys. Rept.}, 104:347, 1984.
\newblock \href {http://dx.doi.org/10.1016/0370-1573(84)90151-0}
  {\path{doi:10.1016/0370-1573(84)90151-0}}.

\bibitem{Hillery:1983ms}
M.~Hillery, R.~F. O'Connell, M.~O. Scully, and Eugene~P. Wigner.
\newblock {Distribution functions in physics: Fundamentals}.
\newblock {\em Phys. Rept.}, 106:121--167, 1984.
\newblock \href {http://dx.doi.org/10.1016/0370-1573(84)90160-1}
  {\path{doi:10.1016/0370-1573(84)90160-1}}.

\bibitem{Lee}
H.~W. Lee.
\newblock {\em Phys. Rep.}, 259:147, 1985.

\bibitem{Vogel:1989zz}
K.~Vogel and H.~Risken.
\newblock {Determination of quasiprobability distributions in terms of
  probability distributions for the rotated quadrature phase}.
\newblock {\em Phys. Rev.}, A40:2847--2849, 1989.
\newblock \href {http://dx.doi.org/10.1103/PhysRevA.40.2847}
  {\path{doi:10.1103/PhysRevA.40.2847}}.

\bibitem{Smithey:1993zz}
D.~T. Smithey, M.~Beck, M.~G. Raymer, and A.~Faridani.
\newblock {Measurement of the Wigner distribution and the density matrix of a
  light mode using optical homodyne tomography: Application to squeezed states
  and the vacuum}.
\newblock {\em Phys. Rev. Lett.}, 70:1244--1247, 1993.
\newblock \href {http://dx.doi.org/10.1103/PhysRevLett.70.1244}
  {\path{doi:10.1103/PhysRevLett.70.1244}}.

\bibitem{Banaszek:1999ya}
K.~Banaszek, C.~Radzewicz, K.~Wodkiewicz, and J.~S. Krasinski.
\newblock {Direct measurement of the Wigner function by photon counting}.
\newblock {\em Phys. Rev.}, A60:674--677, 1999.
\newblock \href {http://arxiv.org/abs/quant-ph/9903027}
  {\path{arXiv:quant-ph/9903027}}, \href
  {http://dx.doi.org/10.1103/PhysRevA.60.674}
  {\path{doi:10.1103/PhysRevA.60.674}}.

\bibitem{Hagiwara:2014iya}
Yoshikazu Hagiwara and Yoshitaka Hatta.
\newblock {Use of the Husimi distribution for nucleon tomography}.
\newblock {\em Nucl. Phys.}, A940:158--166, 2015.
\newblock \href {http://arxiv.org/abs/1412.4591} {\path{arXiv:1412.4591}},
  \href {http://dx.doi.org/10.1016/j.nuclphysa.2015.04.005}
  {\path{doi:10.1016/j.nuclphysa.2015.04.005}}.

\bibitem{Ji:2003ak}
Xiang-dong Ji.
\newblock {Viewing the proton through 'color' filters}.
\newblock {\em Phys. Rev. Lett.}, 91:062001, 2003.
\newblock \href {http://arxiv.org/abs/hep-ph/0304037}
  {\path{arXiv:hep-ph/0304037}}, \href
  {http://dx.doi.org/10.1103/PhysRevLett.91.062001}
  {\path{doi:10.1103/PhysRevLett.91.062001}}.

\bibitem{Belitsky:2003nz}
Andrei~V. Belitsky, Xiang-dong Ji, and Feng Yuan.
\newblock {Quark imaging in the proton via quantum phase space distributions}.
\newblock {\em Phys. Rev.}, D69:074014, 2004.
\newblock \href {http://arxiv.org/abs/hep-ph/0307383}
  {\path{arXiv:hep-ph/0307383}}, \href
  {http://dx.doi.org/10.1103/PhysRevD.69.074014}
  {\path{doi:10.1103/PhysRevD.69.074014}}.

\bibitem{Diehl:2002he}
M.~Diehl.
\newblock {Generalized parton distributions in impact parameter space}.
\newblock {\em Eur. Phys. J.}, C25:223--232, 2002.
\newblock [Erratum: Eur. Phys. J.C31,277(2003)].
\newblock \href {http://arxiv.org/abs/hep-ph/0205208}
  {\path{arXiv:hep-ph/0205208}}, \href
  {http://dx.doi.org/10.1007/s10052-002-1016-9}
  {\path{doi:10.1007/s10052-002-1016-9}}.

\bibitem{Brodsky:2006ku}
S.~J. Brodsky, D.~Chakrabarti, A.~Harindranath, A.~Mukherjee, and J.~P. Vary.
\newblock {Hadron optics in three-dimensional invariant coordinate space from
  deeply virtual compton scattering}.
\newblock {\em Phys. Rev.}, D75:014003, 2007.
\newblock \href {http://arxiv.org/abs/hep-ph/0611159}
  {\path{arXiv:hep-ph/0611159}}, \href
  {http://dx.doi.org/10.1103/PhysRevD.75.014003}
  {\path{doi:10.1103/PhysRevD.75.014003}}.

\bibitem{Brodsky:2006in}
S.~J. Brodsky, D.~Chakrabarti, A.~Harindranath, A.~Mukherjee, and J.~P. Vary.
\newblock {Hadron optics: Diffraction patterns in deeply virtual Compton
  scattering}.
\newblock {\em Phys. Lett.}, B641:440--446, 2006.
\newblock \href {http://arxiv.org/abs/hep-ph/0604262}
  {\path{arXiv:hep-ph/0604262}}, \href
  {http://dx.doi.org/10.1016/j.physletb.2006.08.061}
  {\path{doi:10.1016/j.physletb.2006.08.061}}.

\bibitem{Lorce:2011kd}
C.~Lorc\'e and B.~Pasquini.
\newblock {Quark Wigner Distributions and Orbital Angular Momentum}.
\newblock {\em Phys. Rev.}, D84:014015, 2011.
\newblock \href {http://arxiv.org/abs/1106.0139} {\path{arXiv:1106.0139}},
  \href {http://dx.doi.org/10.1103/PhysRevD.84.014015}
  {\path{doi:10.1103/PhysRevD.84.014015}}.

\bibitem{Meissner:2009ww}
Stephan Meissner, Andreas Metz, and Marc Schlegel.
\newblock {Generalized parton correlation functions for a spin-1/2 hadron}.
\newblock {\em JHEP}, 08:056, 2009.
\newblock \href {http://arxiv.org/abs/0906.5323} {\path{arXiv:0906.5323}},
  \href {http://dx.doi.org/10.1088/1126-6708/2009/08/056}
  {\path{doi:10.1088/1126-6708/2009/08/056}}.

\bibitem{Lorce:2013pza}
C.~Lorc\'e and B.~Pasquini.
\newblock {Structure analysis of the generalized correlator of quark and gluon
  for a spin-1/2 target}.
\newblock {\em JHEP}, 09:138, 2013.
\newblock \href {http://arxiv.org/abs/1307.4497} {\path{arXiv:1307.4497}},
  \href {http://dx.doi.org/10.1007/JHEP09(2013)138}
  {\path{doi:10.1007/JHEP09(2013)138}}.

\bibitem{Martin:1999wb}
Alan~D. Martin, M.~G. Ryskin, and T.~Teubner.
\newblock {Q**2 dependence of diffractive vector meson electroproduction}.
\newblock {\em Phys. Rev.}, D62:014022, 2000.
\newblock \href {http://arxiv.org/abs/hep-ph/9912551}
  {\path{arXiv:hep-ph/9912551}}, \href
  {http://dx.doi.org/10.1103/PhysRevD.62.014022}
  {\path{doi:10.1103/PhysRevD.62.014022}}.

\bibitem{Khoze:2000cy}
Valery~A. Khoze, Alan~D. Martin, and M.~G. Ryskin.
\newblock {Can the Higgs be seen in rapidity gap events at the Tevatron or the
  LHC?}
\newblock {\em Eur. Phys. J.}, C14:525--534, 2000.
\newblock \href {http://arxiv.org/abs/hep-ph/0002072}
  {\path{arXiv:hep-ph/0002072}}, \href
  {http://dx.doi.org/10.1007/s100520000359} {\path{doi:10.1007/s100520000359}}.

\bibitem{Liu:2015eqa}
Tianbo Liu and Bo-Qiang Ma.
\newblock {Quark Wigner distributions in a light-cone spectator model}.
\newblock {\em Phys. Rev.}, D91:034019, 2015.
\newblock \href {http://arxiv.org/abs/1501.07690} {\path{arXiv:1501.07690}},
  \href {http://dx.doi.org/10.1103/PhysRevD.91.034019}
  {\path{doi:10.1103/PhysRevD.91.034019}}.

\bibitem{Lorce:2011ni}
Cedric Lorc\'e, Barbara Pasquini, Xiaonu Xiong, and Feng Yuan.
\newblock {The quark orbital angular momentum from Wigner distributions and
  light-cone wave functions}.
\newblock {\em Phys. Rev.}, D85:114006, 2012.
\newblock \href {http://arxiv.org/abs/1111.4827} {\path{arXiv:1111.4827}},
  \href {http://dx.doi.org/10.1103/PhysRevD.85.114006}
  {\path{doi:10.1103/PhysRevD.85.114006}}.

\bibitem{Hatta:2011ku}
Yoshitaka Hatta.
\newblock {Notes on the orbital angular momentum of quarks in the nucleon}.
\newblock {\em Phys. Lett.}, B708:186--190, 2012.
\newblock \href {http://arxiv.org/abs/1111.3547} {\path{arXiv:1111.3547}},
  \href {http://dx.doi.org/10.1016/j.physletb.2012.01.024}
  {\path{doi:10.1016/j.physletb.2012.01.024}}.

\bibitem{Leader:2013jra}
E.~Leader and C.~Lorc\'e.
\newblock {The angular momentum controversy: What's it all about and does it
  matter?}
\newblock {\em Phys. Rept.}, 541(3):163--248, 2014.
\newblock \href {http://arxiv.org/abs/1309.4235} {\path{arXiv:1309.4235}},
  \href {http://dx.doi.org/10.1016/j.physrep.2014.02.010}
  {\path{doi:10.1016/j.physrep.2014.02.010}}.

\bibitem{Jaffe:1989jz}
R.~L. Jaffe and Aneesh Manohar.
\newblock {The G(1) Problem: Fact and Fantasy on the Spin of the Proton}.
\newblock {\em Nucl. Phys.}, B337:509--546, 1990.
\newblock \href {http://dx.doi.org/10.1016/0550-3213(90)90506-9}
  {\path{doi:10.1016/0550-3213(90)90506-9}}.

\bibitem{Ji:1996nm}
Xiang-Dong Ji.
\newblock {Deeply virtual Compton scattering}.
\newblock {\em Phys. Rev.}, D55:7114--7125, 1997.
\newblock \href {http://arxiv.org/abs/hep-ph/9609381}
  {\path{arXiv:hep-ph/9609381}}, \href
  {http://dx.doi.org/10.1103/PhysRevD.55.7114}
  {\path{doi:10.1103/PhysRevD.55.7114}}.

\bibitem{Wakamatsu:2014zza}
Masashi Wakamatsu.
\newblock {Is gauge-invariant complete decomposition of the nucleon spin
  possible?}
\newblock {\em Int. J. Mod. Phys.}, A29:1430012, 2014.
\newblock \href {http://arxiv.org/abs/1402.4193} {\path{arXiv:1402.4193}},
  \href {http://dx.doi.org/10.1142/S0217751X14300129}
  {\path{doi:10.1142/S0217751X14300129}}.

\bibitem{Hagler:2003jw}
P.~Hagler, A.~Mukherjee, and A.~Schafer.
\newblock {Quark orbital angular momentum in the Wandzura-Wilczek
  approximation}.
\newblock {\em Phys. Lett.}, B582:55--63, 2004.
\newblock \href {http://arxiv.org/abs/hep-ph/0310136}
  {\path{arXiv:hep-ph/0310136}}, \href
  {http://dx.doi.org/10.1016/j.physletb.2003.11.076}
  {\path{doi:10.1016/j.physletb.2003.11.076}}.

\bibitem{Lorce:2015sqe}
C.~Lorc\'e and B.~Pasquini.
\newblock {Multipole decomposition of the nucleon transverse phase space}.
\newblock {\em Phys. Rev.}, D93(3):034040, 2016.
\newblock \href {http://arxiv.org/abs/1512.06744} {\path{arXiv:1512.06744}},
  \href {http://dx.doi.org/10.1103/PhysRevD.93.034040}
  {\path{doi:10.1103/PhysRevD.93.034040}}.

\bibitem{Chakrabarti:2017teq}
D.~Chakrabarti, T.~Maji, C.~Mondal, and A.~Mukherjee.
\newblock {Quark Wigner distributions and spin-spin correlations}.
\newblock {\em Phys. Rev.}, D95(7):074028, 2017.
\newblock \href {http://arxiv.org/abs/1701.08551} {\path{arXiv:1701.08551}},
  \href {http://dx.doi.org/10.1103/PhysRevD.95.074028}
  {\path{doi:10.1103/PhysRevD.95.074028}}.

\bibitem{Chakrabarti:2016yuw}
D.~Chakrabarti, T.~Maji, C.~Mondal, and A.~Mukherjee.
\newblock {Wigner distributions and orbital angular momentum of a proton}.
\newblock {\em Eur. Phys. J.}, C76(7):409, 2016.
\newblock \href {http://arxiv.org/abs/1601.03217} {\path{arXiv:1601.03217}},
  \href {http://dx.doi.org/10.1140/epjc/s10052-016-4258-7}
  {\path{doi:10.1140/epjc/s10052-016-4258-7}}.

\bibitem{Courtoy:2016des}
A.~Courtoy and A.~S. Miramontes.
\newblock {Quark Orbital Angular Momentum in the MIT Bag Model}.
\newblock {\em Phys. Rev.}, D95(1):014027, 2017.
\newblock \href {http://arxiv.org/abs/1611.03375} {\path{arXiv:1611.03375}},
  \href {http://dx.doi.org/10.1103/PhysRevD.95.014027}
  {\path{doi:10.1103/PhysRevD.95.014027}}.

\bibitem{Hagiwara:2016kam}
Yoshikazu Hagiwara, Yoshitaka Hatta, and Takahiro Ueda.
\newblock {Wigner, Husimi, and generalized transverse momentum dependent
  distributions in the color glass condensate}.
\newblock {\em Phys. Rev.}, D94(9):094036, 2016.
\newblock \href {http://arxiv.org/abs/1609.05773} {\path{arXiv:1609.05773}},
  \href {http://dx.doi.org/10.1103/PhysRevD.94.094036}
  {\path{doi:10.1103/PhysRevD.94.094036}}.

\bibitem{Mukherjee:2014nya}
Asmita Mukherjee, Sreeraj Nair, and Vikash~Kumar Ojha.
\newblock {Quark Wigner Distributions and Orbital Angular Momentum in
  Light-front Dressed Quark Model}.
\newblock {\em Phys. Rev.}, D90(1):014024, 2014.
\newblock \href {http://arxiv.org/abs/1403.6233} {\path{arXiv:1403.6233}},
  \href {http://dx.doi.org/10.1103/PhysRevD.90.014024}
  {\path{doi:10.1103/PhysRevD.90.014024}}.

\bibitem{Mukherjee:2015aja}
Asmita Mukherjee, Sreeraj Nair, and Vikash~Kumar Ojha.
\newblock {Wigner distributions for gluons in a light-front dressed quark
  model}.
\newblock {\em Phys. Rev.}, D91(5):054018, 2015.
\newblock \href {http://arxiv.org/abs/1501.03728} {\path{arXiv:1501.03728}},
  \href {http://dx.doi.org/10.1103/PhysRevD.91.054018}
  {\path{doi:10.1103/PhysRevD.91.054018}}.

\bibitem{More:2017zqq}
Jai More, Asmita Mukherjee, and Sreeraj Nair.
\newblock {Quark Wigner Distributions Using Light-Front Wave Functions}.
\newblock {\em Phys. Rev.}, D95(7):074039, 2017.
\newblock \href {http://arxiv.org/abs/1701.00339} {\path{arXiv:1701.00339}},
  \href {http://dx.doi.org/10.1103/PhysRevD.95.074039}
  {\path{doi:10.1103/PhysRevD.95.074039}}.

\bibitem{More:2017zqp}
Jai More, Asmita Mukherjee, and Sreeraj Nair.
\newblock {Wigner Distributions For Gluons}.
\newblock {\em Eur. Phys. J.}, C78(5):389, 2018.
\newblock \href {http://arxiv.org/abs/1709.00943} {\path{arXiv:1709.00943}},
  \href {http://dx.doi.org/10.1140/epjc/s10052-018-5858-1}
  {\path{doi:10.1140/epjc/s10052-018-5858-1}}.

\bibitem{Kanazawa:2014nha}
K.~Kanazawa, C.~Lorc\'e, A.~Metz, B.~Pasquini, and M.~Schlegel.
\newblock {Twist-2 generalized transverse-momentum dependent parton
  distributions and the spin/orbital structure of the nucleon}.
\newblock {\em Phys. Rev.}, D90(1):014028, 2014.
\newblock \href {http://arxiv.org/abs/1403.5226} {\path{arXiv:1403.5226}},
  \href {http://dx.doi.org/10.1103/PhysRevD.90.014028}
  {\path{doi:10.1103/PhysRevD.90.014028}}.

\bibitem{Chakrabarti:2019wjx}
Dipankar Chakrabarti, Narindar Kumar, Tanmay Maji, and Asmita Mukherjee.
\newblock {Sivers and Boer-Mulders GTMDs in Light-front Holographic
  Quark-diquark Model}.
\newblock 2019.
\newblock \href {http://arxiv.org/abs/1902.07051} {\path{arXiv:1902.07051}}.

\bibitem{Mueller:2019gjj}
Niklas Mueller and Raju Venugopalan.
\newblock {Constructing phase space distributions with internal symmetries}.
\newblock {\em Phys. Rev.}, D99(5):056003, 2019.
\newblock \href {http://arxiv.org/abs/1901.10492} {\path{arXiv:1901.10492}},
  \href {http://dx.doi.org/10.1103/PhysRevD.99.056003}
  {\path{doi:10.1103/PhysRevD.99.056003}}.

\bibitem{Bhattacharya:2017bvs}
Shohini Bhattacharya, Andreas Metz, and Jian Zhou.
\newblock {Generalized TMDs and the exclusive double Drell-Yan process}.
\newblock {\em Phys. Lett.}, B771:396--400, 2017.
\newblock \href {http://arxiv.org/abs/1702.04387} {\path{arXiv:1702.04387}},
  \href {http://dx.doi.org/10.1016/j.physletb.2017.05.081}
  {\path{doi:10.1016/j.physletb.2017.05.081}}.

\bibitem{Bhattacharya:2018lgm}
Shohini Bhattacharya, Andreas Metz, Vikash~Kumar Ojha, Jeng-Yuan Tsai, and Jian
  Zhou.
\newblock {Exclusive double quarkonium production and generalized TMDs of
  gluons}.
\newblock 2018.
\newblock \href {http://arxiv.org/abs/1802.10550} {\path{arXiv:1802.10550}}.

\bibitem{Hatta:2016dxp}
Yoshitaka Hatta, Bo-Wen Xiao, and Feng Yuan.
\newblock {Probing the Small- x Gluon Tomography in Correlated Hard Diffractive
  Dijet Production in Deep Inelastic Scattering}.
\newblock {\em Phys. Rev. Lett.}, 116(20):202301, 2016.
\newblock \href {http://arxiv.org/abs/1601.01585} {\path{arXiv:1601.01585}},
  \href {http://dx.doi.org/10.1103/PhysRevLett.116.202301}
  {\path{doi:10.1103/PhysRevLett.116.202301}}.

\bibitem{Hatta:2016aoc}
Yoshitaka Hatta, Yuya Nakagawa, Feng Yuan, Yong Zhao, and Bowen Xiao.
\newblock {Gluon orbital angular momentum at small-$x$}.
\newblock {\em Phys. Rev.}, D95(11):114032, 2017.
\newblock \href {http://arxiv.org/abs/1612.02445} {\path{arXiv:1612.02445}},
  \href {http://dx.doi.org/10.1103/PhysRevD.95.114032}
  {\path{doi:10.1103/PhysRevD.95.114032}}.

\bibitem{Ji:2016jgn}
Xiangdong Ji, Feng Yuan, and Yong Zhao.
\newblock {Hunting the Gluon Orbital Angular Momentum at the Electron-Ion
  Collider}.
\newblock {\em Phys. Rev. Lett.}, 118(19):192004, 2017.
\newblock \href {http://arxiv.org/abs/1612.02438} {\path{arXiv:1612.02438}},
  \href {http://dx.doi.org/10.1103/PhysRevLett.118.192004}
  {\path{doi:10.1103/PhysRevLett.118.192004}}.

\bibitem{Hagiwara:2017fye}
Yoshikazu Hagiwara, Yoshitaka Hatta, Roman Pasechnik, Marek Tasevsky, and Oleg
  Teryaev.
\newblock {Accessing the gluon Wigner distribution in ultraperipheral $pA$
  collisions}.
\newblock {\em Phys. Rev.}, D96(3):034009, 2017.
\newblock \href {http://arxiv.org/abs/1706.01765} {\path{arXiv:1706.01765}},
  \href {http://dx.doi.org/10.1103/PhysRevD.96.034009}
  {\path{doi:10.1103/PhysRevD.96.034009}}.

\end{thebibliography}
\end{document}